\documentclass[12pt]{book}
\usepackage[english]{babel}
\usepackage[utf8]{inputenc}
\usepackage[T1]{fontenc}
\usepackage{indentfirst}
\usepackage{amsmath}
\usepackage{amssymb}
\usepackage{graphicx}
\usepackage{subfigure}
\usepackage{psfrag}

\allowhyphens

\newcommand{\valos}{\mathbb{R}}
\newcommand{\egesz}{\mathbb{N}}

\newcommand{\eps}{\varepsilon}

\newcommand{\ek}{\frac{1}{2}}

\newcommand{\ket}[1]{{\left|#1\right\rangle}}
\newcommand{\bra}[1]{{\left\langle #1\right|}}

\newcommand{\skalarszorzat}[2]{{\langle #1 | #2 \rangle}}
\newcommand{\varhatoertek}[1]{\left\langle #1 \right\rangle}
\newcommand{\vev}{\varhatoertek}


\makeatother

\begin{document}

\thispagestyle{empty}

\begin{center}
~
\vspace{1.5cm}  

{\huge\bf Finite volume form factors and }

\bigskip

{\huge\bf correlation functions at finite}

\bigskip

{\huge\bf temperature}

\vspace{1cm}

{\large Ph.D. thesis}

\vspace{2cm}

{\Large Bal\'azs Pozsgay}

\end{center}

\vspace{2cm}

{\large Supervisor: G\'abor Tak\'acs, D.Sc.

E\"otv\"os University, Budapest

Physics Doctoral School

Particle Physics and Astronomy Program

Doctoral School leader: Zal\'an Horv\'ath

Program leader: Ferenc Csikor
}

\vspace{3.5cm}

\begin{center}
  \large

Theoretical Physics Group of the 

Hungarian Academy of Sciences, 

Theoretical Physics Department

E\"otv\"os University, Budapest

\vspace{0.5cm}

2009
\end{center}

\thispagestyle{empty}

\pagenumbering{roman}

\tableofcontents

\chapter*{Introduction}

\addcontentsline{toc}{chapter}{Introduction}

\pagenumbering{arabic}
\setcounter{page}{1}

Finite size effects play an important  role
in modern statistical physics and quantum field theory. From a
statistical point of view, it is known, that no phase transitions take
place in a finite volume system, and the specific heat
$c(T)$ that is divergent at the critical point in infinite volume,
becomes finite if the system has a finite size. Moreover, it
is only in an interval around the critical temperature $T_c$ that finite
size effects are relevant. Away from this interval, they are negligible
because it is only near $T_c$ that the correlation length is
comparable with the size of the system. An important fact is that
the specific heat and other critical quantities have a scaling
behaviour as a response to varying the size $L$, which is fixed by the 
critical exponents of the infinite volume system  \cite{Fisher:1972zza,Cardy:1986ie}. 

Interesting phenomena occur in quantum field theories as well. The
most prominent example is the Casimir force
between two neutral macroscopic bodies in vacuum \cite{casimir}.  This
force can be determined from the volume dependence of the ground state
energy, which is usually obtained by summing vacuum fluctuations. The
crucial point is that besides the extensive bulk and boundary
contributions (which are proportional to the volume of the domain and
the surface of the boundaries, respectively) there are additional
terms, which (in massive theories) decay exponentially with the
distance of the boundary plates. 
In analogy with the Casimir energy, masses of stable particles also
get exponentially small corrections in a finite volume
\cite{luscher_1particle} and the two-body interaction is also
modified \cite{luscher_2particle}. 


The knowledge of the properties of finite volume QFT is of  central
importance in at least two
ways. On one hand, numerical approaches to QFT 
necessarily presume a finite volume box,
 and in order to interpret
the results correctly a reliable theoretical control of finite size
corrections is needed. Consider for example lattice QCD, where 
in order to approach the continuum limit 
the lattice spacing has to be adjusted  as small as
possible. The number of lattice points is limited by the computational
resources, therefore it is typically not possible to choose a volume
$L^3$, which is large enough to neglect finite size effects.

On the other hand, working in finite volume is not
necessarily a disadvantage. On the contrary, the 
volume dependence of the spectrum can be exploited to obtain (infinite
volume) physical
quantities like the elastic
scattering phase shifts
\cite{luscher_2particle,luscher_szorodas_1+1} or resonance
widths \cite{luscher_unstable,cikk_resonances}.  

\bigskip

In this work we investigate finite size effects in 1+1 dimensional
integrable field theories. The main subjects of interest are finite volume form
factors (matrix elements of local operators on eigenstates of the
finite volume Hamiltonian) and expectation values and correlation
functions at finite temperature.

\bigskip

1+1 dimensional integrable models attracted considerable
interest over the past thirty years. 
This interest is motivated by the fact
that low dimensional integrable models
represent nontrivial interacting field theories that are exactly solvable:
numerous physical quantities can be exactly calculated in these models. 
Phenomena qualitatively known in
higher dimensions, such as universality, duality, etc. can be
quantitatively investigated in 1+1 dimensions.

There is also a second motivation which is of a
more practical nature. Several systems 
of condensed matter physics can be described by 1+1 dimensional effective
field theories that in many cases lead to integrable models. 
Some important examples are the Kondo effect  \cite{Affleck_Kondo,FLS}, and the effect of
impurities in transport processes in general \cite{CS1,CS2,saleur00,saleur98}, behaviour of the edge
states in the fractional quantum Hall effect, problems concerning spin
chains, polymers, carbon nanotubes, etc. 

Besides the applications mentioned above two dimensional field
theories also play a central role in string theory: 
they describe the dynamics of the string on the world sheet.

\bigskip

In statistical physics, two dimensional critical systems can be described
by conformal field theories (CFT) which are fixed points of the
renormalization group flow. In this sense, off-critical theories
correspond to perturbations of CFT's by some of their relevant
operators. Off-critical models do not possess the infinite dimensional
symmetry algebra of the
fixed point CFT, in particular scale invariance is lost. However, it
was observed by A. B. Zamolodchikov
\cite{zam_integrable,zam_int1,zam_int2} that in certain situations there
remains an infinite number of conserved quantities. In this case the
resulting theory is still integrable and can be described by a factorized
scattering theory. The integrals of motion restrict the possible bound
state structure and mass ratios in the theory. Assuming further the
bootstrap principle, ie. that all bound states belong to the same set of
asymptotic particles, it is possible to construct the S-matrix with
only a finite number of physical poles. 

Integrability can be exploited to gain information about the off-shell
physics as well. Applying the ideas of analytic S-matrix theory
and the bootstrap principle to form factors one obtains a rather restrictive set of
equations which they have to obey \cite{smirnov_ff}. These equations can be
considered as axioms for the form factor bootstrap, and supplied with
the principles of maximum analyticity and the cluster property they
contain enough information to determine the form factors  completely
\cite{smirnov_ff,zam_Lee_Yang}. Once the form factors are 
known, it is possible to construct correlation functions trough the
spectral expansion, although the explicit summation of the series is
only possible in some simple cases \cite{Korepin:1998vp,Korepin:1998rj,Oota:1998tr}.

Integrability also offers powerful methods to explore the finite size
properties of these models. It is possible to obtain 
the exact Casimir energy by means of the Thermodynamic Bethe Ansatz (TBA),
both with periodic boundary conditions \cite{zam_tba} or in the presence
of non-trivial (integrable) boundary reflection factors \cite{LeClair:1995uf}. 
A great deal of information is known about the excited state
spectrum as well. Exact methods include the excited state TBA
\cite{excited_TBA,Bazhanov:1996aq} and nonlinear integral equations derived from
lattice regularizations \cite{Destri:1992qk,Destri:1994bv}. 

However, less is known about finite
size corrections to off-shell quantities. 
By Euclidean invariance correlation functions in a finite volume invariance
correspond to the evaluation of thermal correlations.
These objects can be compared directly to
experiments, and it is not surprising that they attracted quite a lot
of interest over the last decade 
 \cite{sachdev,leclair_mussardo,saleurfiniteT,lukyanovfiniteT,delfinofiniteT,mussardodifference,CastroAlvaredo:2002ud,esslerfiniteT,tsvelikfiniteTcorr,Essler:2007jp,konik_heisenberg_spin_chains,james-2009}.
One possible way to approach finite temperature correlation functions
is by establishing an appropriate spectral representation in finite
volume. This constitutes the first motivation for our work, because
finite volume form factors play a central role in this approach.
 Prior to our work only semiclassical results were known
\cite{Smirnov:1998kv,Mussardo:2003ji} and also an exceptional exact
result in the case of a free theory
\cite{Bugrij:2001nf,bugrij-2003-319,Fonseca:2001dc}. 

Besides being a promising tool to obtain correlation functions, finite
volume form factors provide means to verify the bootstrap approach
to form factors. 
The connection between the scattering theory and the Lagrangian (or
perturbed CFT) formulation is rather indirect, however, it is generally
believed, that the solutions of the bootstrap axioms correspond to
the local operators of the field theory.
Evidence includes a direct comparison of the space of the form factor
solutions with the space of primary operators
of the CFT \cite{Koubek:1993ke,Koubek:1994gk,Delfino:2008ia}; the
evaluation of correlation functions both through 
the spectral representation and from perturbation theory around the CFT
\cite{zam_Lee_Yang,Belavin:2003pu}; the evaluation of sum-rules like Zamolodchikov's
c-theorem \cite{Zamolodchikov:1986gt} and the $\Delta$-theorem
\cite{Delfino:1996nf}, both of which can be used 
to express conformal data as spectral sums in terms of form factors.
These tests concern quantities which are constructed from various integrals
over the form factors, whereas the multi-particle matrix elements themselves
have not been accessible. On the other hand, finite volume form factors
can be obtained numerically in the 
perturbed CFT setting by means of the Truncated Conformal Space
Approach (TCSA), and establishing the connection to infinite volume form
factors provides a way to directly 
test the solutions of the form factor bootstrap. This constitutes the
second motivation for our work.

The first part of this thesis (chapters 3, 4 and 5) is devoted to the study of finite volume
form factors in diagonal scattering theories. 
We show that these objects can be obtained in terms of
the infinite volume form factors;
in fact they are related by a proportionality factor which can
be interpreted as a density of states of the finite volume
spectrum.

To our best knowledge, the first result on finite
volume form factors in an interacting QFT was derived by Lellouch and L\"uscher
in \cite{lellouch_luscher}, where they considered the finite size dependence
of kaon decay matrix elements.
It was pointed out by Lin et.al. in \cite{sachrajda} that the final result
of \cite{lellouch_luscher}  can be explained simply by a
non-trivial normalization of states in finite volume; they also
derived a formula for arbitrary two-particle matrix elements.
Our arguments are similar to the ones used in \cite{sachrajda}.
However, in integrable models it is possible to extend the calculations to
arbitrary multi-particle states, because the factorized nature of the S-matrix allows
of an analytic description of the whole finite size spectrum.

Having established the connection to infinite
volume form factors, we can directly access the
form factor functions along certain one-dimensional sections of the
rapidity-space parameterized by the volume $L$. The choice of the
section corresponds
to which multi-particle states we pick from the
finite volume spectrum and it is only limited by the increasing
numerical inaccuracy for higher lying states.
This procedure 
provides a non-trivial check of our analytical calculations on finite
volume form factors, and also 
a direct test of the bootstrap approach to form
factors, as explained above.


In the second part of the work (chapter 6) we develop a method to
evaluate correlation functions at finite temperature:
we introduce finite volume as a regulator of the otherwise ill-defined
Boltzmann sum. We develop a systematic low-temperature expansion using
finite volume form factors and show that the individual terms of this
series have a well defined $L\to\infty$ limit. In fact, they can be
transformed into integral expressions over the infinite volume form
factors. 

There have been previous attempts to attack the problem of finite
temperature correlations.
LeClair and Mussardo proposed an expansion for the one-point and two-point
functions in terms of form factors dressed by appropriate occupation
number factors containing the pseudo-energy function from the thermodynamical
Bethe Ansatz \cite{leclair_mussardo}. It was argued by Saleur \cite{saleurfiniteT}
that their proposal for the two-point function is incorrect; on the
other hand, he gave a proof of the LeClair-Mussardo formula for one-point
functions provided the operator considered is the density of some
local conserved charge. 
In view of the evidence it is now generally
accepted that the conjecture made by LeClair and Mussardo for the
one-point functions is correct; in contrast, the case of two-point
functions (and also higher ones) is not yet fully understood.
We contribute to this issue by comparing the low-temperature expansion
obtained using finite volume form factors to the corresponding terms
in the LeClair-Mussardo proposal.


\bigskip

The thesis is organized as follows. In the next chapter we give a
brief summary of factorized scattering theory and its connection
to CFT, with a section devoted to the finite size properties of these
models. Chapter 2 includes our analytic calculations concerning finite
volume form factors, with two separate sections devoted to form factors
including disconnected pieces. In chapter 3 we present our numerical
studies: we test the analytic results of the previous chapter using
TCSA in the Lee-Yang model and the Ising model in a magnetic
field. In chapter 4 we study the leading exponential corrections (the
$\mu$-term) to scattering states and form factors. 
Chapter 5 deals with the evaluation of a
low-temperature expansion for correlation functions and also with the
comparison of our results to previous proposals. 
The last chapter contains our
conclusions. 




\bigskip
\bigskip

The material presented in this thesis is based on the
papers
\begin{itemize}
\item Balázs Pozsgay and Gábor Takács:
\emph{Form factors in finite volume I: form factor bootstrap and
truncated conformal space} 
Nucl. Phys. \textbf{B788} (2007) 167-208, \texttt{arxiv:0706.1445[hep-th]}
\item Balázs Pozsgay and Gábor Takács:
\emph{Form factors in finite volume II: disconnected terms and finite
temperature correlators} 
Nucl. Phys. \textbf{B788} (2007) 209-251,\\  \texttt{arxiv:0706.3605[hep-th]}
\item Balázs Pozsgay:
\emph{L\"uscher's mu-term and finite volume bootstrap principle for
scattering states and form factors}
Nucl.Phys. {\bf B802} (2008) 435-457,
\texttt{arxiv:0803.4445[hep-th]}
\end{itemize}

\bigskip

\newpage

\addcontentsline{toc}{chapter}{Acknowledgements}

\thispagestyle{empty}

~
\vspace{5cm}

{\bf\Large Acknowledgements}

\vspace{1cm}

First of all, I would like to express my gratitude to my supervisor G\'abor Tak\'acs
for his continuous support over the last four years. He introduced
me to the field of two dimensional integrable models, and his guidance
in the research was of unvaluable help.
I am
especially indebted to him for his personal encouragement during the early
stages of my Ph.D. 

I am also grateful to Zolt\'an Bajnok for illuminating discussions concerning
this work and related ongoing projects.

I am  indebted to Hubert Saleur, who acted as a supervisor during my
visit to the IPhT Saclay, and also to the organizers and lecturers of the
89. Summer School of the \'Ecole de Physique Les Houches. The four
months I spent in France significantly contributed to my enthusiasm
for the field of integrable models and theoretical physics in
general. 

I would like to thank G\'abor Zsolt T\'oth, M\'arton Kormos and
Constantin Candu for lots of useful advice and encouragement during the
writing of this PhD thesis.

Finally, I would like to thank my family and all my friends, who
supported me during the research and pushed me to complete the
dissertation. 


\chapter{Integrable Models and Conformal Field Theories}

\section{Exact S-matrix theories}

In this section we give a brief summary of the basic properties of factorized scattering
theory. For an introduction to this field see the review article of Mussardo
\cite{Mussardo:1992uc}. 

Let us consider a massive scattering theory with $n$ particle types $A_a$ with
masses $m_a$. One-particle states are denoted by
$\ket{\theta}_a$, where $\theta$ is the rapidity variable. We apply
the following Lorentz-invariant normalization:
\begin{equation*}
  _a\skalarszorzat{\theta_a}{\theta_b}_b=2\pi \delta_{ab}\delta(\theta_a-\theta_b)
\end{equation*}

Asymptotic states of the theory are defined as tensor products of
one-particle states and are denoted by
$\ket{\theta_1,\theta_2,\dots,\theta_n}_{a_1a_2\dots a_n}$.
The ordering of the rapidities is defined as
\begin{itemize}
\item $\theta_1>\dots>\theta_n$ for \textit{in} states
\item $\theta_1<\dots<\theta_n$ for \textit{out} states
\end{itemize}


\subsection{Conserved charges and factorization}

In integrable models there are an infinite number of conserved
quantities, which can be represented on the quantum level as mutually
commuting operators $Q_s$, which act diagonally on the one-particle
states:
\begin{equation*}
  Q_s \ket{\theta}_a = \chi_{a}^s e^{s\theta}  \ket{\theta}_a
\end{equation*}
The quantity $\chi_{a}^s$ is the spin-$s$ conserved charge of
particle $A_a$. The spin represents the behavior under
Lorentz-transformations. The charges for $s=1$ and $s=-1$ correspond to a
linear combination of the energy and momentum, therefore $\chi_a^{\pm
  1}=m_a$. 

The conserved charge operators are local in the sense, that they act
additively on multi-particle asymptotic states:
\begin{equation*}
Q_s
\ket{\theta_1,\theta_2,\dots,\theta_n}_{a_1a_2\dots a_n}=
\Big(\sum_{i=1}^n \chi_{a_i}^s  \Big)
\ket{\theta_1,\theta_2,\dots,\theta_n}_{a_1a_2\dots a_n}
\end{equation*}

The conservation of all higher spin charges constrains the S-matrix:
\begin{itemize}
\item All scattering processes are elastic: the number of incoming and
  outgoing particles is the same.
\item The set of conserved charges (including the energy and the
  momentum) is the same for the \textit{in} and \textit{out} states.
\item The S-matrix factorizes: the amplitude for an arbitrary
  scattering process can be obtained by consecutive 
  two-particle scattering processes.
\end{itemize}

The basic object is therefore the two-particle S-matrix, which is
defined for real rapidities by
\begin{equation*}
\ket{\theta_1,\theta_2}_{ab}=S_{ab}^{cd}(\theta_1-\theta_2)\ket{\theta_2,\theta_1}_{cd}
\end{equation*}
The factorization property forces the S-matrix to obey the Yang-Baxter
equations \cite{Yang:1967bm,Baxter:1972hz,zam_zam}
\begin{equation*}
  S_{a_1a_2}^{c_1c_2}(\theta_{12}) S_{c_1c_3}^{b_1b_3}(\theta_{13}) S_{c_2a_3}^{b_2c_3}(\theta_{23})=
 S_{a_1a_3}^{c_1c_3}(\theta_{13}) S_{c_1c_2}^{b_1b_2}(\theta_{12}) S_{a_3c_3}^{c_2b_3}(\theta_{23})
\end{equation*}
where the summation over the repeated indices is understood.

If there are no two particles with the same set of conserved charges,
the S-matrix is diagonal:
\begin{equation*}
  S_{ab}^{cd}(\theta)=S_{ab}(\theta) \delta_{ac}\delta_{bd}
\end{equation*}
In this case the Yang-Baxter equations are automatically satisfied.
In this work we only consider diagonal scattering theories.

The S-matrix can be analytically continued to the whole complex
plane and it satisfies
\begin{eqnarray}
\label{unit1_cross}
  S_{ab}(\theta)S_{ab}^*(\theta)&=&1\\
S_{ab}(i\pi-\theta)&=&S_{a\bar{b}}(\theta)
\end{eqnarray}
due to unitarity and crossing symmetry, where $A_{\bar{b}}$ is the
charge-conjugate of $A_b$. 

The unitarity condition can be expressed as
\begin{equation}
\label{unit2}
   S_{ab}(\theta)S_{ab}(-\theta)=1
\end{equation}
if one assumes real analyticity:
\begin{equation}
\label{real_anal}
  S_{ab}(\theta)=S_{ab}(-\theta^*)=S_{ab}^*(-\theta)
\end{equation}

Equations \eqref{unit1_cross},\eqref{unit2} and \eqref{real_anal} are
rather restrictive and a general solution can be written as
\begin{equation}
  \label{eq:diagonal_S}
  S_{ab}(\theta)= \prod_{\gamma\in \mathcal{G}_{ab}} t_\gamma(\theta)
\end{equation}
with
\begin{equation*}
  t_\gamma(\theta)= \frac{\tanh \ek (\theta+i\pi\gamma)}{\tanh \ek
    (\theta-i\pi\gamma)}=
\frac{\sinh(\theta)+i\sin(\pi\gamma)}{\sinh(\theta)-i\sin(\pi\gamma)}
\end{equation*}
and $\mathcal{G}_{ab}\subset \valos$ is a finite set.

\subsection{Bound states -- The bootstrap principle}

\begin{figure}
   \centering
\psfrag{Aa}{$A_c$}
\psfrag{Ab}{$A_a$}
\psfrag{Ac}{$A_b$}
\psfrag{a1}{$A_1$}
\psfrag{a2}{$A_2$}
\psfrag{a5}{$A_5$}
\psfrag{sim}{$\sim$}
\psfrag{ubca}{$\bar{u}_{ac}^b$}
\psfrag{ucba}{$\bar{u}_{bc}^a$}
\psfrag{mu}{$\mu_{abc}$}
\psfrag{ma}{$m_a$}
\psfrag{mb}{$m_b$}
\psfrag{mc}{$m_c$}
\footnotesize
\subfigure[Fusion angles]{\includegraphics[bb=-80 -20 290 277,scale=0.4]{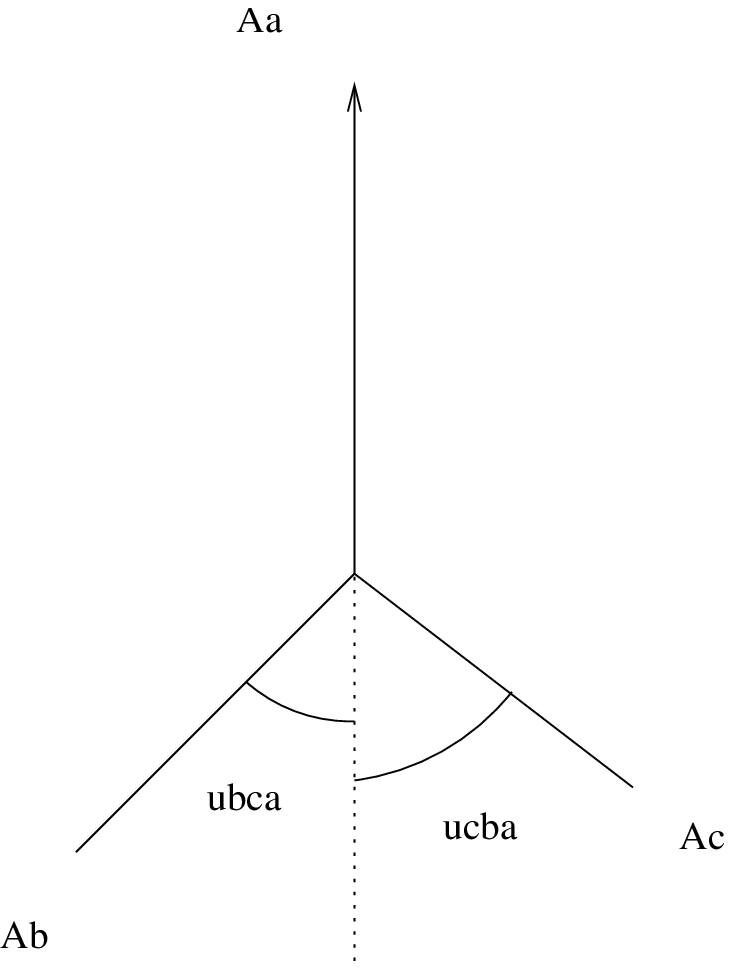}}
\subfigure[The mass triangle]{\includegraphics[bb=-110 -50 227 145,scale=0.4]{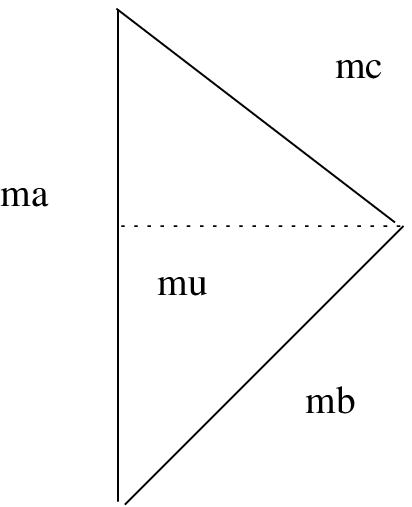}}
\caption{Pictorial representation of particle fusions.\label{fusion}}
 \end{figure}

The S-matrix \eqref{eq:diagonal_S} may contain zeros and simple or multiple poles for
purely imaginary rapidities. In a consistent theory, the interpretation for a simple
pole is a formation of a bound state. \footnote{Higher order poles in
  the S-matrix
are specific to two dimensions and can be explained by the Coleman-Thun mechanism\cite{Coleman:1978kk,Goebel:1986na,Braden:1990wx}. } If particle $A_c$ appears to be
a bound state of $A_a$ and $A_b$, then one has
\begin{equation}
\label{Smatpole}
  \text{Res}_{\theta=iu_{ab}^c}= i(\Gamma_{ab}^c)^2
\end{equation}
with $\Gamma_{ab}^c\in\valos$ and 
\begin{equation*}
  m_c^2=m_a^2+m_b^2+2m_am_b \cos(u_{ab}^c)
\end{equation*}
 The one-particle state
$\ket{\theta}_c$ may be identified with the formal two-particle
state
\begin{equation}
\label{bootstrap_identification}
  \ket{\theta}_c \sim \ket{\theta-i\bar{u}_{ac}^b,\theta+i\bar{u}_{ab}^c}_{ab}
\end{equation}
where the imaginary parts of the rapidities are determined by the
conservation of energy and momentum under the fusion process
(Fig. \ref{fusion}):
\begin{equation*}
  m_a \sin(\bar{u}_{ac}^b)=m_b\sin(\bar{u}_{ab}^c)
\quad \quad m_a \cos(\bar{u}_{ac}^b)+m_b\cos(\bar{u}_{ab}^c)=m_c
\end{equation*}
As a consequence, one also has
$u_{ab}^c=\bar{u}_{ac}^b+\bar{u}_{ab}^c$. For the higher spin
conserved charges the identification
\eqref{bootstrap_identification} requires
\begin{equation}
\label{bootstrap1}
  \chi_c^s=\chi_a^s e^{-is\bar{u}_{ab}^c}  + \chi_b^s e^{is\bar{u}_{ab}^c}
\end{equation}
Scattering processes including $A_c$ can be computed due to the
factorization property of the S-matrix as
\begin{equation}
\label{bootstrap2}
  S_{dc}(\theta)=S_{da}(\theta-i\bar{u}_{ac}^b)S_{db}(\theta+i\bar{u}_{ac}^b)
\end{equation}

Equations \eqref{bootstrap1} and \eqref{bootstrap2} represent strong
restrictions for the possible set of conserved charges and the sets
$\mathcal{G}_{ab}$ describing the scattering processes. The solution
of these equations with a finite number of particles is called the
\textit{bootstrap procedure}.

\section{Integrable theories in finite volume}

\label{integr_in_fv}

Let us consider the theory defined in a finite box of size $L$. 
One of the most characteristic feature of a field theory is the evolution of
its discrete energy levels $E_i(L)$, $i=1\dots\infty$ as a function of
the volume. 

In quantum field theories (including realistic 3+1
dimensional models) low-lying one-particle
 and
two-particle states \cite{luscher_1particle,luscher_2particle,luscher_unstable} can be
described to all orders in $1/L$ in terms of
the infinite volume spectrum (particle types and masses) and the
elastic scattering amplitudes of the theory. 
In non-integrable theories higher lying energy levels are far more difficult to access 
due to the complicated structure of the S-matrix and
only partial results are available \cite{Beane:2007qr,Detmold:2008gh,Luu:2008fg}.
On the other hand,
the S-matrix of integrable models factorizes and there are no
inelastic processes present. These properties enable us to describe 
the whole finite volume spectrum to all orders in $1/L$.

In this section we briefly describe the main characteristics of the
finite volume spectra of diagonal scattering theories. We also
introduce some notations which will be used throughout this work.
We assume periodic boundary conditions for simplicity.

\subsection{One-particle states}

One-particle states of particle $A_a$ in finite volume can be
characterized with the momentum $p_a$, which (due to the periodicity of the
wave function) is constrained to 
\begin{equation*}
  p_a=I\frac{2\pi}{L} \quad \quad I\in\egesz
\end{equation*}
Therefore, the particle type and the momentum quantum number $I$ (the
Lorentz-spin) uniquely determine the one-particle state in
question. For convenience we introduce the notation
\begin{equation*}
  \ket{\{I\}}_{a,L}
\end{equation*}
where the subscript $L$ denotes, that this state is an eigenvector of
the finite volume Hamiltonian.

The energy is given to all orders in $1/L$ by the relativistic formula
\begin{equation}
\label{simple_energy}
  E=\sqrt{m_a^2+p_a^2}
\end{equation}

\begin{figure}
   \centering
   \includegraphics[scale=0.4]{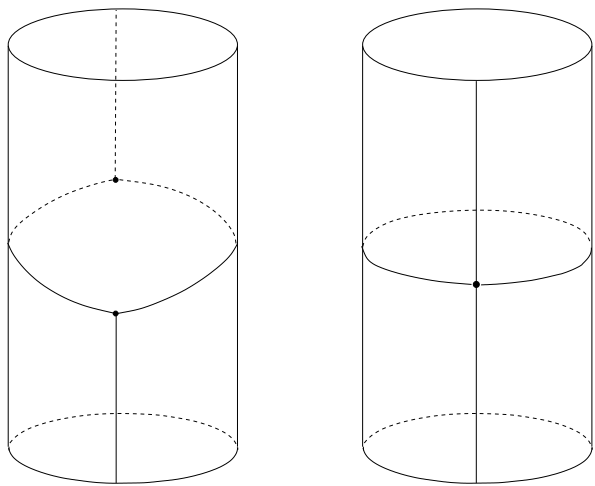}
\caption{The diagram to the left (the $\mu$-term) shows a particle
splitting in two virtual, on-shell particles, traveling around the 
cylinder and recombining. The diagram to the right (the F-term)
shows a virtual particle going around the circumference of the
cylinder.}
\label{mass_corr_diagrams}
 \end{figure}

 There are additional correction terms to
\eqref{simple_energy}, which decay exponentially with the
volume. These terms can be attributed  to virtual processes which are absent in infinite volume.
M. L\"uscher calculated the finite size mass corrections of stable particles (energy
corrections for $I=0$); they consist of the so-called $\mu$-term and
F-term   \cite{luscher_1particle} and are associated to diagrams
depicted in figure \ref{mass_corr_diagrams}. 
The explicit formulas for one dimensional diagonal scattering
theories read  \cite{klassen_melzer}
\begin{eqnarray}
\label{mass_mu_term}
\Delta m_a^{(\mu)}&=&
-\sum_{b,c}\theta(m_a^2-|m_b^2-m_c^2|)\mathcal{M}_{abc}\ \mu_{abc}
\left(\Gamma_{ab}^c\right)^2 e^{-\mu_{abc}L}\\
\label{mass_F_term}
\Delta m_a^{(F)}&=&-\sum_b  \int_{-\infty}^\infty
\frac{d\theta}{2\pi}e^{-m_bL\cosh(\theta)}m_b \cosh(\theta)
\left(S_{ab}(\theta+i\pi/2)-1\right)
\end{eqnarray}
where $\mathcal{M}_{abc}=1$ if $A_c$ is a bound state of $A_a$ and
$A_b$ and zero otherwise;
$\mu_{abc}$ is the altitude of the mass triangle with base
$m_a$ (see figure \ref{fusion}) and $\left(\Gamma_{ab}^c\right)^2$ is the
 residue of $S_{ab}(\theta)$ corresponding to the formation of the
 bound state.

\subsection{Many-particle states}

The Bethe-Yang equations \cite{bethe_original} serve as quantization conditions for a
many-particle state consisting of $n$ particles $A_{i_k}$ with rapidities
$\tilde{\theta}_k$:
\begin{equation}
  \label{Bethe-Yang}
  Q_{k}(\tilde{\theta}_{1},\dots,\tilde{\theta}_{n})_{i_{1}\dots i_{n}}=m_{i_{k}}L\sinh\tilde{\theta}_{k}+\sum_{l\neq k}\delta_{i_{k}i_{l}}(\tilde{\theta}_{k}-\tilde{\theta}_{l})=2\pi I_{k}\quad,\quad k=1,\dots,n
\end{equation}
where $\delta_{ij}(\theta)=-i \log S_{ij}(\theta)$.

For each set of particle types and momentum quantum numbers $I_k$ there is a
unique solution of the equations above.\footnote{The solutions with
  two or more coinciding momenta for a given particle type do not
  correspond to eigenstates of the Hamiltonian
  due to Pauli principle. } The energy of the many-particle
state in question is given to all orders in $1/L$ by the additive
formula
\begin{equation}
\label{simple_energy2}
  E=\sum_{k=1}^n m_{i_k} \cosh(\tilde{\theta_k})
\end{equation}
The effect of the interaction between the particles is the shift of
the rapidities from the values they would assume in a free
theory.

Since many-particle states are completely characterized by the quantum
numbers $I_k$, it is convenient to introduce the following implicit notation for a
$n$-particle state consisting of particles $i_1,i_2,\dots,i_n$ with
rapidities $\tilde{\theta_1},\tilde{\theta_2},\dots,\tilde{\theta_n}$:
\begin{equation*}
  \ket{\{I_1,I_2,\dots,I_n\}}_{i_1i_2,\dots i_n,L}
\end{equation*}
where it is understood that
$\tilde{\theta_1},\tilde{\theta_2},\dots,\tilde{\theta_n}$ are
solutions of the Bethe-equations with quantum numbers $I_1,I_2,\dots,I_n$.

To have an unambiguous definition of the quantum numbers $I_{k}$,
it is convenient
to define phase shift functions $\delta_{ab}$ which are continuous
and odd functions of the rapidity difference $\theta$; we achieve
this using the following convention:\[
S_{ab}(\theta)=S_{ab}(0)\mathrm{e}^{i\delta_{ab}(\theta)}\]
where $\delta_{ab}$ is uniquely specified by continuity and the branch
choice\[
\delta_{ab}(0)=0\]
and it is an odd function of $\theta$ due to the following property
of the scattering amplitude:\[
S_{ab}(\theta)S_{ab}(-\theta)=1\]
(which also implies $S_{ab}(0)=\pm1$). 
For a generic complex rapidity one has
\begin{equation*}
  \delta_{ab}(\theta^*)=\delta_{ab}(\theta)^*
\end{equation*}
Note, that with the convention above the quantum numbers $I_{k}$ can
be both integer and half-integer. Two particle states of the type
$A_aA_a$ have generally half-integer quantum numbers,
because the particles have fermionic statistics ($S_{aa}(0)=-1$) in all known theories\footnote{The only
  exception is the free boson with $S=1$}   .

 Similar to the one-particle energies, there are exponentially
small finite size corrections to \eqref{simple_energy2}. The
$\mu$-term associated to multi-particle energies is investigated in
section \ref{exponential:multi}.


\section{Scattering theories as perturbed CFT's}

Conformal Field Theories are two-dimensional Euclidean field theories, which
possess invariance under conformal transformations, including
scale-invariance. 
In the space of all possible QFT's they represent
fixed points under the renormalization group flow. In statistical
physics, they describe fluctuations of critical systems in the
continuum limit \cite{Polyakov:1970xd}.
Conformal invariance highly constrains the behavior of the
correlation functions, and even the operator content of the theory
\cite{CFT_alap}. In this section we quote some basic properties
of CFT's and introduce two particular models, which will serve as a
testing ground for our results: the scaling Lee-Yang model and the
critical Ising model. For a general
introduction to conformal field theories see \cite{Ginsparg:1988ui,Cardy:2008jc}.

Scale invariance constrains the
energy-momentum tensor $T_{\mu\nu}(x,y)$ to be traceless.
Introducing complex coordinates $z=x+iy$ and $\bar{z}=x-iy$ this
condition can be expressed as
$T_{z\bar{z}}=T_{\bar{z}z}=0$. Conservation of the energy-momentum
tensor on the other hand requires $\partial_{\bar{z}}T_{zz}=\partial_z
T_{\bar{z}\bar{z}}=0$, therefore it is possible to introduce the left- and
right-moving (chiral and anti-chiral) components $T(z) \equiv T_{zz}(z)$ and 
$\bar{T}(\bar{z})\equiv T_{\bar{z}\bar{z}}(\bar{z})$. 

$T(z)$ may be expanded into its Laurent-series around $z=0$ as
\begin{equation*}
  T(z)=\sum_{n\in\egesz} z^{-n-2} L_n
\end{equation*}
The operators $L_n$ satisfy the Virasoro algebra:
\begin{equation*}
  [L_n,L_m]=(n-m)L_{n+m} + \frac{c}{12}(n^3-n)\delta_{n+m,0}
\end{equation*}
which is a central extension of the symmetry algebra of the
classical conformal group, $c$ being the central charge of the
theory. 

The value of $c$ includes a great deal of information about the
theory. It restricts the possible representations of the
Virasoro-algebra, therefore also the operator content and the spectrum of
the theory. The simplest theories are the minimal models \cite{CFT},
which contain a finite number of primary fields and possess no
additional symmetries. They can be
characterized by two coprime integers $p$ and $q$ and the central
charge is given by
\begin{equation*}
  c=1-\frac{6(p-q)^2}{pq}
\end{equation*}
These theories are in addition unitary (there are no negative-norm states in the
spectrum) if $q=p+1$. 

\subsection{CFT in a cylindrical geometry}

A cylinder with spatial circumference $L$ can be mapped to the complex
plane by the conformal transformation
\begin{equation}
\label{eq:exponentialmap}
  z=\exp\Big(\frac{2\pi}{L}(\tau-ix)\Big) \quad\quad
 \bar{z}=\exp\Big(\frac{2\pi}{L}(\tau+ix)\Big)
\end{equation}
where $x$ and $\tau$ are the spatial and the imaginary time
coordinates, respectively. The transformation properties of the
energy-momentum tensor determine the Hamilton-operator (generator of
translations in the time-direction): 
\begin{equation}
\label{conform_H}
  H=\int_0^L dx\ T_{\tau\tau}(x,\tau=0)=\frac{2\pi}{L}\Big(L_0+\bar{L}_0-\frac{c}{12}\Big)
\end{equation}
In minimal models the Hilbert-space is given by
\begin{equation}
\label{conformal_Hilbert_space}
  \mathcal{H}=\bigoplus_{h} \mathcal{V}_h \otimes \bar{\mathcal{V}}_h
\end{equation}
where $\mathcal{V}_h$ ($\bar{\mathcal{V}}_h$) denotes the irreducible
representation of the left (right) Virasoro algebra with highest
weight $h$.

\subsection{Perturbing CFT's}

Conformal field theories represent statistical physical or quantum systems at
criticality. However, they can be also used to approach noncritical
models. 

Let us consider a theory defined by the action
\begin{equation*}
  A=A_{CFT}+\sum_j g_j \int d^2x\   \Phi_j(x)
\end{equation*}
where $A_{CFT}$ is the action of some CFT and $\Phi_j$ are its
relevant operators\footnote{
In the sense of renormalization group flow,
relevant operators 
describe the departure from the critical system. The perturbation leaves the UV behaviour of
the theory unchanged, ie. it is still governed by the CFT. 
If the scaling dimension of the operator
is greater or equal to 1, a renormalization procedure is required to
 properly define the perturbed theory\cite{cardy_perturbing_cfts}
On the other hand, perturbing with irrelevant operators results in an
 ill-defined theory in the UV, with an infinite number of counter-terms occurring
 during renormalization \cite{mussardo_bosonic_S}. }. 
This theory is no longer conformal, the coupling
constants $g_j$ describe the deviation from criticality. However, if
there is only one perturbation present, which only brakes a subset of the
conformal symmetries, the theory may still possess an
infinite number of conservation laws, and it may remain
integrable\footnote{Perturbations with multiple couplings are not
  expected to lead to integrable theories. For a short discussion see section 2.3 in
  \cite{Mussardo:1992uc}. }
\cite{zam_integrable,zam_int1,zam_int2,zam_potts}.  
Let us therefore consider that the theory defined by the action
\begin{equation}
\label{integrable_action}
  A=A_{CFT}+g \int d^2x\ \Phi(x)
\end{equation}
is integrable. Scale-invariance is broken by the perturbation, the energy scale
in the perturbed theory is set by $g^{\frac{1}{2-\Delta_\Phi}}$, where $\Delta_\Phi$ is the
scaling dimension of the field $\Phi$. Depending on the original CFT
and the perturbing operator the action 
\eqref{integrable_action} may define a massive \cite{zam_rsos} or a
massless \cite{zam_massless,saleur_massless2} scattering theory. In
this work we only consider the massive case. 

The question of which perturbed CFT corresponds to which scattering
theory may be answered by different methods. First of all,
Zamolodchikov's counting argument \cite{zam_integrable} provides a
sufficient condition for the existence of higher spin
conservation laws. A second independent method is the Thermodynamic
Bethe Ansatz
\cite{zam_tba,klassen_melzer_tba1,klassen_melzer_tba2}, which provides
integral equations to determine
the finite size dependence of the vacuum energy in terms of the
S-matrix of the theory. It can be used to analytically predict the
central charge of the CFT, the non-universal bulk vacuum energy
density,  and also the scaling dimension of the
perturbing field. 

A third method is the Truncated Conformal Space
Approach  (see \ref{TCSA}), which can be used to numerically determine the
low-lying energy levels of the finite size spectrum. Matching the
numerical data with the predictions of the Bethe-Yang equations \eqref{Bethe-Yang}
one can identify multi-particle states and thus directly test the phase shifts $S_{ab}(\theta)$. For the
details of this procedure see section \ref{identification}.

\subsection{The Lee-Yang model}

The non-unitary minimal model $M_{2,5}$ has central charge $c=-22/5$ and a unique
nontrivial primary field $\Phi$ with scaling weights
$\Delta=\bar{\Delta}=-1/5$. 
The field $\Phi$ is normalized so that it has the following operator
product expansion:\begin{equation}
\Phi(z,\bar{z})\Phi(0,0)=\mathcal{C}(z\bar{z})^{1/5}\Phi(0,0)+(z\bar{z})^{2/5}\mathbb{I}+\dots\label{eq:lyconfope}\end{equation}
where $\mathbb{I}$ is the identity operator and the only nontrivial
structure constant is \[
\mathcal{C}=1.911312699\dots\times i\]
The Hilbert space of the conformal model is given by\[
\mathcal{H}_{LY}=\bigoplus_{h=0,-1/5}\mathcal{V}_{h}\otimes\bar{\mathcal{V}}_{h}\]
where $\mathcal{V}_{h}$ ($\bar{\mathcal{V}}_{h}$) denotes the irreducible
representation of the left (right) Virasoro algebra with highest weight
$h$. 

The off-critical Lee-Yang model is defined by the Hamiltonian
\begin{equation}
H^{SLY}=H_{CFT}+i\lambda\int_{0}^{L}dx\Phi(0,x)\label{eq:lypcftham}\end{equation}
where $H_{CFT}$
is the conformal Hamiltonian. This theory is related to the Lee-Yang edge
singularity of the Ising model in an imaginary magnetic
field \cite{Yang:1952be,Lee:1952ig,Cardy:1985yy}. Despite 
its lack of unitarity, it is also relevant in condensed matter
physics: it is related to the theory of non-intersecting branched
polymers in 4 dimensions \cite{Parisi:1980ia}.

When $\lambda>0$ the theory above has
a single particle in its spectrum with mass $m$ that can be related
to the coupling constant as \cite{zam_tba} \begin{equation}
\lambda=0.09704845636\dots\times m^{12/5}\label{eq:lymassgap}\end{equation}
and the bulk energy density is given by\begin{equation}
\mathcal{B}=-\frac{\sqrt{3}}{12}m^{2}\label{eq:lybulk}\end{equation}
The $S$-matrix reads \cite{Cardy_Mussardo__Lee_Yang}\begin{equation}
S_{LY}(\theta)=\frac{\sinh\theta+i\sin\frac{2\pi}{3}}{\sinh\theta-i\sin\frac{2\pi}{3}}\label{eq:Smatly}\end{equation}
and the particle occurs as a bound state of itself at $\theta=2\pi i/3$
with the three-particle coupling given by\[
\Gamma^{2}=-2\sqrt{3}\]
where the negative sign is due to the non-unitarity of the
model. According to the convention introduced in section \ref{integr_in_fv}
we define the phase-shift via the relation\begin{equation}
S_{LY}(\theta)=-\mathrm{e}^{i\delta(\theta)}\label{eq:lydeltachoice}\end{equation}
so that $\delta(0)=0$. 

\subsection{The Ising-model in a magnetic field}

The critical Ising model is the described by the unitary conformal
field theory  $M_{3,4}$
with $c=1/2$ and has two nontrivial primary fields: the spin operator
$\sigma$ with $\Delta_{\sigma}=\bar{\Delta}_{\sigma}=1/16$ and the
energy density $\epsilon$ with $\Delta_{\epsilon}=\bar{\Delta}_{\epsilon}=1/2$.
The magnetic perturbation\[
H=H_{0}^{I}+h\int_{0}^{L}dx\sigma(0,x)\]
is massive (and its physics does not depend on the sign of the external
magnetic field $h$). The spectrum and the exact $S$ matrix is described
by the famous $E_{8}$ factorized scattering theory \cite{zamE8}, which
contains eight particles $A_{i},\; i=1,\dots,8$ with mass ratios
given by \begin{eqnarray*}
m_{2} & = & 2m_{1}\cos\frac{\pi}{5}\\
m_{3} & = & 2m_{1}\cos\frac{\pi}{30}\\
m_{4} & = & 2m_{2}\cos\frac{7\pi}{30}\\
m_{5} & = & 2m_{2}\cos\frac{2\pi}{15}\\
m_{6} & = & 2m_{2}\cos\frac{\pi}{30}\\
m_{7} & = & 2m_{4}\cos\frac{\pi}{5}\\
m_{8} & = & 2m_{5}\cos\frac{\pi}{5}\end{eqnarray*}
and the mass gap relation is \cite{phonebook}\[
m_{1}=(4.40490857\dots)|h|^{8/15}\]
or\begin{equation}
h=\kappa_{h}m_{1}^{15/8}\qquad,\qquad\kappa_{h}=0.06203236\dots\label{eq:ising_massgap}\end{equation}
The bulk energy density is given by\begin{equation}
B=-0.06172858982\dots\times m_{1}^{2}\label{eq:isingbulk}\end{equation}
We also quote the scattering phase shift of two $A_{1}$ particles:
\begin{equation}
S_{11}(\theta)=\left\{ \frac{1}{15}\right\} _{\theta}\left\{ \frac{1}{3}\right\} _{\theta}\left\{ \frac{2}{5}\right\} _{\theta}\quad,\quad\left\{ x\right\} _{\theta}=\frac{\sinh\theta+i\sin\pi x}{\sinh\theta-i\sin\pi x}\label{eq:s11_ising}\end{equation}
All other amplitudes $S_{ab}$ are determined by the $S$ matrix bootstrap
\cite{zamE8}; the only one we need later is that of the $A_{1}-A_{2}$
scattering, which takes the form\[
S_{12}(\theta)=\left\{ \frac{1}{5}\right\} _{\theta}\left\{ \frac{4}{15}\right\} _{\theta}\left\{ \frac{2}{5}\right\} _{\theta}\left\{ \frac{7}{15}\right\} _{\theta}\]

Our interest in the Ising model is motivated by the fact that this
is the simplest model in which form factors of an operator different
from the perturbing one are known, and also its spectrum and bootstrap
structure is rather complex, both of which stands in contrast with
the much simpler case of scaling Lee-Yang model.

\subsection{Finite size spectrum from Conformal Field Theory -- the
  Truncated Conformal Space approach}

\label{TCSA}

Let us assume that the infinite volume scattering theory corresponds
to a perturbed conformal theory defined by the action 
\begin{equation}
\label{integrable_action2}
  A=A_{CFT}+g \int d^2x\ \Phi(x)
\end{equation}
If this connection is already established, one can use the conformal
data to obtain the finite size spectrum of the theory. The perturbed
Hamiltonian is obtained by mapping the
cylinder with circumference $L$ to the complex plane. Using the
transformation properties of the primary field $\Phi(x)$ one has
\begin{equation}
\label{perturbed_conformal_H}
  H=H_{CFT}+2\pi g \left(\frac{2\pi}{L}\right)^{\Delta_\Phi-1} \Phi(0,0)
\end{equation}
where $H_{CFT}$ is the conformal Hamiltonian \eqref{conform_H}. Both
the spectrum $H_{CFT}$ and the matrix elements of $\Phi$ between its
eigenstates can be calculated, therefore the complete and exact finite size spectrum
can be obtained in principle by diagonalizing
\eqref{perturbed_conformal_H} in the infinite dimensional
Hilbert-space \eqref{conformal_Hilbert_space}. 

Instead of the impossible task of diagonalizing an infinite matrix
Yurov and Zamolodchikov proposed the Truncated Conformal Space 
Approach (TCSA) \cite{yurovzam} as a numerical approximation
scheme. The method consists of truncating the
Hilbert-space  \eqref{conformal_Hilbert_space} to the states which
have scaling dimension (eigenvalue under $L_0+\bar{L}_0$) smaller than
some threshold $e_{cut}$. In this finite Hilbert-space the
diagonalization procedure can be performed on a computer, and one
obtains eigenvalues $E_i(L,e_{cut})$, which are approximations
to the exact energy levels $E_i(L)$, and
\begin{equation*}
  \lim_{e_{cut}\to\infty} E_i(L,e_{cut})=E_i(L)
\end{equation*}
The convergence is faster for the low-lying states.

Besides being an effective tool to obtain finite size quantities, TCSA
can also serve as a regulator of the UV 
divergences of the vacuum energy if $\Delta_\Phi \ge 1$.

In this work (chapter \ref{s:numerics}) we use TCSA to determine the finite volume form factors
of local operators in the scaling Lee-Yang and Ising models. The
numerical data is then used to confirm the analytic results of chapter
2. The details of the TCSA methods we used are explained in section
\ref{numerics_howto}.

\chapter{Finite Volume Form Factors -- Analytic results}

In this chapter we determine the finite volume matrix elements in
terms of the infinite volume form factors and the S-matrix of the
theory.

In section \ref{fv_ff:bootstrap} we briefly summarize the ingredients
of the so-called form factor bootstrap program, which leads to
explicit solutions for the infinite volume form factor functions. In
section \ref{fv_ff} we present our results for finite volume form
factors which do not include disconnected pieces. 

Disconnected terms occur in the case of diagonal form
factors and matrix  elements including zero-momentum particles. These
two special situations are discussed in sections \ref{diagonal} and
\ref{zero-momentum}, respectively.  

\section{The form factor bootstrap program}

\label{fv_ff:bootstrap}

The (infinite volume) form factors of a local operator $\mathcal{O}(t,x)$ are defined
as\begin{equation}
F_{mn}^{\mathcal{O}}(\theta_{1}^{'},\dots,\theta_{m}^{'}|\theta_{1},\dots,\theta_{n})_{j_{1}\dots j_{m};i_{1}\dots i_{n}}=\,_{j_{1}\dots j_{m}}\langle\theta_{1}^{'},\dots,\theta_{m}^{'}\vert\mathcal{O}(0,0)\vert\theta_{1},\dots,\theta_{n}\rangle_{i_{1}\dots i_{n}}\label{eq:genff}\end{equation}
In first place they are defined for real rapidities, however they can
be analytically continued to the whole complex plane. The properties
of the form factor functions are explained in detail in Smirnov's
review \cite{smirnov_ff}. 

With the help of the crossing relations\begin{eqnarray}
 &  & F_{mn}^{\mathcal{O}}(\theta_{1}^{'},\dots,\theta_{m}^{'}|\theta_{1},\dots,\theta_{n})_{j_{1}\dots j_{m};i_{1}\dots i_{n}}=\nonumber \\
 &  & \qquad F_{m-1n+1}^{\mathcal{O}}(\theta_{1}^{'},\dots,\theta_{m-1}^{'}|\theta_{m}^{'}+i\pi,\theta_{1},\dots,\theta_{n})_{j_{1}\dots j_{m-1};j_{m}i_{1}\dots i_{n}}\nonumber \\
 &  & \qquad+\sum_{k=1}^{n}\Big(2\pi\delta_{j_{m}i_{k}}\delta(\theta_{m}^{'}-\theta_{k})\prod_{l=1}^{k-1}S_{i_{l}i_{k}}(\theta_{l}-\theta_{k})\times\nonumber \\
 &  & \qquad F_{m-1n-1}^{\mathcal{O}}(\theta_{1}^{'},\dots,\theta_{m-1}^{'}|\theta_{1},\dots,\theta_{k-1},\theta_{k+1}\dots,\theta_{n})_{j_{1}\dots j_{m-1};j_{m}i_{1}\dots i_{k-1}i_{k+1}\dots i_{n}}\Big)\label{eq:ffcrossing}\end{eqnarray}
all form factors can be expressed in terms of the elementary form
factors\[
F_{n}^{\mathcal{O}}(\theta_{1},\dots,\theta_{n})_{i_{1}\dots i_{n}}=\langle0\vert\mathcal{O}(0,0)\vert\theta_{1},\dots,\theta_{n}\rangle_{i_{1}\dots i_{n}}\]
which satisfy the following axioms \cite{Karowski:1978vz}: 

I. Exchange:

\begin{center}\begin{eqnarray}
 &  & F_{n}^{\mathcal{O}}(\theta_{1},\dots,\theta_{k},\theta_{k+1},\dots,\theta_{n})_{i_{1}\dots i_{k}i_{k+1}\dots i_{n}}=\nonumber \\
 &  & \qquad S_{i_{k}i_{k+1}}(\theta_{k}-\theta_{k+1})F_{n}^{\mathcal{O}}(\theta_{1},\dots,\theta_{k+1},\theta_{k},\dots,\theta_{n})_{i_{1}\dots i_{k+1}i_{k}\dots i_{n}}\label{eq:exchangeaxiom}\end{eqnarray}
\par\end{center}

II. Cyclic permutation: \begin{equation}
F_{n}^{\mathcal{O}}(\theta_{1}+2i\pi,\theta_{2},\dots,\theta_{n})=F_{n}^{\mathcal{O}}(\theta_{2},\dots,\theta_{n},\theta_{1})\label{eq:cyclicaxiom}\end{equation}

III. Kinematical singularity\begin{equation}
-i\mathop{\textrm{Res}}_{\theta=\theta^{'}}F_{n+2}^{\mathcal{O}}(\theta+i\pi,\theta^{'},\theta_{1},\dots,\theta_{n})_{i\, j\, i_{1}\dots i_{n}}=\left(1-\delta_{i\, j}\prod_{k=1}^{n}S_{i\, i_{k}}(\theta-\theta_{k})\right)F_{n}^{\mathcal{O}}(\theta_{1},\dots,\theta_{n})_{i_{1}\dots i_{n}}\label{eq:kinematicalaxiom}\end{equation}

IV. Dynamical singularity \begin{equation}
-i\mathop{\textrm{Res}}_{\theta=\theta^{'}}F_{n+2}^{\mathcal{O}}(\theta+i\bar{u}_{jk}^{i}/2,\theta^{'}-i\bar{u}_{ik}^{j}/2,\theta_{1},\dots,\theta_{n})_{i\, j\, i_{1}\dots i_{n}}=\Gamma_{ij}^{k}F_{n+1}^{\mathcal{O}}(\theta,\theta_{1},\dots,\theta_{n})_{k\, i_{1}\dots i_{n}}\label{eq:dynamicalaxiom}\end{equation}
whenever $k$ occurs as the bound state of the particles $i$ and
$j$.

The essence of the form factor bootstrap program is to 
 obtain explicit analytical solution to the above set of recursive
 equations \cite{Karowski:1978vz,karowski_weisz_elso_S,zam_Lee_Yang,delfino_mussardo}.
Axioms I-IV are supplemented by the assumption of maximum analyticity
(i.e. that the form factors are meromorphic functions which only have
the singularities prescribed by the axioms) and possible further conditions
expressing properties of the particular operator whose form factors
are sought.

\section{Finite volume form factors without disconnected pieces}

\label{fv_ff}

\subsection{Elementary form factors}

\label{elementary}

Here we determine finite volume form factors of the form
\begin{equation*}
 \bra{0}O \vert\{ I_{1},I_{2},\dots,I_{n}\}\rangle_{i_{1}\dots i_{n},L}
\end{equation*}
by comparing the  Euclidean two-point functions of arbitrary local operators
\begin{equation}
\label{2point_funct_kulonbseg}
  \langle\mathcal{O}(\bar{x})\mathcal{O}'(0,0)\rangle
  \quad\text{and}\quad
\langle\mathcal{O}(\bar{x})\mathcal{O}'(0,0)\rangle_{L}
\end{equation}
defined in the infinite volume theory, and in a finite but large volume $L$,
respectively. Given that one uses the same renormalization
prescriptions for the operators, the finite size correction to the
correlation function decays exponentially with $L$:
\begin{equation}
\langle\mathcal{O}(\bar{x})\mathcal{O}'(0)\rangle-\langle\mathcal{O}(\bar{x})\mathcal{O}'(0)\rangle_{L}\sim
O(\mathrm{e}^{-\mu L})\label{eq:corrfinvoldiff}
\end{equation}
where $\mu$ is some characteristic mass scale. In
\ref{estimate_of_residual} we discuss this relation in detail; here we
only use the fact, that the finite size correction decays faster than
any power of $1/L$.

Choosing $\bar{x}=(\tau,0)$ and inserting a complete set of states one
obtains the spectral representations
\begin{eqnarray}
\langle\mathcal{O}(\bar{x})\mathcal{O}'(0,0)\rangle & = & \sum_{n=0}^{\infty}\sum_{i_{1}\dots i_{n}}\left(\prod_{k=1}^{n}\int_{-\infty}^{\infty}\frac{d\theta_{k}}{2\pi}\right)F_{n}^{\mathcal{O}}(\theta_{1},\theta_{2},\dots,\theta_{n})_{i_{1}\dots i_{n}}\times\nonumber \\
 &  & \quad F_{n}^{\mathcal{O}'}(\theta_{1},\theta_{2},\dots,\theta_{n})_{i_{1}\dots i_{n}}^{+}\exp\left(-\tau\sum_{k=1}^{n}m_{i_{k}}\cosh\theta_{k}\right)\label{eq:spectralrepr}\end{eqnarray}
and
\begin{eqnarray}
\label{eq:finitevolspectralrepr}
\langle\mathcal{O}(\tau,0)\mathcal{O}'(0,0)\rangle_{L} & 
= & \sum_{n=0}^{\infty}\sum_{i_{1}\dots i_{n}}\sum_{I_{1}\dots I_{n}}\langle0\vert\mathcal{O}(0,0)\vert\{ I_{1},I_{2},\dots,I_{n}\}\rangle_{i_{1}\dots i_{n},L}\times \\
\nonumber 
 &  & \quad_{i_{1}\dots i_{n}}\langle\{
 I_{1},I_{2},\dots,I_{n}\}|\mathcal{O}'(0,0)|0\rangle_{L}
\exp\left(-\tau\sum_{k=1}^{n}m_{i_{k}}\cosh\tilde{\theta}_{k}\right)
\end{eqnarray}
where we used multi-particle states and form factors of
the infinite volume and finite volume theories, respectively, and in
\eqref{eq:finitevolspectralrepr}  it is
understood that the rapidities $\tilde{\theta}_k$ are solutions of the
Bethe-Yang equations 
\begin{equation}
Q_{k}(\tilde{\theta}_{1},\dots,\tilde{\theta}_{n})_{i_{1}\dots i_{n}}=m_{i_{k}}L\sinh\tilde{\theta}_{k}+\sum_{l\neq k}\delta_{i_{k}i_{l}}(\tilde{\theta}_{k}-\tilde{\theta}_{l})=2\pi I_{k}\quad,\quad k=1,\dots,n\label{eq:betheyang}\end{equation}
The second form factor in
\eqref{eq:spectralrepr} is defined as
\[
F_{n}^{\mathcal{O}'}(\theta_{1},\theta_{2},\dots,\theta_{n})_{i_{1}\dots i_{n}}^{+}=\,_{i_{1}\dots i_{n}}\langle\theta_{1},\dots,\theta_{n}|\mathcal{O}'(0,0)|0\rangle=F_{n}^{\mathcal{O}'}(\theta_{1}+i\pi,\theta_{2}+i\pi,\dots,\theta_{n}+i\pi)_{i_{1}\dots i_{n}}\],
which is just the complex conjugate of $F_{n}^{\mathcal{O}'}$ for
unitary theories.



To relate the finite and infinite volume form factors a further step
is necessary, because the integrals in the spectral representation
(\ref{eq:spectralrepr}) must also be discretized. Let us consider
this problem first for the case of free particles:\[
\left(\prod_{k=1}^{n}\int_{-\infty}^{\infty}\frac{d\theta_{k}}{2\pi}\right)f(\theta_{1},\dots,\theta_{n})=\left(\prod_{k=1}^{n}\int_{-\infty}^{\infty}\frac{dp_{k}}{2\pi
    E_{k}}\right)f(p_{1},\dots,p_{n})\]
where \[
p_{k}=m_{i_{k}}\sinh\theta_{k}\quad,\quad E_{k}=m_{i_{k}}\cosh\theta_{k}\]
are the momenta and energies of the particles. In finite volume\[
p_{k}=\frac{2\pi I_{k}}{L}\]
and it is well-known (as a consequence of the Poisson summation formula,
cf. \cite{luscher_2particle}) that\begin{equation}
\sum_{I_{1},\dots I_{n}}g\left(\frac{2\pi I_{1}}{L},\dots,\frac{2\pi I_{n}}{L}\right)=\left(\frac{L}{2\pi}\right)^{n}\left(\prod_{k=1}^{n}\int_{-\infty}^{\infty}dp_{k}\right)g(p_{1},\dots,p_{n})+O(L^{-N})\label{eq:intsumcorr}\end{equation}
provided the function $g$ and its first $N$ derivatives are integrable.
Recalling that form factors are analytic functions for real momenta,
in our case this is true for derivatives of any order, due to the
exponential suppression factor in the spectral integrals, provided
the form factors grow at most polynomially in the momentum, i.e.\[
\left|F_{n}(\theta_{1}+\theta,\theta_{2}+\theta,\dots,\theta_{n}+\theta)\right|\sim\mathrm{e}^{x\left|\theta\right|}\quad\mathrm{as}\quad\left|\theta\right|\rightarrow\infty\]
This is true if we only consider operators which have a power-like
short distance singularity in their two-point functions \cite{delfino_mussardo}:\[
\langle0\vert\mathcal{O}(\bar{x})\mathcal{O}(0)\vert0\rangle=\frac{1}{r^{2\Delta}}\]

Therefore the discrete sum differs from the continuum integral only
by terms decaying faster than any power in $1/L$, i.e. by terms exponentially
suppressed in $L$. Taking into account that \eqref{2point_funct_kulonbseg}
is valid for any pair $\mathcal{O}$, $\mathcal{O}'$ of scaling fields,
we obtain \begin{equation}
\langle0\vert\mathcal{O}(0,0)\vert\{ I_{1},\dots,I_{n}\}\rangle_{i_{1}\dots i_{n},L}=\frac{1}{\sqrt{\rho_{i_{1}\dots i_{n}}^{(0)}(\tilde{\theta}_{1},\dots,\tilde{\theta}_{n})}}F_{n}^{\mathcal{O}}(\tilde{\theta}_{1},\dots,\tilde{\theta}_{n})_{i_{1}\dots i_{n}}+O(\mathrm{e}^{-\mu'L})\label{eq:freepartffrelation}\end{equation}
where\[
\sinh\tilde{\theta}_{k}=\frac{2\pi I_{k}}{m_{i_{k}}L}\]
and\begin{equation}
\rho_{i_{1}\dots i_{n}}^{(0)}(\tilde{\theta}_{1},\dots,\tilde{\theta}_{n})=\prod_{k=1}^{n}m_{i_{k}}L\cosh\tilde{\theta}_{k}\label{eq:freejacobidet}\end{equation}
$\rho_{n}^{(0)}$ is nothing else than the Jacobi determinant corresponding
to changing from the variables $2\pi I_{k}$ to the rapidities $\tilde{\theta}_{k}$.
The term $O(\mathrm{e}^{-\mu'L})$ signifies that our considerations
are valid to all orders in $1/L$, although our argument does not
tell us the value of $\mu'$: to do that, we would need more information
about the correction term in the discretization (\ref{eq:intsumcorr}). 

In the case of interacting particles a more careful analysis is necessary.
To generalize (\ref{eq:freepartffrelation}) one has to
employ the Jacobian of the mapping $\{I_1,I_2,\dots,I_n\}\to
\{\theta_1,\theta_2,\dots,\theta_n\}$ described by the Bethe-Yang
equations. Therefore we define
\begin{eqnarray}
\rho_{i_{1}\dots i_{n}}(\theta_{1},\dots,\theta_{n}) & = & \det\mathcal{J}^{(n)}(\theta_{1},\dots,\theta_{n})_{i_{1}\dots i_{n}}\label{eq:byjacobian}\\
 &  & \mathcal{J}_{kl}^{(n)}(\theta_{1},\dots,\theta_{n})_{i_{1}\dots i_{n}}=\frac{\partial Q_{k}(\theta_{1},\dots,\theta_{n})_{i_{1}\dots i_{n}}}{\partial\theta_{l}}\quad,\quad k,l=1,\dots,n\nonumber \end{eqnarray}
The finite volume form factors are then given by
\begin{equation}
\langle0\vert\mathcal{O}(0,0)\vert\{ I_{1},\dots,I_{n}\}\rangle_{i_{1}\dots i_{n},L}=\frac{1}{\sqrt{\rho_{i_{1}\dots i_{n}}(\tilde{\theta}_{1},\dots,\tilde{\theta}_{n})}}F_{n}^{\mathcal{O}}(\tilde{\theta}_{1},\dots,\tilde{\theta}_{n})_{i_{1}\dots i_{n}}+O(\mathrm{e}^{-\mu'L})\label{eq:ffrelation}\end{equation}
where $\tilde{\theta}_{k}$ are the solutions of the Bethe-Yang equations
(\ref{eq:betheyang}) corresponding to the state with the specified
quantum numbers $I_{1},\dots,I_{n}$ at the given volume $L$. Note,
that the Bethe-Yang equations are correct to all orders in $1/L$,
therefore there can be no error terms in \eqref{eq:ffrelation}
analytic in $1/L$. In subsection \ref{estimate_of_residual} we consider the error
exponent $\mu'$ in detail.

The quantity $\rho_{i_{1}\dots i_{n}}(\theta_{1},\dots,\theta_{n})$
is nothing else than the density of states in rapidity space. It is
also worthwhile to mention that relation (\ref{eq:ffrelation}) can
be interpreted as an expression for the finite volume multi-particle
state in terms of the corresponding infinite volume state as follows\begin{equation}
\vert\{ I_{1},\dots,I_{n}\}\rangle_{i_{1}\dots i_{n},L}=\frac{1}{\sqrt{\rho_{i_{1}\dots i_{n}}(\tilde{\theta}_{1},\dots,\tilde{\theta}_{n})}}\vert\tilde{\theta}_{1},\dots,\tilde{\theta}_{n}\rangle_{i_{1}\dots i_{n}}\label{eq:staterenorm}\end{equation}
This relation between the density and the normalization of states
is a straightforward application of the ideas put forward by Saleur
in \cite{saleurfiniteT}.

To get an idea of the structure of $\rho_{i_{1}\dots
  i_{n}}(\theta_{1},\dots,\theta_{n})$  it is instructive to consider
the simplest cases. The quantization rule for the one-particle states
is identical in the free and interacting cases, therefore 
\begin{equation*}
  \rho_i(\theta)=E_iL \quad\text{with}\quad E_i=m_i\cosh(\theta)
\end{equation*}
In the two-particle case one has
\begin{equation*}
  \mathcal{J}^{(2)}(\theta_1,\theta_2)_{i_1i_2}=
  \begin{pmatrix}
    E_1L+\varphi_{i_1i_2}(\theta_1-\theta_2) & -\varphi_{i_1i_2}(\theta_1-\theta_2) \\
-\varphi_{i_1i_2}(\theta_1-\theta_2) & E_2L+\varphi_{i_1i_2}(\theta_1-\theta_2) 
  \end{pmatrix}
\end{equation*}
where we used the symmetric function
$\varphi_{i_1i_2}(\theta)=d/d\theta\ \delta_{i_1i_2}(\theta)$ and
$E_{1,2}=m_{i_{1,2}}\cosh(\theta_{1,2})$. The two-particle density is
therefore given by 
\begin{equation*}
  \rho_{i_1i_2}(\theta_1,\theta_2)=E_1E_2L^2+(E_1+E_2)L\ \varphi_{i_1i_2}(\theta_1-\theta_2) 
\end{equation*}
The leading term in the expression above is simply the 
product of two one-particle densities, just like in the free
case. This is a general rule: the leading $\mathcal{O}(L^n)$ term in
the n-particle density is given by the product of one-particle
densities. The effect of the interaction is the appearance of sub-leading terms proportional to
$\varphi_{i_1i_2}(\theta)$. These terms turn out to be crucial in the calculations
of thermal correlations (see section \ref{low_T_expansion}) and in
direct numerical tests of the form factors. 
The
necessity of introducing the full densities \eqref{eq:byjacobian} was first shown 
in \cite{cikk_resonances} and it is demonstrated in great detail in
section \ref{s:numerics}.

Note, that there is no preferred way to
order the rapidities on the circle, since there are no genuine asymptotic
\emph{in/out} particle configurations. This means that in relation
(\ref{eq:ffrelation}) there is no preferred way to order the rapidities
inside the infinite volume form factor function $F_{n}^{\mathcal{O}}$.
Different orderings are related by $S$-matrix factors according to
the exchange axiom (\ref{eq:exchangeaxiom}), which are indeed phases.
Such phases do not contribute to correlation functions (cf. the spectral
representation (\ref{eq:spectralrepr})), nor to any physically meaningful
quantity derived from them. In section \ref{numerics} we show
that relations like (\ref{eq:ffrelation}) must always be understood
to hold only up to physically irrelevant phase factors.

We also remark, that there are no finite volume states for which the
quantum numbers of any two of the particles are identical. The reason
is that \[
S_{ii}(0)=-1\]
(with the exception of free bosonic theories) and so the wave function
corresponding to the appropriate solution of the Bethe-Yang equations
(\ref{eq:betheyang}) vanishes. We can express this in terms of form
factors as follows:\[
\langle0\vert\mathcal{O}(0,0)\vert\{ I_{1},I_{2},\dots,I_{n}\}\rangle_{i_{1}\dots i_{n},L}=0\]
whenever $I_{k}=I_{l}$ and $i_{k}=i_{l}$ for some $k$ and $l$.
Using this convention we can assume that the summation in (\ref{eq:finitevolspectralrepr})
runs over all possible values of the quantum numbers without exclusions.
Note that even in this case the relation (\ref{eq:ffrelation}) can
be maintained since due to the exchange axiom (\ref{eq:exchangeaxiom})\[
F_{n}^{\mathcal{O}}(\tilde{\theta}_{1},\dots,\tilde{\theta}_{n})_{i_{1}\dots i_{n}}=0\]
whenever $\tilde{\theta}_{k}=\tilde{\theta}_{l}$ and $i_{k}=i_{l}$
for some $k$ and $l$.

\subsection{Generic form factors without disconnected pieces}

\label{fv_ff:generic}

Using the crossing formula (\ref{eq:ffcrossing}),
eqn. (\ref{eq:staterenorm}) allows us to construct the general form
factor functions (\ref{eq:genff}) in finite volume as follows:\begin{eqnarray}
 &  & \,_{j_{1}\dots j_{m}}\langle\{ I_{1}',\dots,I_{m}'\}\vert\mathcal{O}(0,0)\vert\{ I_{1},\dots,I_{n}\}\rangle_{i_{1}\dots i_{n},L}=\nonumber \\
 &  & \qquad\frac{F_{m+n}^{\mathcal{O}}(\tilde{\theta}_{m}'+i\pi,\dots,\tilde{\theta}_{1}'+i\pi,\tilde{\theta}_{1},\dots,\tilde{\theta}_{n})_{j_{m}\dots j_{1}i_{1}\dots i_{n}}}{\sqrt{\rho_{i_{1}\dots i_{n}}(\tilde{\theta}_{1},\dots,\tilde{\theta}_{n})\rho_{j_{1}\dots j_{m}}(\tilde{\theta}_{1}',\dots,\tilde{\theta}_{m}')}}+O(\mathrm{e}^{-\mu L})\label{eq:genffrelation}\end{eqnarray}
provided that there are no rapidities that are common between the
left and the right states i.e. the sets $\left\{ \tilde{\theta}_{1},\dots,\tilde{\theta}_{n}\right\} $
and $\left\{ \tilde{\theta}_{1}',\dots,\tilde{\theta}_{m}'\right\} $
are disjoint. The latter condition is necessary to eliminate disconnected
pieces. The rapidities entering \eqref{eq:genffrelation} are
determined by the Bethe-Yang equations \eqref{Bethe-Yang};
due to the presence of the scattering terms containing the phase shift
functions $\delta$, equality of two quantum numbers $I_{k}$ and
$I_{l}'$ does not mean that the two rapidities themselves are equal
in finite volume $L$. It is easy to see that there are only two cases
when exact equality of some rapidities can occur:

\begin{enumerate}
\item The two states are identical, i.e. $n=m$ and \begin{eqnarray*}
\{ j_{1}\dots j_{m}\} & = & \{ i_{1}\dots i_{n}\}\\
\{ I_{1}',\dots,I_{m}'\} & = & \{ I_{1},\dots,I_{n}\}\end{eqnarray*}
in which case all the rapidities are pairwise equal. The discussion of
such matrix elements is presented in section \ref{diagonal}.
\item Both states are parity symmetric states in the spin zero sector, i.e.
\begin{eqnarray*}
\{ I_{1},\dots,I_{n}\} & \equiv & \{-I_{n},\dots,-I_{1}\}\\
\{ I_{1}',\dots,I'_{m}\} & \equiv & \{-I'_{m},\dots,-I'_{1}\}\end{eqnarray*}
and the particle species labels are also compatible with the symmetry,
i.e. $i_{n+1-k}=i_{k}$ and $j_{m+1-k}=j_{k}$. Furthermore, both
states must contain one (or possibly more, in a theory with more than
one species) particle of quantum number $0$, whose rapidity is then
exactly $0$ for any value of the volume $L$ due to the symmetric
assignment of quantum numbers. The discussion of such matrix elements is presented in section
\ref{zero-momentum}. 
\end{enumerate}

We stress that eqns. (\ref{eq:ffrelation}, \ref{eq:genffrelation})
are exact to all orders of powers in $1/L$; we refer to the corrections
non-analytic in $1/L$ (eventually decaying exponentially as indicated)
as \emph{residual finite size effects}, following the terminology
introduced in \cite{cikk_resonances}.

We would like to remark, that similar arguments that led us to \eqref{eq:genffrelation}
were previously used to obtain the finite size dependence
of kaon decay matrix elements by Lin et al. \cite{sachrajda}. The idea
of normalizing finite volume form factors with particle densities also
appeared in \cite{ising_ff2}; however, they used the unsatisfactory
free-theory densities.

\subsection{Estimation of the residual finite size corrections}

\label{estimate_of_residual}

There are two sources of error terms which may contribute to
\eqref{eq:ffrelation}:
\begin{itemize}
\item Finite size corrections to the correlator
  $\langle\mathcal{O}(\tau,0)\mathcal{O}'(0,0)\rangle_{L}$.
\item Discretization errors introduced in \eqref{eq:intsumcorr}.
\end{itemize}
Here we show, that both types of error terms behave as
$\mathcal{O}(e^{-\mu L})$ where $\mu$ is some characteristic mass
scale of the theory.

One can use L\"uscher's finite volume expansion introduced in
\cite{luscher_1particle} to study finite size effects to the
correlator.
According to L\"uscher's classification of finite volume Feynman graphs,
the difference between the finite and infinite volume correlation
function is given by contributions from graphs of nontrivial gauge
class, i.e. graphs in which some propagator has a nonzero winding
number around the cylinder. Such graphs always carry an exponential
suppression factor in $L$, whose exponent can be determined by analyzing
the singularities of the propagators and vertex functions entering
the expressions. In a massive theory, all such singularities lie away
from the real axis of the Mandelstam variables, and the one with the
smallest imaginary part determines $\mu$. It turns out that the value
of $\mu$ is determined by the exact mass spectrum of the particles
and also the bound state fusions between them \cite{luscher_2particle,klassen_melzer}.
Therefore it is universal, which means that it is independent of the
correlation function considered. In general $\mu\leq m$ where
$m$ is the lightest particle mass (the mass gap of the theory),
because there are always corrections in which the lightest particle
loops around the finite volume $L$, and so the mass shell pole of
the corresponding exact propagator is always present. Contributions
from such particle loops to the vacuum expectation value are evaluated
in subsection \ref{numerics} (for a graphical representation see figure \ref{fig:vevLcorr}),
while an example of a finite volume correction corresponding to a
bound state fusion (a so-called $\mu$-term) is discussed in subsection
\ref{numerics_1p_ising} (figure \ref{fig:ising3pcorr}). The $\mu$-terms associated to
scattering states and form factors are analyzed in chapter \ref{exponential}.

The estimation
\begin{equation}
\langle\mathcal{O}(\tau,0)\mathcal{O}'(0)\rangle-\langle\mathcal{O}(\tau,0)\mathcal{O}'(0)\rangle_{L}\sim
O(\mathrm{e}^{-\mu L})
\end{equation}
is in fact a pessimistic one. 
Making use of
the Euclidean invariance of the theory and performing the rotation
$(\tau,x)\to (-x,\tau)$, the finite-size effects in
the relation above can be obtained by calculating finite temperature corrections to
the correlations function with the temperature given by $T=1/L$. 
In section \ref{lowT_2p} we consider such contributions; they are of
order $e^{-mL}$, ie. $\mu=m$ with $m$ being the mass gap of the theory. The
absence of correction terms with $\mu<m$ can also be explained by the
fact, that it is impossible to draw a relevant finite volume diagram
(possibly including a particle fusion) which would carry the suppression factor
$e^{-\mu L}$.

Having discussed finite size effects to the the correlator, we now
prove that the discretization procedure introduces error terms which
behave as $e^{-\mu' L}$, where in fact $\mu'=\mu$.

Recall that the Poisson formula gives the discrete sum in terms of
a Fourier transform: the leading term is the Fourier transform of
the summand evaluated at wave number $0$ (i.e. the integral) and
the corrections are determined by the decay of the Fourier transform
at large wave numbers. The function we need to consider is\begin{eqnarray}
h(p_{1},\dots,p_{n}) & = & \sum_{i_{1}\dots i_{n}}\rho_{i_{1}\dots i_{n}}(\theta_{1},\dots,\theta_{n})^{-1}F_{n}^{\mathcal{O}}(\theta_{1},\theta_{2},\dots,\theta_{n})_{i_{1}\dots i_{n}}F_{n}^{\mathcal{O}'}(\theta_{1},\theta_{2},\dots,\theta_{n})_{i_{1}\dots i_{n}}^{+}\times\nonumber \\
 &  & \exp\left(-r\sum_{k=1}^{n}m_{i_{k}}\cosh\theta_{k}\right)\label{eq:hfunction}\end{eqnarray}
where $p_{k}=m_{i_{k}}\sinh\theta_{k}$ are the momentum variables.
Due to the analyticity of the form factors for real rapidities, this
function is analytic for physical (real) momenta, and together with
all of its derivatives decays more rapidly than any power at infinity.
Therefore its Fourier transform taken in the momentum variables has
the same asymptotic property, i.e. it (and its derivatives) decay
more rapidly than any power:\[
\tilde{h}(\kappa_{1},\dots,\kappa_{n})\sim\mathrm{e}^{-\mu'|\kappa|}\]
for large $\kappa$. As a result, discretization introduces an error
of order $\mathrm{e}^{-\mu'L}$. The asymptotic exponent $\mu'$
of the Fourier transform can be generally determined by shifting the
contour of the integral transform and is given by the position of
the singularity closest to the real momentum domain (this is essentially
the procedure that L\"uscher uses in \cite{luscher_1particle}). Singularities
of the form factors are given by the same analytic structure as that
of the amplitudes which determine the exponent $\mu$ in
eqn. (\ref{eq:corrfinvoldiff}). Thus we find that $\mu'=\mu$. 

This argument is just an intuitive reasoning, although it can be made
a little more precise. First of all, we must examine whether the determinant
$\rho_{i_{1}\dots i_{n}}(\theta_{1},\dots,\theta_{n})$ can have any
zeros. It can always be written in the form\[
\rho_{i_{1}\dots i_{n}}(\theta_{1},\dots,\theta_{n})=\left(\prod_{k=1}^{n}m_{i_{k}}L\cosh\theta_{k}\right)\left(1+O(1/L)\right)\]
The leading factor can only be zero when $\theta_{k}=\frac{i\pi}{2}$
for some $k$, which corresponds to $p_{k}=im_{k}$, giving $\mu'=m_{k}$
in case this is the closest singularity. That gives a correction\[
\mathrm{e}^{-m_{k}L}\]
which is the same as the contribution by an on-shell propagator wound
around the finite volume, and such corrections are already included
in the $\mathrm{e}^{-\mu L}$ term of (\ref{eq:corrfinvoldiff}).
Another possibility is that some phase-shift function $\delta(\theta)$
in the $O(1/L)$ terms contributes a large term, which balances the
$1/L$ pre-factor. For that its argument must be close to a singularity,
and then according to eqn. (\ref{Smatpole}) we can write\[
\delta(\theta)\sim\log(\theta-u)\sim O(L)\]
where $u$ is the position of the singularity in the phase-shift%
. This requires that the singularity is approached exponentially close
(as a function of the volume $L$), but the positions of all these
singularities are again determined by singularities of the vertex
functions, so this gives no new possibilities for the exponent $\mu'$. 

A further issue that can be easily checked is whether the Fourier
integral is convergent for large momenta; the function (\ref{eq:hfunction})
is cut off at the infinities by the factor\[
\prod_{k=1}^{n}\exp(-m_{i_{k}}r\cosh\theta_{k})\]
which can only go wrong if for some $k$\[
\mathrm{\Re e}\,\cosh\theta_{k}<0\]
but that requires \[
\mathrm{\Im m}\,\theta_{k}>\frac{\pi}{2}\]
which is already farther from the real momentum domain then the position
of the on-shell propagator singularity.

\subsection{Comparison with an exact result}

As far as we know, the only exact result in the literature  was obtained in \cite{Fonseca:2001dc} and
independently in \cite{Bugrij:2001nf,bugrij-2003-319}, where the
authors considered the finite volume form
factors of the spin-field operator $\sigma$ in the Ising-model at zero
external field.  
Although the model in question is a free theory (it can be described in
terms of free massive Majorana fermions), it is interesting
to compare the results with the general rule presented in the
previous subsections. On one hand, this confirms our calculations. On
the other hand, it might be used in further work as a starting point
to study sub-leading exponential corrections (for the leading term, see
section \ref{exponential:fv_ff}).  

The Hilbert space of the theory consists of two sectors: the
Neveu-Schwarz (NS) and the Ramond (R) sectors. The fermionic
fields  are antiperiodic or periodic, respectively,
leading to the quantization rule for the particle momenta
\begin{eqnarray}
\label{NSq}
  p&=&m\sinh(\theta)=\frac{2\pi}{L} k\quad\quad k\in \egesz +\frac{1}{2}  \quad\quad
  \text{(NS)}\\
\label{Rq}
  p&=&m\sinh(\theta)=\frac{2\pi}{L} n\quad\quad n\in \egesz   \quad\quad
  \text{(R)}
\end{eqnarray}
Multi-particle states are denoted by
\begin{equation*}
  \ket{k_1,\dots,k_N}_{\text{NS}} \quad\quad\text{and}\quad\quad  \ket{n_1,\dots,n_N}_{\text{R}}
\end{equation*}
for states of the NS and R sectors, respectively. For simplicity we
only consider the $T>T_c$ phase of the theory, where spin-flip
symmetry is unbroken. In this phase the allowed excitations have an
even number of particles in both the NS and R sectors.

The spin operator has non-vanishing matrix elements only between states of
different sectors. The result for a generic finite volume matrix element reads (eq. 2.12
in \cite{Fonseca:2001dc})
\begin{eqnarray}
\nonumber
_{\text{NS}}  \langle k_1, k_2, \cdots, k_K |\sigma(0,0)| n_1, n_2, \cdots,
n_N \rangle_{\text{R}}
=&\\
S(L)\,\prod_{j=1}^{K}\,
{\tilde g}(\theta_{k_j})\,\prod_{i=1}^{N}\,g(\theta_{n_i})\,
&F_{K,N}(\theta_{k_1}, \cdots \theta_{k_K}| \theta_{n_1}, \cdots, 
\theta_{n_N})\,,
\label{exact_fermionic_ff}
\end{eqnarray}
where $\theta_n$ ($\theta_k$) stand for the finite-size rapidities 
related to the integers $n$ (half-integers $k$) by the equations
\eqref{NSq}-\eqref{Rq}. $F_{K,N}$ is the spin-field 
form factor in infinite-space \cite{karowski_weisz_elso_S}; the
explicit form of this function is not needed for the present
purposes. The overall factor $S(L)$ is given by
\begin{equation}
  S(L) = 
\exp\left\{{{(mL)^2}\over 2}
\!\!\!\int\!\!\!\!\int_{-\infty}^\infty
{{d\theta_1 d\theta_2}\over{(2\pi)^2}}
{{\sinh\theta_1\,\sinh\theta_2}\over{\sinh(mL\cosh\theta_1)\,
\sinh(mL\cosh\theta_2)}}\log\left|
\coth{{\theta_1 - \theta_2}\over 2}\right|\right\}
\end{equation}
The momentum-dependent leg factors $g$ and 
${\tilde g}$ are
\begin{equation}
g(\theta)=e^{\kappa(\theta)}/\sqrt{mL\cosh\theta}\,, \qquad 
{\tilde g}(\theta)=e^{-\kappa(\theta)}/\sqrt{mL\cosh\theta}  
\end{equation}
where 
\begin{equation}
\kappa(\theta) = \int_{-\infty}^{\infty}\,{{d\theta'}\over{2\pi}}\,
{1\over{\cosh(\theta-\theta')}}
\log\left({{1-e^{-mL\cosh\theta'}}\over 
{1+e^{-mL\cosh\theta'}}}\right)  
\end{equation}

It is straightforward to show, that the exact result
\eqref{exact_fermionic_ff} coincides with our formula
\eqref{eq:genffrelation} to all orders in $1/L$; the error terms are
of order $e^{-mL}$. First of all, the normalization factor $S(L)$ does
not depend on the form factor in question and for large $L$ its
behaviour is given by
\begin{equation*}
  S(L)\approx 1+\mathcal{O}(e^{-2mL})
\end{equation*}
The function $\kappa(\theta)$ decays exponentially with
$L$, therefore one has
\begin{equation*}
  g(\theta)=\frac{1}{\sqrt{\rho_1(\theta)}}+O(e^{-mL})
\qquad
\tilde{g}(\theta)=\frac{1}{\sqrt{\rho_1(\theta)}}+O(e^{-mL})
\end{equation*}
where $\rho_1(\theta)=EL=m\cosh(\theta)L$ is the one-particle density
derived from eqs. \eqref{NSq}-\eqref{Rq}. Being a free theory, the
products of one-particle densities equals the multi-particle densities
entering \eqref{eq:genffrelation} and one has indeed
\begin{eqnarray*}
_{\text{NS}}  \langle k_1, k_2, \cdots, k_K |\sigma(0,0)| n_1, n_2, \cdots,
n_N \rangle_{\text{R}}&
\\
=\frac{F_{K,N}(\theta_{k_1}, \cdots \theta_{k_K}| \theta_{n_1}, \cdots, 
\theta_{n_N})}{\sqrt{\rho_K(\theta_{k_1}, \cdots \theta_{k_K})\rho_N(\theta_{n_1}, \cdots, 
\theta_{n_N})}}&
+\mathcal{O}(e^{-mL})
\end{eqnarray*}

It was shown in \ref{estimate_of_residual} the leading correction term
to \eqref{eq:genffrelation}  is at most of order $e^{-\mu L}$ where $\mu\le m$
with $m$ being the mass of the lightest particle. In chapter
\ref{exponential} we show that the leading term (the $\mu$-term) is
connected with the inner structure of the particles under the
bootstrap procedure. In fact, in most theories $\mu<m$,
whereas in the present case the leading exponential correction is
only of order 
$e^{-mL}$. The reason for this is simple: the model in question is a
free theory, there are no particle fusions. Therefore in this case there is
no $\mu$-term present.

\section{Diagonal matrix elements}

\label{diagonal}

In this section we consider diagonal matrix elements of the form
\begin{equation*}
  \,_{i_{1}\dots i_{n}}\langle\{ I_{1}\dots I_{n}\}|\Psi|\{ I_{1}\dots I_{n}\}\rangle_{i_{1}\dots i_{n},L}
\end{equation*}
First we derive the first two ($n=1$ and $n=2$) cases using form
factor perturbation theory \cite{nonintegrable}. Built on these
rigorous results we conjecture the general formula in
\ref{diagonal:general}. 

Two additional subsections are devoted to the
discussion of our results. In \ref{symm_conn} we examine the connection between
two evaluation methods of the infinite volume diagonal form factors.
In \ref{diagonal:saleur} we prove that a conjecture made by Saleur in \cite{saleurfiniteT} 
exactly coincides with ours. 

\subsection{Form factor perturbation theory and disconnected contributions}

In the framework of conformal perturbation theory, we consider a model
with the action \begin{equation}
\mathcal{A}(\mu,\lambda)=\mathcal{A}_{\mathrm{CFT}}-\mu\int dtdx\Phi(t,x)-\lambda\int dtdx\Psi(t,x)\label{eq:nonint_Lagrangian}\end{equation}
such that in the absence of the coupling $\lambda$, the model defined
by the action $\mathcal{A}(\mu,\lambda=0)$ is integrable. The two
perturbing fields are taken as scaling fields of the ultraviolet limiting
conformal field theory, with left/right conformal weights $h_{\Phi}=\bar{h}_{\Phi}<1$
and $h_{\Psi}=\bar{h}_{\Psi}<1$, i.e. they are relevant and have
zero conformal spin, resulting in a Lorentz-invariant field theory. 

The integrable limit $\mathcal{A}(\mu,\lambda=0)$ is supposed to
define a massive spectrum, with the scale set by the dimensionful
coupling $\mu$. The exact spectrum in this case consists of 
massive particles, forming a factorized scattering theory with known
$S$ matrix amplitudes, and characterized by a mass scale $M$ (which
we take as the mass of the fundamental particle generating the bootstrap),
which is related to the coupling $\mu$ via the mass gap relation\[
\mu=\kappa M^{2-2h_{\Phi}}\]
where $\kappa$ is a (non-perturbative) dimensionless constant. 

Switching on a second independent coupling $\lambda$ in general spoils
integrability, deforms the mass spectrum and the $S$ matrix, and
in particular allows decay of the particles which are stable at the
integrable point. One way to approach the dynamics of the model is
the form factor perturbation theory proposed in \cite{nonintegrable}.
Let us denote the form factors of the operator $\Psi$ in the $\lambda=0$
theory by\[
F_{n}^{\Psi}\left(\theta_{1},\dots,\theta_{n}\right)_{i_{1}\dots i_{n}}=\langle0|\Psi(0,0)|\theta_{1}\dots\theta_{n}\rangle_{i_{1}\dots i_{n}}^{\lambda=0}\]
Using perturbation theory to first order in $\lambda$, the following
quantities can be calculated \cite{nonintegrable}:

\begin{enumerate}
\item The vacuum energy density is shifted by an amount\begin{equation}
\delta\mathcal{E}_{vac}=\lambda\left\langle 0\right|\Psi\left|0\right\rangle _{\lambda=0}.\label{vac_energy_shift}\end{equation}

\item The mass (squared) matrix $M_{ab}^{2}$ gets a correction\begin{equation}
\delta M_{ab}^{2}=2\lambda F_{2}^{\Psi}\left(i\pi\,,\,0\right)_{a\bar{b}}\delta_{m_{a},m_{b}}\label{mass_correction}\end{equation}
(where the bar denotes the antiparticle) supposing that the original
mass matrix was diagonal and of the form $M_{ab}^{2}=m_{a}^{2}\delta_{ab}\:.$
\item The scattering amplitude for the four particle process $a+b\,\rightarrow\, c+d$
is modified by \begin{equation}
\delta S_{ab}^{cd}\left(\theta,\lambda\right)=-i\lambda\frac{F_{4}^{\Psi}\left(i\pi,\,\theta+i\pi,\,0,\,\theta\right)_{\bar{c}\bar{d}ab}}{m_{a}m_{b}\sinh\theta}\quad,\quad\theta=\theta_{a}-\theta_{b}\:.\label{smatr_corr}\end{equation}
It is important to stress that the form factor amplitude in the above
expression must be defined as the so-called {}``symmetric'' evaluation\[
\lim_{\epsilon\rightarrow0}F_{4}^{\Psi}\left(i\pi+\epsilon,\,\theta+i\pi+\epsilon,\,0,\,\theta\right)_{\bar{c}\bar{d}ab}\]
(see eqn. (\ref{eq:Fs_definition}) below). It is also necessary to
keep in mind that eqn. (\ref{smatr_corr}) gives the variation of
the scattering phase when the center-of-mass energy (or, the Mandelstam
variable $s$) is kept fixed \cite{nonintegrable}. Therefore, in
terms of rapidity variables, this variation corresponds to the following:\[
\delta S_{ab}^{cd}\left(\theta,\lambda\right)=\frac{\partial S_{ab}^{cd}\left(\theta,\lambda=0\right)}{\partial\theta}\delta\theta+\lambda\left.\frac{\partial S_{ab}^{cd}\left(\theta,\lambda\right)}{\partial\lambda}\right|_{\lambda=0}\]
where \[
\delta\theta=-\frac{m_{a}\delta m_{a}+m_{a}\delta m_{a}+(m_{b}\delta m_{a}+m_{a}\delta m_{b})\cosh\theta}{m_{a}m_{b}\sinh\theta}\]
 is the shift of the rapidity variable induced by the mass corrections
given by eqn. (\ref{mass_correction}).
\end{enumerate}
It is also possible to calculate the (partial) decay width of particles
\cite{resonances}, but we do not need it here. 

We can use the above results to calculate diagonal matrix elements
involving one particle. For simplicity we present the derivation for
a theory with a single particle species. Let us start with the one-particle
case. The variation of the energy of a stationary one-particle state
with respect to the vacuum (i.e. the finite volume particle mass)
can be expressed as the difference between the first order perturbative
results for the one-particle and vacuum states in volume $L$:\begin{equation}
\Delta m(L)=\lambda L\left(\langle\{0\}|\Psi|\{0\}\rangle_{L}-\langle0|\Psi|0\rangle_{L}\right)\label{eq:finvoldm}\end{equation}
On the other hand, using L\"uscher's results \cite{luscher_2particle} it
only differs from the infinite volume mass in terms exponentially
falling with $L$. Using eqn. (\ref{mass_correction})\[
\Delta m(L)=\frac{\lambda}{m}F^{\Psi}(i\pi,0)+O\left(\mathrm{e}^{-\mu L}\right)\]
Similarly, the vacuum expectation value receives only corrections
falling off exponentially with $L$. Therefore we obtain\[
\langle\{0\}|\Psi|\{0\}\rangle_{L}=\frac{1}{mL}\left(F^{\Psi}(i\pi,0)+mL\langle0|\Psi|0\rangle\right)+\dots\]
with the ellipsis denoting residual finite size corrections. Note
that the factor $mL$ is just the one-particle Bethe-Yang Jacobian
$\rho_{1}(\theta)=mL\cosh\theta$ evaluated for a stationary particle
$\theta=0$. 

We can extend the above result to moving particles in the following
way. Up to residual finite size corrections, the one-particle energy
is given by \[
E(L)=\sqrt{m^{2}+p^{2}}\]
 with \[
p=\frac{2\pi I}{L}\]
where $I$ is the Lorentz spin (which is identical to the particle
momentum quantum number). Therefore \[
E\Delta E=m\Delta m\]
whereas perturbation theory gives: \[
\Delta E=\lambda L\left(\langle\{ I\}|\Psi|\{ I\}\rangle_{L}-\langle0|\Psi|0\rangle_{L}\right)\]
and so we obtain \begin{equation}
\langle\{ I\}|\Psi|\{ I\}\rangle_{L}=\frac{1}{\rho_{1}(\tilde{\theta})}\left(F^{\Psi}(i\pi,0)+\rho_{1}(\tilde{\theta})\langle0|\Psi|0\rangle\right)+\dots\label{eq:d1formula}\end{equation}
where \[
\sinh\tilde{\theta}=\frac{2\pi I}{mL}\,\Rightarrow\,\rho_{1}(\tilde{\theta})=\sqrt{m^{2}L^{2}+4\pi^{2}I^{2}}\]

One can use a similar argument to evaluate diagonal two-particle matrix
elements in finite volume. 
The energy levels
can be calculated from the relevant Bethe-Yang equations\begin{eqnarray*}
m_{i_{1}}L\sinh\tilde{\theta}_{1}+\delta(\tilde{\theta}_{1}-\tilde{\theta}_{2}) & = & 2\pi I_{1}\\
m_{i_{2}}L\sinh\tilde{\theta}_{2}+\delta(\tilde{\theta}_{2}-\tilde{\theta}_{1}) & = & 2\pi I_{2}\end{eqnarray*}
and (up to residual finite size corrections) \[
E_{2}(L)=E_{2pt}(L)-E_{0}(L)=m_{i_{1}}\cosh\tilde{\theta}_{1}+m_{i_{2}}\cosh\tilde{\theta}_{2}\]
where $i_{1}$ and $i_{2}$ label the particle species. After a somewhat
tedious, but elementary calculation the variation of this energy difference
with respect to $\lambda$ can be determined, using (\ref{mass_correction})
and (\ref{smatr_corr}):\begin{eqnarray*}
\Delta E_{2}(L) & = & \lambda\frac{L}{\rho_{i_{1}i_{2}}\left(\tilde{\theta}_{1},\tilde{\theta}_{2}\right)}\Big(F_{4}^{\Psi}\left(\tilde{\theta}_{2}+i\pi,\tilde{\theta}_{1}+i\pi,\tilde{\theta}_{1},\tilde{\theta}_{2}\right)_{i_{2}i_{1}i_{1}i_{2}}+m_{i_{1}}L\cosh\tilde{\theta}_{1}F_{2}^{\Psi}(i\pi,0)_{i_{2}i_{2}}\\
 &  & +m_{i_{2}}L\cosh\tilde{\theta}_{2}F^{\Psi}_2(i\pi,0)_{i_{1}i_{1}}\Big)\end{eqnarray*}
where all quantities (such as Bethe-Yang rapidities $\tilde{\theta}_{i}$,
masses $m_{i}$ and the two-particle state density $\rho_{2}$) are
in terms of the $\lambda=0$ theory. This result expresses the fact
that there are two sources for the variation of two-particle energy
levels: one is the mass shift of the individual particles, and the
second is due to the variation in the interaction. On the other hand,
in analogy with (\ref{eq:finvoldm}) we have\[
\Delta E_{2}(L)=\lambda L\left({}_{i_{1}i_{2}}\langle\{ I_{1},I_{2}\}|\Psi|\{ I_{1},I_{2}\}\rangle_{i_{1}i_{2},L}-\langle0|\Psi|0\rangle_{L}\right)\]
and so we obtain the following relation:\begin{eqnarray}
{}_{i_{1}i_{2}}\langle\{ I_{1},I_{2}\}|\Psi|\{ I_{1},I_{2}\}\rangle_{i_{1}i_{2},L} & = & \frac{1}{\rho_{i_{1}i_{2}}\left(\tilde{\theta}_{1},\tilde{\theta}_{2}\right)}\Big(F_{4}^{\Psi}\left(\tilde{\theta}_{2}+i\pi,\tilde{\theta}_{1}+i\pi,\tilde{\theta}_{1},\tilde{\theta}_{2}\right)_{i_{2}i_{1}i_{1}i_{2}}\nonumber \\
 &  & +m_{i_{1}}L\cosh\tilde{\theta}_{1}F_{2}^{\Psi}(i\pi,0)_{i_{2}i_{2}}\nonumber \\
 &  & +m_{i_{2}}L\cosh\tilde{\theta}_{2}F_{2}^{\Psi}(i\pi,0)_{i_{1}i_{1}}+\langle0|\Psi|0\rangle\Big)+\dots\label{eq:d2formula}\end{eqnarray}
where the ellipsis again indicate residual finite size effects. The
above argument is a generalization of the derivation of the mini-Hamiltonian
coefficient $C$ in Appendix C of \cite{cikk_resonances}. 

We wish to remark that the result \eqref{eq:d2formula} extends to two-particle states in
non-integrable models below the inelastic
threshold, as the Bethe-Yang
equations remain valid in this case  \cite{luscher_2particle}.

\subsection{Generalization to higher number of particles}

\label{diagonal:general}

Let us now introduce some more convenient notations. Given a state
\[
|\{ I_{1}\dots I_{n}\}\rangle_{i_{1}\dots i_{n}}\]
we denote \begin{equation}
\rho(\{ k_{1},\dots,k_{r}\})_{L}=\rho_{i_{k_{1}}\dots i_{k_{r}}}(\tilde{\theta}_{k_{1}},\dots,\tilde{\theta}_{k_{r}})\label{eq:rhonotation}\end{equation}
where $\tilde{\theta}_{l}$, $l=1,\dots,n$ are the solutions of the
$n$-particle Bethe-Yang equations (\ref{eq:betheyang}) at volume
$L$ with quantum numbers $I_{1},\dots,I_{n}$ and $\rho(\{ k_{1},\dots,k_{r}\},L)$
is the $r$-particle Bethe-Yang Jacobi determinant (\ref{eq:byjacobian})
involving only the $r$-element subset $1\leq k_{1}<\dots<k_{r}\leq n$
of the $n$ particles, evaluated with rapidities $\tilde{\theta}_{k_{1}},\dots,\tilde{\theta}_{k_{r}}$.
Let us further denote\[
\mathcal{F}(\{ k_{1},\dots,k_{r}\})_{L}=F_{2r}^{s}(\tilde{\theta}_{k_{1}},\dots,\tilde{\theta}_{k_{r}})_{i_{k_{1}}\dots i_{k_{r}}}\]
where \begin{equation}
F_{2n}^{s}(\theta_{1},\dots,\theta_{n})_{i_{1}\dots i_{n}}=\lim_{\epsilon\rightarrow0}F_{2n}^{\Psi}(\theta_{n}+i\pi+\epsilon,\dots,\theta_{1}+i\pi+\epsilon,\theta_{1},\dots,\theta_{n})_{i_{1}\dots i_{n}i_{n}\dots i_{1}}\label{eq:Fs_definition}\end{equation}
is the so-called symmetric evaluation of diagonal $n$-particle matrix
elements, which we analyze more closely in the next subsection. Note
that the exclusion property mentioned at the end of subsection 2.1
carries over to the symmetric evaluation too: (\ref{eq:Fs_definition})
vanishes whenever the rapidities of two particles of the same species
coincide. 

Based on the above results, we conjecture that the general rule for
a diagonal matrix element takes the form of a sum over all bipartite
divisions of the set of the $n$ particles involved (including the
trivial ones when $A$ is the empty set or the complete set $\{1,\dots,n\}$):\begin{eqnarray}
\,_{i_{1}\dots i_{n}}\langle\{ I_{1}\dots I_{n}\}|\Psi|\{ I_{1}\dots I_{n}\}\rangle_{i_{1}\dots i_{n},L} & = & \frac{1}{\rho(\{1,\dots,n\})_{L}}\times\label{eq:diaggenrule}\\
 &  & \sum_{A\subset\{1,2,\dots n\}}\mathcal{F}(A)_{L}\rho(\{1,\dots,n\}\setminus A)_{L}+O(\mathrm{e}^{-\mu L})\nonumber \end{eqnarray}

\subsection{Diagonal matrix elements in terms of connected form factors}

\label{symm_conn}

Here we discuss diagonal matrix elements in terms of connected
form factors. The results will be used in the following subsection to
prove that a conjecture made by Saleur in \cite{saleurfiniteT} 
exactly coincides with our eqn. (\ref{eq:diaggenrule}). To simplify
notations we omit the particle species labels; they can be restored
easily if needed.

\subsubsection{Relation between connected and symmetric matrix elements}

The purpose of this discussion is to give a treatment of the ambiguity
inherent in diagonal matrix elements. Due to the existence of kinematical
poles (\ref{eq:kinematicalaxiom}) the expression \[
F_{2n}(\theta_{1}+i\pi,\theta_{2}+i\pi,...,\theta_{n}+i\pi,\theta_{n},...,\theta_{2},\theta_{1})\]
which is relevant for diagonal multi-particle matrix elements, is
not well-defined. Let us consider the regularized version\[
F_{2n}(\theta_{1}+i\pi+\epsilon_{1},\theta_{2}+i\pi+\epsilon_{2},...,\theta_{n}+i\pi+\epsilon_{n},\theta_{n},...,\theta_{2},\theta_{1})\]
It was first observed in \cite{nonintegrable} that the singular parts
of this expression drop when taking the limits $\epsilon_{i}\rightarrow0$
simultaneously; however, the end result depends on the direction of
the limit, i.e. on the ratio of the $\epsilon_{i}$ parameters. The
terms that are relevant in the limit can be written in the following
general form: \begin{eqnarray}
F_{2n}(\theta_{1}+i\pi+\epsilon_{1},\theta_{2}+i\pi+\epsilon_{2},...,\theta_{n}+i\pi+\epsilon_{n},\theta_{n},...,\theta_{2},\theta_{1})=\label{mostgeneral}\\
\prod_{i=1}^{n}\frac{1}{\epsilon_{i}}\cdot\sum_{i_{1}=1}^{n}\sum_{i_{2}=1}^{n}...\sum_{i_{n}=1}^{n}a_{i_{1}i_{2}...i_{n}}(\theta_{1},\dots,\theta_{n})\epsilon_{i_{1}}\epsilon_{i_{2}}...\epsilon_{i_{n}}+\dots\nonumber \end{eqnarray}
 where $a_{i_{1}i_{2}...i_{n}}$ is a completely symmetric tensor
of rank $n$ and the ellipsis denote terms that vanish when taking
$\epsilon_{i}\rightarrow0$ simultaneously.

In our previous considerations we used the symmetric limit, which
is defined by taking all $\epsilon_{i}$ equal: \[
F_{2n}^{s}(\theta_{1},\theta_{2},...,\theta_{n})=\lim_{\epsilon\to0}F_{2n}(\theta_{1}+i\pi+\epsilon,\theta_{2}+i\pi+\epsilon,...,\theta_{n}+i\pi+\epsilon,\theta_{n},...,\theta_{2},\theta_{1})\]
It is symmetric in all the variables $\theta_{1},\dots,\theta_{n}$.
There is another evaluation with this symmetry property, namely the
so-called connected form factor, which is defined as the $\epsilon_{i}$
independent part of eqn. (\ref{mostgeneral}), i.e. the part which
does not diverge whenever any of the $\epsilon_{i}$ is taken to zero:
\begin{equation}
F_{2n}^{c}(\theta_{1},\theta_{2},...,\theta_{n})=n!\, a_{12...n}\label{eq:connected}\end{equation}
where the appearance of the factor $n!$ is simply due to the permutations
of the $\epsilon_{i}$.

\subsubsection{The relation for $n\leq3$}

We now spell out the relation between the symmetric and connected
evaluations for $n=1$, $2$ and $3$. 

The $n=1$ case is simple, since the two-particle form factor $F_{2}(\theta_{1},\theta_{2})$
has no singularities at $\theta_{1}=\theta_{2}+i\pi$ and therefore
\begin{equation}
F_{2}^{s}(\theta)=F_{2}^{c}(\theta)=F_{2}(i\pi,0)\label{eq:fs2fc2}\end{equation}
It is independent of the rapidities and will be denoted $F_{2}^{c}$
in the sequel. 

For $n=2$ we need to consider\begin{equation}
F_{4}(\theta_{1}+i\pi+\epsilon_{1},\theta_{2}+i\pi+\epsilon_{2},\theta_{2},\theta_{1})\approx\frac{a_{11}\epsilon_{1}^{2}+2a_{12}\epsilon_{1}\epsilon_{2}+a_{22}\epsilon_{2}^{2}}{\epsilon_{1}\epsilon_{2}}\label{fourresidue}\end{equation}
which gives\begin{eqnarray*}
F_{4}^{s}(\theta_{1},\theta_{2}) & = & a_{11}+2a_{12}+a_{22}\\
F_{4}^{c}(\theta_{1},\theta_{2}) & = & 2a_{12}\end{eqnarray*}
The terms $a_{11}$ and $a_{22}$ can be expressed using the two-particle
form factor. Taking an infinitesimal, but fixed $\epsilon_{2}\ne0$
\[
\mathop{\mathrm{Res}}_{\epsilon_{1}=0}F_{4}(\theta_{1}+i\pi+\epsilon_{1},\theta_{2}+i\pi+\epsilon_{2},\theta_{2},\theta_{1})=a_{22}\epsilon_{2}\]
 whereas according to (\ref{eq:dynamicalaxiom})\[
\mathop{\mathrm{Res}}_{\epsilon_{1}=0}F_{4}(\theta_{1}+i\pi+\epsilon_{1},\theta_{2}+i\pi+\epsilon_{2},\theta_{2},\theta_{1})=i\left(1-S(\theta_{1}-\theta_{2})S(\theta_{1}-\theta_{2}-i\pi-\epsilon_{2})\right)F_{2}(\theta_{2}+i\pi+\epsilon_{2},\theta_{2})\]
To first order in $\epsilon_{2}$\[
S(\theta_{1}-\theta_{2}-i\pi-\epsilon_{2})=S(\theta_{2}-\theta_{1}+\epsilon_{2})=S(\theta_{2}-\theta_{1})(1+i\varphi(\theta_{2}-\theta_{1})\epsilon_{2}+\dots)\]
where \[
\varphi(\theta)=-i\frac{d}{d\theta}\log S(\theta)\]
is the derivative of the two-particle phase shift defined before.
Therefore we obtain \[
a_{22}=\varphi(\theta_{2}-\theta_{1})F_{2}^{c}\]
and similarly \[
a_{11}=\varphi(\theta_{1}-\theta_{2})F_{2}^{c}\]
and so \begin{equation}
F_{4}^{s}(\theta_{1},\theta_{2})=F_{4}^{c}(\theta_{1},\theta_{2})+2\varphi(\theta_{1}-\theta_{2})F_{2}(i\pi,0)\label{cs_four}\end{equation}
In the case of the trace of the energy-momentum tensor $\Theta$ the
following expressions are known \cite{mussardodifference} \begin{eqnarray*}
F_{2}^{\Theta} & = & 2\pi m^{2}\\
F_{4}^{\Theta,s} & = & 8\pi m^{2}\varphi(\theta_{1}-\theta_{2})\cosh^{2}\left(\frac{\theta_{1}-\theta_{2}}{2}\right)\\
F_{4}^{\Theta,c} & = & 4\pi m^{2}\varphi(\theta_{1}-\theta_{2})\cosh(\theta_{1}-\theta_{2})\end{eqnarray*}
and they are in agreement with (\ref{cs_four}).

For $n=3$, a procedure similar to the above gives the following relation:\begin{eqnarray}
F_{6}^{s}(\theta_{1},\theta_{2},\theta_{3}) & = & F_{6}^{c}(\theta_{1},\theta_{2},\theta_{3})+\left[F_{4}^{c}(\theta_{1},\theta_{2})(\varphi(\theta_{1}-\theta_{3})+\varphi(\theta_{2}-\theta_{3}))+\mathrm{permutations}\right]\nonumber \\
 &  & +3F_{2}^{c}\left[\varphi(\theta_{1}-\theta_{2})\varphi(\theta_{1}-\theta_{3})+\mathrm{permutations}\right]\label{eq:f6sa}\end{eqnarray}
where we omitted terms that only differ by permutation of the particles.

\subsubsection{Relation between the connected and symmetric evaluation in the general
case}

Our goal is to compute the general expression \begin{equation}
F_{2n}(\theta_{1},\dots,\theta_{n}|\epsilon_{1},\dots,\epsilon_{n})=F_{2n}(\theta_{1}+i\pi+\epsilon_{1},\theta_{2}+i\pi+\epsilon_{2},...,\theta_{n}+i\pi+\epsilon_{n},\theta_{n},...,\theta_{2},\theta_{1})\label{f2n_eztkellene}\end{equation}
Let us take $n$ vertices labeled by the numbers $1,2,\dots,n$ and
let $G$ be the set of the directed graphs $G_{i}$ with the following
properties: 

\begin{itemize}
\item $G_{i}$ is tree-like. 
\item For each vertex there is at most one outgoing edge. 
\end{itemize}
For an edge going from $i$ to $j$ we use the notation $E_{ij}$.

\paragraph{Theorem 1\label{par:Theorem-1}}

(\ref{f2n_eztkellene}) can be evaluated as a sum over all graphs
in $G$, where the contribution of a graph $G_{i}$ is given by the
following two rules: 

\begin{itemize}
\item Let $A_{i}=\{ a_{1},a_{2},\dots,a_{m}\}$ be the set of vertices from
which there are no outgoing edges in $G_{i}$. The form factor associated
to $G_{i}$ is \begin{equation}
F_{2m}^{c}(\theta_{a_{1}},\theta_{a_{2}},\dots,\theta_{a_{m}})\label{egygrafformfaktora}\end{equation}
 
\item For each edge $E_{jk}$ the form factor above has to be multiplied
by \[
\frac{\epsilon_{j}}{\epsilon_{k}}\varphi(\theta_{j}-\theta_{k})\]
 
\end{itemize}
Note that since cannot contain cycles, the product of the $\epsilon_{i}/\epsilon_{j}$
factors will never be trivial (except for the empty graph with no
edges).

\paragraph*{Proof}

The proof goes by induction in $n$. For $n=1$ we have \[
F_{2}^{s}(\theta_{1})=F_{2}^{c}(\theta_{1})=F_{2}(i\pi,0)\]
This is in accordance with the theorem, because for $n=1$ there is
only the trivial graph which contains no edges and a single node. 

Now assume that the theorem is true for $n-1$ and let us take the
case of $n$ particles. Consider the residue of the matrix element
(\ref{f2n_eztkellene}) at $\epsilon_{n}=0$ while keeping all the
$\epsilon_{i}$ finite \[
R=\mathop{\mathrm{Res}}_{\epsilon_{n}=0}F_{2n}(\theta_{1}..\theta_{n}|\epsilon_{1}..\epsilon_{n})\]
According to the theorem the graphs contributing to this residue are
exactly those for which the vertex $n$ has an outgoing edge and no
incoming edges. Let $R_{j}$ be sum of the diagrams where the outgoing
edge is $E_{nj}$ for some $j=1,\dots,n-1$, and so \[
R=\sum_{j=1}^{n-1}R_{j}\]
The form factors appearing in $R_{j}$ do not depend on $\theta_{n}$.
Therefore we get exactly the diagrams that are needed to evaluate
$F_{2(n-1)}(\theta_{1}..\theta_{n-1}|\epsilon_{1}..\epsilon_{n-1})$,
apart from the proportionality factor associated to the link $E_{nj}$
and so \[
R_{j}=\frac{\epsilon_{j}}{\epsilon_{n}}\varphi(\theta_{j}-\theta_{n})F_{2(n-1)}(\theta_{1}..\theta_{n-1}|\epsilon_{1}..\epsilon_{n-1})\]
and summing over $j$ gives \begin{equation}
R=(\epsilon_{1}\varphi(\theta_{1}-\theta_{n})+\epsilon_{2}\varphi(\theta_{2}-\theta_{n})+\dots+\epsilon_{n-1}\varphi(\theta_{n-1}-\theta_{n}))F_{2(n-1)}(\theta_{1}..\theta_{n-1}|\epsilon_{1}..\epsilon_{n-1})\label{erremitlepsz}\end{equation}
In order to prove the theorem, we only need to show that the residue
indeed takes this form. On the other hand, the kinematical residue
axiom (\ref{eq:kinematicalaxiom}) gives \[
R=i\left(1-\prod_{j=1}^{n-1}S(\theta_{n}-\theta_{j})S(\theta_{n}-\theta_{j}-i\pi-\epsilon_{j})\right)F_{2(n-1)}(\theta_{1}..\theta_{n-1}|\epsilon_{1}..\epsilon_{n-1})\]
 which is exactly the same as eqn. (\ref{erremitlepsz}) when expanded
to first order in $\epsilon_{j}$.

We thus checked that the theorem gives the correct result for the
terms that include a $1/\epsilon_{n}$ singularity. Using symmetry
in the rapidity variables this is true for all the terms that include
at least one $1/\epsilon_{i}$ for an arbitrary $i$. There is only
one diagram that cannot be generated by the inductive procedure, namely
the empty graph. However, there are no singularities ($1/\epsilon_{i}$
factors) associated to it, and it gives $F_{2n}^{c}(\theta_{1},\dots,\theta_{n})$
by definition. \emph{Qed}.

\begin{figure}
\noindent \begin{centering}\includegraphics[scale=1.2]{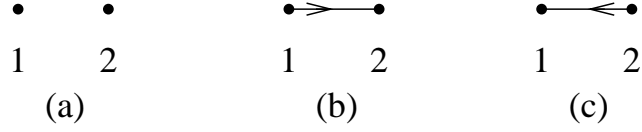}\par\end{centering}

\caption{\label{fig:neq2gr} The graphs relevant for $n=2$}
\end{figure}

\begin{figure}
\noindent \begin{centering}\includegraphics[scale=1.2]{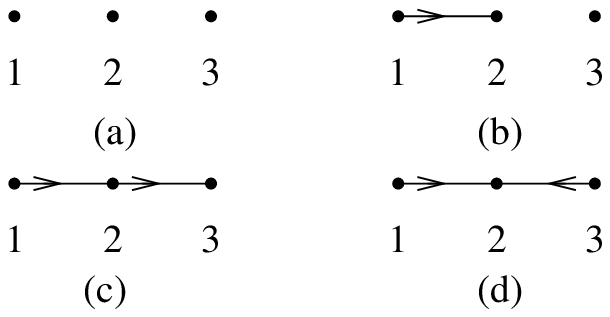}\par\end{centering}

\caption{\label{fig:neq3gr} The graphs relevant for $n=3$}
\end{figure}

We now illustrate how the theorem works. For $n=2$, there are only
three graphs, depicted in figure \ref{fig:neq2gr}. Applying the rules
yields \[
F_{4}(\theta_{1},\theta_{2}|\epsilon_{1},\epsilon_{2})=F_{4}^{c}(\theta_{1},\theta_{2})+\varphi(\theta_{1}-\theta_{2})\left(\frac{\epsilon_{1}}{\epsilon_{2}}+\frac{\epsilon_{2}}{\epsilon_{1}}\right)F_{2}^{c}\]
which gives back (\ref{cs_four}) upon putting $\epsilon_{1}=\epsilon_{2}$.
For $n=3$ there are $4$ different kinds of graphs, the representatives
of which are shown in figure \ref{fig:neq3gr}; all other graphs can
be obtained by permuting the node labels $1,2,3$. The contributions
of these graphs are

\begin{eqnarray*}
(a) & : & F_{6}^{c}(\theta_{1},\theta_{2},\theta_{3})\\
(b) & : & \frac{\epsilon_{2}}{\epsilon_{1}}\varphi(\theta_{1}-\theta_{2})F_{4}^{c}(\theta_{2},\theta_{3})\\
(c) & : & \frac{\epsilon_{2}}{\epsilon_{1}}\frac{\epsilon_{3}}{\epsilon_{2}}\varphi(\theta_{1}-\theta_{2})\varphi(\theta_{2}-\theta_{3})F_{2}^{c}=\frac{\epsilon_{3}}{\epsilon_{1}}\varphi(\theta_{1}-\theta_{2})\varphi(\theta_{2}-\theta_{3})F_{2}^{c}\\
(d) & : & \frac{\epsilon_{2}}{\epsilon_{1}}\frac{\epsilon_{2}}{\epsilon_{3}}\varphi(\theta_{1}-\theta_{2})\varphi(\theta_{3}-\theta_{2})F_{2}^{c}\end{eqnarray*}
Adding up all the contributions and putting $\epsilon_{1}=\epsilon_{2}=\epsilon_{3}$
we recover eqn. (\ref{eq:f6sa}).

\subsection{Consistency with Saleur's proposal}

\label{diagonal:saleur}

Saleur proposed an expression for diagonal matrix elements in terms
of connected form factors in \cite{saleurfiniteT}, which is partially
based on earlier work by Balog \cite{balogtba} and also on the determinant
formula for normalization of states in the framework of algebraic
Bethe Ansatz, derived by Gaudin, and also by Korepin (see \cite{qism}
and references therein). To describe it, we must extend the normalization
of finite volume states defined previously to the case when
the particle rapidities form a proper subset of some multi-particle
Bethe-Yang solution. 

It was shown in subsection \ref{fv_ff:generic} that the normalization of a finite volume
state is given by\[
\vert\{ I_{1},\dots,I_{n}\}\rangle_{L}=\frac{1}{\sqrt{\rho_{n}(\tilde{\theta}_{1},\dots,\tilde{\theta}_{n})}}\vert\tilde{\theta}_{1},\dots,\tilde{\theta}_{n}\rangle\]
We again omit the particle species labels, and also denote the $n$-particle
determinant by $\rho_{n}$. Let us take a subset of particle indices
$A\in\{1,\dots,n\}$ and define the corresponding sub-determinant
by\[
\tilde{\rho}_{n}(\tilde{\theta}_{1},\dots,\tilde{\theta}_{n}|A)=\det\mathcal{J}_{A}^{(n)}\]
where $\mathcal{J}_{A}^{(n)}$ is the sub-matrix of the matrix $\mathcal{J}^{(n)}$
defined in eqn. (\ref{eq:byjacobian}) which is given by choosing
the elements whose indices belong to $A$. The full matrix can be
written explicitly as\[
\mathcal{J}^{(n)}=\begin{pmatrix}E_{1}L+\varphi_{12}+\dots+\varphi_{1n} & -\varphi_{12} & \dots & -\varphi_{1n}\\
-\varphi_{12} & E_{2}L+\varphi_{21}+\varphi_{23}+\dots+\varphi_{2n} & \dots & -\varphi_{2n}\\
\vdots & \vdots & \ddots & \vdots\\
-\varphi_{1n} & -\varphi_{2n} & \dots & E_{n}L+\varphi_{1n}+\dots+\varphi_{n-1,n}\end{pmatrix}\]
where the following abbreviations were used: $E_{i}=m_{i}\cosh\theta_{i}$,
$\varphi_{ij}=\varphi_{ji}=\varphi(\theta_{i}-\theta_{j})$. Note
that $\tilde{\rho}_{n}$ depends on all the rapidities, not just those
which correspond to elements of $A$. It is obvious that\[
\rho_{n}(\tilde{\theta}_{1},\dots,\tilde{\theta}_{n})\equiv\tilde{\rho}_{n}(\tilde{\theta}_{1},\dots,\tilde{\theta}_{n}|\{1,\dots,n\})\]
Saleur proposed the definition\begin{equation}
\langle\{\tilde{\theta}_{k}\}_{k\in A}\vert\{\tilde{\theta}_{k}\}_{k\in A}\rangle_{L}=\tilde{\rho}_{n}(\tilde{\theta}_{1},\dots,\tilde{\theta}_{n}|A)\label{eq:detnorm}\end{equation}
where \[
\vert\{\tilde{\theta}_{k}\}_{k\in A}\rangle_{L}\]
is a {}``partial state'' which contains only the particles with
index in $A$, but with rapidities that solve the Bethe-Yang equations
for the full $n$-particle state. Note that this is not a proper state
in the sense that it is not an eigenstate of the Hamiltonian since
the particle rapidities do not solve the Bethe-Yang equations relevant
for a state consisting of $|A|$ particles (where $|A|$ denotes the
cardinal number -- i.e. number of elements -- of the set $A$). The
idea behind this proposal is that the density of these partial states
in rapidity space depends on the presence of the other particles which
are not included, and indeed it is easy to see that it is given by
$\tilde{\rho}_{n}(\tilde{\theta}_{1},\dots,\tilde{\theta}_{n}|A)$.

In terms of the above definitions, Saleur's conjecture for the diagonal
matrix element is\begin{eqnarray}
\,_{i_{1}\dots i_{n}}\langle\{ I_{1}\dots I_{n}\}|\Psi|\{ I_{1}\dots I_{n}\}\rangle_{i_{1}\dots i_{n},L} & = & \frac{1}{\rho_{n}(\tilde{\theta}_{1},\dots,\tilde{\theta}_{n})}\times\label{eq:diaggenrulesaleur}\\
 &  & \sum_{A\subset\{1,2,\dots n\}}F_{2|A|}^{c}(\{\tilde{\theta}_{k}\}_{k\in A})\tilde{\rho}(\tilde{\theta}_{1},\dots,\tilde{\theta}_{n}|A)+O(\mathrm{e}^{-\mu L})\nonumber \end{eqnarray}
which is just the standard representation of the full matrix element
as the sum of all the connected contributions provided we accept eqn.
(\ref{eq:detnorm}). The full amplitude is obtained by summing over
all possible bipartite divisions of the particles, where the division
is into particles that are connected to the local operator, giving
the connected form factor $F^{c}$ and into those that simply go directly
from the initial to the final state which contribute the norm of the
corresponding partial multi-particle state. 

Using the results of the previous subsection, it is easy to check explicitly
 that our rule for the diagonal matrix
elements as given in eqn. (\ref{eq:diaggenrule}) is equivalent to
eqn. (\ref{eq:diaggenrulesaleur}). We now give a complete proof for
the general case.

\paragraph{Theorem 2}

\begin{equation}
\sum_{A\subset N}F_{2|A|}^{c}(\{\theta_{k}\}_{k\in A})\tilde{\rho}(\theta_{1},\dots,\theta_{n}|A)=\sum_{A\subset N}F_{2|A|}^{s}(\{\theta_{k}\}_{k\in A})\rho(\{\theta_{k}\}_{k\in N\setminus A})\label{ketfelekepp}\end{equation}
where we denoted $N=\{1,2,\dots,n\}$.

\paragraph{Proof}

The two sides of eqn. (\ref{ketfelekepp}) differ in two ways: 

\begin{itemize}
\item The form factors on the right hand side are evaluated according to
the ,,symmetric'' prescription, and in addition to the connected
part also they contain extra terms, which are proportional to connected
form factors with fewer particles. 
\item The densities $\tilde{\rho}$ on the left hand side are not determinants
of the form (\ref{eq:byjacobian}) written down in terms of the particles
contained in $N\setminus A$: they contain additional terms due to
the presence of the particles in $A$ as well. 
\end{itemize}
Here we show that eqn. (\ref{ketfelekepp}) is merely a reorganization
of these terms. 

For simplicity consider first the term on the left hand side which
corresponds to $A=\{ m+1,m+2,\dots,n\}$, i.e. \[
F_{2m}^{c}(\theta_{m+1},\dots,\theta_{n})\tilde{\rho}(\theta_{1},\dots,\theta_{n}|A)\]
We expand $\tilde{\rho}$ in terms of the physical multi-particle
densities $\rho$. In order to accomplish this, it is useful to rewrite
the sub-matrix $\mathcal{J}_{N\setminus A}^{n}{}$ as \[
\mathcal{J}^{(n)}|_{N\setminus A}=\mathcal{J}^{m}(\theta_{1},\dots,\theta_{m})+\begin{pmatrix}\sum\limits _{i=m+1}^{n}\varphi_{1i}\\
 &  & \sum\limits _{i=m+1}^{n}\varphi_{2i}\\
 &  &  &  & \ddots\\
 &  &  &  &  &  & \sum\limits _{i=m+1}^{n}\varphi_{mi}\end{pmatrix}\]
where $\mathcal{J}^{m}$ is the $m$-particle Jacobian matrix which
does not contain any terms depending on the particles in $A$. The
determinant of $\mathcal{J}_{N\setminus A}^{n}{}$ can be written
as a sum over the subsets of $N\setminus A$. For a general subset
$B\subset N\setminus A$ let us use the notation $B=\{ b_{1},b_{2},\dots,b_{|B|}\}$.
We can then write \begin{equation}
\tilde{\rho}(\theta_{1},\dots,\theta_{n}|A)=\text{det}\mathcal{J}^{(n)}|_{N\setminus A}=\sum_{B}\left[\rho(N\setminus(A\cup B))\prod_{i=1}^{|B|}\left(\sum_{c_{i}=m+1}^{n}\varphi_{b_{i},c_{i}}\right)\right]\label{egyfajta}\end{equation}
 where $\rho(N\setminus(A\cup B))$ is the $\rho$-density (\ref{eq:byjacobian})
written down with the particles in $N\setminus(A\cup B)$. 

Applying a suitable permutation of variables we can generalize eqn.
(\ref{egyfajta}) to an arbitrary subset $A\subset N$: \begin{equation}
\tilde{\rho}(\theta_{1},\dots,\theta_{n}|A)=\text{det}\mathcal{J}^{(n)}|_{N\setminus A}=\sum_{B}\rho(N\setminus(A\cup B))\sum_{C}(\prod_{i=1}^{|B|}\varphi_{b_{i},c_{i}})\label{egyfajta2}\end{equation}
where the second summation goes over all the sets $C=\{ c_{1},c_{2},\dots,c_{|B|}\}$
with $|C|=|B|$ and $c_{i}\in A$. The left hand side of eqn. (\ref{ketfelekepp})
can thus be written as \begin{eqnarray}
\sum_{A\subset N}F_{2|A|}^{c}(\{\theta_{k}\}_{k\in A})\tilde{\rho}(\theta_{1},\dots,\theta_{n}|A) & = & \sum_{\begin{array}{c}
A,B\subset N\\
A\cap B=\emptyset\end{array}}\rho(N\setminus(A\cup B))\sum_{C}F_{(A,B,C)}\label{ketfelekepp2}\\
\mbox{where} &  & F_{(A,B,C)}=F_{2|A|}^{c}(\{\theta_{k}\}_{k\in A})\prod_{i=1}^{|B|}\varphi_{b_{i},c_{i}}\nonumber \end{eqnarray}
We now show that there is a one-to-one correspondence between all
the terms in (\ref{ketfelekepp2}) and those on the right hand side
of (\ref{ketfelekepp}) if the symmetric evaluations $F_{2k}^{s}$
are expanded according to Theorem 1. To each triplet $(A,B,C)$ let
us assign the graph $G_{(A,B,C)}$ defined as follows: 

\begin{itemize}
\item The vertices of the graph are the elements of the set $A\cup B$. 
\item There are exactly $|B|$ edges in the graph, which start at $b_{i}$
and end at $c_{i}$ with $i=1,\dots,|B|$. 
\end{itemize}
The contribution of $G_{(A,B,C)}$ to $F_{2(|A|+|B|)}^{s}(\{\theta_{k}\}_{k\in A\cup B})$
is nothing else than $F_{(A,B,C)}$ which can be proved by applying
the rules of Theorem 1. Note that all the possible diagrams with at
most $n$ vertices are contained in the above list of the $G_{(A,B,C)}$,
because a general graph $G$ satisfying \textit{\emph{the conditions}}
in Theorem 1 can be characterized by writing down the set of vertices
with and without outgoing edges (in this case $B$ and $A$) and the
endpoints of the edges (in this case $C$).

It is easy to see that the factors $\rho(N\setminus(A\cup B))$ multiplying
the $F_{(A,B,C)}$ in (\ref{ketfelekepp2}) are also the correct ones:
they are just the density factors multiplying $F_{2(|A|+|B|)}^{s}(\{\theta_{k}\}_{k\in A\cup B})$
on the right hand side of (\ref{ketfelekepp}). \emph{Qed}.

\section{Zero-momentum particles}

\label{zero-momentum}

\subsection{Theories with one particle species}

In theories with only a single particle species, there
can only be a single particle of zero momentum in a multi-particle
state due to the exclusion principle. For the momentum to be exactly
zero in finite volume it is necessary that all the other particles
should come with quantum numbers in pairs of opposite sign, which
means that the state must have $2n+1$ particles in a configuration\[
|\{ I_{1},\dots,I_{n},0,-I_{n},\dots,-I_{1}\}\rangle_{L}\]
Therefore we consider matrix elements of the form\[
\langle\{ I_{1}',\dots,I_{k}',0,-I_{k}',\dots,-I_{1}'\}|\Phi|\{ I_{1},\dots,I_{l},0,-I_{l},\dots,-I_{1}\}\rangle_{L}\]
(with $k=0$ or $l=0$ corresponding to a state containing a single
stationary particle). We also suppose that the two sets $\{ I_{1},\dots,I_{k}\}$
and $\{ I_{1}',\dots,I_{l}'\}$ are not identical, otherwise we have
the case of diagonal matrix elements treated in section \ref{diagonal}.

We need to examine form factors of the form\[
F_{2k+2l+2}(i\pi+\theta_{1}',\dots,i\pi+\theta_{k}',i\pi-\theta_{k}',\dots,i\pi-\theta_{1}',i\pi+\theta,0,\theta_{1},\dots,\theta_{l},-\theta_{l},\dots,-\theta_{1})\]
where the particular ordering of the rapidities was chosen to ensure
that no additional $S$ matrix factors appear in the disconnected
terms of the crossing relation (\ref{eq:ffcrossing}). Using the singularity
axiom (\ref{eq:kinematicalaxiom}), plus unitarity and crossing symmetry
of the $S$-matrix it is easy to see that the residue of the above
function at $\theta=0$ vanishes, and so it has a finite limit as
$\theta\rightarrow0$. However, this limit depends on direction just
as in the case of the diagonal matrix elements considered in section
4. Therefore we must specify the way it is taken, and just as previously
we use a prescription that is maximally symmetric in all variables:
we choose to shift all rapidities entering the left hand state with
the same amount to define\begin{eqnarray}
 &  & \mathcal{F}_{k,l}(\theta_{1}',\dots,\theta_{k}'|\theta_{1},\dots,\theta_{l})=\nonumber \\
 &  & \lim_{\epsilon\rightarrow0}F_{2k+2l+2}(i\pi+\theta_{1}'+\epsilon,\dots,i\pi+\theta_{k}'+\epsilon,i\pi-\theta_{k}'+\epsilon,\dots,i\pi-\theta_{1}'+\epsilon,\nonumber \\
 &  & i\pi+\epsilon,0,\theta_{1},\dots,\theta_{l},-\theta_{l},\dots,-\theta_{1})\label{eq:oddoddlimitdef}\end{eqnarray}
Using the above definition, by analogy to (\ref{eq:diaggenrule})
we conjecture that
\begin{eqnarray}
&&\langle\{ I_{1}',\dots,I_{k}',0,-I_{k}',\dots,-I_{1}'\}|\Phi|\{ I_{1},\dots,I_{l},0,-I_{l},\dots,-I_{1}\}\rangle_{L}\label{eq:oddoddlyrule}
  = \\ & &\frac{1}{\sqrt{\rho_{2k+1}(\tilde{\theta}_{1}',\dots,\tilde{\theta}_{k}',0,-\tilde{\theta}_{k}',\dots,-\tilde{\theta}_{1}')\rho_{2l+1}(\tilde{\theta}_{1},\dots,\tilde{\theta}_{l},0,-\tilde{\theta}_{l},\dots,-\tilde{\theta}_{1})}}\times\nonumber \\
 &  & \Big(\mathcal{F}_{k,l}(\tilde{\theta}_{1}',\dots,\tilde{\theta}_{k}'|\tilde{\theta}_{1},\dots,\tilde{\theta}_{l})+mL\, F_{2k+2l}(i\pi+\tilde{\theta}_{1}',\dots,i\pi+\tilde{\theta}_{k}',\nonumber \\
 &  & i\pi-\tilde{\theta}_{k}',\dots,i\pi-\tilde{\theta}_{1}',\tilde{\theta}_{1},\dots,\tilde{\theta}_{l},-\tilde{\theta}_{l},\dots,-\tilde{\theta}_{1})\Big)+O(\mathrm{e}^{-\mu L})\nonumber \end{eqnarray}
where $\tilde{\theta}$ denote the solutions of the appropriate Bethe-Yang
equations at volume $L$, $\rho_{n}$ is a shorthand notation for
the $n$-particle Bethe-Yang density (\ref{eq:byjacobian}) and equality
is understood up to phase factors. We recall from section
\ref{elementary} that relative phases of multi-particle states are in
general fixed differently in the form factor bootstrap and in
numerical methods (TCSA). Also
note that reordering particles gives phase factors on the right hand
side according to the exchange axiom (\ref{eq:exchangeaxiom}). This
issue is obviously absent in the case of diagonal matrix elements
treated in the previous section, since any such phase factor cancels out
between the state and its conjugate. Such phases do not affect correlation
functions, or as a consequence, any physically relevant quantities
since they can all be expressed in terms of correlators. 

There is some argument that can be given in support of eqn. (\ref{eq:oddoddlyrule}).
Note that the zero-momentum particle occurs in both the left and right
states, which actually makes it unclear how to define a density similar
to $\tilde{\rho}$ in (\ref{eq:detnorm}). Such a density would take
into account the interaction with the other particles. However, the
nonzero rapidities entering of the two states are different and therefore
there is no straightforward way to apply Saleur's recipe (\ref{eq:diaggenrulesaleur})
here. Using the maximally symmetric definition (\ref{eq:oddoddlimitdef})
the shift $\epsilon$ can be equally put on the right hand side rapidities
as well, and therefore we expect that the density factor multiplying
the term $F_{2k+2l}$ in (\ref{eq:oddoddlyrule}) would be the one-particle
state density in which none of the other rapidities appear, which
is exactly $mL$ for a stationary particle. This is a natural guess
from eqn. (\ref{eq:diaggenrule}) which states that when diagonal
matrix elements are expressed using the symmetric evaluation, only
densities of the type $\rho$ appear. 

Another argument can be formulated using the observation that eqn.
(\ref{eq:oddoddlyrule}) is only valid if $\mathcal{F}_{k,l}$ is
defined as in (\ref{eq:oddoddlimitdef}); all other possible ways
to take the limit can be related in a simple way to this definition
and so the rule (\ref{eq:oddoddlyrule}) can be rewritten appropriately.
Let us consider two other natural choices\begin{eqnarray*}
 &  & \mathcal{F}_{k,l}^{+}(\theta_{1}',\dots,\theta_{k}'|\theta_{1},\dots,\theta_{l})=\\
 &  & \lim_{\epsilon\rightarrow0}F_{2k+2l+2}(i\pi+\theta_{1}',\dots,i\pi+\theta_{k}',i\pi-\theta_{k}',\dots,i\pi-\theta_{1}',i\pi,\epsilon,\theta_{1},\dots,\theta_{l},-\theta_{l},\dots,-\theta_{1})\\
 &  & \mathcal{F}_{k,l}^{-}(\theta_{1}',\dots,\theta_{k}'|\theta_{1},\dots,\theta_{l})=\\
 &  & \lim_{\epsilon\rightarrow0}F_{2k+2l+2}(i\pi+\theta_{1}',\dots,i\pi+\theta_{k}',i\pi-\theta_{k}',\dots,i\pi-\theta_{1}',i\pi+\epsilon,0,\theta_{1},\dots,\theta_{l},-\theta_{l},\dots,-\theta_{1})\end{eqnarray*}
in which the shift is put only on the zero-momentum particle on the
right/left, respectively. Using the kinematical residue axiom (\ref{eq:kinematicalaxiom}),
$\mathcal{F}^{\pm}$ can be related to $\mathcal{F}$ via\begin{eqnarray*}
 &  & \mathcal{F}_{k,l}(\theta_{1}',\dots,\theta_{k}'|\theta_{1},\dots,\theta_{l})=\mathcal{F}_{k,l}^{+}(\theta_{1}',\dots,\theta_{k}'|\theta_{1},\dots,\theta_{l})\\
 &  & +2\sum_{i=1}^{l}\varphi(\theta_{i})F_{2k+2l}(i\pi+\theta_{1}',\dots,i\pi+\theta_{k}',i\pi-\theta_{k}',\dots,i\pi-\theta_{1}',\theta_{1},\dots,\theta_{l},-\theta_{l},\dots,-\theta_{1})\\
 &  & \mathcal{F}_{k,l}(\theta_{1}',\dots,\theta_{k}'|\theta_{1},\dots,\theta_{l})=\mathcal{F}_{k,l}^{-}(\theta_{1}',\dots,\theta_{k}'|\theta_{1},\dots,\theta_{l})\\
 &  & -2\sum_{i=1}^{k}\varphi(\theta_{i}')F_{2k+2l}(i\pi+\theta_{1}',\dots,i\pi+\theta_{k}',i\pi-\theta_{k}',\dots,i\pi-\theta_{1}',\theta_{1},\dots,\theta_{l},-\theta_{l},\dots,-\theta_{1})\end{eqnarray*}
With the help of the above relations eqn. (\ref{eq:oddoddlyrule})
can also be rewritten in terms of $\mathcal{F}^{\pm}$. The way $\mathcal{F}$
and therefore also eqn. (\ref{eq:oddoddlyrule}) are expressed in
terms of $\mathcal{F^{\pm}}$ shows a remarkable and natural symmetry
under the exchange of the left and right state (and correspondingly
$\mathcal{F}^{+}$ with $\mathcal{F}^{-}$), which provides a further
support to our conjecture.

The above two arguments cannot be considered as a proof; we do not
have a proper derivation of relation (\ref{eq:oddoddlyrule}). On the
other hand, as we show in section \ref{numerics:zero} it agrees very
well with 
numerical data which would be impossible if there were some additional
$\varphi$ terms present; such terms, as shown in section
\ref{elementary} would contribute corrections of order $1/l$ in terms 
of the dimensionless volume parameter $l=mL$. 

\subsection{Generalization to more than one particle species }

\label{zero_generalized}

In a general diagonal scattering theory, there can be more than one
particles of zero momentum in a multi-particle state. This can occur
if the following two conditions hold:
\begin{itemize}
\item The sets of the rapidities are ``parity-symmetric''  for each particle
species separately, ie. for every particle with non-zero rapidity $\theta$ there
is a particle of the same species with rapidity
$-\theta$. The same holds for the momentum quantum numbers. It is
convenient to denote such a state with $2n+m$ 
particles by
\begin{equation*}
  \ket{\{I_1,-I_1,I_2,-I_2,\dots,I_n,-I_n,0,0,\dots,0\}}_{i_1i_1i_2i_2\dots i_{n}i_{n}\ j_{1}\dots j_{m},L}
\end{equation*}
\item The scattering phases between the zero-momentum particles satisfy
  \begin{equation*}
    \mathop{\prod_{l=1\dots m}}_{l\ne k} S_{j_kj_l}(0)=1 \quad\quad k=1\dots m
  \end{equation*}
This condition is necessary for the Bethe-Yang equations to hold.
\end{itemize}

Here we consider matrix elements between two different states
$\ket{\varphi}$ and $\ket{\varphi'}$ which satisfy the previous two conditions.
The generalization of \eqref{eq:oddoddlyrule} is straightforward: one
has to include a disconnected term for every proper subset of those
zero-momentum particles, which are present in both states. To
precisely formulate this rule, once again ({\small and in this section for the
last time }    $\ddot\smile$ ) we need to introduce new notations. We
wish to remark that the result presented below is unpublished material. 

Let $\ket{\psi}$ and $\ket{\psi'}$ be a scattering state with
$n+o$ and $n'+o$ particles, respectively, where $o$ denotes the number
of those zero-momentum
particles which are present in both states.
The sets of the remaining
particles consist of ``parity-symmetric'' pairs or they may also contain 
zero-momentum particles that are not present in the other state. For
the rapidities of these particles we use the standard notation
$\theta_1\dots \theta_n$ and $\theta'_1\dots \theta'_{n'}$ and in the
following we do
not distinguish whether a particular rapidity belongs to a pair or is zero.
Particle types of these first $n$ and $n'$ particles are denoted by
$i_1,\dots i_n$ and $i'_1,\dots i'_{n'}$, whereas for the remaining
$o$ particles we use $j_1,\dots j_o$.

For every subset $A\subset \{j_1,\dots j_o\}$ we define
\begin{eqnarray*}
\mathcal{F}^o_{nn'}(A)=  \lim_{\eps\to 0} 
F_{n+n'+2|A|}(i\pi+\theta_1+\eps,\dots,i\pi+\theta_n+\eps,\underbrace{i\pi+\eps,\dots,i\pi+\eps}_{|A|
\text{ times}},\underbrace{0,\dots,0}_{|A|
\text{ times}},\theta'_1,\dots,\theta'_{n'})_{B}
\end{eqnarray*}
where the label of particle types is given by
\begin{equation*}
 B= \{i_1\dots i_n\} A \bar{A} \{i'_1\dots i'_{n'}\}
\end{equation*}
with $\bar{A}$ consisting of the elements of $A$ listed in reverse
order.

With these notations, the conjectured general formula reads
\begin{eqnarray}
  \label{eq:zero_mom_most_general}
\nonumber
  \bra{\psi}\Phi\ket{\psi'}_L=
\frac{1}{\sqrt{
\rho_{n+o}(\theta_1,\dots,\theta_n,0,\dots,0)
\rho_{n'+o}(\theta'_1,\dots,\theta'_{n'},0,\dots,0)
}}\times\\
\sum_{A \subset \{j_1,\dots j_o\}} \mathcal{F}^o_{nn'}(A) 
\prod_{j\in B} (m_jL)
+\mathcal{O}(e^{-\mu L})
\end{eqnarray}
where $B= \{j_1,\dots j_o\} \setminus A$.
Note, that an exchange of any two rapidities in $\theta_1\dots
\theta_n$ or $\theta'_1\dots \theta'_{n'}$ yields the same phase 
factor in each term; the expression \eqref{eq:zero_mom_most_general}
is therefore well-defined and 
is to be understood up to
an overall phase factor,
similar to \eqref{eq:oddoddlyrule} and \eqref{eq:genffrelation}.

\chapter{Finite Volume Form Factors -- Numerical analysis}

\label{s:numerics}

Here we compare the analytic predictions of the previous
chapter to numerical data obtained by the Truncated Conformal Space
Approach (TCSA). 

In \ref{numerics_howto} we explain in detail the methods we used to
numerically determine the finite volume matrix elements in the
Lee-Yang model and in the Ising model in a magnetic field. In sections
\ref{numerics}-\ref{numerics:general}
we present results on form factors without disconnected terms. The
special cases of diagonal matrix elements and form factors with
zero-momentum particles are investigated in \ref{numerics:diagonal}
and \ref{numerics:zero}, respectively.

\section{Form factors from truncated conformal space}

\label{numerics_howto}

\subsection{Truncated conformal space approach for scaling Lee-Yang model}

We use the truncated conformal space approach (TCSA) developed by
Yurov and Zamolodchikov in \cite{yurovzam}. 

Due to translational invariance of the Hamiltonian (\ref{eq:lypcftham}),
the conformal Hilbert space $\mathcal{H}$ can be split into sectors
characterized by the eigenvalues of the total spatial momentum \[
P=\frac{2\pi}{L}\left(L_{0}-\bar{L}_{0}\right)\]
the operator $L_{0}-\bar{L}_{0}$ generates Lorentz transformations
and its eigenvalue is called Lorentz spin. For a numerical evaluation
of the spectrum, the Hilbert space is truncated by imposing a cut
in the conformal energy. The truncated conformal space corresponding
to a given truncation and fixed value $s$ of the Lorentz spin reads\[
\mathcal{H}_{\mathrm{TCS}}(s,e_{\mathrm{cut}})=\left\{ |\psi\rangle\in\mathcal{H}\:|\;\left(L_{0}-\bar{L}_{0}\right)|\psi\rangle=s|\psi\rangle,\;\left(L_{0}+\bar{L}_{0}-\frac{c}{12}\right)|\psi\rangle=e|\psi\rangle\,:\, e\leq e_{\mathrm{cut}}\right\} \]
On this subspace, the dimensionless Hamiltonian matrix can be written
as\begin{equation}
h_{ij}=\frac{2\pi}{l}\left(L_{0}+\bar{L}_{0}-\frac{c}{12}+i\frac{\kappa l^{2-2\Delta}}{(2\pi)^{1-2\Delta}}G^{(s)-1}B^{(s)}\right)\label{eq:dimlesstcsaham}\end{equation}
where energy is measured in units of the particle mass $m$, $l=mL$
is the dimensionless volume parameter, \begin{equation}
G_{ij}^{(s)}=\langle i|j\rangle\label{eq:Gs}\end{equation}
is the conformal inner product matrix and \begin{equation}
B_{ij}^{(s)}=\left.\langle i|\Phi(z,\bar{z})|j\rangle\right|_{z=\bar{z}=1}\label{eq:Bs}\end{equation}
is the matrix element of the operator $\Phi$ at the point $z=\bar{z}=1$
on the complex plane between vectors $|i\rangle$, $|j\rangle$ from
$\mathcal{H}_{\mathrm{TCS}}(s,e_{\mathrm{cut}})$. The natural basis
provided by the action of Virasoro generators is not orthonormal and
therefore $G^{(s)-1}$ must be inserted to transform the left vectors
to the dual basis. The Hilbert space and the matrix elements are constructed
using an algorithm developed by Kausch et al. and first used in \cite{takacs_watts}.

Diagonalizing the matrix $h_{ij}$ we obtain the energy levels as
functions of the volume, with energy and length measured in units
of $m$. The maximum value of the cutoff $e_{\mathrm{cut}}$ we used
was $30$, in which case the Hilbert space contains around one thousand
vectors, slightly depending on the spin.

\subsection{Exact form factors of the primary field $\Phi$ in the
  Lee-Yang model}

Form factors of the trace of the stress-energy tensor $\Theta$ were
computed by Al.B. Zamolodchikov in \cite{zam_Lee_Yang}, and using the relation\begin{equation}
\Theta=i\lambda\pi(1-\Delta)\Phi\label{eq:thetaphirel}\end{equation}
we can rewrite them in terms of $\Phi$. They have the form\begin{equation}
F_{n}(\theta_{1},\dots,\theta_{n})=\langle\Phi\rangle H_{n}Q_{n}(x_{1},\dots,x_{n})\prod_{i=1}^{n}\prod_{j=i+1}^{n}\frac{f(\theta_{i}-\theta_{j})}{x_{i}+x_{j}}\label{eq:lyff}\end{equation}
with the notations\begin{eqnarray*}
f(\theta) & = & \frac{\cosh\theta-1}{\cosh\theta+1/2}v(i\pi-\theta)v(-i\pi+\theta)\\
v(\theta) & = & \exp\left(2\int_{0}^{\infty}dt\frac{\sinh\frac{\pi t}{2}\sinh\frac{\pi t}{3}\sinh\frac{\pi t}{6}}{t\sinh^{2}\pi t}\mathrm{e}^{i\theta t}\right)\\
x_{i} & = & \mathrm{e}^{\theta_{i}}\qquad,\qquad H_{n}=\left(\frac{3^{1/4}}{2^{1/2}v(0)}\right)^{n}\end{eqnarray*}
The mass-gap of the theory is related to the coupling constant 
as
 \begin{equation}
\lambda=0.09704845636\dots\times
m^{12/5}\label{eq:lymassgap2}\end{equation}
The exact vacuum expectation value of the field $\Phi$ is \[
\langle\Phi\rangle=1.239394325\dots\times i\, m^{-2/5}\]
which can be readily obtained using (\ref{eq:lymassgap2}, \ref{eq:thetaphirel})
and also the known vacuum expectation value of $\Theta$ \cite{zam_Lee_Yang}\[
\langle\Theta\rangle=-\frac{\pi m^{2}}{4\sqrt{3}}\]
The functions $Q_{n}$ are symmetric polynomials in the variables
$x_{i}$. Defining the elementary symmetric polynomials of $n$ variables
by the relations\[
\prod_{i=1}^{n}(x+x_{i})=\sum_{i=0}^{n}x^{n-i}\sigma_{i}^{(n)}(x_{1},\dots,x_{n})\qquad,\qquad\sigma_{i}^{(n)}=0\mbox{ for }i>n\]
they can be constructed as\begin{eqnarray*}
Q_{1} & = & 1\qquad,\qquad Q_{2}=\sigma_{1}^{(2)}\qquad,\qquad Q_{3}=\sigma_{1}^{(3)}\sigma_{2}^{(3)}\\
Q_{n} & = & \sigma_{1}^{(n)}\sigma_{n-1}^{(n)}P_{n}\quad,\qquad n>3\\
P_{n} & = & \det\mathcal{M}^{(n)}\quad\mbox{where}\quad\mathcal{M}_{ij}^{(n)}=\sigma_{3i-2j+1}^{(n)}\quad,\quad i,j=1,\dots,n-3\end{eqnarray*}
Note that the one-particle form factor is independent of the rapidity:\begin{equation}
F_{1}^{\Phi}=1.0376434349\dots\times
im^{-2/5}\label{eq:ly1pff}\end{equation}

\subsection{Truncated fermionic space approach for the Ising model}

The conformal Ising model can be represented as the theory of a massless
Majorana fermion with the action\[
\mathcal{A}_{Ising}=\frac{1}{2\pi}\int d^{2}z\left(\bar{\psi}\partial\bar{\psi}+\psi\bar{\partial}\psi\right)\]
On the conformal plane the model has two sectors, with the mode expansions\[
\psi(z)=\begin{cases}
\sum_{r\in\mathbb{Z}+\frac{1}{2}}b_{r}z^{-r-1/2} & \mbox{ Neveu-Schwarz (NS) sector}\\
\sum_{r\in\mathbb{Z}}b_{r}z^{-r-1/2} & \mbox{ Ramond (R) sector}\end{cases}\]
and similarly for the anti-holomorphic field $\bar{\psi}$. The Hilbert
space is the direct sum of a certain projection of the NS and R sectors,
with the Virasoro content\[
\mathcal{H}_{Ising}=\bigoplus_{h=0,\frac{1}{2},\frac{1}{16}}\mathcal{V}_{h}\otimes\bar{\mathcal{V}}_{h}\]
The spin field $\sigma$ connects the NS and R sectors, and its matrix
elements $B_{ij}^{(s)}$ in the sector with a given conformal spin
$s$ (cf. eqn. (\ref{eq:Bs})) can be most conveniently computed in
the fermionic basis using the work of Yurov and Zamolodchikov \cite{yurovzam_fermionic_TFCSA},
who called this method the truncated fermionic space approach. The
fermionic basis can easily be chosen orthonormal, and thus in this
case the metrics $G^{(s)}$ on the spin subspaces (cf. eqn. (\ref{eq:Gs}))
are all given by unit matrices of appropriate dimension. Apart from
the choice of basis all the calculation proceeds very similarly to
the case of the Lee-Yang model. Energy and volume is measured in units
of the lowest particle mass $m=m_{1}$ and using relation (\ref{eq:ising_massgap})
one can write the dimensionless Hamiltonian in the form (\ref{eq:dimlesstcsaham}).
The highest cutoff we use is $e_{\mathrm{cut}}=30$, in which case
the Hilbert space contains around three thousand vectors (slightly
depending on the value of the spin chosen).

We remark that the energy density operator can be represented in the
fermionic language as \[
\epsilon=\bar{\psi}\psi\]
which makes the evaluation of its matrix elements in the fermionic
basis extremely simple.

\subsection{Ising model --  Form factors of the energy density operator}

The form factors of the operator $\epsilon$ in the $E_{8}$ model
were first calculated in \cite{ising_ff1} and their determination
was carried further in \cite{resonances}. The exact vacuum expectation
value of the field $\epsilon$ is given by \cite{vevs}\[
\langle\epsilon\rangle=\epsilon_{h}|h|^{8/15}\qquad,\qquad\epsilon_{h}=2.00314\dots\]
or in terms of the mass scale $m=m_{1}$\begin{equation}
\langle\epsilon\rangle=0.45475\dots\times m\label{eq:epsilonexactvev}\end{equation}
The form factors are not known for the general $n$-particle case
in a closed form, i.e. no formula similar to that in (\ref{eq:lyff})
exists. They can be evaluated by solving the appropriate polynomial
recursion relations derived from the form-factor axioms. We do not
present explicit formulae here; instead we refer to the above papers.
For practical calculations we used the results computed by Delfino,
Grinza and Mussardo, which can be downloaded from the Web in \texttt{Mathematica}
format \cite{isingff}.

\subsection{Evaluating matrix elements of a local operator $\mathcal{O}$ in
TCSA}

\subsubsection{Identification of multi-particle states}

\label{identification}

Diagonalizing the TCSA Hamiltonian (\ref{eq:dimlesstcsaham}) yields
a set of eigenvalues and eigenvectors at each value of the volume,
but it is not immediately obvious how to select the same state at
different values of the volume. Therefore in order to calculate form
factors it is necessary to identify the states with the corresponding
many-particle interpretation.

Finding the vacuum state is rather simple since it is the lowest lying
state in the spin-0 sector and its energy is given by \[
E_{0}(L)=BL+\dots\]
where the ellipsis indicate residual finite size effects decaying
exponentially fast with volume $L$ and $B$ is the bulk energy density
which in the models we consider is exactly known (\ref{eq:lybulk},
\ref{eq:isingbulk}). One-particle states can be found using that
their energies can be expressed as\[
E_{i}^{(s)}(L)=BL+\sqrt{\left(\frac{2\pi s}{L}\right)^{2}+m_{i}^{2}}+\dots\]
again up to residual finite size effects where $s$ is the spin of
the sector considered and $i$ is the species label (every sector
contains a single one-particle state for each species). 

Higher multi-particle states can be identified by comparing the measured
eigenvalues to the levels predicted by the Bethe-Yang equations. Fixing
species labels $i_{1},\dots,i_{n}$ and momentum quantum numbers $I_{1},\dots,I_{n}$,
eqns. (\ref{eq:betheyang}) can be solved to give the rapidities $\tilde{\theta}_{1},\dots,\tilde{\theta}_{n}$
of the particles as function of the dimensionless volume parameter
$l=mL$. Then the energy of the multi-particle state in question is\[
E_{i_{1}\dots i_{n}}^{(I_{1}\dots I_{n})}(L)=BL+\sum_{k=1}^{n}m_{i_{k}}\cosh\tilde{\theta}_{k}+\dots\]
which can be compared to the spectrum. 

For each state there exists a range of the volume, called the \emph{scaling
region}, where $L$ is large enough so that the omitted residual finite
size effects can be safely neglected and small enough so that the
truncation errors are also negligible. More precisely, the scaling
region for any quantity depending on the volume can be defined as
the volume range in which the residual finite size corrections and
the truncation errors are of the same order of magnitude; since both
sources of error show a dependence on the state and the particular
quantity considered (as well as on the value of the cutoff), so does
the exact position of the scaling region itself.

\begin{figure}
\begin{centering}\psfrag{el}{$e(l)$}\psfrag{l}{$l$}\includegraphics[scale=0.75]{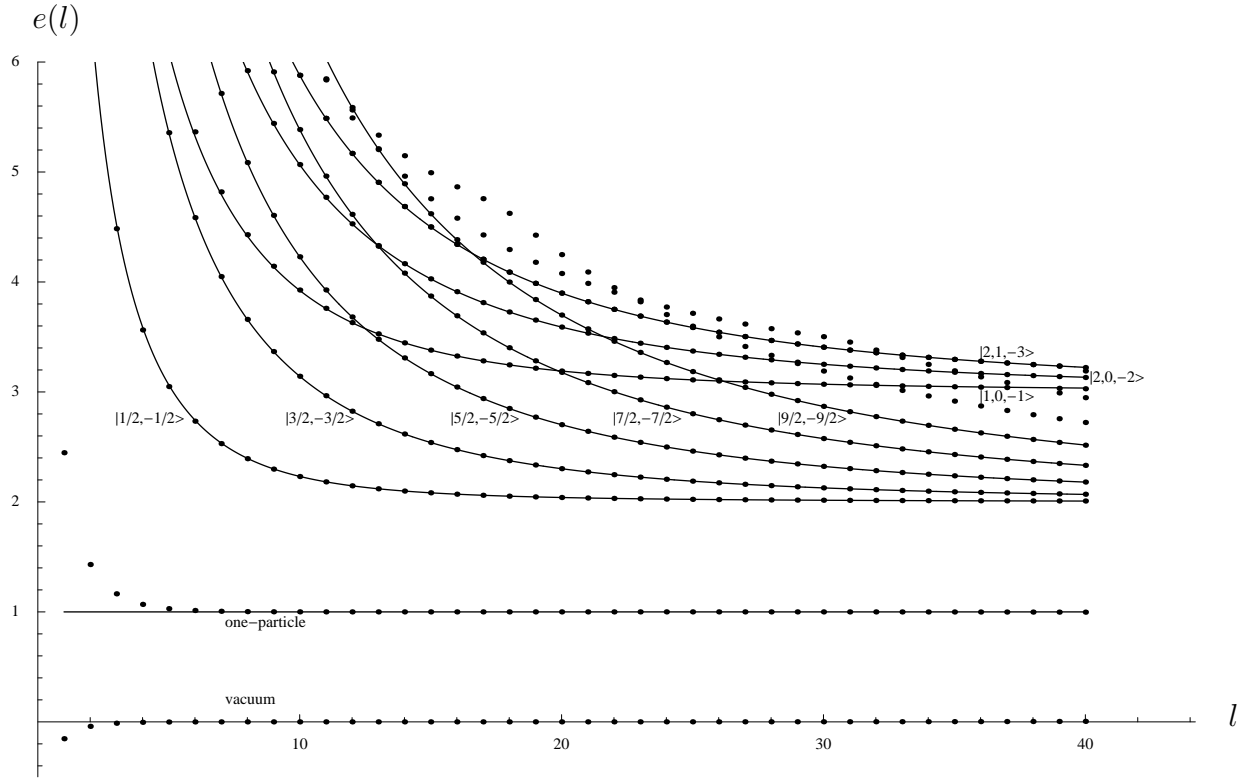}\par\end{centering}

\caption{\label{fig:lyfinvol}The first $13$ states in the finite volume
spectrum of scaling Lee-Yang model. We plot the energy in units of
$m$ (with the bulk subtracted): $e(l)=(E(L)-BL)/m$, against the
dimensionless volume variable $l=mL$. $n$-particle states are labeled
by $|I_{1},\dots,I_{n}\rangle$, where the $I_{k}$ are the momentum
quantum numbers. The state labeled $|2,1,-3\rangle$ is actually two-fold
degenerate because of the presence of $|-2,-1,3\rangle$ (up to a
splitting which vanishes as $\mathrm{e}^{-l}$, cf. the discussion
in subsection 3.3). The dots are the TCSA results and the continuous
lines are the predictions of the Bethe-Yang equations (\ref{eq:betheyang}).
The points not belonging to any of the Bethe-Yang lines drawn are
two- and three-particle states which are only partly contained in
the first $13$ levels due to line crossings, whose presence is a
consequence of the integrability of the model. }
\end{figure}

In the scaling region, we can use a comparison between the Bethe-Yang
predictions and the numerical energy levels to sort the states and
label them by multi-particle quantum numbers. An example is shown
in figure \ref{fig:lyfinvol}, where we plot the first few states
in the spin-0 sector of the scaling Lee-Yang model and their identification
in terms of multi-particle states is given. In this case, the agreement
with the predicted bulk energy density and the Bethe-Yang levels in
the scaling region is better than one part in $10^{4}$ for every
state shown (with the TCSA cutoff taken at $e_{\mathrm{cut}}=30$).

\subsubsection{\label{sub:matrixeval}Evaluation of matrix elements}

Suppose that we computed two Hamiltonian eigenvectors as functions
of the volume $L$ (labeled by their quantum numbers in the Bethe-Yang
description (\ref{eq:betheyang}), omitting the particle species labels
for brevity):\begin{eqnarray*}
|\{ I_{1},\dots,I_{n}\}\rangle_{L} & = & \sum_{i}\Psi_{i}(I_{1},\dots,I_{n};L)|i\rangle\\
|\{ I_{1}',\dots,I_{k}'\}\rangle_{L} & = & \sum_{j}\Psi_{j}(I_{1}',\dots,I_{k}';L)|j\rangle\end{eqnarray*}
in the sector with spin $s$ and spin $s'$, respectively. Let the
inner products of these vectors with themselves be given by\begin{eqnarray*}
\mathcal{N} & = & \sum_{i,j}\Psi_{i}(I_{1},\dots,I_{n};L)G_{ij}^{(s)}\Psi_{j}(I_{1},\dots,I_{n};L)\\
\mathcal{N}' & = & \sum_{i,j}\Psi_{i}(I_{1}',\dots,I_{k}';L)G_{ij}^{(s')}\Psi_{j}(I_{1}',\dots,I_{k}';L)\end{eqnarray*}
It is important that the components of the left eigenvector are not
complex conjugated. In the Ising model we work in a basis where all
matrix and vector components are naturally real. In the Lee-Yang model,
the TCSA eigenvectors are chosen so that all of their components $\Psi_{i}$
are either purely real or purely imaginary depending on whether the
basis vector $|i\rangle$ is an element of the $h=\bar{h}=0$ or the
$h=\bar{h}=-1/5$ component in the Hilbert space. It is well-known
that the Lee-Yang model is non-unitary, which is reflected in the
presence of complex structure constants as indicated in (\ref{eq:lyconfope}).
This particular convention for the structure constants forces upon
us the above inner product, because it is exactly the one under which
TCSA eigenvectors corresponding to different eigenvalues are orthogonal.
We remark that by redefining the structure constants and the conformal
inner product it is also possible to use a manifestly real representation
for the Lee-Yang TCSA (up to some truncation effects that lead to
complex eigenvalues in the vicinity of level crossings \cite{yurovzam}).
Note that the above conventions mean that the phases of the eigenvectors
are fixed up to a sign.

Let us consider a spinless primary field $\mathcal{O}$ with scaling
weights $\Delta_{\mathcal{O}}=\bar{\Delta}_{\mathcal{O}}$, which
can be described as the matrix \[
O_{ij}^{(s',s)}=\left.\langle i|\mathcal{O}(z,\bar{z})|j\rangle\right|_{z=\bar{z}=1}\quad,\quad|i\rangle\in\mathcal{H}_{\mathrm{TCS}}(s',e_{\mathrm{cut}})\quad,\quad|j\rangle\in\mathcal{H}_{\mathrm{TCS}}(s,e_{\mathrm{cut}})\]
between the two truncated conformal space sectors. Then the matrix
element of $\mathcal{O}$ can be computed as\begin{eqnarray}
 &  & m^{-2\Delta_{\mathcal{O}}}\langle\{ I_{1}',\dots,I_{k}'\}\vert\mathcal{O}(0,0)\vert\{ I_{1},\dots,I_{n}\}\rangle_{L}=\nonumber \\
 &  & \qquad\left(\frac{2\pi}{mL}\right)^{2\Delta_{\mathcal{O}}}\frac{1}{\sqrt{\mathcal{N}}}\frac{1}{\sqrt{\mathcal{N}'}}\sum_{j,l}\Psi_{j}(I_{1}',\dots,I_{k}';L)O_{jl}^{(s',s)}\Psi_{l}(I_{1},\dots,I_{n};L)\label{eq:fftcsaevaluation}\end{eqnarray}
where the volume dependent pre-factor comes from the transformation
of the primary field $\mathcal{O}$ under the exponential map (\ref{eq:exponentialmap})
and we wrote the equation in a dimensionless form using the mass scale
$m$. The above procedure is a generalization of the one used by Guida
and Magnoli to evaluate vacuum expectation values in \cite{guidamagnoli};
it was extended to one-particle form factors in the context of the
tricritical Ising model by Fioravanti et al. in \cite{onepff}.

\section{Vacuum expectation values and one-particle form factors}

\label{numerics}

\subsection{\label{sub:opffly} Scaling Lee-Yang model}

Before the one-particle form factor we discuss the vacuum expectation
value. Let us define the dimensionless function\[
\phi(l)=-im^{2/5}\langle0|\Phi|0\rangle_{L}\]
where the finite volume expectation value is evaluated from TCSA using
(\ref{eq:fftcsaevaluation}).%
\begin{figure}
\begin{centering}\psfrag{l}{$l$}
\psfrag{vev}{$\phi$}\includegraphics[scale=1.3]{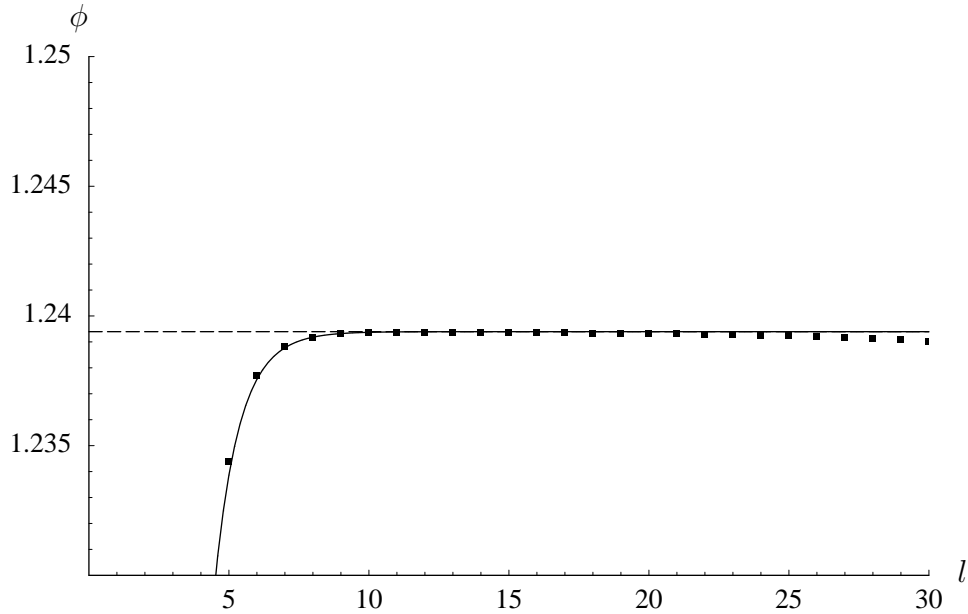}\par\end{centering}

\caption{\label{fig:lyvev}The vacuum expectation value of $\Phi$ in finite
volume. The dashed line shows the exact infinite volume value, while
the continuous line corresponds to eqn. (\ref{eq:vevLcorr}). }
\end{figure}
We performed measurement of $\phi$ as a function of both the cutoff
$e_{\mathrm{cut}}=21\dots30$ and the volume $l=1\dots30$ and then
extrapolated the cutoff dependence fitting a function\[
\phi(l,e_{\mathrm{cut}})=\phi(l)+A(l)e_{\mathrm{cut}}^{-12/5}\]
(where the exponent was chosen by verifying that it provides an optimal
fit to the data). The data corresponding to odd and even values of
the cutoff must be extrapolated separately \cite{cikk_resonances},
therefore one gets two estimates for the result, but they only differ
by a very small amount (of order $10^{-5}$ at $l=30$ and even less
for smaller volumes). The theoretical prediction for $\phi(l)$ is\[
\phi(l)=1.239394325\dots+O(\mathrm{e}^{-l})\]
The numerical result (after extrapolation) is shown in figure \ref{fig:lyvev}
from which it is clear that there is a long scaling region. Estimating
the infinite volume value from the flattest part of the extrapolated
curve (at $l$ around $12$) we obtain the following measured value\[
\phi(l=\infty)=1.23938\dots\]
where the numerical errors from TCSA are estimated to affect only
the last displayed digit, which corresponds to an agreement within
one part in $10^{5}$.

\begin{figure}
\noindent \begin{centering}\psfrag{O}{$\Phi$}
\psfrag{sum}{\Large $+\,\displaystyle\sum_i$}
\psfrag{i}{$i$}\includegraphics{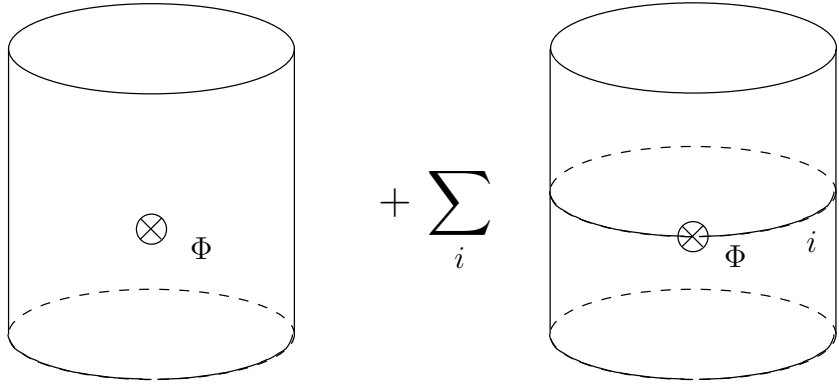}\par\end{centering}

\caption{\label{fig:vevLcorr} Graphical representation of eqn. (\ref{eq:vevLcorr}).}
\end{figure}

There is also a way to compute the leading exponential correction,
which was derived by Delfino \cite{delfinofiniteT}:\begin{equation}
\langle\Phi\rangle_{L}=\langle\Phi\rangle+\frac{1}{\pi}\sum_{i}F_{2}(i\pi,0)_{ii}K_{0}(m_{i}r)+\dots\label{eq:vevLcorr}\end{equation}
where \[
K_{0}(x)=\int_{0}^{\infty}d\theta\,\cosh\theta\,\mathrm{e}^{-x\cosh\theta}\]
is the modified Bessel-function, and the summation is over the particle
species $i$ (there is only a single term in the scaling Lee-Yang
model). This agrees very well with the numerical data, as demonstrated
in table \ref{tab:vevLcorr} and also in figure \ref{fig:lyvev}.
Using L\"uscher's finite-volume perturbation theory introduced in \cite{luscher_1particle},
the correction term can be interpreted as the sum of Feynman diagrams
where there is exactly one propagator that winds around the cylinder,
and therefore eqn. (\ref{eq:vevLcorr}) can be represented graphically
as shown in figure \ref{fig:vevLcorr}. On the other hand, by
Euclidean invariance this finite size correction coincides with the
first term in the low-T expansion for one-point functions at finite
temperature 
(see subsections \ref{ss:leclair_mussardo} and \ref{e-2mr}).

\begin{table}
\noindent \begin{centering}\begin{tabular}{|c|c|c|}
\hline 
$l$&
$\phi(l)$ (predicted)&
$\phi(l)$ (TCSA)\tabularnewline
\hline
\hline 
2&
1.048250&
1.112518\tabularnewline
\hline 
3&
1.184515&
1.195345\tabularnewline
\hline 
4&
1.222334&
1.224545\tabularnewline
\hline 
5&
1.233867&
1.234396\tabularnewline
\hline 
6&
1.237558&
1.237698\tabularnewline
\hline 
7&
1.238774&
1.238811\tabularnewline
\hline 
8&
1.239182&
1.239189\tabularnewline
\hline 
9&
1.239321&
1.239317\tabularnewline
\hline 
10&
1.239369&
1.239360\tabularnewline
\hline 
11&
1.239385&
1.239373\tabularnewline
\hline
12&
1.239391&
1.239375\tabularnewline
\hline
\end{tabular}\par\end{centering}

\caption{\label{tab:vevLcorr} Comparison of eqn. (\ref{eq:vevLcorr}) to
TCSA data}
\end{table}

\begin{figure}
\begin{centering}\psfrag{f1}{$\tilde{f}_1$}\psfrag{l}{$l$}
\psfrag{s=0}{$s=0$}
\psfrag{s=1}{$s=1$}
\psfrag{s=2}{$s=2$}
\psfrag{Exact}{exact}\includegraphics[scale=1.3]{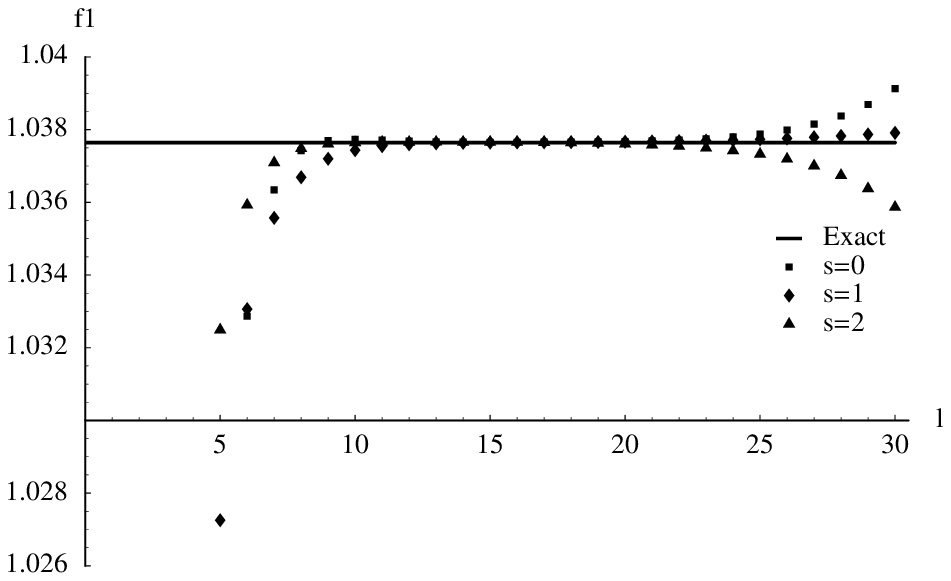}\par\end{centering}

\caption{\label{fig:oply}One-particle form factor from sectors with spin
$s=0,1,2$. The continuous line shows the exact infinite volume prediction.}
\end{figure}

To measure the one-particle form factor we use the correspondence
(\ref{eq:ffrelation}) between the finite and infinite volume form
factors to define the dimensionless function\[
\tilde{f}_{1}^{s}(l)=-im^{2/5}\left(l^{2}+\left(2\pi s\right)^{2}\right)^{1/4}\langle0|\Phi|\{ s\}\rangle_{L}\]
where $|\{ s\}\rangle_{L}$ is the finite volume one-particle state
with quantum number $I=s$ i.e. from the spin-$s$ sector. The theoretical
prediction for this quantity is\begin{equation}
\tilde{f}_{1}^{s}(l)=1.0376434349\dots+O(\mathrm{e}^{-l})\label{eq:lyopffpred}\end{equation}
The numerical results (after extrapolation in the cutoff) are shown
in figure \ref{fig:oply}. The scaling region gives the following
estimates for the infinite volume limit:\begin{eqnarray*}
\tilde{f}_{1}^{0}(l=\infty) & = & 1.037654\dots\\
\tilde{f}_{1}^{1}(l=\infty) & = & 1.037650\dots\\
\tilde{f}_{1}^{2}(l=\infty) & = & 1.037659\dots\end{eqnarray*}
which show good agreement with eqn. (\ref{eq:lyopffpred}) (the relative
deviation is again around $10^{-5}$, as for the vacuum expectation
value).

\subsection{Ising model in magnetic field}

\label{numerics_1p_ising}

\begin{figure}
\noindent \begin{centering}\psfrag{igaziertek}{$\phi(l)$}
\psfrag{l}{$l$}
\psfrag{extrapolated}{extrapolated}
\psfrag{ecut30}{$e_{\mathrm{cut}}=30$}
\psfrag{ecut28}{$e_{\mathrm{cut}}=28$}
\psfrag{ecut26}{$e_{\mathrm{cut}}=26$}\includegraphics[scale=1.2]{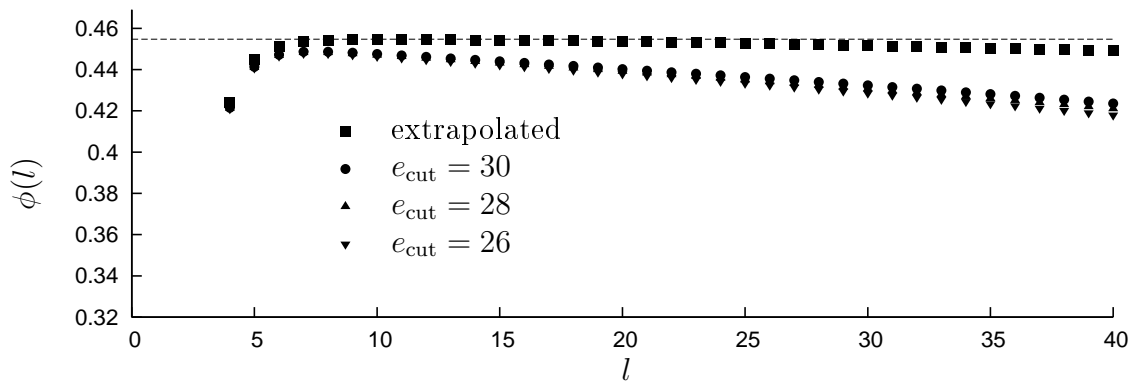}\par\end{centering}

\caption{\label{fig:vevisingraw} Measuring the vacuum expectation value of
$\epsilon$ in the Ising model}
\end{figure}

For the Ising model, we again start with checking the dimensionless
vacuum expectation value for which, using eqn. (\ref{eq:epsilonexactvev})
we have the prediction \[
\phi(l)=\frac{1}{m}\langle\epsilon\rangle_{L}=0.45475\dots+O\left(\mathrm{e}^{-l}\right)\]
where $m=m_{1}$ is the mass of the lightest particle and $l=mL$
as before. The TCSA data are shown in figure \ref{fig:vevisingraw}.
Note that there is substantial dependence on the cutoff $e_{\mathrm{cut}}$
and also that extrapolation in $e_{\mathrm{cut}}$ is really required
to achieve good agreement with the infinite volume limit. Reading
off the plateau value from the extrapolated data gives the estimate\[
\frac{1}{m}\langle\epsilon\rangle=0.4544\dots\]
for the infinite volume vacuum expectation value, which has $8\cdot10^{-4}$
relative deviation from the exact result. Our first numerical comparison
thus already tells us that we can expect much larger truncation errors
than in the Lee-Yang case. It is also clear from figure \ref{fig:vevisingraw}
that in order to attain suitable precision in the Ising model extrapolation
in the cutoff is very important.

\begin{figure}
\noindent \begin{centering}\psfrag{f0}{$\bar{\phi}(l)$}
\psfrag{l}{$l$}
\psfrag{ecut-fitted}{extrapolated}
\psfrag{predicted}{predicted}
\psfrag{infinite-volume}{infinite volume}\includegraphics[scale=1.3]{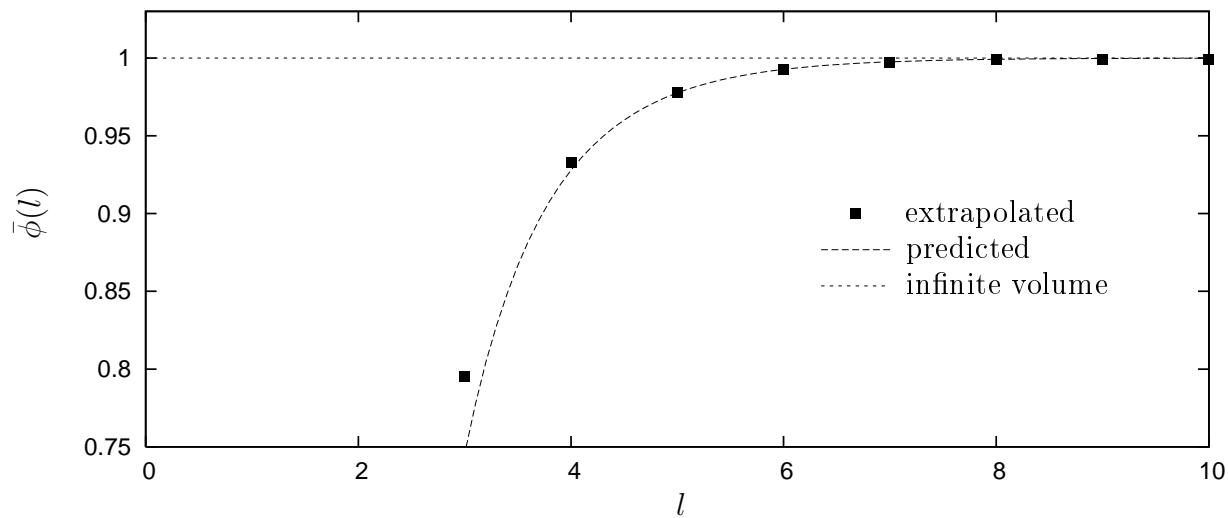}\par\end{centering}

\caption{\label{fig:vevisingLdep}The volume dependence of the vacuum expectation
value of $\epsilon$ in the Ising model, showing the extrapolated
value and the prediction from eqn. (\ref{eq:vevLcorr}), normalized
by the value in the infinite volume limit.}
\end{figure}

Defining the function \[
\bar{\phi}(l)=\langle\epsilon\rangle_{L}/\langle\epsilon\rangle\]
we can calculate the leading exponential correction using eqn. (\ref{eq:vevLcorr})
and the exact two-particle form factors from \cite{isingff}. It only
makes sense to include particles $i=1,2,3$ since the contribution
of the fourth particle is sub-leading with respect to two-particle
terms from the lightest particle due to $m_{4}>2m_{1}$. The result
is shown in figure \ref{fig:vevisingLdep}; we do not give the data
in numerical tables, but we mention that the relative deviation between
the predicted and measured value is better than $10^{-3}$ in the
range $5<l<10$. 

\begin{figure}
\noindent \begin{centering}\psfrag{f1}{${\tilde f}_1$}
\psfrag{f2}{${\tilde f}_2$}
\psfrag{f3}{${\tilde f}_3$}
\psfrag{l}{$l$}
\psfrag{s=0}{$\scriptstyle s=0$}
\psfrag{s=1}{$\scriptstyle s=1$}
\psfrag{s=2}{$\scriptstyle s=2$}
\psfrag{s=3}{$\scriptstyle s=3$}
\psfrag{exact}{\small exact}\\
\subfigure[$A_1$]{\includegraphics{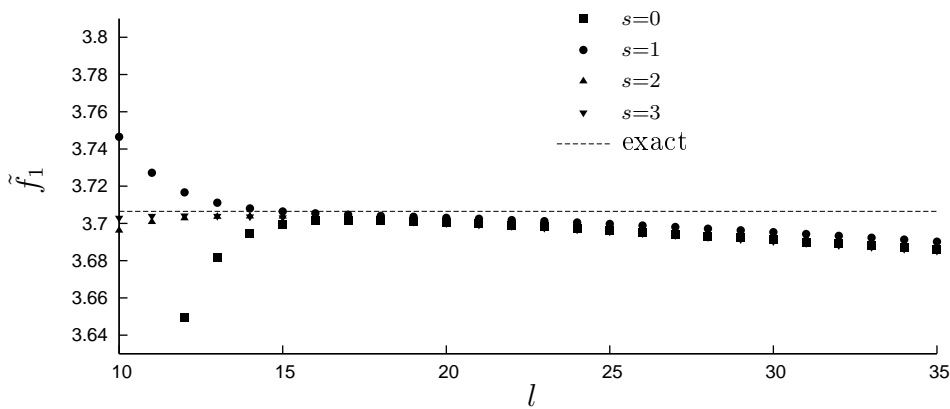}}\\
\subfigure[$A_2$]{\includegraphics{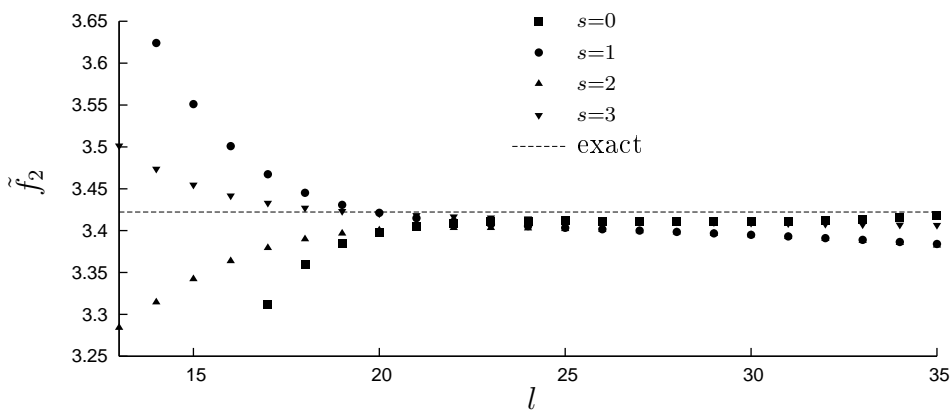}}\\
\subfigure[$A_3$]{\includegraphics{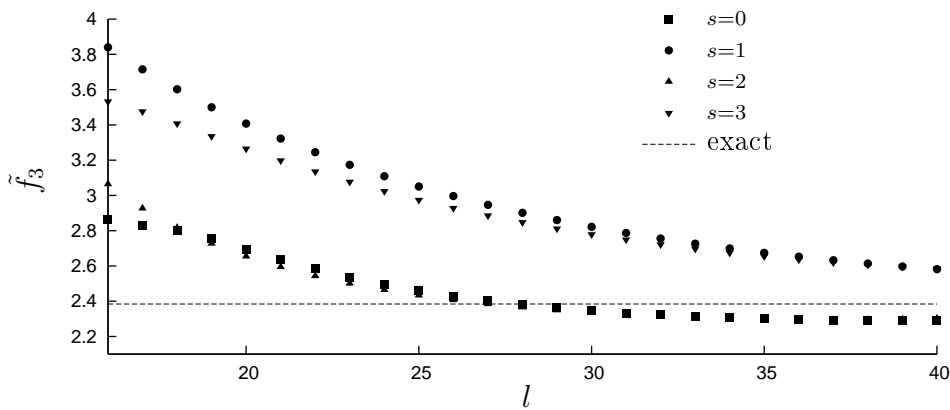}}\par\end{centering}

\caption{\label{fig:onepffising}One-particle form factors measured from TCSA
(dots) compared to the infinite-volume prediction from exact form
factors. All numerical data have been extrapolated to $e_{\mathrm{cut}}=\infty$
and $s$ denotes the Lorentz spin of the state considered. The relative
deviation in the scaling region is around $10^{-3}$ for $A_{1}$
and $A_{2}$; there is no scaling region for $A_{3}$ (see the discussion
in the main text).}
\end{figure}

From now on we normalize all form factors of the operator $\epsilon$
by the infinite volume vacuum expectation value (\ref{eq:epsilonexactvev}),
i.e. we consider form factors of the operator\begin{equation}
\Psi=\epsilon/\langle\epsilon\rangle\label{eq:psidef}\end{equation}
which conforms with the conventions used in \cite{resonances,isingff}.
We define the dimensionless one-particle form factor functions as\[
\tilde{f}_{i}^{s}(l)=\left(\left(\frac{m_{i}l}{m_{1}}\right)^{2}+\left(2\pi s\right)^{2}\right)^{1/4}\langle0|\Psi|\{ s\}\rangle_{i,\, L}\]
In the plots of figure \ref{fig:onepffising} we show how these functions
measured from TCSA compare to predictions from the exact form factors
for particles $i=1,2,3$ and spins $s=0,1,2,3$. 

It is evident that the scaling region sets in much later than for
the Lee-Yang model; therefore for the Ising model we do not plot data
for low values of the volume (all plots start from $l\sim10...15$).
This also means that truncation errors in the scaling region are also
much larger than in the scaling Lee-Yang model; we generally found
errors larger by an order of magnitude after extrapolation in the
cutoff. We remark that extrapolation improves the precision by an
order of magnitude compared to the raw data at the highest value of
the cutoff. 

\begin{figure}
\noindent \begin{centering}\includegraphics{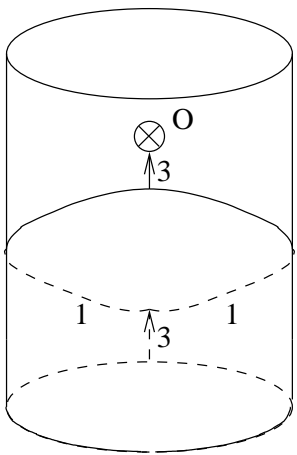}\par\end{centering}

\caption{\label{fig:ising3pcorr}Leading finite size correction (a so-called
$\mu$-term) to the one-particle form factor of $A_{3}$, which results
from the process of splitting up into two copies of $A_{1}$ which
then wind around the cylinder once before recombining into $A_{3}$
again.}
\end{figure}

Note the rather large finite size correction in the case of $A_{3}$.
This can be explained rather simply as the presence of a
$\mu$-term. Based on L\"uscher's finite-volume perturbation
theory \cite{luscher_1particle} it is expected, that similar to the
$\mu$-term for the masses of stable particles \eqref{mass_mu_term}, the leading contribution 
 (which is associated to the diagram depicted in figure \ref{fig:ising3pcorr}) has 
the volume
dependence 
\[
\mathrm{e}^{-\mu_{311}L}\qquad,
\qquad\mu_{311}=\sqrt{m_{1}^{2}-\frac{m_{3}^{2}}{4}}=0.10453\dots\times m_{1}\]
Therefore we can expect a contribution suppressed only by $\mathrm{e}^{-0.1l}$.
A numerical fit of the $l$-dependence in the $s=0$ case is perfectly
consistent with this expectation. As a result, no scaling region can
be found, because truncation errors are too large in the volume range
where the exponential correction is suitably small. We do not elaborate
on this issue further here:
chapter \ref{exponential} is devoted to the study of exponential corrections.

\section{Two-particle form factors}

\subsection{\label{sub:sly2pff} Scaling Lee-Yang model}

Following the ideas in the previous subsection, we can again define
a dimensionless function for each two-particle state as follows:\[
f_{2}(l)_{I_{1}I_{2}}=-im^{2/5}\langle0|\Phi|\{ I_{1},I_{2}\}\rangle_{L}\quad,\quad l=mL\]
Relation (\ref{eq:ffrelation}) gives the following prediction in
terms of the exact form-factors:\begin{equation}
f_{2}(l)_{I_{1}I_{2}}=\frac{-im^{2/5}}{\sqrt{\rho_{11}(\tilde{\theta}_{1}(l),\tilde{\theta}_{2}(l))}}F_{2}^{\Phi}(\tilde{\theta}_{1}(l),\tilde{\theta}_{2}(l))+O(\mathrm{e}^{-l})\label{eq:lyf2prediction}\end{equation}
where $\tilde{\theta}_{1}(l),\,\tilde{\theta}_{2}(l)$ solve the Bethe-Yang
equations\begin{eqnarray*}
l\sinh\tilde{\theta}_{1}+\delta(\tilde{\theta}_{1}-\tilde{\theta}_{2}) & = & 2\pi I_{1}\\
l\sinh\tilde{\theta}_{2}+\delta(\tilde{\theta}_{2}-\tilde{\theta}_{1}) & = & 2\pi I_{2}\end{eqnarray*}
and the density of states is given by\[
\rho_{11}(\theta_{1},\theta_{2})=l^{2}\cosh\theta_{1}\cosh\theta_{2}+l\cosh\theta_{1}\varphi(\theta_{2}-\theta_{1})+l\cosh\theta_{2}\varphi(\theta_{1}-\theta_{2})\]
the phase shift $\delta$ is defined according to eqn. (\ref{eq:lydeltachoice})
and \[
\varphi(\theta)=\frac{d\delta(\theta)}{d\theta}\]
There is a further issue to take into account: the relative phases
of the multi-particle states are a matter of convention and the choice
made in subsection \ref{sub:matrixeval} for the TCSA eigenvectors
may differ from the convention adapted in the form factor bootstrap.
Therefore in the numerical work we compare the absolute values of
the functions $f_{2}(l)$ computed from TCSA with those predicted
from the exact form factors. Note that this issue is present for any
non-diagonal matrix element, and was in fact tacitly dealt with in
the case of one-particle matrix elements treated in subsection \ref{sub:opffly}.

\begin{figure}
\begin{centering}\psfrag{f2}{$|f_2(l)|$}\psfrag{l}{$l$}
\psfrag{tcsa11}{$\scriptstyle\langle 0|\Phi|\{ 1/2,-1/2\}\rangle$}
\psfrag{tcsa33}{$\scriptstyle\langle 0|\Phi|\{ 3/2,-3/2\}\rangle$}
\psfrag{tcsa55}{$\scriptstyle\langle 0|\Phi|\{ 5/2,-5/2\}\rangle$}\includegraphics[scale=1.3]{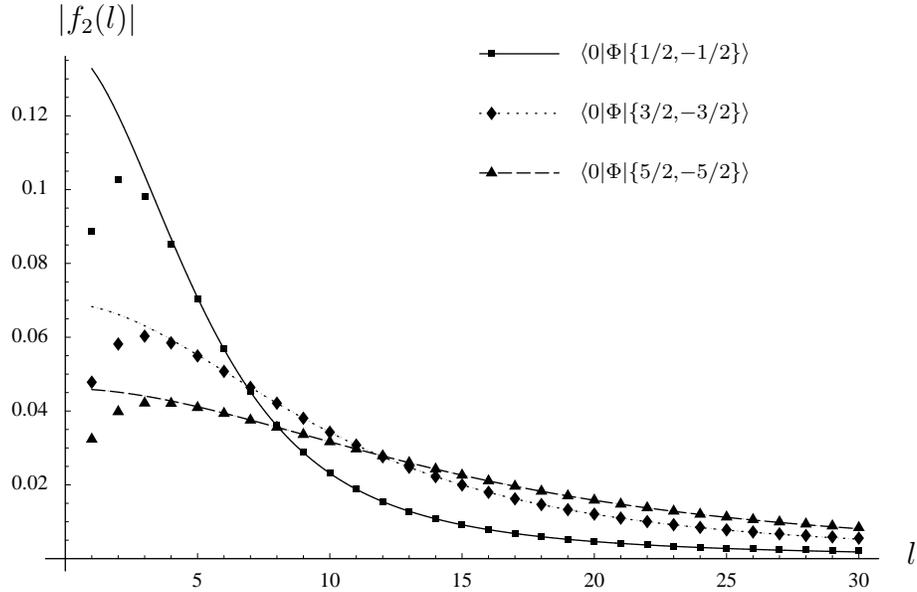}\par\end{centering}

\caption{\label{fig:lytps0}Two-particle form factors in the spin-$0$ sector.
Dots correspond to TCSA data, while the lines show the corresponding
form factor prediction.}
\end{figure}

\begin{figure}
\begin{centering}\psfrag{f2}{$|f_2(l)|$}\psfrag{l}{$l$}
\psfrag{tcsa13}{$\scriptstyle\langle 0|\Phi|\{ 3/2,-1/2\}\rangle$}
\psfrag{tcsa35}{$\scriptstyle\langle 0|\Phi|\{ 5/2,-3/2\}\rangle$}
\psfrag{tcsa15}{$\scriptstyle\langle 0|\Phi|\{ 5/2,-1/2\}\rangle$}
\psfrag{tcsa37}{$\scriptstyle\langle 0|\Phi|\{ 7/2,-3/2\}\rangle$}\includegraphics[scale=1.3]{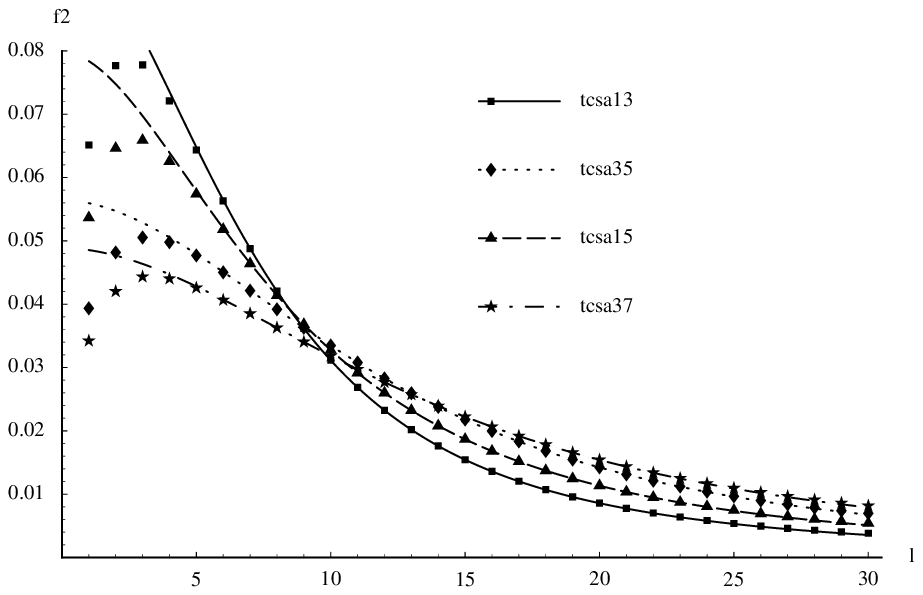}\par\end{centering}

\caption{\label{fig:lytps12}Two-particle form factors in the spin-$1$ and
spin-$2$ sectors. Dots correspond to TCSA data, while the lines show
the corresponding form factor prediction.}
\end{figure}

\begin{figure}
\begin{centering}\psfrag{logdevf2}{$\log |f_2^{FF}(l)-f_2^{TCSA}(l)|$}
\psfrag{l}{$l$}\includegraphics{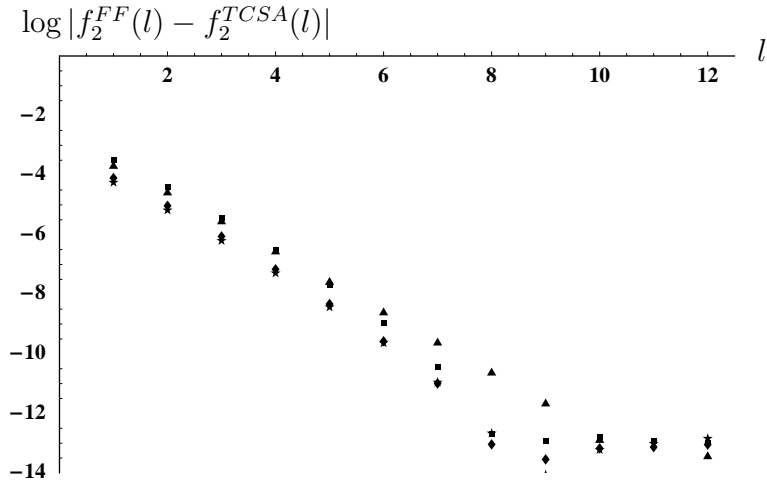}\par\end{centering}

\caption{\label{fig:errorterm}Estimating the error term in (\ref{eq:lyf2prediction})
using the data in table \ref{tab:lytps12}. The various plot symbols
correspond to the same states as specified in figure \ref{fig:lytps12}.}
\end{figure}

The prediction (\ref{eq:lyf2prediction}) for the finite volume two-particle
form factors is compared with spin-$0$ states graphically in figure
\ref{fig:lytps0} and numerically in table \ref{tab:lytps0}, while
the spin-$1$ and spin-$2$ case is presented in figure \ref{fig:lytps12}
and in table \ref{tab:lytps12}. These contain no more than a representative
sample of our data: we evaluated similar matrix elements for a large
number of two-particle states for values of the volume parameter $l$
running from $1$ to $30$. The behaviour of the relative deviation
is consistent with the presence of a correction of $e^{-l}$ type
up to $l\sim9\dots10$ (i.e. the logarithm of the deviation is very
close to being a linear function of $l$), and after $l\sim16\dots18$
it starts to increase due to truncation errors. This is demonstrated
in figure \ref{fig:errorterm} using the data presented in table \ref{tab:lytps12}
for spin-$1$ and spin-$2$ states%
\footnote{Note that the dependence of the logarithm of the deviation on the
volume is not exactly linear because the residual finite size correction
can also contain a factor of some power of $l$, and so it is expected
that a $\log l$ contribution is also present in the data plotted
in figure \ref{fig:errorterm}.%
}, but it is equally valid for all the other states we examined. In
the intermediate region $l\sim10\dots16$ the two sources of numerical
deviation are of the same order, and so that range can be considered
as the optimal scaling region: according to the data in the tables
agreement there is typically around $10^{-4}$ (relative deviation).
It is also apparent that scaling behaviour starts at quite low values
of the volume (around $l\sim4$ the relative deviation is already
down to around 1\%). 

It can be verified by explicit evaluation that in the scaling region
the Bethe-Yang density of states ($\rho$) given in (\ref{eq:byjacobian})
differs by corrections of relative magnitude $10^{-1}-10^{-2}$ (analytically:
of order $1/l$) from the free density of states ($\rho^{0}$) in
(\ref{eq:freejacobidet}), and therefore without using the proper
interacting density of states it is impossible to obtain the precision
agreement we demonstrated. In fact the observed $10^{-4}$ relative
deviation corresponds to corrections of order $l^{-4}$ at $l=10$,
but it is of the order of estimated truncation errors%
\footnote{Truncation errors can be estimated by examining the dependence of
the extracted data on the cutoff $e_{\mathrm{cut}}$, as well as by
comparing TCSA energy levels to the Bethe-Yang predictions.%
}. 

These results are very strong evidence for the main statement in (\ref{eq:lyf2prediction})
(and thus also (\ref{eq:ffrelation})), namely, that all $1/L$ corrections
are accounted by the proper interacting state density factor and that
all further finite size corrections are just residual finite size
effects decaying exponentially in $L$. In section \ref{sub:manypff}
we show that data from higher multi-particle form factors fully support
the above conclusions drawn from the two-particle form factors.

\begin{table}
\begin{centering}\begin{tabular}{|r||c|c||c|c||c|c|}
\hline 
&
\multicolumn{2}{c|}{$I_{1}=1/2$ , $I_{2}=-1/2$}&
\multicolumn{2}{c|}{$I_{1}=3/2$ , $I_{2}=-3/2$}&
\multicolumn{2}{c|}{$I_{1}=5/2$ , $I_{2}=-5/2$}\tabularnewline
\hline 
$l$&
TCSA&
FF&
TCSA&
FF&
TCSA&
FF\tabularnewline
\hline
\hline 
2&
0.102780&
0.120117&
0.058158&
0.066173&
0.039816&
0.045118\tabularnewline
\hline 
4&
0.085174&
0.086763&
0.058468&
0.059355&
0.042072&
0.042729\tabularnewline
\hline 
6&
0.056828&
0.056769&
0.050750&
0.050805&
0.039349&
0.039419\tabularnewline
\hline 
8&
0.036058&
0.035985&
0.042123&
0.042117&
0.035608&
0.035614\tabularnewline
\hline 
10&
0.023168&
0.023146&
0.034252&
0.034248&
0.031665&
0.031664\tabularnewline
\hline 
12&
0.015468&
0.015463&
0.027606&
0.027604&
0.027830&
0.027828\tabularnewline
\hline
14&
0.010801&
0.010800&
0.022228&
0.022225&
0.024271&
0.024267\tabularnewline
\hline
16&
0.007869&
0.007867&
0.017976&
0.017972&
0.021074&
0.021068\tabularnewline
\hline
18&
0.005950&
0.005945&
0.014652&
0.014645&
0.018268&
0.018258\tabularnewline
\hline
20&
0.004643&
0.004634&
0.012061&
0.012050&
0.015844&
0.015827\tabularnewline
\hline
\end{tabular}\par\end{centering}

\caption{\label{tab:lytps0}Two-particle form factors $\left|f_{2}(l)\right|$
in the spin-$0$ sector}
\end{table}

\begin{table}
\begin{centering}\begin{tabular}{|r||c|c||c|c||c|c||c|c|}
\hline 
&
\multicolumn{2}{c|}{$I_{1}=3/2$ , $I_{2}=-1/2$}&
\multicolumn{2}{c|}{$I_{1}=5/2$ , $I_{2}=-3/2$}&
\multicolumn{2}{c|}{$I_{1}=5/2$ , $I_{2}=-1/2$}&
\multicolumn{2}{c|}{$I_{1}=7/2$ , $I_{2}=-3/2$}\tabularnewline
\hline 
$l$&
TCSA&
FF&
TCSA&
FF&
TCSA&
FF&
TCSA&
FF\tabularnewline
\hline
\hline 
2&
0.077674&
0.089849&
0.048170&
0.054711&
0.064623&
0.074763&
0.042031&
0.047672\tabularnewline
\hline 
4&
0.072104&
0.073571&
0.049790&
0.050566&
0.062533&
0.063932&
0.044034&
0.044716\tabularnewline
\hline 
6&
0.056316&
0.056444&
0.045031&
0.045100&
0.051828&
0.052009&
0.040659&
0.040724\tabularnewline
\hline 
8&
0.042051&
0.042054&
0.039191&
0.039193&
0.041370&
0.041394&
0.036284&
0.036287\tabularnewline
\hline 
10&
0.031146&
0.031144&
0.033469&
0.033467&
0.032757&
0.032759&
0.031850&
0.031849\tabularnewline
\hline 
12&
0.023247&
0.023245&
0.028281&
0.028279&
0.026005&
0.026004&
0.027687&
0.027684\tabularnewline
\hline
14&
0.017619&
0.017616&
0.023780&
0.023777&
0.020802&
0.020799&
0.023941&
0.023936\tabularnewline
\hline
16&
0.013604&
0.013599&
0.019982&
0.019977&
0.016808&
0.016802&
0.020659&
0.020652\tabularnewline
\hline
18&
0.010717&
0.010702&
0.016831&
0.016822&
0.013735&
0.013724&
0.017835&
0.017824\tabularnewline
\hline
20&
0.008658&
0.008580&
0.014249&
0.014227&
0.011357&
0.011337&
0.015432&
0.015413\tabularnewline
\hline
\end{tabular}\par\end{centering}

\caption{\label{tab:lytps12}Two-particle form factors $\left|f_{2}(l)\right|$
in the spin-$1$ and spin-$2$ sectors}
\end{table}

\subsection{Ising model in magnetic field}

\begin{figure}
\noindent \begin{centering}\psfrag{f11}{$|\bar{f}_{11}|$}\psfrag{l}{$l$}
\psfrag{---a-0.5a0.5}{$\scriptstyle\langle 0|\Psi|\{1/2,-1/2\}\rangle_{11}$}
\psfrag{---a-1.5a1.5}{$\scriptstyle\langle 0|\Psi|\{3/2,-3/2\}\rangle_{11}$}
\psfrag{---a-0.5a1.5}{$\scriptstyle\langle 0|\Psi|\{3/2,-1/2\}\rangle_{11}$}\subfigure[$A_1A_1$]{\includegraphics[scale=1.2]{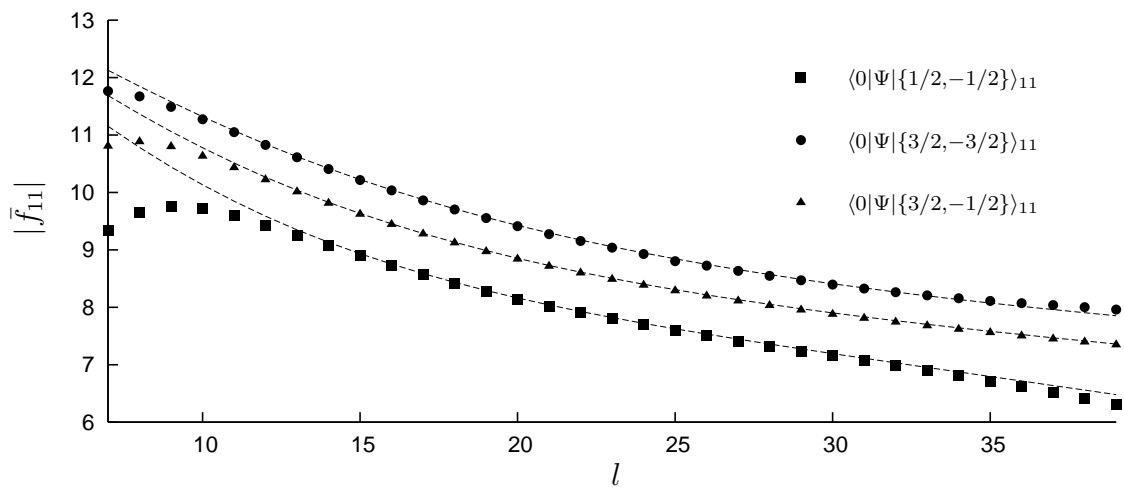}}\\
\psfrag{f12}{$|\bar{f}_{12}|$}\psfrag{l}{$l$}
\psfrag{---a-1b1}{$\scriptstyle\langle 0|\Psi|\{-1,1\}\rangle_{12}$}
\psfrag{---a0b1}{$\scriptstyle\langle 0|\Psi|\{0,1\}\rangle_{12}$}
\psfrag{---a1b1}{$\scriptstyle\langle 0|\Psi|\{1,1\}\rangle_{12}$}\subfigure[$A_1A_2$]{\includegraphics[scale=1.2]{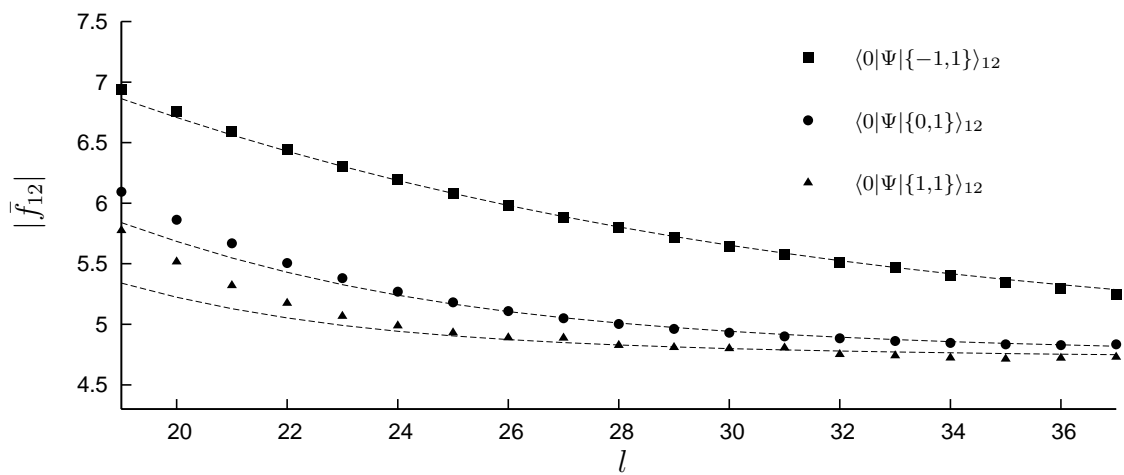}}\par\end{centering}

\caption{\label{fig:ising2pt} Two-particle form factors in the Ising model.
Dots correspond to TCSA data, while the lines show the corresponding
form factor prediction.}
\end{figure}

In this case, there is some further subtlety to be solved before proceeding
to the numerical comparison. Namely, there are spin-$0$ states which
are parity reflections of each other, but are degenerate according
to the Bethe-Yang equations. An example is the state $|\{1,-1\}\rangle_{12}$
in figure \ref{fig:ising2pt} (b), which is degenerate with $|\{-1,1\}\rangle_{12}$
to all orders in $1/L$. In general the degeneracy of these states
is lifted by residual finite size effects (more precisely by quantum
mechanical tunneling -- a detailed discussion of this mechanism was
given in the framework of the $k$-folded sine-Gordon model in \cite{takacs_extrapolation}).
Since the finite volume spectrum is parity symmetric, the TCSA eigenvectors
correspond to the states\[
|\{1,-1\}\rangle_{12,\, L}^{\pm}=\frac{1}{\sqrt{2}}\left(|\{1,-1\}\rangle_{12,\, L}\pm|\{-1,1\}\rangle_{12,\, L}\right)\]
Because the Hilbert space inner product is positive definite, the
TCSA eigenvectors $|\{1,-1\}\rangle_{12}^{\pm}$ can be chosen orthonormal
and the problem can be resolved by calculating the form factor matrix
element using the two-particle state vectors\[
\frac{1}{\sqrt{2}}\left(|\{1,-1\}\rangle_{12,\, L}^{+}\pm|\{1,-1\}\rangle_{12,\, L}^{-}\right)\]
Because the Ising spectrum is much more complicated than that of the
scaling Lee-Yang model (and truncation errors are larger as well),
we only identified two-particle states containing two copies of $A_{1}$
, or an $A_{1}$ and an $A_{2}$. The numerical results are plotted
in figures \ref{fig:ising2pt} (a) and (b), respectively. The finite
volume form factor functions of the operator $\Psi$ (\ref{eq:psidef})
are defined as\[
\bar{f}_{11}\left(l\right)_{I_{1}I_{2}}=\sqrt{\rho_{11}(\tilde{\theta}_{1}(l),\tilde{\theta}_{2}(l))}\langle0|\Psi|\{ I_{1},I_{2}\}\rangle_{11}\]
where\begin{eqnarray*}
 &  & l\sinh\tilde{\theta}_{1}+\delta_{11}(\tilde{\theta}_{1}-\tilde{\theta}_{2})=2\pi I_{1}\\
 &  & l\sinh\tilde{\theta}_{2}+\delta_{11}(\tilde{\theta}_{2}-\tilde{\theta}_{1})=2\pi I_{2}\\
 &  & \rho_{11}(\theta_{1},\theta_{2})=l^{2}\cosh\theta_{1}\cosh\theta_{2}+l\cosh\theta_{1}\varphi_{11}(\theta_{2}-\theta_{1})+l\cosh\theta_{2}\varphi_{11}(\theta_{1}-\theta_{2})\\
 &  & \varphi_{11}(\theta)=\frac{d\delta_{11}(\theta)}{d\theta}\end{eqnarray*}
and\[
\bar{f}_{12}\left(l\right)_{I_{1}I_{2}}=\sqrt{\rho_{11}(\tilde{\theta}_{1},\tilde{\theta}_{2})}\langle0|\Psi|\{ I_{1},I_{2}\}\rangle_{12}\]
with\begin{eqnarray*}
 &  & l\sinh\tilde{\theta}_{1}+\delta_{12}(\tilde{\theta}_{1}-\tilde{\theta}_{2})=2\pi I_{1}\\
 &  & \frac{m_{2}}{m_{1}}l\sinh\tilde{\theta}_{2}+\delta_{12}(\tilde{\theta}_{2}-\tilde{\theta}_{1})=2\pi I_{2}\\
 &  & \rho_{12}(\theta_{1},\theta_{2})=\frac{m_{2}}{m_{1}}l^{2}\cosh\theta_{1}\cosh\theta_{2}+l\cosh\theta_{1}\varphi_{12}(\theta_{2}-\theta_{1})+\frac{m_{2}}{m_{1}}l\cosh\theta_{2}\varphi_{12}(\theta_{1}-\theta_{2})\\
 &  & \varphi_{12}(\theta)=\frac{d\delta_{12}(\theta)}{d\theta}\end{eqnarray*}
and are compared against the form factor functions\[
F_{2}^{\Psi}(\tilde{\theta}_{1}(l),\tilde{\theta}_{2}(l))_{11}\]
and \[
F_{2}^{\Psi}(\tilde{\theta}_{1}(l),\tilde{\theta}_{2}(l))_{12}\]
respectively. 

Although (as we already noted) truncation errors in the Ising model
are much larger than in the Lee-Yang case, extrapolation in the cutoff
improves them by an order of magnitude compared to the evaluation
at the highest cutoff (in our case $30$). After extrapolation, deviations
in the scaling region become less than $1$\% (with a minimum of around
$10^{-3}$ in the $A_{1}A_{1}$, and $10^{-4}$ in the $A_{1}A_{2}$
case), and even better for states with nonzero total spin. As noted
in the previous subsection this means that the numerics is really
sensitive to the dependence of the particle rapidities and state density
factors on the interaction between the particles; generally the truncation
errors in the extrapolated data are about two orders of magnitude
smaller than the interaction corrections.

It is a general tendency that the agreement is better in the sectors
with nonzero spin, and the scaling region starts at smaller values
of the volume. This is easy to understand for the energy levels, since
for low-lying states nonzero spin generally means higher particle
momenta. The higher the momenta of the particles, the more the Bethe-Yang
contributions dominate over the residual finite size effects. This
is consistent with the results of Rummukainen and Gottlieb in \cite{Rummukainen:1995vs}
where it was found that resonance phase shifts can be more readily
extracted from sectors with nonzero momentum; our data show that this
observation carries over to general matrix elements as well. 

\section{\label{sub:manypff} Many-particle form factors}

\subsection{Scaling Lee-Yang model}

\begin{figure}
\begin{centering}\psfrag{f3}{$|\tilde{f}_3(l)|$}\psfrag{l}{$l$}
\psfrag{tcsa101}{$\scriptstyle\langle 0|\Phi|\{ 1,0,-1\}\rangle$}
\psfrag{tcsa202}{$\scriptstyle\langle 0|\Phi|\{ 2,0,-2\}\rangle$}
\psfrag{tcsa303}{$\scriptstyle\langle 0|\Phi|\{ 3,0,-3\}\rangle$}
\psfrag{tcsa312}{$\scriptstyle\langle 0|\Phi|\{ 3,-1,-2\}\rangle$}\includegraphics[scale=1.6]{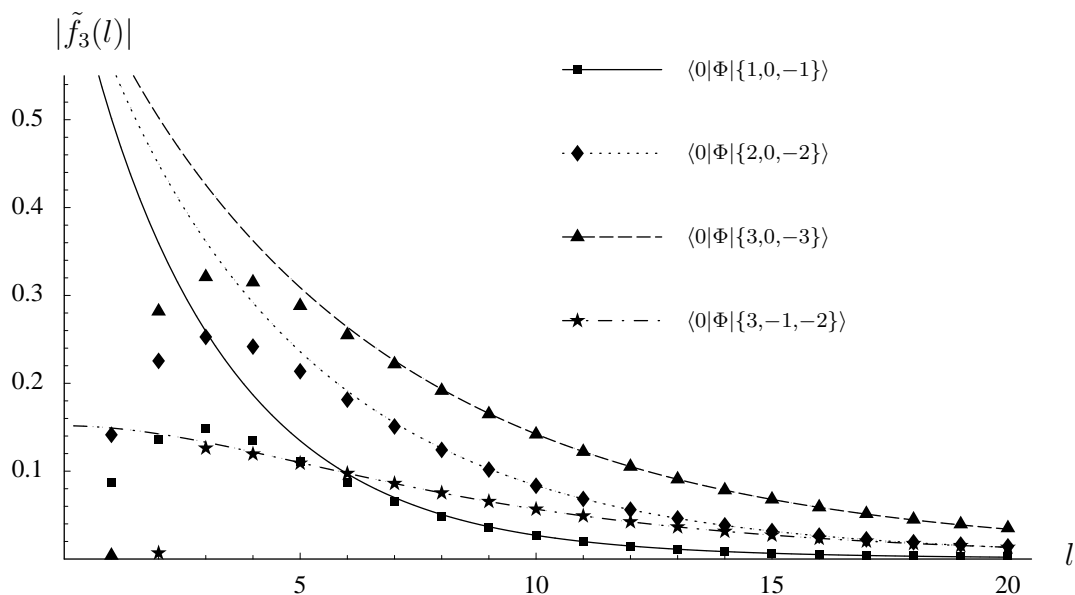}\par\end{centering}

\caption{\label{fig:lythps0}Three-particle form factors in the spin-$0$
sector. Dots correspond to TCSA data, while the lines show the corresponding
form factor prediction.}
\end{figure}

\begin{figure}
\begin{centering}\psfrag{f4}{$|\tilde{f}_4(l)|$}\psfrag{l}{$l$}
\psfrag{tcsa3113}{$\scriptstyle\langle 0|\Phi|\{ 3/2,1/2,-1/2,-3/2\}\rangle$}
\psfrag{tcsa5115}{$\scriptstyle\langle 0|\Phi|\{ 5/2,1/2,-1/2,-5/2\}\rangle$}
\psfrag{tcsa7117}{$\scriptstyle\langle 0|\Phi|\{ 7/2,1/2,-1/2,-7/2\}\rangle$}
\psfrag{tcsa7135}{$\scriptstyle\langle 0|\Phi|\{ 7/2,1/2,-3/2,-5/2\}\rangle$}\includegraphics[scale=1.6]{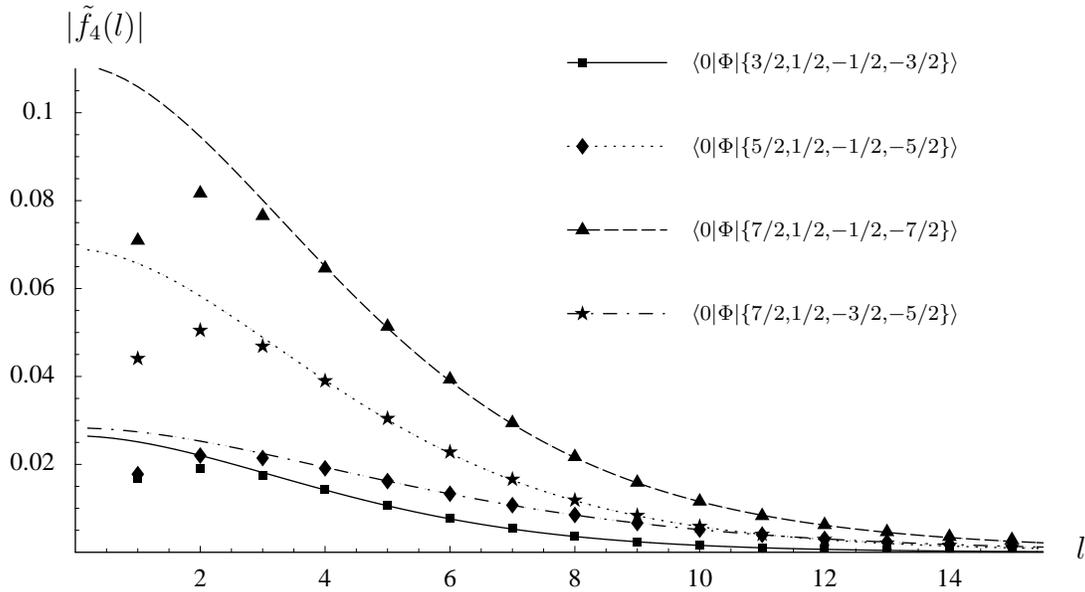}\par\end{centering}

\caption{\label{fig:lyfps0}Four-particle form factors in the spin-$0$ sector.
Dots correspond to TCSA data, while the lines show the corresponding
form factor prediction.}
\end{figure}

We also performed numerical evaluation of three and four-particle
form factors in the scaling Lee-Yang model; some of the results are
presented in figures \ref{fig:lythps0} and \ref{fig:lyfps0}, respectively.
For the sake of brevity we refrain from presenting explicit numerical
tables; we only mention that the agreement between the numerical TCSA
data and the prediction from the exact form factor solution is always
better than $10^{-3}$ in the scaling region. For better visibility
we plotted the functions\[
\tilde{f}_{k}(l)_{I_{1}\dots I_{k}}=-im^{2/5}\sqrt{\rho_{k}(\tilde{\theta}_{1},\dots,\tilde{\theta}_{k})}\langle0|\Phi|\{ I_{1},\dots,I_{k}\}\rangle_{L}\quad,\quad l=mL\]
for which relation (\ref{eq:ffrelation}) gives:\begin{equation}
\tilde{f}_{k}(l)_{I_{1}\dots I_{k}}=-im^{2/5}F_{k}^{\Phi}(\tilde{\theta}_{1},\dots,\tilde{\theta}_{k})+O(\mathrm{e}^{-l})\label{eq:lyfkprediction}\end{equation}
Due to the fact that in the Lee-Yang model there is only a single
particle species, we introduced the simplified notation $\rho_{n}$
for the $n$-particle Jacobi determinant.

The complication noted in subsection 3.2.2 for the Ising state $|\{1,-1\}\rangle_{12}$
is present in the Lee-Yang model as well. The Bethe-Yang equations
give degenerate energy values for the states $|\{ I_{1},\dots,I_{k}\}\rangle_{L}$
and $|\{-I_{k},\dots,-I_{1}\}\rangle_{L}$ (as noted before, the degeneracy
is lifted by quantum mechanical tunneling). For states with nonzero
spin this causes no problem, because these two states are in sectors
of different spin (their spins differ by a sign) and similarly there
is no difficulty when the two quantum number sets are identical, i.e.\[
\{ I_{1},\dots,I_{k}\}=\{-I_{1},\dots,-I_{k}\}\]
since then there is a single state. However, there are states in the
zero spin sector (i.e. with $\sum_{k}I_{k}=0$) for which \[
\{ I_{1},\dots,I_{k}\}\neq\{-I_{1},\dots,-I_{k}\}\]
We use two such pairs of states in our data here: the three-particle
states $|\{3,-1,-2\}\rangle_{L}$, $|\{2,1,-3\}\rangle_{L}$ and the
four-particle states $|\{7/2,1/2,-3/2,-5/2\}\rangle_{L}$, $|\{5/2,3/2,-1/2,-7/2\}\rangle_{L}$.
Again, the members of such pairs are related to each other by spatial
reflection, which is a symmetry of the exact finite-volume Hamiltonian
and therefore (supposing that the eigenvectors are orthonormal) the
finite volume eigenstates correspond to \[
|\{ I_{1},\dots,I_{k}\}\rangle_{L}^{\pm}=\frac{1}{\sqrt{2}}\left(|\{ I_{1},\dots,I_{k}\}\rangle_{L}\pm|\{-I_{k},\dots,-I_{1}\}\rangle_{L}\right)\]
and this must be taken into account when evaluating the form factor
matrix elements. In the Lee-Yang case, however, the inner product
is not positive definite (and some nonzero vectors may have zero {}``length'',
although this does not happen for TCSA eigenvectors, because they
are orthogonal to each other and the inner product is non-degenerate),
but there is a simple procedure that can be used in the general case.
Suppose the two TCSA eigenvectors corresponding to such a pair are
$v_{1}$ and $v_{2}$. Then we can define their inner product matrix
as \[
g_{ij}=v_{i}G^{(0)}v_{j}\]
using the TCSA inner product (\ref{eq:Gs}). The appropriate basis
vectors of this two-dimensional subspace, which can be identified
with $|\{ I_{1},\dots,I_{k}\}\rangle_{L}$ and $|\{-I_{k},\dots,-I_{1}\}\rangle_{L}$,
can be found by solving the two-dimensional generalized eigenvalue
problem\[
g\cdot w=\lambda P\cdot w\]
for the vector $(w_{1},w_{2})$ describing orientation in the subspace,
with \[
P=\left(\begin{array}{cc}
0 & 1\\
1 & 0\end{array}\right)\]
This procedure has the effect of rotating from the basis of parity
eigenvectors to basis vectors which are taken into each other by spatial
reflection.


\subsection{Ising model in a magnetic field}

\begin{figure}
\noindent \begin{centering}\psfrag{f111}{$|\tilde{f}_{111}|$}\psfrag{l}{$l$}
\psfrag{---a0a0a0}{$\scriptstyle\langle 0|\Psi|\{1,0,-1\}\rangle_{111}$}
\psfrag{---a-1a0a2}{$\scriptstyle\langle 0|\Psi|\{2,0,-1\}\rangle_{111}$}
\psfrag{---a0a1a2}{$\scriptstyle\langle 0|\Psi|\{2,1,0\}\rangle_{111}$}\includegraphics[scale=1.3]{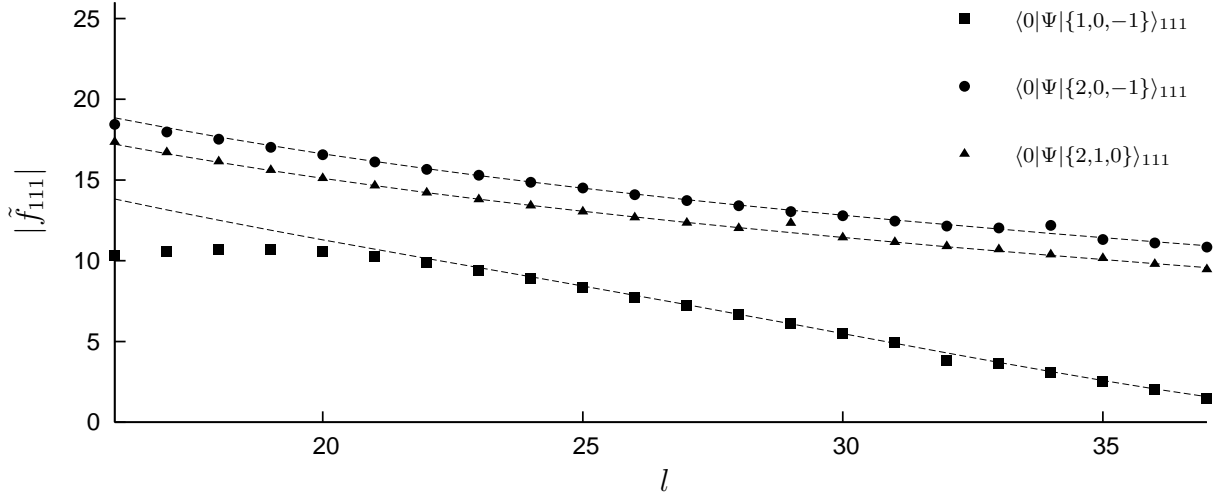}\par\end{centering}

\caption{\label{fig:ising3pt} Three-particle form factors in the Ising model.
Dots correspond to TCSA data, while the lines show the corresponding
form factor prediction.}
\end{figure}

As we already noted, it is much harder to identify%
\footnote{To identify $A_{1}A_{1}A_{1}$ states it is necessary to use at least
$e_{\mathrm{cut}}=22$ or $24$ and even then the agreement with the
Bethe-Yang prediction is still only within $20\%$, but the identification
can be made for the first few $A_{1}A_{1}A_{1}$ states using data
up to $e_{\mathrm{cut}}=30$. Truncation errors are substantially
decreased by extrapolation to $e_{\mathrm{cut}}=\infty$. %
} higher states in the Ising model due to the complexity of the spectrum,
and so we only performed an analysis of states containing three $A_{1}$
particles. We define\[
\tilde{f}_{111}\left(l\right)_{I_{1}I_{2}I_{3}}=\sqrt{\rho_{111}(\tilde{\theta}_{1}(l),\tilde{\theta}_{2}(l),\tilde{\theta}_{3}(l))}\langle0|\Psi|\{ I_{1},I_{2},I_{3}\}\rangle_{111}\]
where $\tilde{\theta}_{i}(l)$ are the solutions of the three-particle
Bethe-Yang equations in (dimensionless) volume $l$ and $\rho_{111}$
is the appropriate $3$-particle determinant. The results of the comparison
can be seen in figure \ref{fig:ising3pt}. The numerical precision
indicated for two-particle form factors at the end of subsection 3.2.2,
as well as the remarks made there on the spin dependence apply here
as well; we only wish to emphasize that for $A_{1}A_{1}A_{1}$ states
with nonzero total spin the agreement between the extrapolated TCSA
data and the form factor prediction in the optimal part of the scaling
region is within $2\times10^{-4}$. 

\section{General form factors without disconnected pieces}

\label{numerics:general}

\subsection{Scaling Lee-Yang model}

In this model there is a single particle species, so we can introduce
the following notations:\[
f_{kn}(l)_{I_{1},\dots,I_{n}}^{I_{1}',\dots,I_{k}'}=-im^{2/5}\langle\{ I_{1}',\dots,I_{k}'\}\vert\Phi(0,0)\vert\{ I_{1},\dots,I_{n}\}\rangle_{L}\]
and also \[
\tilde{f}_{kn}(l)_{I_{1},\dots,I_{n}}^{I_{1}',\dots,I_{k}'}=-im^{2/5}\sqrt{\rho_{k}(\tilde{\theta}_{1}',\dots,\tilde{\theta}_{k}')}\sqrt{\rho_{n}(\tilde{\theta}_{1},\dots,\tilde{\theta}_{n})}\langle\{ I_{1}',\dots,I_{k}'\}\vert\Phi(0,0)\vert\{ I_{1},\dots,I_{n}\}\rangle_{L}\]
for which relation (\ref{eq:genffrelation}) yields\begin{eqnarray}
f_{kn}(l)_{I_{1},\dots,I_{n}}^{I_{1}',\dots,I_{k}'} & = & -im^{2/5}\frac{F_{k+n}^{\Phi}(\tilde{\theta}_{k}'+i\pi,\dots,\tilde{\theta}_{1}'+i\pi,\tilde{\theta}_{1},\dots,\tilde{\theta}_{n})}{\sqrt{\rho_{n}(\tilde{\theta}_{1},\dots,\tilde{\theta}_{n})\rho_{k}(\tilde{\theta}_{1}',\dots,\tilde{\theta}_{m}')}}+O(\mathrm{e}^{-l})\nonumber \\
\tilde{f}_{kn}(l)_{I_{1},\dots,I_{n}}^{I_{1}',\dots,I_{k}'} & = & -im^{2/5}F_{k+n}^{\Phi}(\tilde{\theta}_{k}'+i\pi,\dots,\tilde{\theta}_{1}'+i\pi,\tilde{\theta}_{1},\dots,\tilde{\theta}_{n})+O(\mathrm{e}^{-l})\label{eq:fmnrels}\end{eqnarray}
For the plots we chose to display $f$ or $\tilde{f}$ depending on
which one gives a better visual picture. The numerical results shown
here are just a fraction of the ones we actually obtained, but all
of them show an agreement with precision $10^{-4}-10^{-3}$ in the
scaling region (the volume range corresponding to the scaling region
typically varies depending on the matrix element considered due to
variation in the residual finite size corrections and truncation effects).

\begin{figure}
\noindent \begin{centering}\psfrag{f11}{$|\tilde{f}_{11}(l)|$}\psfrag{l}{$l$}
\psfrag{tcsa01}{$\scriptstyle\langle\{ 0\}|\Phi|\{ 1\}\rangle$}
\psfrag{tcsa02}{$\scriptstyle\langle\{ 0\}|\Phi|\{ 2\}\rangle$}
\psfrag{tcsa1m1}{$\scriptstyle\langle\{ -1\}|\Phi|\{ 1\}\rangle$}
\psfrag{tcsa12}{$\scriptstyle\langle\{ 1\}|\Phi|\{ 2\}\rangle$}
\psfrag{tcsam12}{$\scriptstyle\langle\{ -1\}|\Phi|\{ 2\}\rangle$}
\includegraphics[scale=1.4]{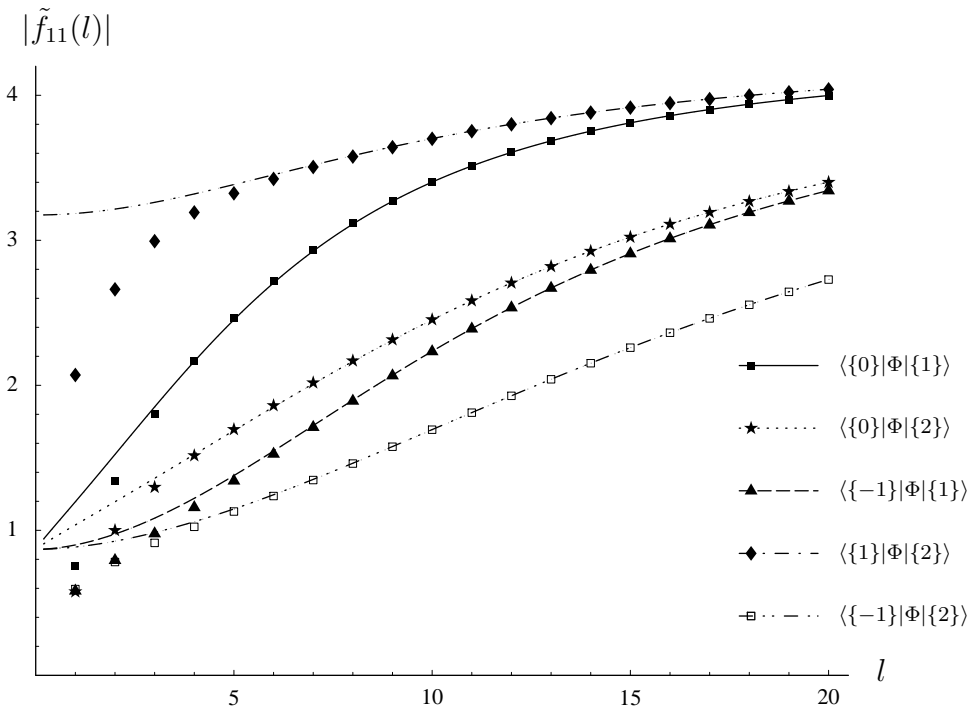}\par\end{centering}

\caption{\label{fig:ly1r1r} One-particle--one-particle form factors in Lee-Yang
model. Dots correspond to TCSA data, while the lines show the corresponding
form factor prediction.}
\end{figure}

\begin{figure}
\noindent \begin{raggedright}\psfrag{f12}{$|{f}_{12}(l)|$}\psfrag{l}{$l$}
\psfrag{tcsa01m1}{$\scriptstyle\langle\{ 0\}|\Phi|\{ 1/2,-1/2\}\rangle$}
\psfrag{tcsa03m3}{$\scriptstyle\langle\{ 0\}|\Phi|\{ 3/2,-3/2\}\rangle$}
\psfrag{tcsa01m3}{$\scriptstyle\langle\{ 0\}|\Phi|\{ 1/2,-3/2\}\rangle$}
\psfrag{tcsa21m3}{$\scriptstyle\langle\{ 2\}|\Phi|\{ 1/2,-3/2\}\rangle$}

\includegraphics[scale=1.4]{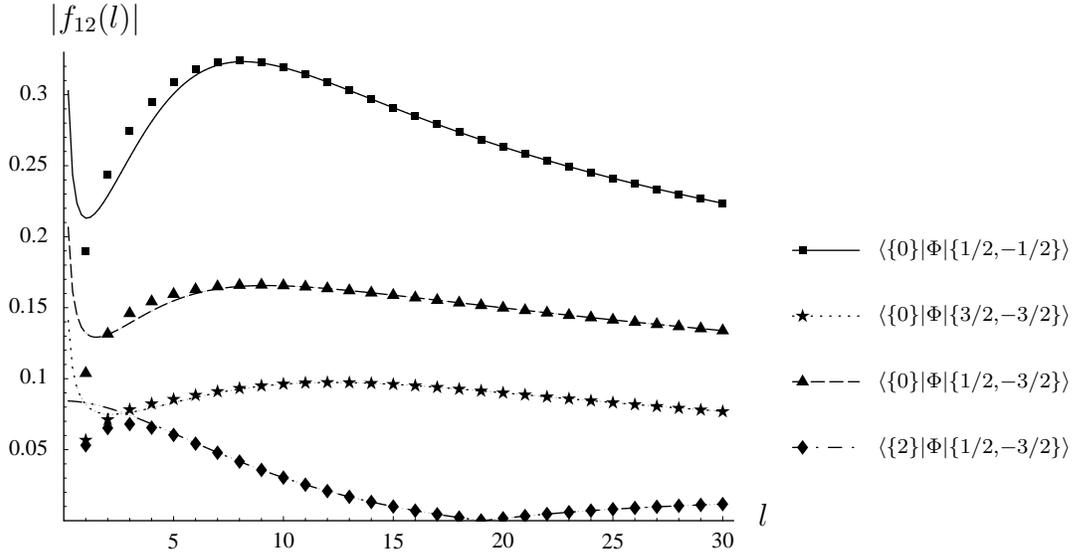}\par\end{raggedright}

\caption{\label{fig:ly1r2r} One-particle--two-particle form factors in Lee-Yang
model. Dots correspond to TCSA data, while the lines show the corresponding
form factor prediction.}
\end{figure}

The simplest cases involve one and two-particle states: the one-particle--one-particle
data in figure \ref{fig:ly1r1r} actually test the two-particle form
factor $F_{2}^{\Phi}$, while the one-particle--two-particle plot
\ref{fig:ly1r2r} corresponds to $F_{3}^{\Phi}$ (we obtained similar
results on $F_{4}^{\Phi}$ using matrix elements $f_{22}$). Note
that in contrast to the comparisons performed in subsections 3.2 and
3.3, these cases involve the form factor solutions (\ref{eq:lyff})
at complex values of the rapidities. In general, all tests performed
with TCSA can test form factors at rapidity arguments with imaginary
parts $0$ or $\pi$, which are the only parts of the complex rapidity
plane where form factors eventually correspond to physical matrix
elements. 


\begin{figure}
\noindent \begin{raggedright}\psfrag{f13}{$|{f}_{13}(l)|$}\psfrag{l}{$l$}
\psfrag{tcsa110m1}{$\scriptstyle\langle\{ 1\}|\Phi|\{ 1,0,-1\}\rangle$}
\psfrag{tcsa120m2}{$\scriptstyle\langle\{ 1\}|\Phi|\{ 2,0,-2\}\rangle$}
\psfrag{tcsa03m1m2}{$\scriptstyle\langle\{ 0\}|\Phi|\{ 3,-1,-2\}\rangle$}
\psfrag{tcsa13m1m2}{$\scriptstyle\langle\{ 1\}|\Phi|\{ 3,-1,-2\}\rangle$}
\psfrag{tcsa1m312}{$\scriptstyle\langle\{ 1\}|\Phi|\{ -3,1,2\}\rangle$}
\includegraphics[scale=1.4]{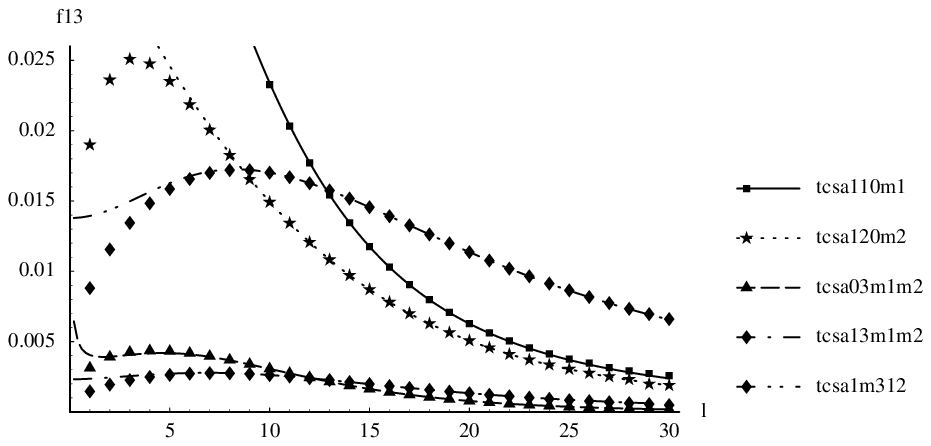}\par\end{raggedright}

\caption{\label{fig:ly1r3r} One-particle--three-particle form factors in
Lee-Yang model. Dots correspond to TCSA data, while the lines show
the corresponding form factor prediction.}
\end{figure}

\begin{figure}
\noindent \begin{raggedright}\psfrag{f14}{$|{f}_{14}(l)|$}\psfrag{l}{$l$}
\psfrag{tcsa01}{$\scriptstyle\langle\{ 0\}|\Phi|\{ 3/2,1/2,-1/2,-3/2\}\rangle$}
\psfrag{tcsa02}{$\scriptstyle\langle\{ 0\}|\Phi|\{ 5/2,1/2,-1/2,-5/2\}\rangle$}
\psfrag{tcsa12}{$\scriptstyle\langle\{ 1\}|\Phi|\{ 3/2,1/2,-1/2,-3/2\}\rangle$}
\psfrag{tcsa14}{$\scriptstyle\langle\{ 1\}|\Phi|\{ 5/2,3/2,-1/2,-7/2\}\rangle$}
\psfrag{tcsa15}{$\scriptstyle\langle\{ 1\}|\Phi|\{ 7/2,1/2,-3/2,-5/2\}\rangle$}
\includegraphics[scale=1.4]{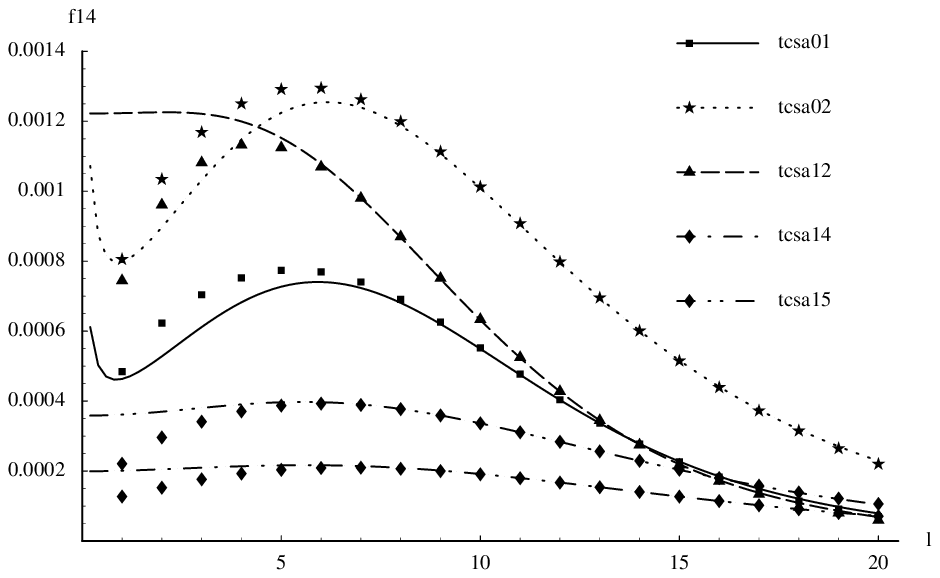}\par\end{raggedright}

\caption{\label{fig:ly1r4r} One-particle--four-particle form factors in Lee-Yang
model. Dots correspond to TCSA data, while the lines show the corresponding
form factor prediction.}
\end{figure}

\begin{figure}
\noindent \begin{raggedright}\psfrag{f23}{$|{f}_{23}(l)|$}\psfrag{l}{$l$}
\psfrag{tcsa11}{$\scriptstyle\langle\{ 1/2,-1/2\}|\Phi|\{ 1,0,-1\}\rangle$}
\psfrag{tcsa12}{$\scriptstyle\langle\{ 3/2,-3/2\}|\Phi|\{ 1,0,-1\}\rangle$}
\psfrag{tcsa21}{$\scriptstyle\langle\{ 1/2,-1/2\}|\Phi|\{ 2,0,-2\}\rangle$}
\psfrag{tcsas14}{$\scriptstyle\langle\{ 3/2,-1/2\}|\Phi|\{ 3,-1,-2\}\rangle$}
\psfrag{tcsas15}{$\scriptstyle\langle\{ 3/2,-1/2\}|\Phi|\{ -3,1,2\}\rangle$}
\includegraphics[scale=1.4]{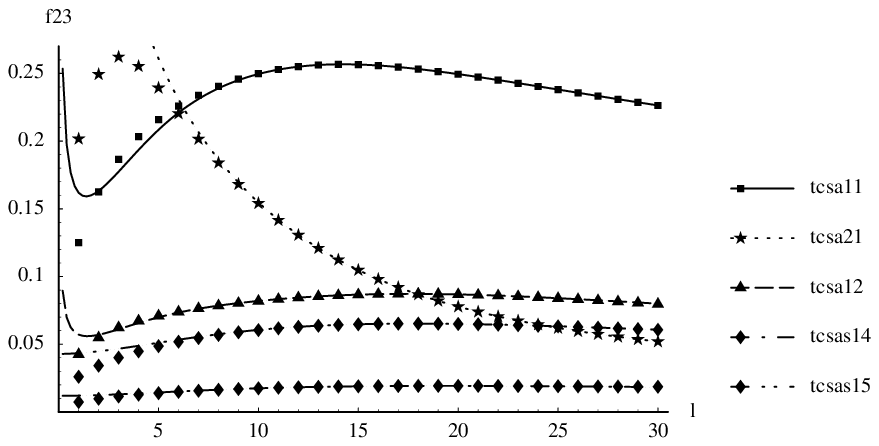}\par\end{raggedright}

\caption{\label{fig:ly2r3r} Two-particle--three-particle form factors in
Lee-Yang model. Dots correspond to TCSA data, while the lines show
the corresponding form factor prediction.}
\end{figure}

One-particle--three-particle and one-particle--four-particle matrix
elements $f_{13}$ and $f_{14}$ contribute another piece of useful
information. We recall that there are pairs of parity-related states
in the spin-0 factors which we cannot distinguish in terms of their
elementary form factors. In subsection 3.3 we showed the example of
the three-particle states\[
|\{3,-1,-2\}\rangle_{L}\mbox{ and }|\{2,1,-3\}\rangle_{L}\]
and the four-particle states\[
|\{7/2,1/2,-3/2,-5/2\}\rangle_{L}\mbox{ and }|\{5/2,3/2,-1/2,-7/2\}\rangle_{L}\]
In fact it is only true that they cannot be distinguished if the left
state is parity-invariant. However, using a one-particle state of
nonzero spin on the left it is possible to distinguish and appropriately
label the two states, as shown in figures \ref{fig:ly1r3r} and \ref{fig:ly1r4r}.
This can also be done using matrix elements with two-particle states
of nonzero spin: the two-particle--three-particle case $f_{23}$ is
shown in \ref{fig:ly2r3r} (similar results were obtained for $f_{24}$).
Examining the data in detail shows that the identifications provided
using different states on the left are all consistent with each other. 


\begin{figure}
\noindent \begin{raggedright}\psfrag{f3344}{$|{f}_{33}(l)|$ or $|{f}_{44}(l)|$}\psfrag{l}{$l$}
\psfrag{tcsa3314}{$\scriptstyle\langle\{ 1,0,-1\}|\Phi|\{ 3,-1,-2\}\rangle$}
\psfrag{tcsa3324}{$\scriptstyle\langle\{ 1,0,-1\}|\Phi|\{ 3,-1,-2\}\rangle$}
\psfrag{tcsa3334}{$\scriptstyle\langle\{ 1,0,-1\}|\Phi|\{ 3,-1,-2\}\rangle$}
\psfrag{tcsas4412}{$\scriptstyle\langle\{ \frac{\scriptstyle 3}{\scriptstyle 2},\frac{\scriptstyle 1}{\scriptstyle 2},-\frac{\scriptstyle 1}{\scriptstyle 2},-\frac{\scriptstyle 3}{\scriptstyle 2}\}|\Phi|\{\frac{\scriptstyle 5}{\scriptstyle 2},\frac{\scriptstyle 1}{\scriptstyle 2},-\frac{\scriptstyle 1}{\scriptstyle 2},-\frac{\scriptstyle 5}{\scriptstyle 2}\}\rangle$}
\psfrag{tcsas4413}{$\scriptstyle\langle\{ \frac{\scriptstyle 3}{\scriptstyle 2},\frac{\scriptstyle 1}{\scriptstyle 2},-\frac{\scriptstyle 1}{\scriptstyle 2},-\frac{\scriptstyle 3}{\scriptstyle 2}\}|\Phi|\{ \frac{\scriptstyle 7}{\scriptstyle 2},\frac{\scriptstyle 1}{\scriptstyle 2},-\frac{\scriptstyle 1}{\scriptstyle 2},-\frac{\scriptstyle 7}{\scriptstyle 2}\}\rangle$}
\includegraphics[scale=1.4]{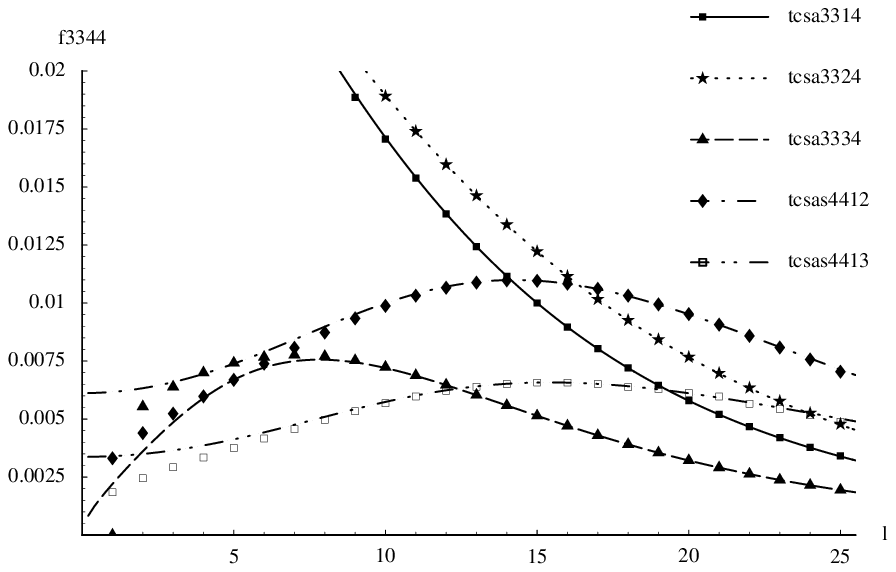}\par\end{raggedright}

\caption{\label{fig:ly3r3rand4r4r} Three-particle--three-particle and four-particle--four-particle
form factors in Lee-Yang model. Dots correspond to TCSA data, while
the lines show the corresponding form factor prediction.}
\end{figure}

\begin{figure}
\noindent \begin{raggedright}\psfrag{f34}{$|{f}_{34}(l)|$}\psfrag{l}{$l$}
\psfrag{tcsa11}{$\scriptstyle\langle\{ 1,0,-1\}|\Phi|\{ \frac{\scriptstyle 3}{\scriptstyle 2},\frac{\scriptstyle 1}{\scriptstyle 2},-\frac{\scriptstyle 1}{\scriptstyle 2},-\frac{\scriptstyle 3}{\scriptstyle 2}\}\rangle$}
\psfrag{tcsa12}{$\scriptstyle\langle\{ 1,0,-1\}|\Phi|\{ \frac{\scriptstyle 5}{\scriptstyle 2},\frac{\scriptstyle 1}{\scriptstyle 2},-\frac{\scriptstyle 1}{\scriptstyle 2},-\frac{\scriptstyle 5}{\scriptstyle 2}\}\rangle$}
\psfrag{tcsa22}{$\scriptstyle\langle\{ 2,0,-2\}|\Phi|\{ \frac{\scriptstyle 5}{\scriptstyle 2},\frac{\scriptstyle 1}{\scriptstyle 2},-\frac{\scriptstyle 1}{\scriptstyle 2},-\frac{\scriptstyle 5}{\scriptstyle 2}\}\rangle$}
\psfrag{tcsa44}{$\scriptstyle\langle\{ 3,-1,-2\}|\Phi|\{\frac{\scriptstyle 7}{\scriptstyle 2},\frac{\scriptstyle 1}{\scriptstyle 2},-\frac{\scriptstyle 3}{\scriptstyle 2},-\frac{\scriptstyle 5}{\scriptstyle 2}\}\rangle$}
\psfrag{tcsa45}{$\scriptstyle\langle\{ 2,1,-3\}|\Phi|\{ \frac{\scriptstyle 7}{\scriptstyle 2},\frac{\scriptstyle 1}{\scriptstyle 2},-\frac{\scriptstyle 3}{\scriptstyle 2},-\frac{\scriptstyle 5}{\scriptstyle 2}\}\rangle$}
\includegraphics[scale=1.4]{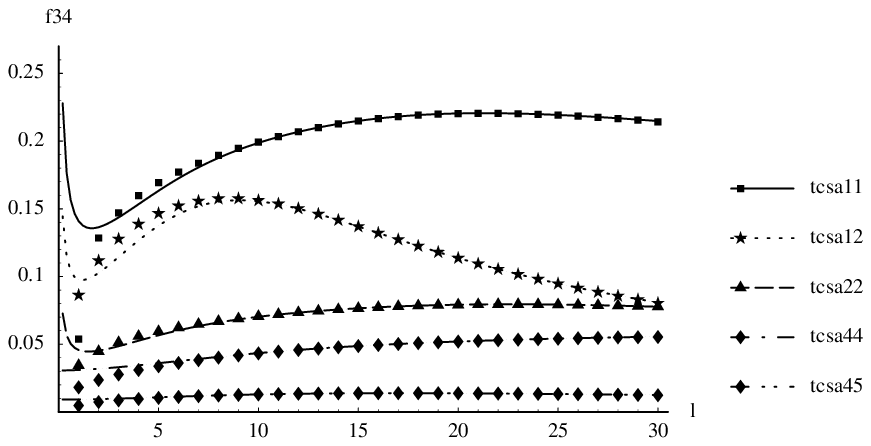}\par\end{raggedright}

\caption{\label{fig:ly3r4r} Three-particle--four-particle form factors in
Lee-Yang model. Dots correspond to TCSA data, while the lines show
the corresponding form factor prediction.}
\end{figure}

It is also interesting to note that the $f_{14}$ (figure \ref{fig:ly1r4r})
and $f_{23}$ data (figure \ref{fig:ly2r3r}) provide a test for the
five-particle form factor solutions $F_{5}$. This is important since
it is progressively harder to identify many-particle states in the
TCSA spectrum for two reasons. First, the spectrum itself becomes
more and more dense as we look for higher levels; second, the truncation
errors grow as well. Both of these make the identification of the
energy levels by comparison with the predictions of the Bethe-Yang
equations more difficult; in the Lee-Yang case we stopped at four-particle
levels. However, using general matrix elements and the relations (\ref{eq:fmnrels})
we can even get data for form factors up to $8$ particles, a sample
of which is shown in figures \ref{fig:ly3r3rand4r4r} ($f_{33}$ and
$f_{44}$, corresponding to $6$ and $8$ particle form-factors) and
\ref{fig:ly3r4r} ($f_{34}$ which corresponds to $7$ particle form
factors).


\subsection{Ising model in magnetic field}

In the case of the Ising model, we define the functions\[
\tilde{f}_{j_{1}\dots j_{m};i_{1}\dots i_{n}}(l)=\sqrt{\rho_{i_{1}\dots i_{n}}(\tilde{\theta}_{1},\dots,\tilde{\theta}_{n})\rho_{j_{1}\dots j_{m}}(\tilde{\theta}_{1}',\dots,\tilde{\theta}_{m}')}\times\,_{j_{1}\dots j_{m}}\langle\{ I_{1}',\dots,I_{m}'\}\vert\Psi\vert\{ I_{1},\dots,I_{n}\}\rangle_{i_{1}\dots i_{n},L}\]
which are compared against form factors\[
F_{m+n}^{\Psi}(\tilde{\theta}_{m}'+i\pi,\dots,\tilde{\theta}_{1}'+i\pi,\tilde{\theta}_{1},\dots,\tilde{\theta}_{n})_{j_{m}\dots j_{1}i_{1}\dots i_{n}}\]
where $\tilde{\theta}_{i}$ and $\tilde{\theta}_{j}'$ denote the
rapidities obtained as solutions of the appropriate Bethe-Yang equations
at the given value of the volume. We chose states for which the necessary
form factor solution was already known (and given in \cite{isingff})
i.e. we did not construct new form factor solutions ourselves. %
\begin{figure}
\noindent \begin{centering}\psfrag{f11}{$|\tilde{f}_{1;1}|$}\psfrag{l}{$l$}
\psfrag{a-1---a1}{$\scriptstyle{}_{1}\langle \{-1\}|\Psi|\{1\}\rangle_{1}$}
\psfrag{a-2---a2}{$\scriptstyle{}_{1}\langle \{-2\}|\Psi|\{2\}\rangle_{1}$}
\psfrag{a-3---a3}{$\scriptstyle{}_{1}\langle \{-3\}|\Psi|\{3\}\rangle_{1}$}\subfigure[$A_1-A_1$ matrix elements]{\includegraphics[scale=1.2]{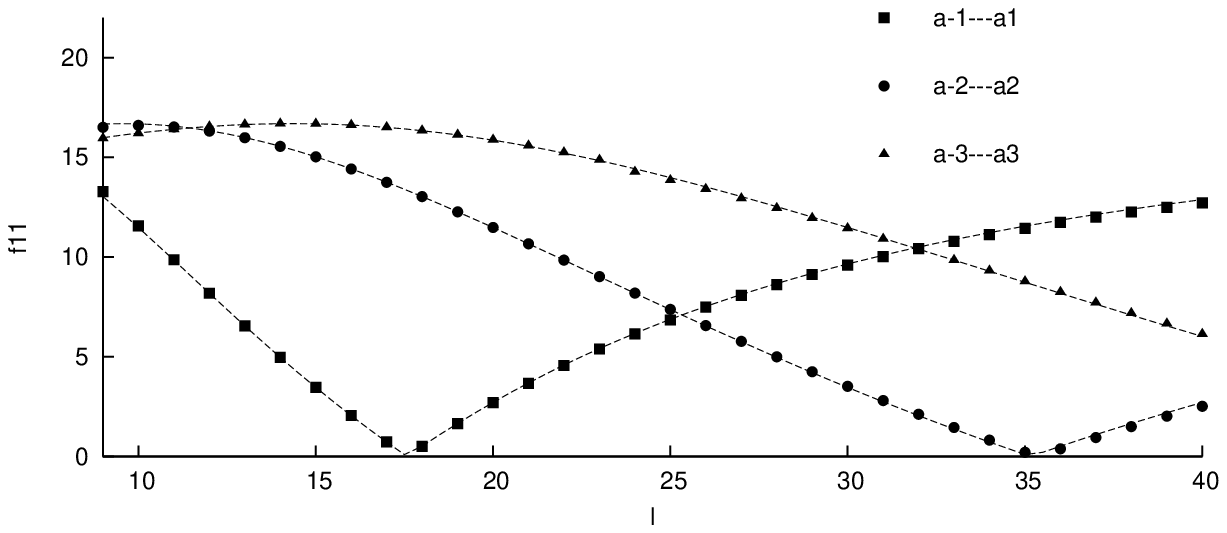}}\\
\psfrag{f12}{$|\tilde{f}_{1;2}|$}\psfrag{l}{$l$}
\psfrag{a0---b0}{$\scriptstyle{}_{1}\langle \{0\}|\Psi|\{0\}\rangle_{2}$}
\psfrag{a-1---b2}{$\scriptstyle{}_{1}\langle \{-1\}|\Psi|\{2\}\rangle_{2}$}
\psfrag{a-2---b2}{$\scriptstyle{}_{1}\langle \{-2\}|\Psi|\{2\}\rangle_{2}$}\subfigure[$A_1-A_2$ matrix elements]{\includegraphics[scale=1.2]{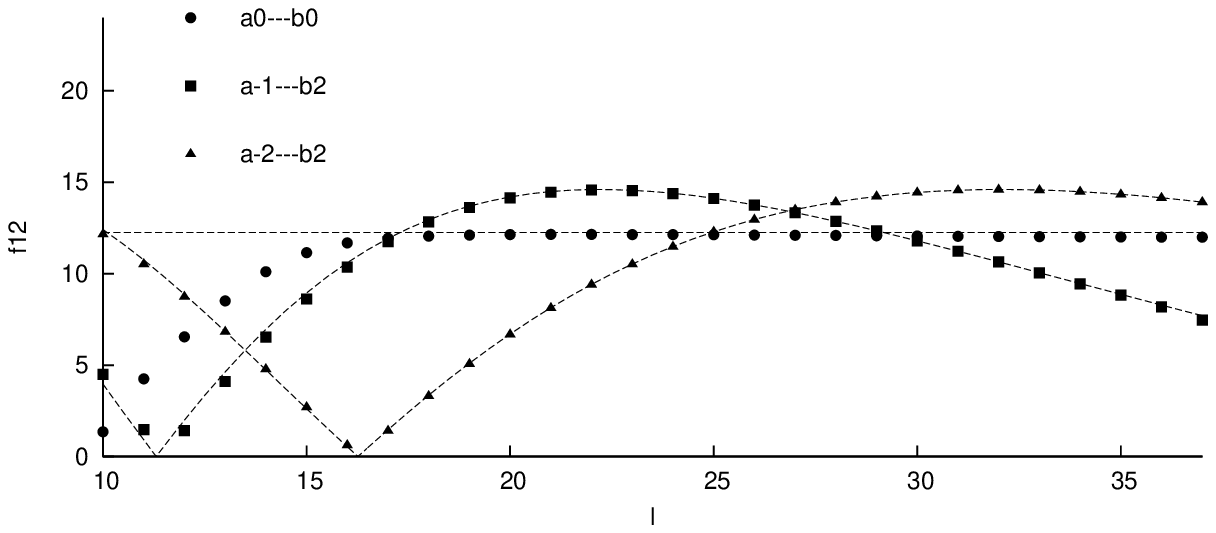}}\par\end{centering}

\caption{\label{fig:ising1p1p} One-particle--one-particle form factors in
the Ising model. Dots correspond to TCSA data, while the lines show
the corresponding form factor prediction.}
\end{figure}

\begin{figure}
\noindent \begin{centering}\psfrag{f111}{$|\tilde{f}_{1;11}|$}\psfrag{l}{$l$}
\psfrag{a0---a-0.5a0.5}{$\scriptstyle{}_{1}\langle \{0\}|\Psi|\{1/2,-1/2\}\rangle_{11}$}
\psfrag{a1---a-0.5a0.5}{$\scriptstyle{}_{1}\langle \{1\}|\Psi|\{1/2,-1/2\}\rangle_{11}$}
\psfrag{a2---a-0.5a0.5}{$\scriptstyle{}_{1}\langle \{2\}|\Psi|\{1/2,-1/2\}\rangle_{11}$}
\psfrag{a-1---a-0.5a1.5}{$\scriptstyle{}_{1}\langle \{-1\}|\Psi|\{3/2,-1/2\}\rangle_{11}$}\subfigure[$A_1-A_1A_1$ matrix elements]{\includegraphics[scale=1.3]{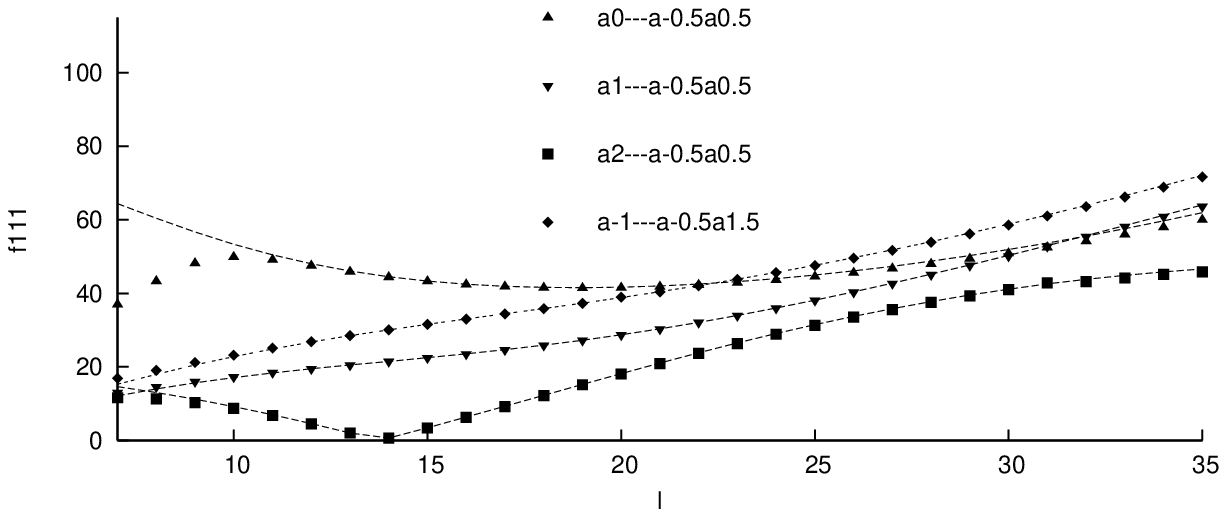}}\\
\psfrag{f112}{$|\tilde{f}_{1;12}|$}\psfrag{l}{$l$}
\psfrag{a0---a0b1}{$\scriptstyle{}_{1}\langle \{0\}|\Psi|\{1,0\}\rangle_{21}$}
\psfrag{a-1--a0b1}{$\scriptstyle{}_{1}\langle \{-1\}|\Psi|\{1,0\}\rangle_{21}$}
\psfrag{a0---a-1b1}{$\scriptstyle{}_{1}\langle \{0\}|\Psi|\{-1,1\}\rangle_{12}$}\subfigure[$A_1-A_1A_2$ matrix elements]{\includegraphics[scale=1.3]{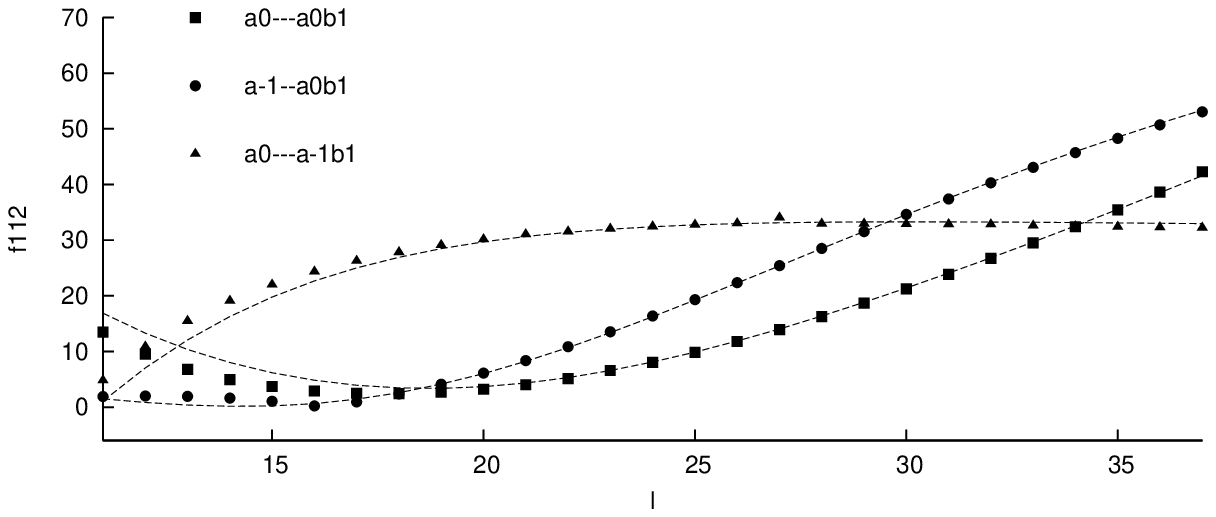}}\par\end{centering}

\caption{\label{fig:ising1p2p} One-particle--two-particle form factors in
the Ising model. Dots correspond to TCSA data, while the lines show
the corresponding form factor prediction.}
\end{figure}

One-particle--one-particle form factors are shown in figure \ref{fig:ising1p1p};
these provide another numerical test for the two-particle form factors
examined previously in subsection 3.2.2. One-particle--two-particle
form factors, besides testing again the three-particle form factor
$A_{1}A_{1}A_{1}$ (figure \ref{fig:ising1p2p} (a)) also provide
information on $A_{1}A_{1}A_{2}$ (figure \ref{fig:ising1p2p} (b)). 

\begin{figure}
\noindent \begin{centering}\psfrag{f1111}{$|\tilde{f}_{1;111}|$}\psfrag{l}{$l$}
\psfrag{a0---a-1a0a2}{$\scriptstyle{}_{1}\langle \{0\}|\Psi|\{2,0,-1\}\rangle_{111}$}
\psfrag{a-1---a-1a0a2}{$\scriptstyle{}_{1}\langle \{-1\}|\Psi|\{2,0,-1\}\rangle_{111}$}
\psfrag{a-1---a-1a1a2}{$\scriptstyle{}_{1}\langle \{-1\}|\Psi|\{2,1,-1\}\rangle_{111}$}\subfigure[$A_1-A_1A_1A_1$ matrix elements]{\includegraphics[scale=1.2]{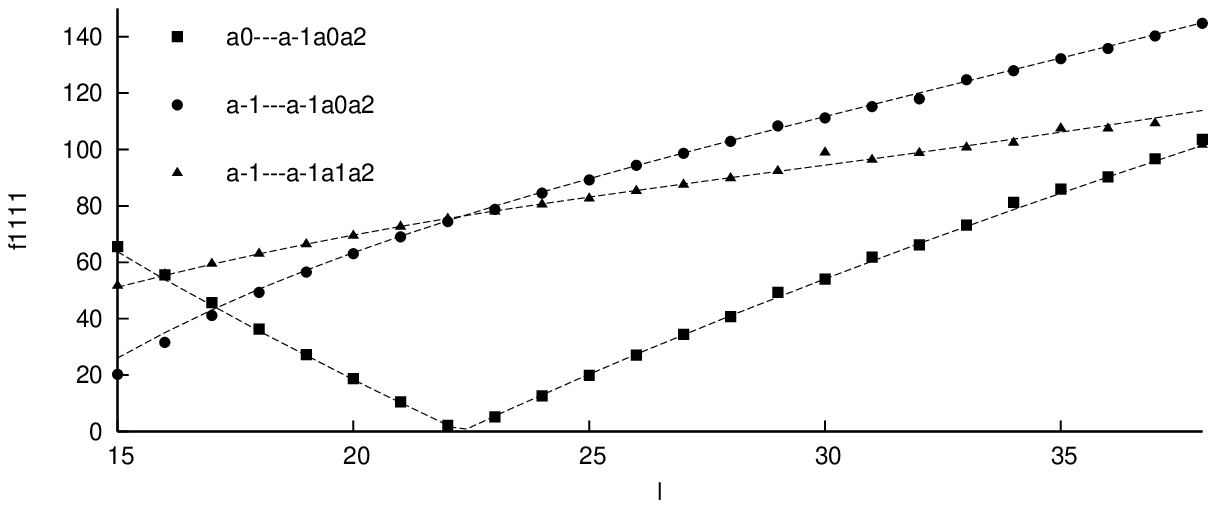}}\\
\psfrag{f1111}{$|\tilde{f}_{11;11}|$}\psfrag{l}{$l$}
\psfrag{a-1.5a0.5---a-0.5a1.5}{$\scriptstyle{}_{11}\langle \{1/2,-3/2\}|\Psi|\{3/2,-1/2\}\rangle_{11}$}
\psfrag{a-0.5a-1.5---a0.5a1.5}{$\scriptstyle{}_{11}\langle \{-1/2,-3/2\}|\Psi|\{3/2,1/2\}\rangle_{11}$}
\psfrag{a-0.5a1.5---a0.5a1.5}{$\scriptstyle{}_{11}\langle \{-1/2,3/2\}|\Psi|\{3/2,1/2\}\rangle_{11}$}\subfigure[$A_1A_1-A_1A_1$ matrix elements]{\includegraphics[scale=1.2]{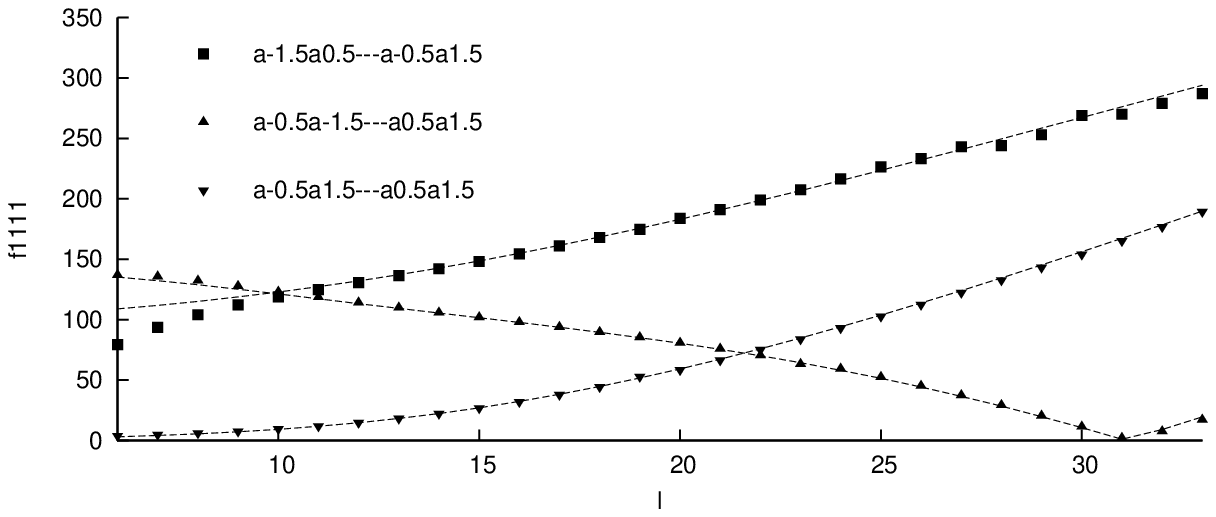}}\par\end{centering}

\caption{\label{fig:ising1p3pand2p2p}One-particle--three-particle and two-particle--two-particle
form factors in the Ising model. Dots correspond to TCSA data, while
the lines show the corresponding form factor prediction.}
\end{figure}

Finally, one-particle--three-particle and two-particle--two-particle
matrix elements can be compared to the $A_{1}A_{1}A_{1}A_{1}$ form
factor, which again shows that by considering general matrix elements
we can go substantially higher in the form factor tree than using
only elementary form factors. 

We remark that the cusps on the horizontal axis in the form factor
plots correspond to zeros where the form factors change sign; they
are artifacts introduced by taking the absolute value of the matrix
elements. The pattern of numerical deviations between TCSA data and
exact form factor predictions is fully consistent with the discussion
in the closing paragraphs of subsections 3.2.2 and 3.3.2. The deviations
in the scaling region are around $1$\% on average, with agreement
of the order of $10^{-3}$ in the optimal range. 

\vspace{5cm}

\section{Diagonal matrix elements}

\label{numerics:diagonal}

Here we compare the predictions of section \ref{diagonal} for the diagonal form
factors to the numerical data obtained from TCSA. 

\subsection{Diagonal one-particle and two-particle form factors}

\begin{figure}
\begin{centering}\psfrag{fd1}{$f_{11}$}
\psfrag{l}{$l$}
\psfrag{tcsa0}{$\langle\{0\}|\Phi|\{0\}\rangle$}
\psfrag{tcsa1}{$\langle\{1\}|\Phi|\{1\}\rangle$}
\psfrag{tcsa2}{$\langle\{2\}|\Phi|\{2\}\rangle$}\includegraphics[scale=1.2]{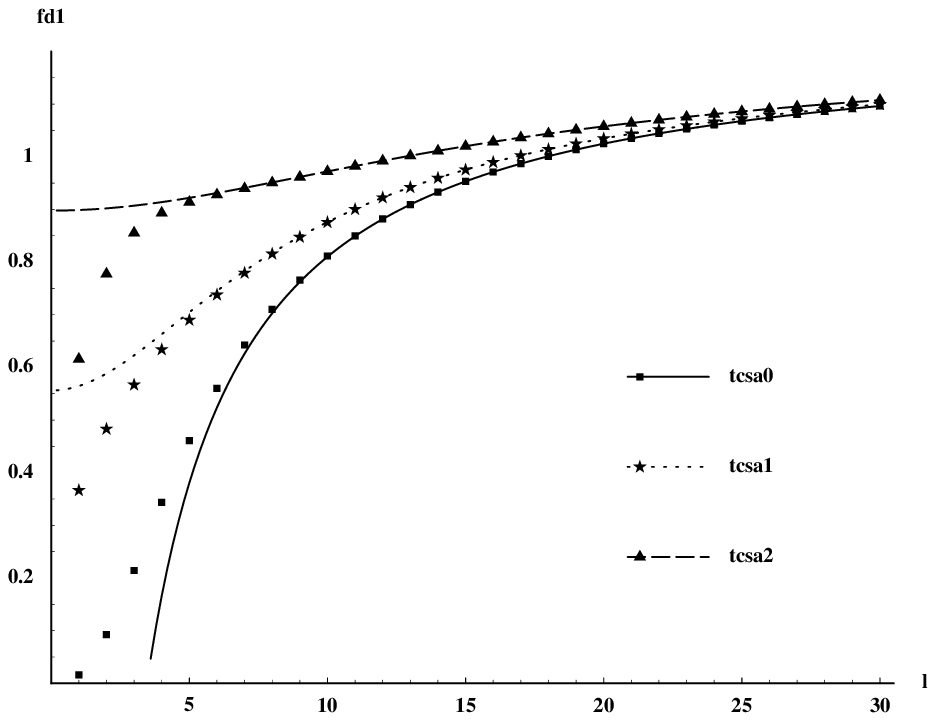}\par\end{centering}

\caption{\label{fig:d1ly}Diagonal $1$-particle matrix elements in the scaling
Lee-Yang model. The discrete points correspond to the TCSA data, while
the continuous line corresponds to the prediction from exact form
factors.}
\end{figure}

\begin{figure}
\noindent \begin{centering}\psfrag{l}{$l$}
\psfrag{ffff11}{$f_{1,1}$}
\psfrag{a0---a0}{${}_{1}\langle\{0\}|\Psi|\{0\}\rangle_{1}$}
\psfrag{a1---a1}{${}_{1}\langle\{1\}|\Psi|\{1\}\rangle_{1}$}
\psfrag{a2---a2}{${}_{1}\langle\{2\}|\Psi|\{2\}\rangle_{1}$}
\psfrag{a3---a3}{${}_{1}\langle\{3\}|\Psi|\{3\}\rangle_{1}$}\subfigure[$A_1$--$A_1$]{\includegraphics[scale=1.2]{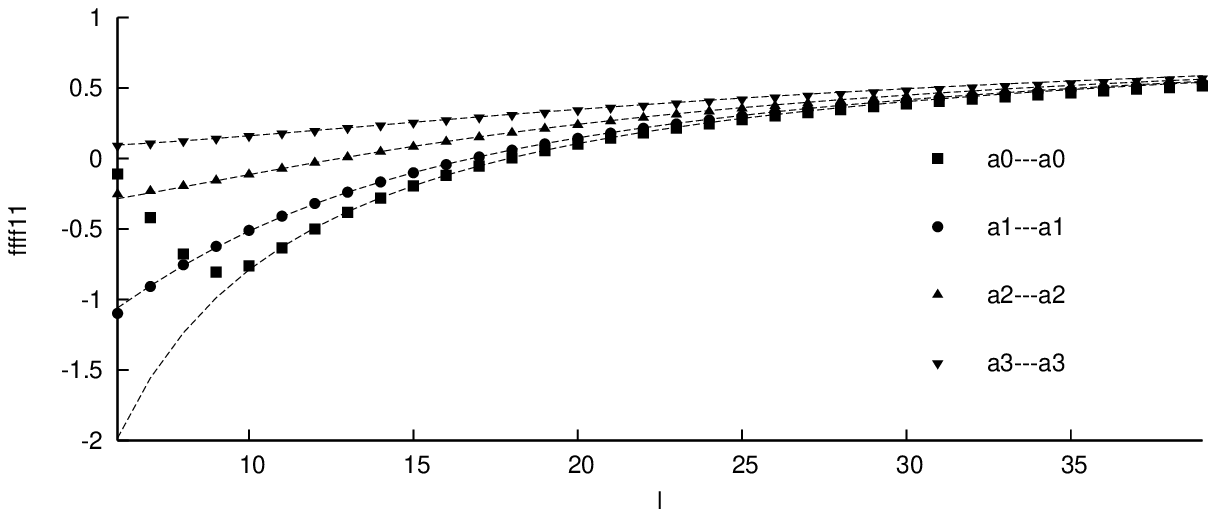}}\\
\psfrag{l}{$l$}
\psfrag{ffff22}{$f_{2,2}$}
\psfrag{b0---b0}{${}_{2}\langle\{0\}|\Psi|\{0\}\rangle_{2}$}
\psfrag{b1---b1}{${}_{2}\langle\{1\}|\Psi|\{1\}\rangle_{2}$}
\psfrag{b2---b2}{${}_{2}\langle\{2\}|\Psi|\{2\}\rangle_{2}$}
\psfrag{b3---b3}{${}_{2}\langle\{3\}|\Psi|\{3\}\rangle_{2}$}\subfigure[$A_2$--$A_2$]{\includegraphics[scale=1.2]{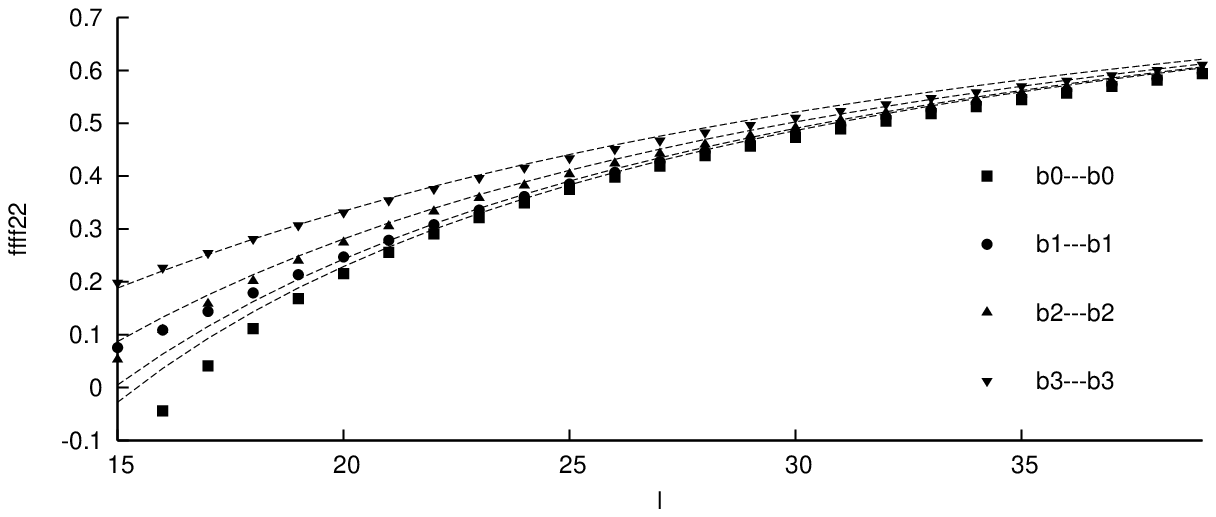}}\par\end{centering}

\caption{\label{fig:d1ising}Diagonal 1-particle matrix elements in the Ising
model. The discrete points correspond to the TCSA data, while the
continuous line corresponds to the prediction from exact form factors.}
\end{figure}

Figure (\ref{fig:d1ly}) shows the comparison of eqn. (\ref{eq:d1formula})
to numerical data obtained from Lee-Yang TCSA: the matching is spectacular,
especially in the so-called scaling region (the volume range where
residual finite size corrections are of the order of truncation errors,
cf. \cite{fftcsa1}) where the relative deviation is less than $10^{-4}$.
Diagonal one-particle matrix elements
for the Ising model are shown in figure \ref{fig:d1ising}.

\begin{figure}
\begin{centering}\psfrag{fd2}{$f_{22}$}
\psfrag{l}{$l$}
\psfrag{tcsa11}{$\langle\{\frac12,-\frac12\}|\Phi|\{\frac12,-\frac12\}\rangle$}
\psfrag{tcsa33}{$\langle\{\frac12,-\frac12\}|\Phi|\{\frac32,-\frac32\}\rangle$}
\psfrag{tcsa31}{$\langle\{\frac32,-\frac12\}|\Phi|\{\frac32,-\frac12\}\rangle$}
\psfrag{tcsa53}{$\langle\{\frac52,-\frac32\}|\Phi|\{\frac52,-\frac32\}\rangle$}\includegraphics[scale=1.2]{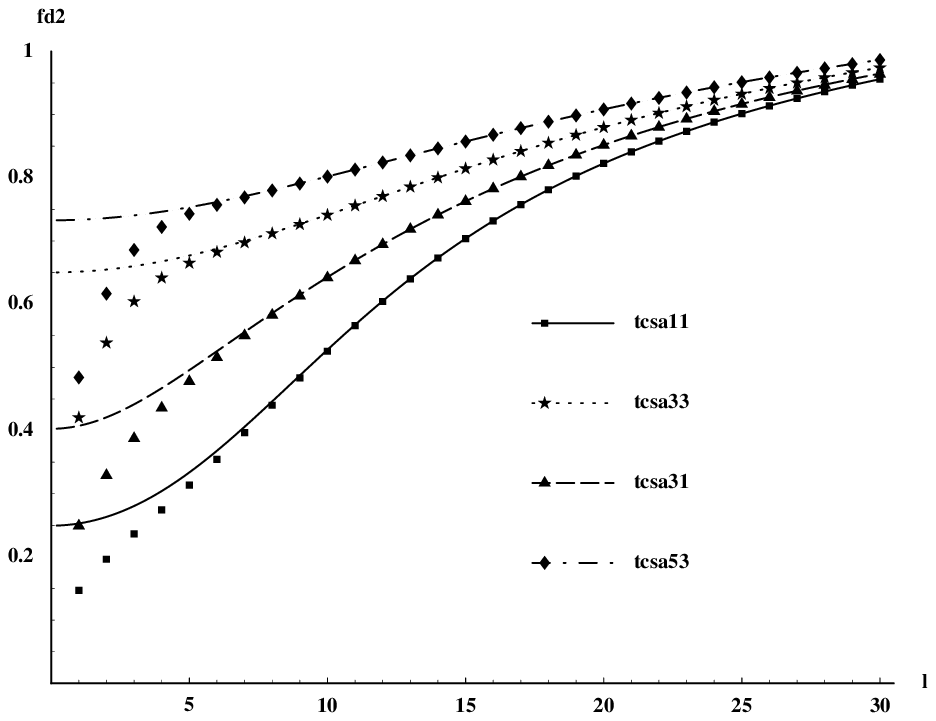}\par\end{centering}

\caption{\label{fig:d2ly}Diagonal $2$-particle matrix elements in the scaling
Lee-Yang model. The discrete points correspond to the TCSA data, while
the continuous line corresponds to the prediction from exact form
factors.}
\end{figure}

\begin{figure}
\noindent \begin{centering}\psfrag{l}{$l$}
\psfrag{ff1111}{$f_{11,11}$}
\psfrag{a-0.5a0.5---a-0.5a0.5}{${}_{11}\langle\{1/2,-1/2\}|\Psi|\{1/2,-1/2\}\rangle_{11}$}
\psfrag{a-0.5a1.5---a-0.5a1.5}{${}_{11}\langle\{3/2,-1/2\}|\Psi|\{3/2,-1/2\}\rangle_{11}$}
\psfrag{a0.5a1.5---a0.5a1.5}{${}_{11}\langle\{3/2,1/2\}|\Psi|\{3/2,1/2\}\rangle_{11}$}
\psfrag{a0.5a2.5---a0.5a2.5}{${}_{11}\langle\{5/2,1/2\}|\Psi|\{5/2,1/2\}\rangle_{11}$}\includegraphics[scale=1.1]{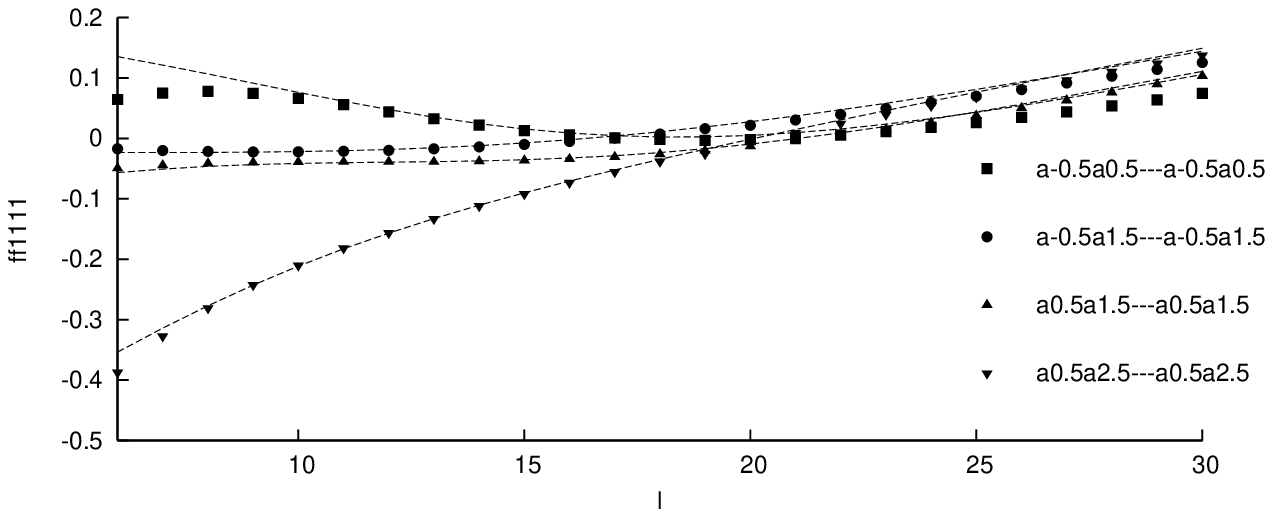}\par\end{centering}

\caption{\label{fig:d2ising}Diagonal 2-particle matrix elements in the Ising
model. The discrete points correspond to the TCSA data, while the
continuous line corresponds to the prediction from exact form factors.}
\end{figure}

Formula \eqref{eq:d2formula} describing diagonal two-particle matrix elements 
is tested against numerical data in the Lee-Yang model in figure \ref{fig:d2ly},
and the agreement is as precise as it was for the one-particle case.
Similar results can be found in the Ising case; they are shown in
figure \ref{fig:d2ising}.

\subsection{The general result}

\begin{figure}
\begin{centering}\psfrag{fd3}{$f_{33}$}
\psfrag{l}{$l$}
\psfrag{tcsa1}{$\langle\{1,0,-1\}|\Phi|\{1,0,-1\}\rangle$}
\psfrag{tcsa2}{$\langle\{2,0,-2\}|\Phi|\{2,0,-2\}\rangle$}
\psfrag{tcsa3}{$\langle\{3,0,-3\}|\Phi|\{3,0,-3\}\rangle$}
\psfrag{tcsa4}{$\langle\{3,-1,-2\}|\Phi|\{3,-1,-2\}\rangle$}\includegraphics[scale=1.2]{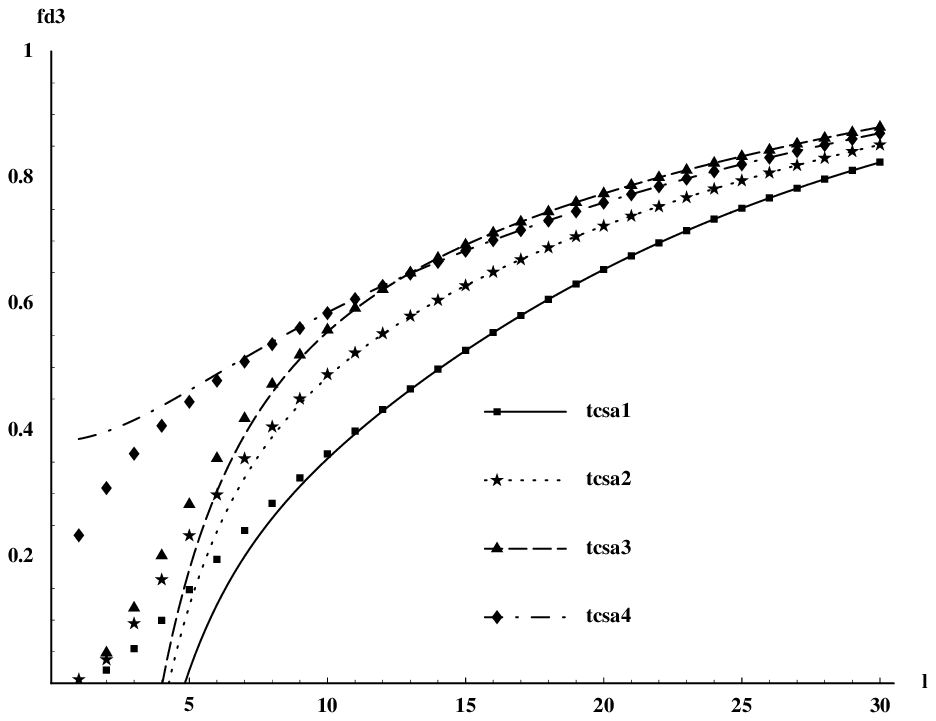}\par\end{centering}

\caption{\label{fig:d3ly}Diagonal $3$-particle matrix elements in the scaling
Lee-Yang model. The discrete points correspond to the TCSA data, while
the continuous line corresponds to the prediction from exact form
factors.}
\end{figure}

\begin{figure}
\begin{centering}\psfrag{fd4}{$f_{44}$}
\psfrag{l}{$l$}
\psfrag{tcsa1}{$\langle\{\frac32,\frac12,-\frac12,-\frac32\}|\Phi|\{\frac32,\frac12,-\frac12,-\frac32\}\rangle$}
\psfrag{tcsa2}{$\langle\{\frac52,\frac12,-\frac12,-\frac52\}|\Phi|\{\frac52,\frac12,-\frac12,-\frac52\}\rangle$}
\psfrag{tcsa3}{$\langle\{\frac72,\frac12,-\frac12,-\frac72\}|\Phi|\{\frac72,\frac12,-\frac12,-\frac72\}\rangle$}
\psfrag{tcsa4}{$\langle\{\frac72,\frac12,-\frac32,-\frac52\}|\Phi|\{\frac72,\frac12,-\frac32,-\frac52\}\rangle$}\includegraphics[scale=1.2]{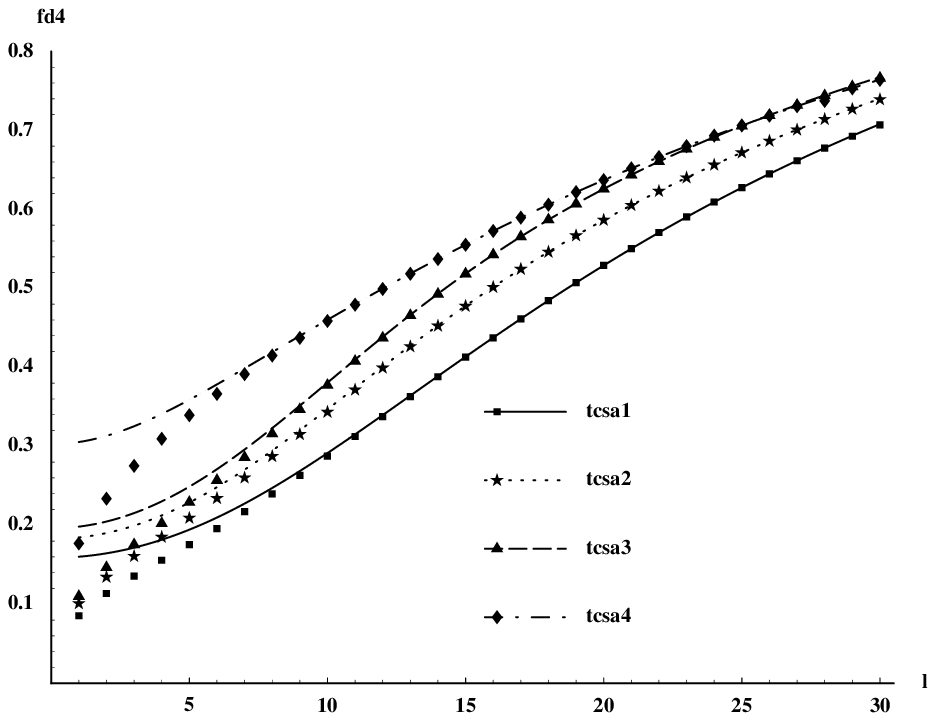}\par\end{centering}

\caption{\label{fig:d4ly}Diagonal $4$-particle matrix elements in the scaling
Lee-Yang model. The discrete points correspond to the TCSA data, while
the continuous line corresponds to the prediction from exact form
factors.}
\end{figure}

Formula \ref{eq:diaggenrule} can be tested against matrix elements with $n=3$ and $n=4$
in the Lee-Yang model, which are displayed in figures \ref{fig:d3ly}
and \ref{fig:d4ly}, respectively. The agreement is excellent as before,
with the relative deviation in the scaling region being of the order
of $10^{-4}$.


\section{Zero-momentum particles}

\label{numerics:zero}

\begin{figure}
\begin{centering}\psfrag{fd13}{$|f_{13}|$}
\psfrag{l}{$l$}
\psfrag{tcsa1}{$\langle\{0\}|\Phi|\{1,0,-1\}\rangle$}
\psfrag{tcsa2}{$\langle\{0\}|\Phi|\{2,0,-2\}\rangle$}
\psfrag{tcsa3}{$\langle\{0\}|\Phi|\{3,0,-3\}\rangle$}\includegraphics[scale=1.2]{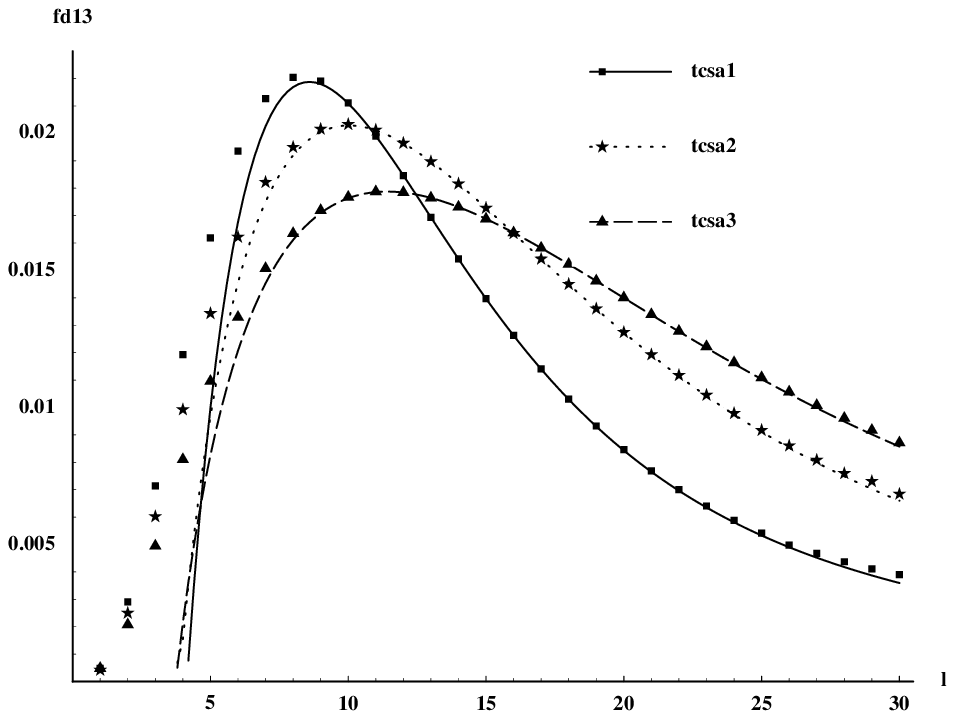}\par\end{centering}

\caption{\label{fig:fd13ly}$1$-particle--$3$-particle matrix elements in
the scaling Lee-Yang model. The discrete points correspond to the
TCSA data, while the continuous line corresponds to the prediction
from exact form factors.}
\end{figure}
\begin{figure}
\begin{centering}\psfrag{fd33}{$|f_{33}|$}
\psfrag{l}{$l$}
\psfrag{tcsa1}{$\langle\{1,0,-1\}|\Phi|\{2,0,-2\}\rangle$}
\psfrag{tcsa2}{$\langle\{1,0,-1\}|\Phi|\{3,0,-3\}\rangle$}
\psfrag{tcsa3}{$\langle\{2,0,-2\}|\Phi|\{3,0,-3\}\rangle$}\includegraphics[scale=1.2]{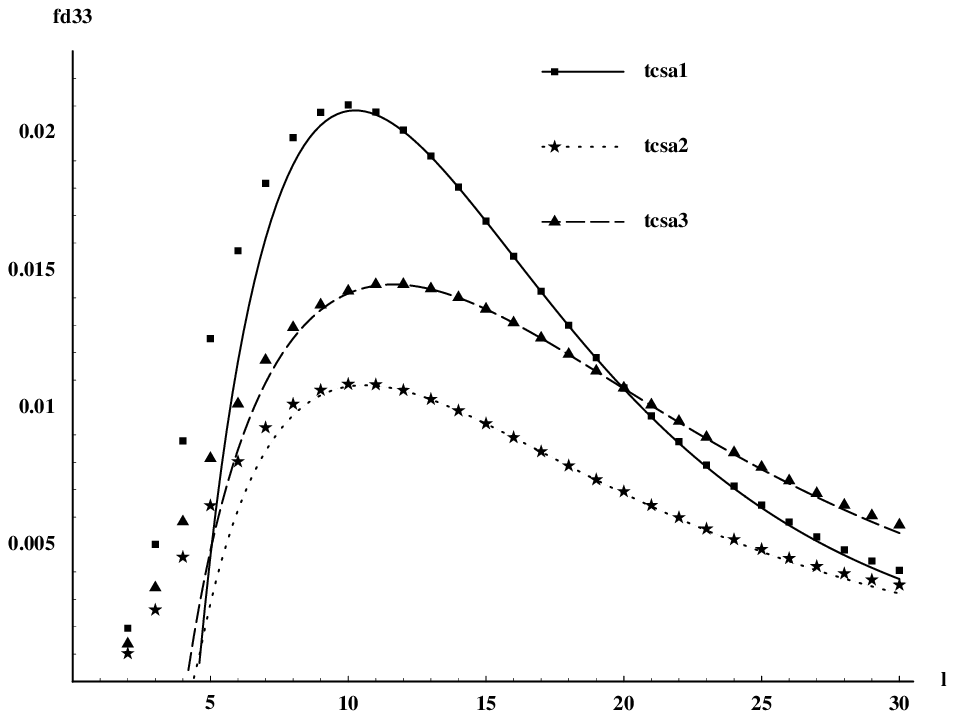}\par\end{centering}

\caption{\label{fig:fd33ly}$3$-particle--$3$-particle matrix elements in
the scaling Lee-Yang model. The discrete points correspond to the
TCSA data, while the continuous line corresponds to the prediction
from exact form factors.}
\end{figure}

In this section we test the predictions of sec. \ref{zero-momentum}
for disconnected pieces associated to zero-momentum particles. 

\subsection{Lee-Yang model}

First we test eqn. (\ref{eq:oddoddlyrule}) using low-lying symmetric
states with a zero-momentum particle present. Data for 
matrix elements of the type
\begin{equation*}
  \bra{\{0\}}\Phi\ket{\{-I,0,I\}}_L \quad \text{and} \quad 
\bra{\{-I',0,I'\}}\Phi\ket{\{-I,0,I\}}_L
\end{equation*}
are shown
in figures \ref{fig:fd13ly} and \ref{fig:fd33ly}, respectively. The
agreement is precise as in all previous cases.

\begin{figure}
\psfrag{f25}{$|f_{25}|$}\psfrag{f35}{$|f_{35}|$}
\psfrag{l}{$l$}
\psfrag{tcsa33}{$\langle\{\frac32,-\frac32\}|\Phi|\{2,1,0,-1,-2\}\rangle$}
\psfrag{tcsa53}{$\langle\{\frac52,-\frac32\}|\Phi|\{2,1,0,-1,-2\}\rangle$}
\psfrag{tcsathp}{$\langle\{3,-1,-2\}|\Phi|\{2,1,0,-1,-2\}\rangle$}\includegraphics[scale=0.8]{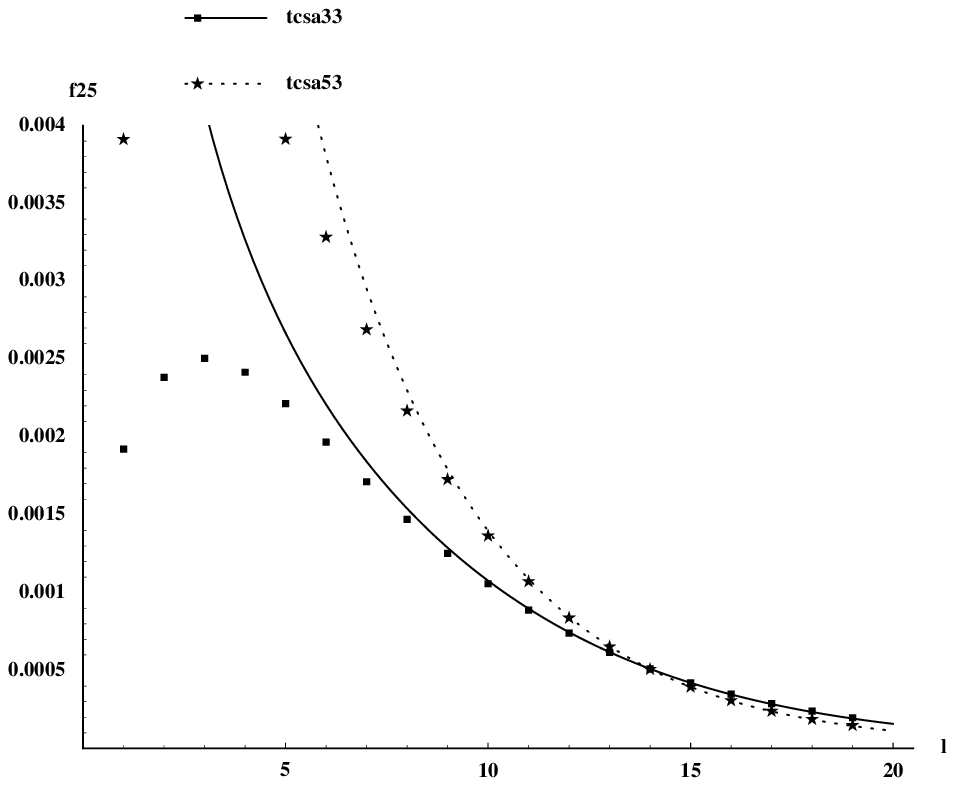}~~\includegraphics[scale=0.8]{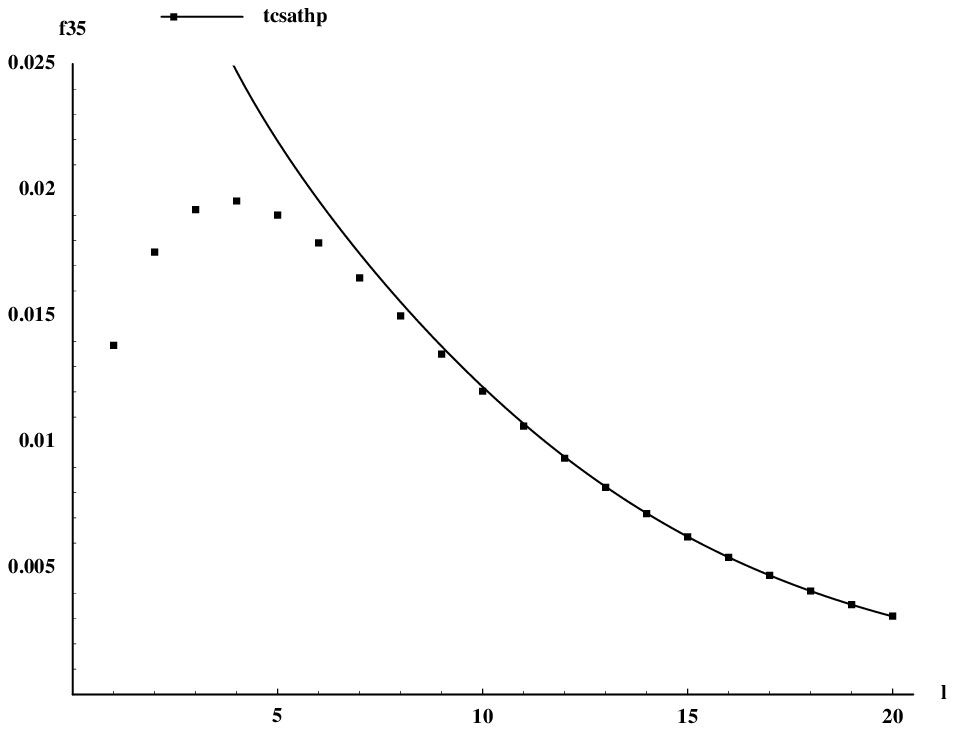}

\caption{\label{fig:5ptident} Identifying the $5$-particle state using form
factors. The discrete points correspond to the TCSA data, while the
continuous line corresponds to the prediction from exact form factors.}
\end{figure}

The support for eqn. (\ref{eq:oddoddlyrule}) can be strengthened using
$5$-particle states. It is not easy to find them because they are
high up in the spectrum, and identification using the process of matching
against Bethe-Yang predictions (as described in subsection \ref{identification}) becomes
ambiguous. We could identify the first $5$-particle state by combining
the Bethe-Yang matching with predictions for matrix elements with
no disconnected pieces given by eqn. (\ref{eq:genffrelation}), as
shown in figure \ref{fig:5ptident}. Some care must be taken in choosing
the other state because many choices give matrix elements that are
too small to be measured reliably in TCSA: since vector components
and TCSA matrices are mostly of order $1$ or slightly less, getting
a result of order $10^{-4}$ or smaller involves a lot of cancellation
between a large number of individual contributions, which inevitably
leads to the result being dominated by truncation errors. Despite
these difficulties, combining Bethe-Yang level matching with form
factor evaluation we could identify the first five-particle level
up to $l=20$.

\begin{figure}
\begin{centering}\psfrag{fd13}{$|f_{35}|$}
\psfrag{l}{$l$}
\psfrag{tcsa1}{$\langle\{1,0,-1\}|\Phi|\{2,1,0,-1,-2\}\rangle$}
\psfrag{tcsa2}{$\langle\{2,0,-2\}|\Phi|\{2,1,0,-1,-2\}\rangle$}
\psfrag{tcsa3}{$\langle\{3,0,-3\}|\Phi|\{2,1,0,-1,-2\}\rangle$}\includegraphics[scale=1.2]{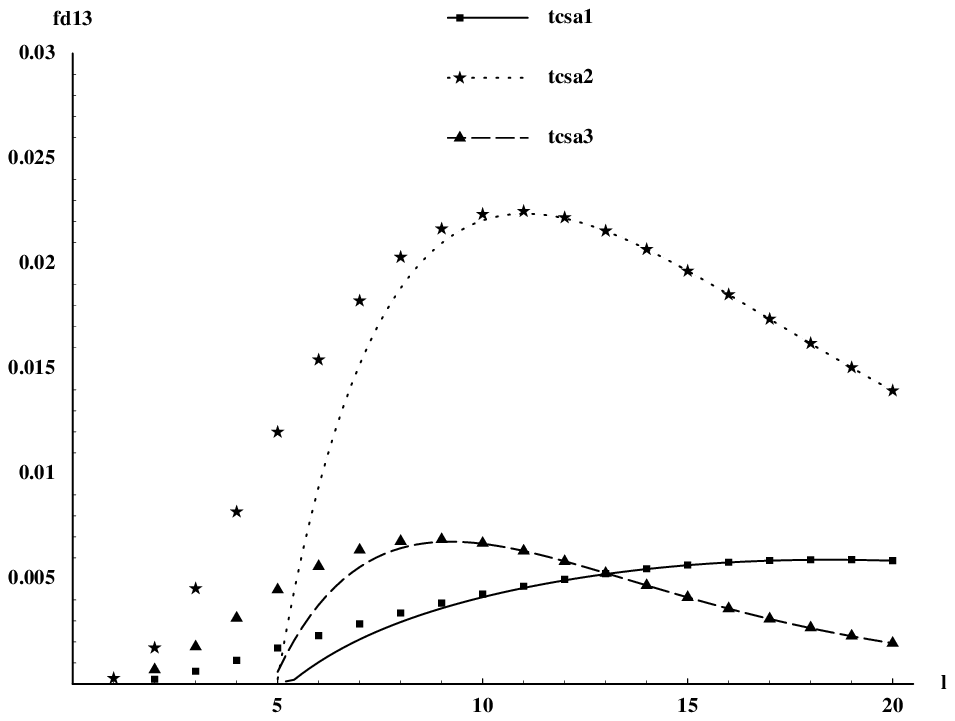}\par\end{centering}

\caption{\label{fig:fd35ly}$3$-particle--$5$-particle matrix elements in
the scaling Lee-Yang model. The discrete points correspond to the
TCSA data, while the continuous line corresponds to the prediction
from exact form factors.}
\end{figure}

The simplest matrix element involving a five-particle state and zero-momentum
disconnected pieces is the $1$-$5$ one, but the prediction of eqn.
(\ref{eq:oddoddlyrule}) turns out to be too small to be usefully
compared to TCSA. However, it is possible to find $3$-$5$ matrix
elements that are sufficiently large, and the data shown in figure
\ref{fig:fd35ly} confirm our conjecture with a relative precision
of somewhat better than $10^{-3}$ in the scaling region.

We close by noting that since the agreement is better than one part
in $10^{3}$ in the scaling region, which is typically found in the
range of volume $l\sim10\dots20$, and also this precision holds for
quite a large number of independent matrix elements, the presence
of additional $\varphi$ terms in eqn. (\ref{eq:oddoddlyrule}) can
be confidently excluded.

\subsection{Ising model in magnetic field}

\begin{figure}
\noindent \begin{centering}\psfrag{l}{$l$}
\psfrag{ff1111}{$|f_{1,111}|$}
\psfrag{a0---a0a0a0}{${}_{1}\langle\{0\}|\Psi|\{1,0,-1\}\rangle_{111}$}\includegraphics{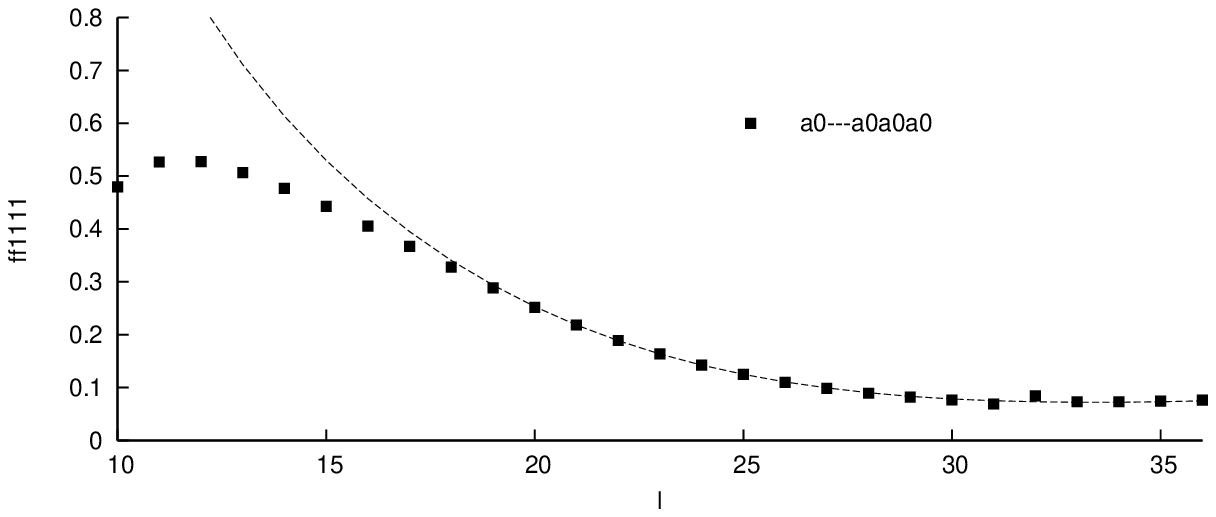}\par\end{centering}

\caption{\label{fig:fd13ising}$A_{1}-A_{1}A_{1}A_{1}$ matrix element in
Ising model with a zero-momentum particle}
\end{figure}

We first test our analytic results on the example of the matrix element
\begin{equation*}
_1\bra{\{0\}}\Phi\ket{\{-1,0,1\}}_{111,L}  
\end{equation*}
Since all
particles are of species $A_{1}$, formula \eqref{eq:oddoddlyrule} is
applicable here. In figure \ref{fig:fd13ising} we 
plot the numerical results against the analytic prediction.

\begin{figure}
\noindent \begin{centering}\psfrag{l}{$l$}
\psfrag{ff112}{$|f_{1,12}|$}
\psfrag{a0---a0b0}{${}_{1}\langle\{0\}|\Psi|\{0,0\}\rangle_{12}$}\includegraphics{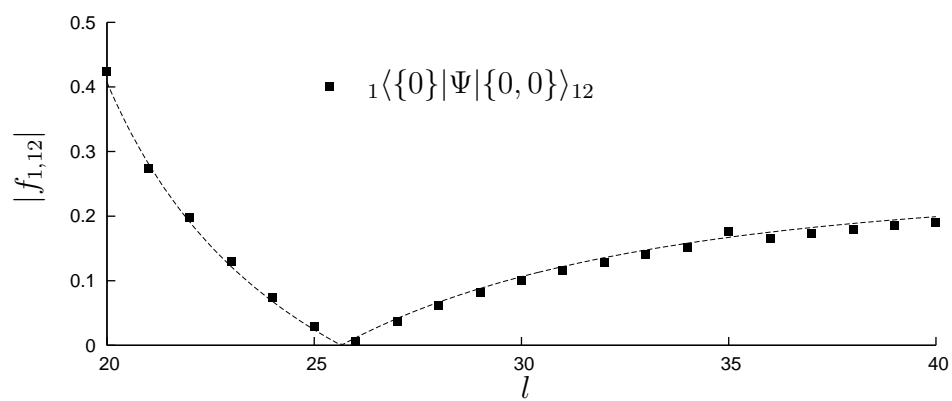}\par\end{centering}

\caption{\label{fig:fd12ising}$A_{1}-A_{1}A_{2}$ matrix element in Ising
model with zero-momentum particle}
\end{figure}

Due to the fact that the Ising model has more than one particle
species, it is possible to have more than one stationary particles
in the same state. Our TCSA data allow us to locate one such state,
with a stationary $A_{1}$ and $A_{2}$ particle, for which we have the prediction\[
f_{1,12}={}_{1}\langle\{0\}|\Psi|\{0,0\}\rangle_{12}=\frac{1}{m_{1}L\sqrt{m_{2}L}}\left(\lim_{\epsilon\rightarrow0}F_{3}(i\pi+\epsilon,0,0)_{112}+m_{1}L\, F_{1}(0)_{2}\right)\]
where $F_{1}(0)_{2}$ is the one-particle form factor corresponding
to $A_{2}$. This is compared to TCSA data in figure \ref{fig:fd12ising}
and a convincing agreement is found.

Note that in both of figures \ref{fig:fd13ising} and \ref{fig:fd12ising}
there is a point which obviously deviates from the prediction. This
is a purely technical issue, and is due to the presence of a line
crossing close to this particular value of the volume which makes
the cutoff dependence more complicated and so slightly upsets the
extrapolation in the cutoff. 

%

\chapter{Residual finite size effects}

\label{exponential}

The Bethe-Yang equations describe multi-particle energies in finite
volume to all orders in $1/L$. The same holds for the results
presented in section 2: formulas \eqref{eq:genffrelation},
\eqref{eq:diaggenrule} and \eqref{eq:oddoddlyrule} determine the
finite volume form factors to all orders in $1/L$. 
The aim of this
chapter is to derive the $\mu$-term (which is the leading exponential
correction) associated to moving one-particle states, generic
multi-particle scattering states, and finite volume form factors. 

The analytic results are derived using the principle of bound state
quantization in finite volume. To confirm the calculations, 
all our formulas are tested against TCSA data.

We wish to remark, that there has been a recent interest in exponential corrections also in the
framework of AdS/CFT correspondence. The $\mu$-term and F-term for
moving one-particle states was derived in \cite{Janik:2007wt} and
\cite{Heller:2008at}, for multi-particle states in
\cite{Bajnok:2008bm} and \cite{Hatsuda:2008na}. For the $\mu$-term these results
coincide with ours, but they can be considered as independent;
they use completely different methods: generalizations 
of L\"uscher's approach, summation of the vacuum fluctuations and
calculations based on the TBA for excited states. 


\section{One-particle states}

\label{exp:1particle}


The energy of a moving one-particle state $\ket{\{I\}}_{c,L}$ is
given to all orders by
\begin{equation}
  \label{eq:1particle_ujra}
E=\sqrt{m_c^2+p_c^2} \quad \text{with}  \quad p_c=\frac{2\pi I}{L}
\end{equation}
Here we determine the leading $\mu$-term associated to \eqref{eq:1particle_ujra}
 Based on the description of 
mass corrections it is expected that this contribution is associated 
to the fusion $A_aA_b\to A_c$ with the smallest $\mu_{ab}^c$.
We assume that $a=b$, ie. the fusion in question is a symmetric
one.
This happens to be true for the lightest particle in models with the ''$\Phi^3$-property'' 
and for other low lying states in most known models. 
A possible extension to non-symmetric fusions has not yet been carried
out, it is left for further research.

The (infinite volume) bootstrap principle for a symmetric fusion consists of the identification
\begin{equation}
\label{infi_bootstrap}
  \ket{\theta}_c \sim \ket{\theta+i\bar{u}_{ac}^a,\theta-i\bar{u}_{ac}^a}_{aa}
\end{equation}
resulting in
\begin{equation*}
  m_c=2m_a\cos(\bar{u}_{ac}^a) \quad\quad \left(\mu_{aa}^c\right)^2=m_a^2-\frac{m_c^2}{4}
\end{equation*}
Smallness of $\mu_{aa}^c$ means that $m_c$ is close to $2m_a$, in
other words the
binding energy is small. 

For a moment let us lay aside the framework of QFT
and consider quantum mechanics with an
attractive potential. Bound states are described by 
wave functions 
\begin{equation*}
\Psi(x_1,x_2)=e^{iP(x_1+x_2)}\psi(x_1-x_2)
\end{equation*}
where $P$ is the total momentum and $\psi(x)$ is the appropriate
solution of the Schr\"odinger equation in the relative coordinate. It is localized around $x=0$ and
shows exponential decay at infinity. Except for the
region $x_1\approx x_2$, the  wave function can be approximated
with a product of plane waves with imaginary momenta $p_{1,2}=P\pm ik$. The interaction
results in the quantization of the allowed values of $k$.

The theory in finite volume is described along the same lines. There
are however two differences:
\begin{itemize}
\item The total momentum gets quantized.
\item $\psi(x)$ (and therefore $k$) obtains finite volume corrections.
\end{itemize}

This picture also applies to relativistic integrable
theories. We consider $A_c$ as a simple quantum mechanical
bound state of two elementary particles and  use the infinite volume
scattering data to describe the interaction between the constituents.
To develop these ideas, let us consider the spectrum of the theory
defined on a circle with circumference $L$. We state the identification 
\begin{equation}
  \label{finiteL_azonositas}
\ket{A_c(\theta)}_L\sim  \ket{A_a(\theta_1)A_a(\theta_2)}_L
\end{equation}
where the $\theta_{1,2}$ are complex to describe a
bound-state; this idea also appeared in \cite{takacs_watts,bpt1}. 
Relation \eqref{finiteL_azonositas} can be regarded
as  the finite volume realization of \eqref{infi_bootstrap}. 
 The total energy and momentum of the bound state have to be purely real,
constraining the rapidities to take the form
\begin{equation}
\label{finiteL_rapidities}
  \theta_1=\theta+iu,\quad\quad
  \theta_2=\theta-iu
\end{equation}
where the dependence on $L$ is suppressed. Energy and momentum
are calculated as
\begin{equation}
  E=2m_a\cos(u)\cosh(\theta)\quad\quad p=2m_a\cos(u)\sinh(\theta)
\end{equation}

The quantization
condition for an $n$-particle state is given by the Bethe-Yang equations 
\begin{equation*}
  e^{ip_jL}\mathop{\prod_{k=1}^n}_{k\ne j}
S_{i_ji_k}(\theta_j-\theta_k)=1,\quad\quad
j=1\dots n
\end{equation*}
To
quantize the bound state in finite volume, an appropriate analytic
continuation of the above equations with $n=2$ can be applied.
This procedure is justified
by the same reasoning that leads to original Bethe-Yang equations: one
assumes plane waves (with imaginary momenta) except for
the localized interaction, which is described by the S-matrix of the
infinite volume theory.
Inserting \eqref{finiteL_rapidities} and separating the real and imaginary
parts 
\begin{eqnarray}
\label{kepzetes_Bethe_Yang_1}
e^{im_a \cos(u)\sinh(\theta)L} e^{-m_a\sin(u)\cosh(\theta)L}
S_{aa}(2iu)&=&1\\
\label{kepzetes_Bethe_Yang_2}
e^{im_a \cos(u)\sinh(\theta)L} e^{m_a\sin(u)\cosh(\theta)L}
S_{aa}(-2iu)&=&1
\end{eqnarray}
Multiplying the two equations and making use of $S(2iu)=S(-2iu)^{-1}$ one arrives at
\begin{equation}
\label{kvantalasi_1}
  e^{2im_a \cos(u)\sinh(\theta)L}=1\quad \textrm{or}\quad  2m_a \cos(u)\sinh(\theta)=\frac{2\pi I}{L}
\end{equation}
which is the quantization condition
for the total momentum. $I$ is to be identified with the momentum quantum number
of $A_c$.  The quantization condition for $u$ is found by eliminating
$\theta$ from \eqref{kepzetes_Bethe_Yang_1}:
\begin{equation}
  \label{u_egyenlete}
  e^{-m_aL\sin(u)\sqrt{1+\left(\frac{\pi I}{m_aL\cos(u)}\right)^2}}
S_{aa}(2iu)=(-1)^I
\end{equation}
The exponential factor forces $u$ to be close to the pole of
the S-matrix associated to the formation of the bound-state. For the
case at hand it reads
\begin{equation}
\label{S-matrix_polus}
S_{aa}(\theta\sim iu_{aa}^c)\sim \frac{i\left(\Gamma_{aa}^c\right)^2}{\theta-iu_{aa}^c}
\end{equation}
with $u_{aa}^c=2\bar{u}_{ac}^a$.
Note the appearance of $(-1)^I$ on the rhs. of \eqref{u_egyenlete}, which is a natural
consequence of the quantization of the total momentum. This sign
determines the direction from which the pole is approached.

The exact solution of \eqref{u_egyenlete} can be developed into a power series in
$e^{-\mu_{aa}^c L}$, where the first term is found by replacing $u$
with $\bar{u}_{ac}^a$ in the exponent:
\begin{equation}
\label{uminusubar}
  u-\bar{u}_{ac}^a=(-1)^I \frac{1}{2}\left(\Gamma_{aa}^c\right)^2
  e^{-\mu_{aa}^cL \sqrt{1+\left(\frac{2\pi I}{m_cL}\right)^2}}+
O(e^{-2\mu_{aa}^cL})
\end{equation}
First order corrections to the energy
are readily evaluated to give
\begin{eqnarray}
\label{elso_korrekcio}
  E&=&E_0- (-1)^I  \left(\Gamma_{aa}^c\right)^2
\frac{\mu_{aa}^c m_c}{E_0} e^{-\frac{\mu_{aa}^cE_0}{m_c} L }+O(e^{-2\mu_{aa}^cL})
\end{eqnarray}
where $E_0$ is the ordinary  one-particle energy
\begin{equation*}
  E_0=\sqrt{m_c^2+\left(\frac{2\pi I}{L}\right)^2}
\end{equation*}
In the case of zero momentum the former result simplifies to 
the leading term in \eqref{mass_mu_term}. 
For large volumes we
recover
\begin{equation*}
u\to \bar{u}_{ac}^a\quad\quad  \theta\to \textrm{arsh} \frac{2\pi I}{m_cL}
\end{equation*}

Having established the quantization procedure we now turn to the
question of momentum quantum numbers inside 
the bound state. 
For the phase shift we adopt the convention that was used throughout
this work
\begin{equation*}
  S_{ab}(\theta)=S_{ab}(0)e^{i\delta_{ab}(\theta)}
\end{equation*}
Note that $\delta_{ab}(iu)$ is purely imaginary.

With this choice of the phase shift the Bethe-Yang equations in their logarithmic form 
\begin{eqnarray*}
  l \sinh(\theta+iu) +\delta_{11}(2iu)&=&2\pi I_1\\
  l \sinh(\theta-iu) +\delta_{11}(-2iu)&=&2\pi I_2
\end{eqnarray*}
imply $I_1=I_2$. Quantization of the total momentum on the other hand requires
$I_1=I_2=I/2$. Note that different conventions for $\delta_{ab}$ would result in a less
transparent rule for dividing $I$ among the two constituents. The only
disadvantage of our choice is the appearance of the unphysical half-integer quantum
numbers.  
With this convention
the bound-state quantization can be written in
short-hand notation as
\begin{equation*}
  \ket{\{I\}}_{c,L}\sim\ket{\{I/2,I/2\}}_{aa,L}
\end{equation*}




\subsection{Numerical analysis}

It seems plausible that  by
exactly solving the bound state quantization condition \eqref{u_egyenlete} one obtains all higher order
corrections that go as $e^{-n\mu_{aa}^cL}$ with $n\in \mathbb{N}$.  In this
subsection we present numerical evidence to support this claim.

We investigate 
 the Ising model in the presence of a magnetic
field. 
The first three particles lie below the two-particle threshold and they all show up as $A_1A_1$
bound states. These fusions are responsible for the leading $\mu$-term.
The corresponding parameters (in units of $m_1$)
are listed in the table below. The exponent of the next-to-leading
correction (the error exponent) is denoted by $\mu'$.

\begin{center}
\begin{tabular}{|c|ccc|cc|}
\hline
a& $m_a$ &  $\mu_{11}^a$ & $\left(\Gamma_{11}^a\right)^2$ & $\mu'$ & \\
\hline
1 & 1 & 0.86603 & 205.14 & $1$ & ($m_1$)\\

2 & 1.6180 & 0.58779 & 120.80 & 0.95106 & ($\mu_{12}^2$) \\

 3&  1.9890 & 0.10453& 1.0819 &  0.20906 &  ($2\mu_{11}^3$) \\
\hline
\end{tabular}
\end{center}

In \cite{klassen_melzer} Klassen and Melzer performed the numerical
analysis of mass corrections. The analytic predictions were compared to TCSA data
and to transfer matrix results. They
observed the expected behavior of mass corrections of $A_1$ and
$A_2$; in the former case they were also able to verify the F-term. On
the other hand, the precision of their TCSA data was not sufficient to
reach volumes where the $\mu$-term for $A_3$ could have been tested. 
This limitation is a natural consequence of the unusually small
exponent $\mu_{11}^3$: the next-to-leading contribution is of order
$e^{-2\mu_{11}^3L}$, still very slowly decaying.

We employ
the TFCSA 
routines that were successfully used in the previous chapter.
Calculations are performed for $I=0,1,2,3$ at different values of
the volume. One-particle states of $A_1$, $A_2$ and $A_3$ are easily
identified: they are the lowest lying levels in the spectrum, except
for $I=0$ where the lowest state is the vacuum. 
We use the dimensionless quantities $l=m_1L$ and $e=E/m_1$.  

The results are extrapolated from $e_{cut}=20..30$ to
$e_{cut}=\infty$.
Experience from the previous results shows
that this extrapolation technique reduces the numerical
errors by an order of magnitude. 
We resign here from quantitatively monitoring the TCSA errors and
constrain ourselves to a range
of the volume parameter where it is safe to neglect truncation effects.

\subsubsection{$A_3$}

We begin our analysis with the most interesting case of $A_3$. At each
value of $l$ and $I$ the following procedure is performed.
\begin{itemize}
\item The energy correction is calculated according to
  \eqref{elso_korrekcio}
\item The quantization condition \eqref{u_egyenlete} is solved for
  $u$ and the energy correction is calculated by
  \begin{equation*}
 \Delta e=2\cosh(\theta)\cos(u)-e_0
\end{equation*}
where $\theta$ is determined by the total momentum quantization
\eqref{kvantalasi_1} and $e_0$ is the  ordinary one-particle energy.
\item The exact correction is calculated numerically by $\Delta e=e^{TCSA}-e_0$.
\end{itemize}

The choice for the range of the volumes is limited in two ways. On one hand,
$l$ has to be sufficiently large in order to reduce the contribution of the
F terms and other higher order
finite size corrections. On the other hand, numerical errors grow with the volume and
eventually become
comparable with the finite size corrections, resulting in an upper bound
on $l$.  The window $l=30..40$ is suitable for
our purposes.

\begin{figure}
  \centering
\subfigure[$I=0$]{  \includegraphics[scale=0.5,angle=-90]{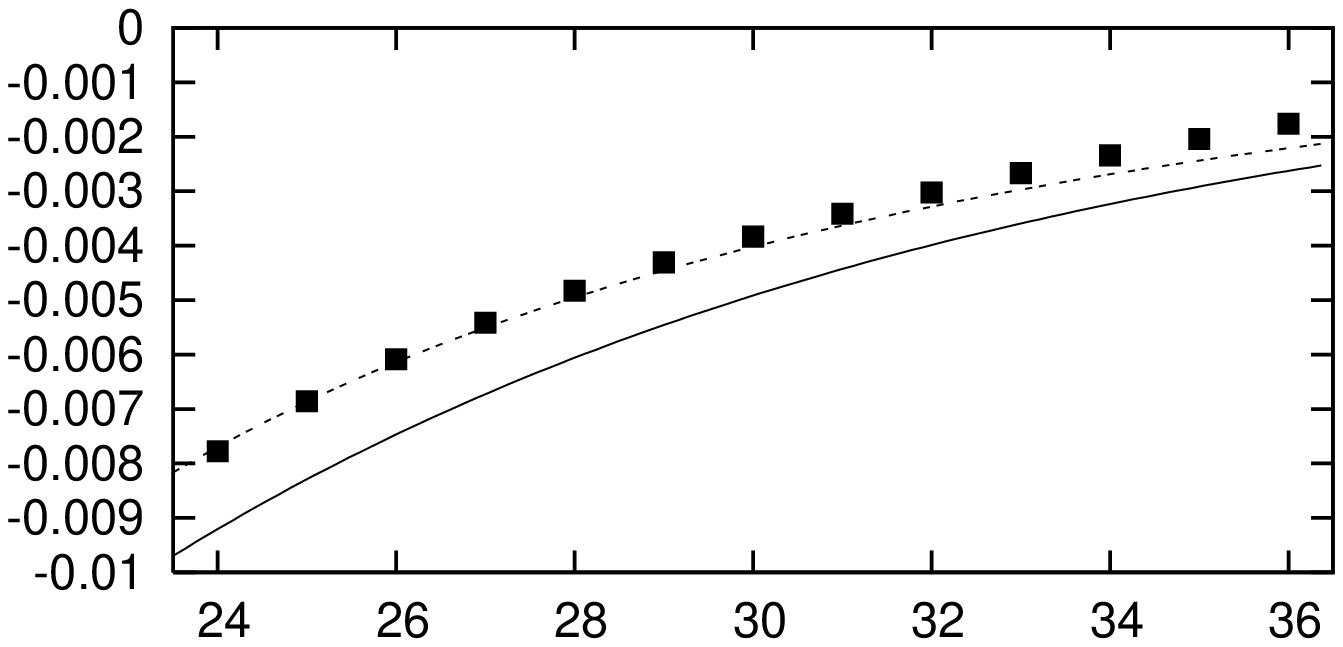}}
\subfigure[$I=1$]{ \includegraphics[scale=0.5,angle=-90]{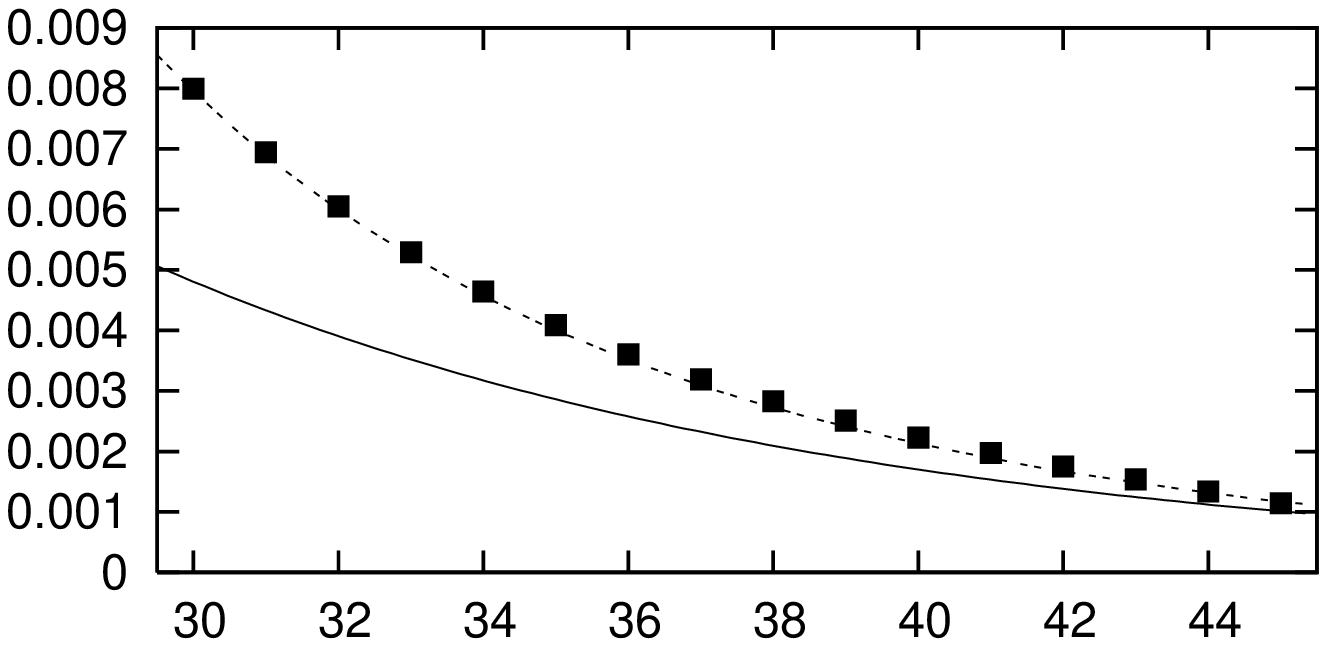}}

\subfigure[$I=2$]{ \includegraphics[scale=0.5,angle=-90]{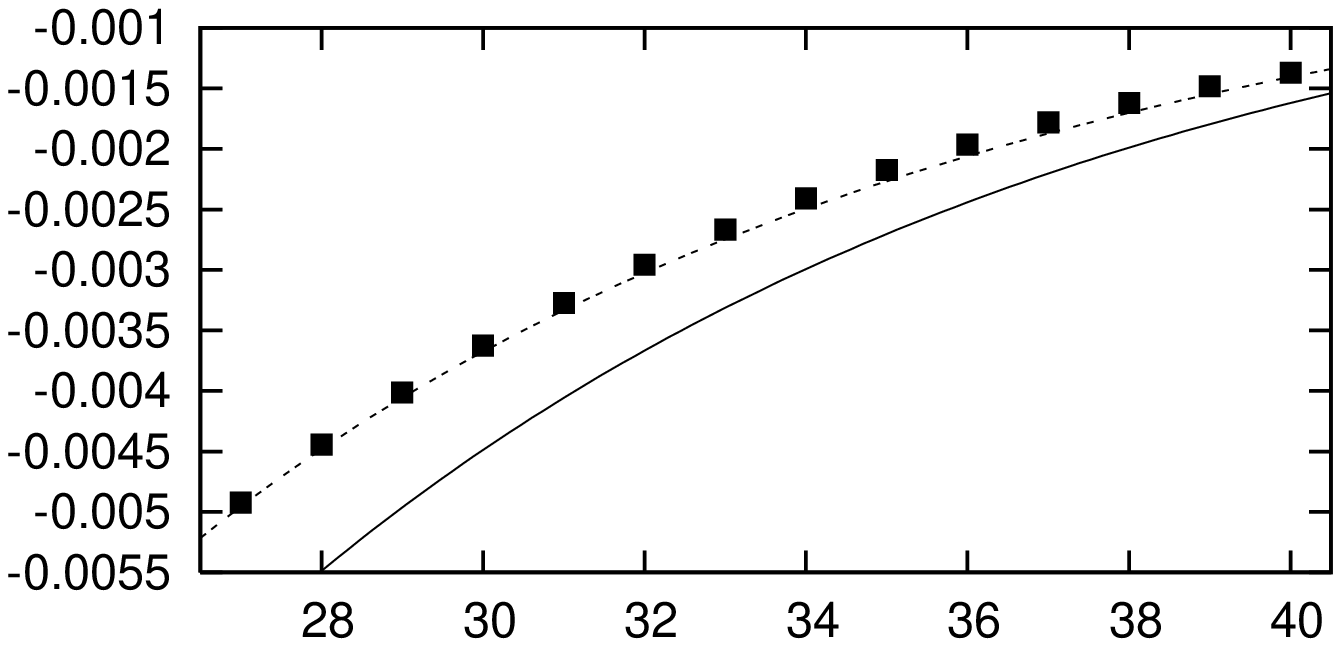}}
\subfigure[$I=3$]{\includegraphics[scale=0.5,angle=-90]{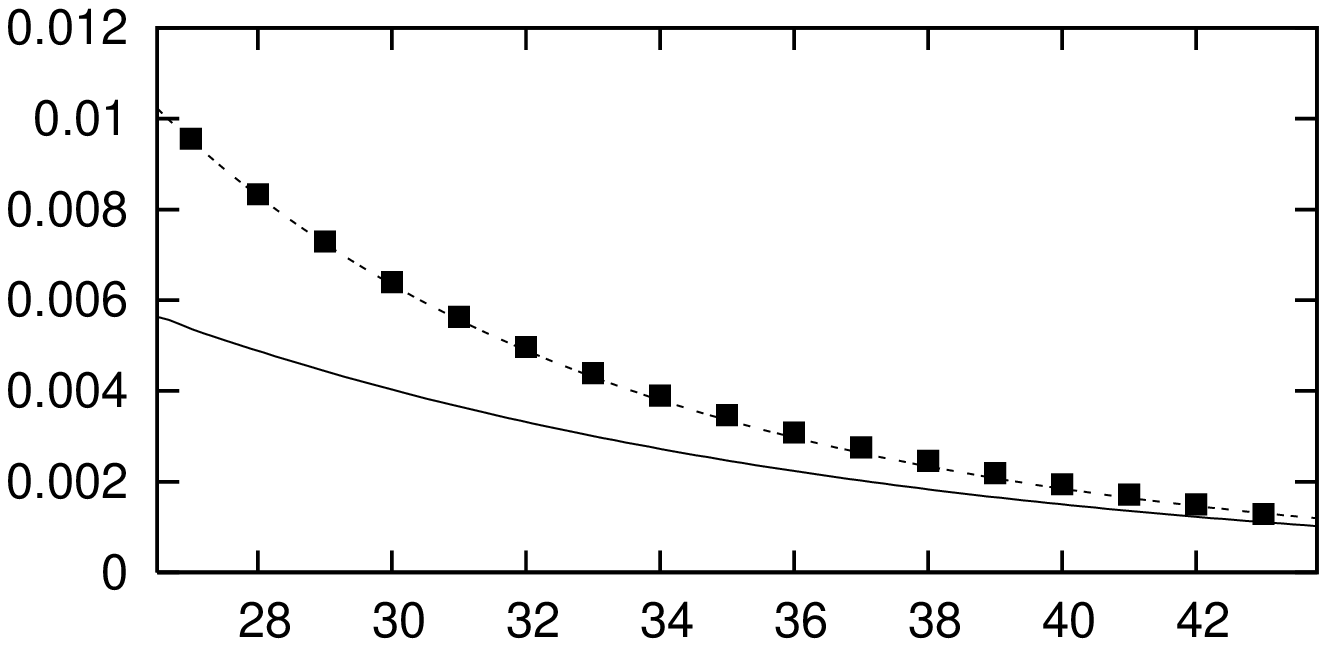}}
\caption{
Finite size corrections to $A_3$ one-particle levels (in sectors
$I=0\dots 3$) as a function of
the volume. The TCSA data
are plotted against theoretical predictions of the single
$\mu$-term associated to the $A_1A_1\to A_3$ fusion (solid curve) and
the exact solution of the bound-state quantization (dotted curve). 
\label{fig:A3_nagy_terfogat}}
\end{figure}

The results are shown in figure \ref{fig:A3_nagy_terfogat}. It is
clear that the $\mu$-term yields the correct prediction in the
$L\to\infty$ limit. However,  higher order terms cause a
significant deviation for $l<40$, which is in turn accurately described
by the bound state prediction.
The sign of the correction depends on
the parity of $I$ as predicted by \eqref{elso_korrekcio}.

Based on the success of this first numerical test we also explored the
region $l<30$.
Inspecting the behavior of $u$ as a function of $l$ reveals an interesting
phenomenon. It is obvious from \eqref{uminusubar} that $u(l)$ is
monotonously increasing if $I$ is odd, with the infinite volume
limit fixed to $\bar{u}_{ac}^a$.  However, the complex
conjugate pair $\theta_{1,2}$ approaches the real axis as $l$ is
decreased and they collide at a critical volume $l=l_c$. For $l<l_c$ 
they separate again but stay on the real line, providing a unique
solution with two distinct purely real rapidities. The same behavior
was also observed in \cite{takacs_watts,bpt1}.

The interpretation of this phenomenon is evident: if the volume is comparable to
the characteristic size of the bound-state, there is enough energy in
the system for the constituents to become unbound. 
Therefore the $A_3$ one-particle level becomes an $A_1A_1$
scattering state for $l<l_c$. We call this phenomenon the
``dissociation of the bound state''. The same result was
obtained also in the boundary sine-Gordon model by a semiclassical analysis \cite{bpt_semiclass}. 

The value of $l_c$ can be found
by exploiting the fact that the Jacobian of the Bethe-Yang equations (viewed as a mapping from
$(\theta_1,\theta_2)$ to $(I_1,I_2)$) vanishes at the critical
point. A straightforward calculation yields
\begin{equation*}
  l_c=\sqrt{4\varphi_{11}(0)^2-I^2\pi^2}
\end{equation*}
where $\varphi_{11}(\theta)=\delta_{11}'(\theta)$. 
The numerical values for the case at hand
are 
\begin{equation*}
  l_c=27.887\quad (I=1) \quad\textrm{ and }  \quad
l_c=26.434 \quad (I=3)
\end{equation*}

\begin{figure}
  \centering
\Large
\psfrag{spin0}{$I=0$}
\psfrag{spin1}{$I=1$}
\psfrag{spin2}{$I=2$}
\psfrag{spin3}{$I=3$}
\psfrag{l}{$l$}
\psfrag{e}{$e$}
\psfrag{igazi}{$m_3$}
\psfrag{magy1}{$A_1A_1\to A_3$}
\psfrag{magy2}{$(I=3)\quad (I=1)$}
\includegraphics[angle=-90,scale=0.5]{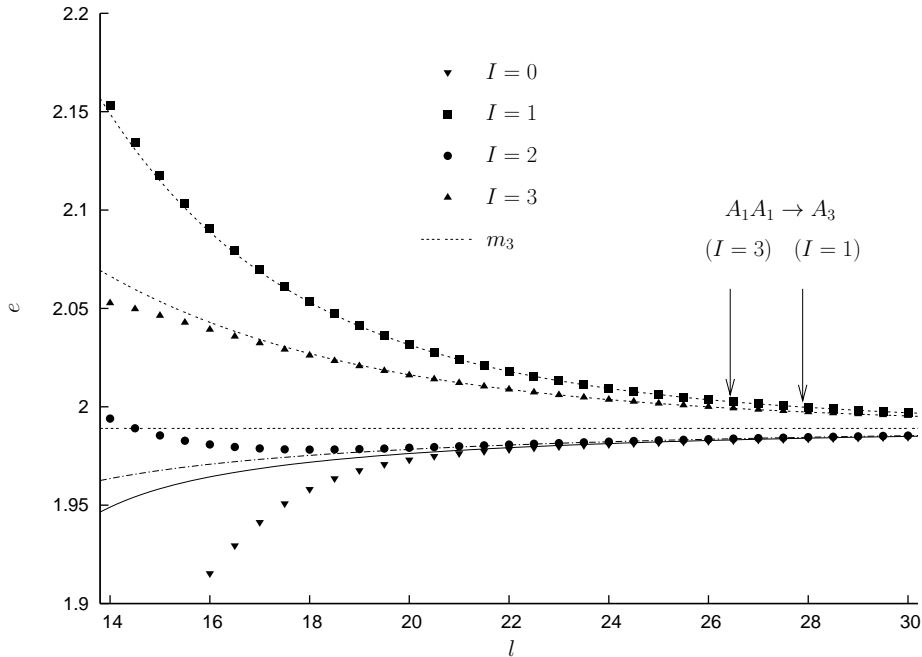}  
\caption{
$A_3$ one-particle levels in sectors $I=0\dots 3$  as a function of
the volume. Dots represent TCSA data, while the lines show the corresponding
prediction of the $A_1A_1$ bound state quantization. In sectors $I=1$ and
$I=3$ the bound state dissociates at $l_c$ and for $l<l_c$
 a conventional $A_1A_1$ scattering state replaces $A_3$ in the spectrum. 
The values of $l_c$ are shown by the two arrows.
\label{A3_full}}
\end{figure}

We are now in the position to complete the numerical analysis.
The Bethe-equations are solved at each value of $l$, 
providing two distinct real rapidities for $l<l_c$ (with $I$ being odd),
and a complex conjugate pair otherwise. The energy is calculated in
either case as
\begin{equation*}
  e=\cosh(\theta_1)+\cosh(\theta_2)
\end{equation*}
which is compared to TCSA data. The results are exhibited in figure 
\ref{A3_full}.

The agreement for the upper two curves ($I=1$ and $I=3$) is not as surprising as it may
seem because what one sees here are conventional $A_1A_1$ scattering
states. The Bethe-Yang equation determining their energy is exact up to $O(e^{-\mu'L})$ where 
$\mu'=\mu_{11}^1$
is the smallest exponent that occurs in the sequence of finite volume
corrections of $A_1$.
On the other hand, the energy levels are analytic functions of L, which
leads to the conclusion that the prediction of \eqref{u_egyenlete} is correct up to
$O(e^{-\mu_{11}^1L})$ even for $L>L_c$.  Comparing the numerical
values one finds $\mu_{11}^1>8\mu_{11}^3$. We conclude
 that the bound state picture  indeed accounts for
finite volume corrections up to the first few orders in
$e^{-\mu_{aa}^cL}$ (the first 8 orders in the case at hand).

\subsubsection{$A_1$ and $A_2$}

Particles $A_1$ and $A_2$ also appear as $A_1A_1$ bound states.
However, there is no point in applying the complete bound state
quantization to them, because the error terms dominate over the higher
order contributions from \eqref{u_egyenlete}: the exponents of the
sub-leading finite size corrections $m_1$ and $\mu_{12}^2$ are smaller than $2\mu_{11}^1$ and $2\mu_{11}^2$. 
Nevertheless, the leading $\mu$-term can be verified by choosing suitable windows in $l$.

\begin{figure}
  \centering
\subfigure[$I=0$]{\includegraphics[scale=0.3,angle=-90]{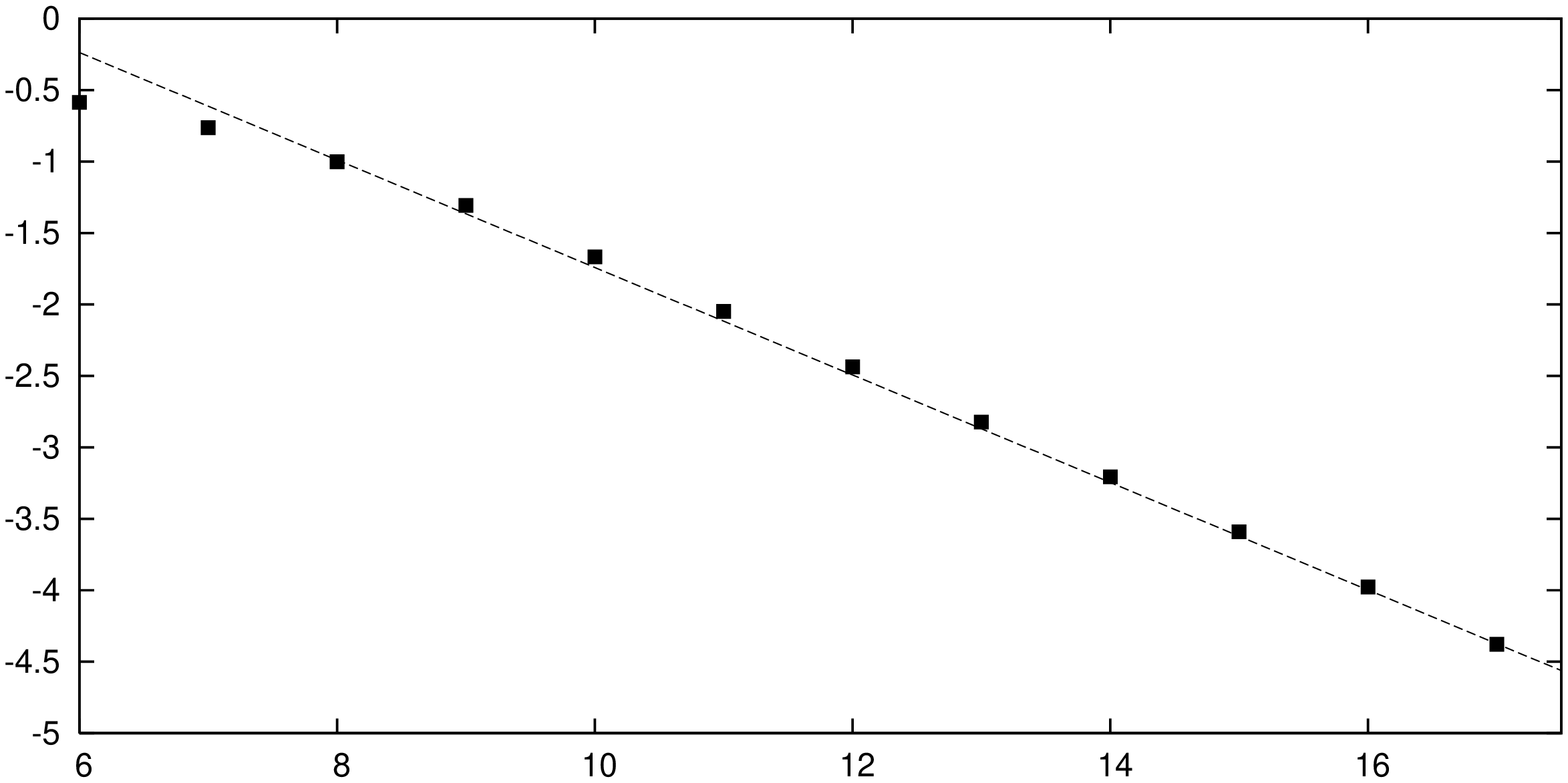}}
\subfigure[$I=1$]{ \includegraphics[scale=0.3,angle=-90]{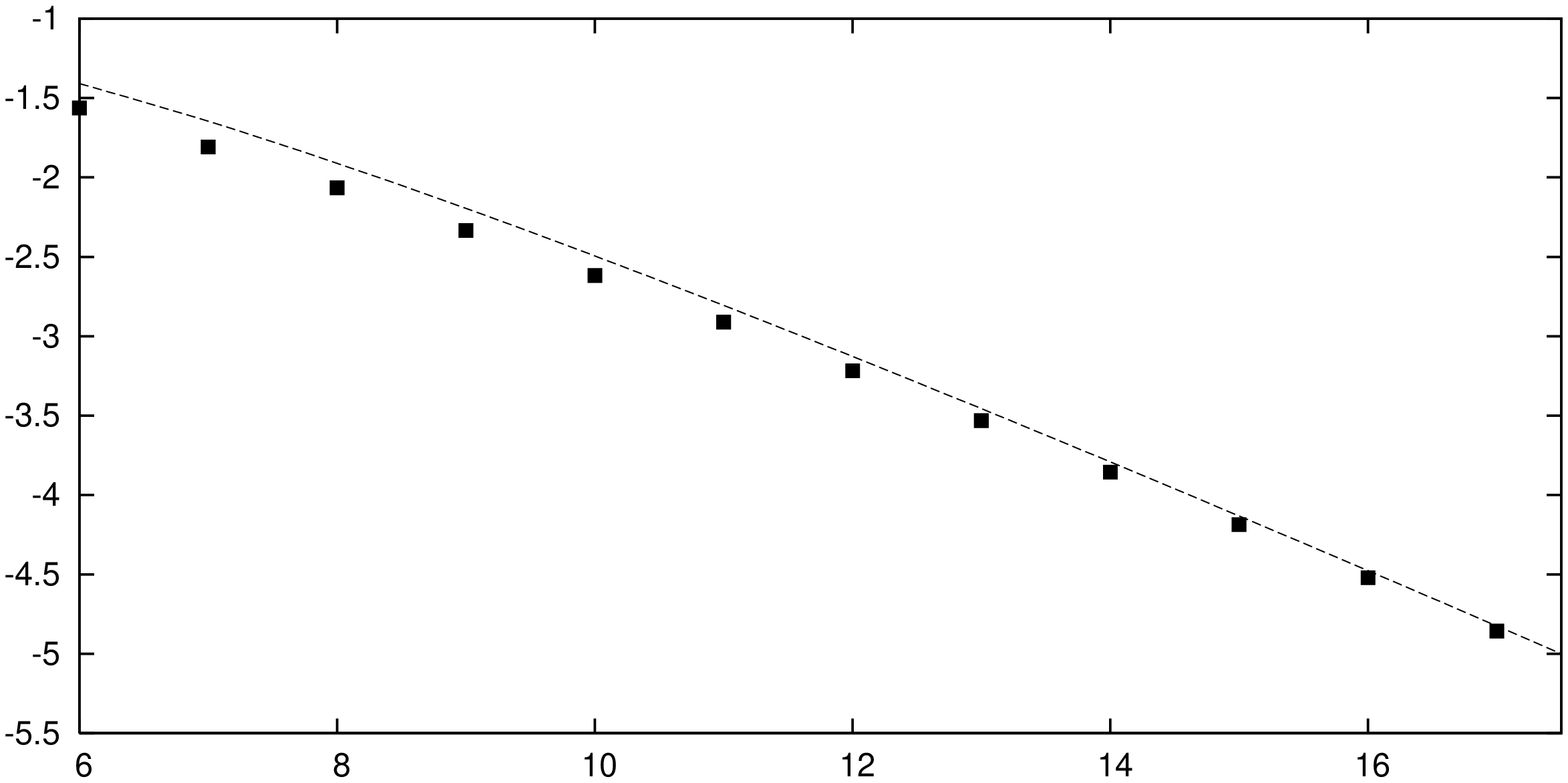}}

\subfigure[$I=2$]{ \includegraphics[scale=0.3,angle=-90]{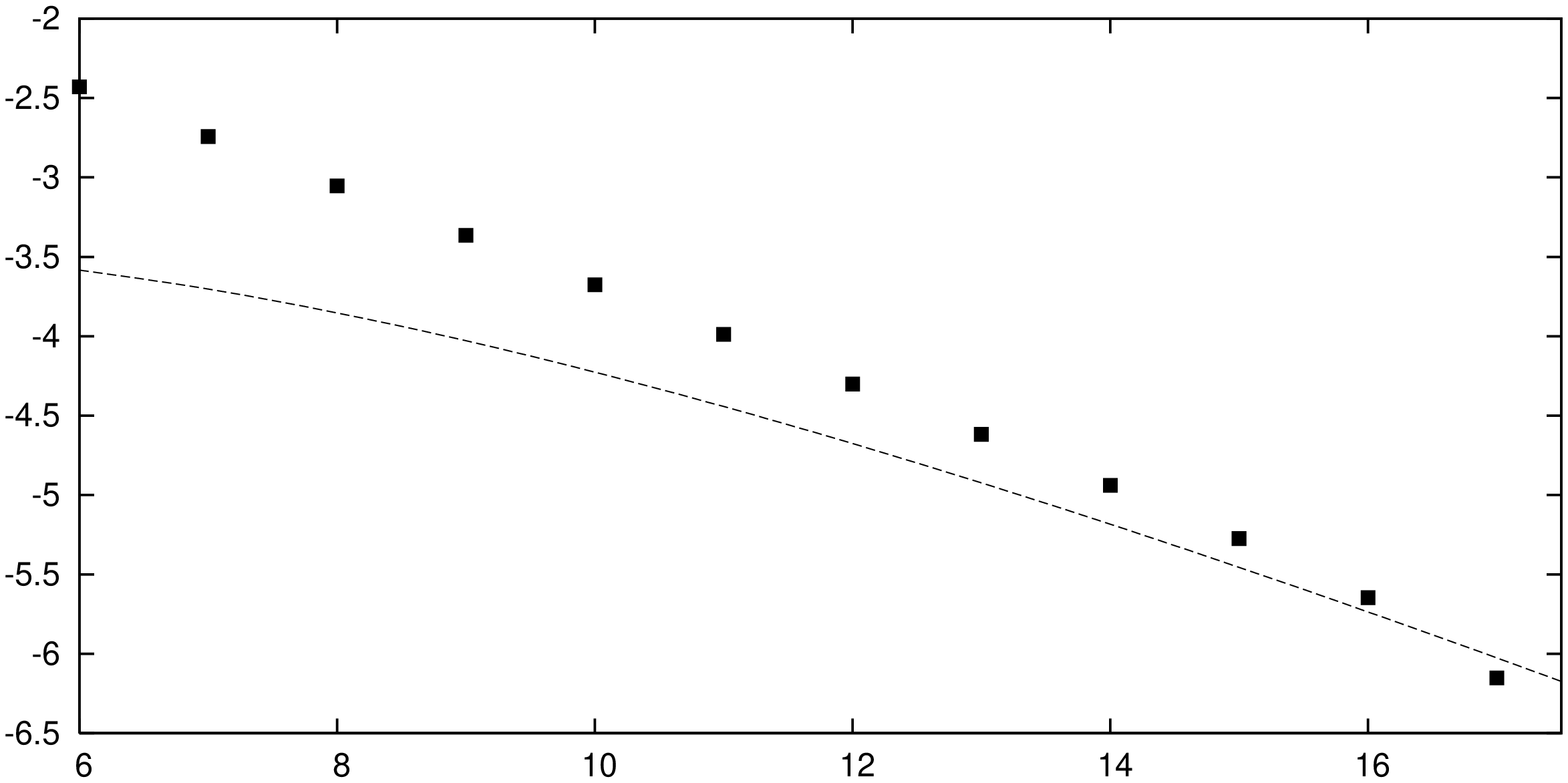}}
\subfigure[$I=3$]{\includegraphics[scale=0.3,angle=-90]{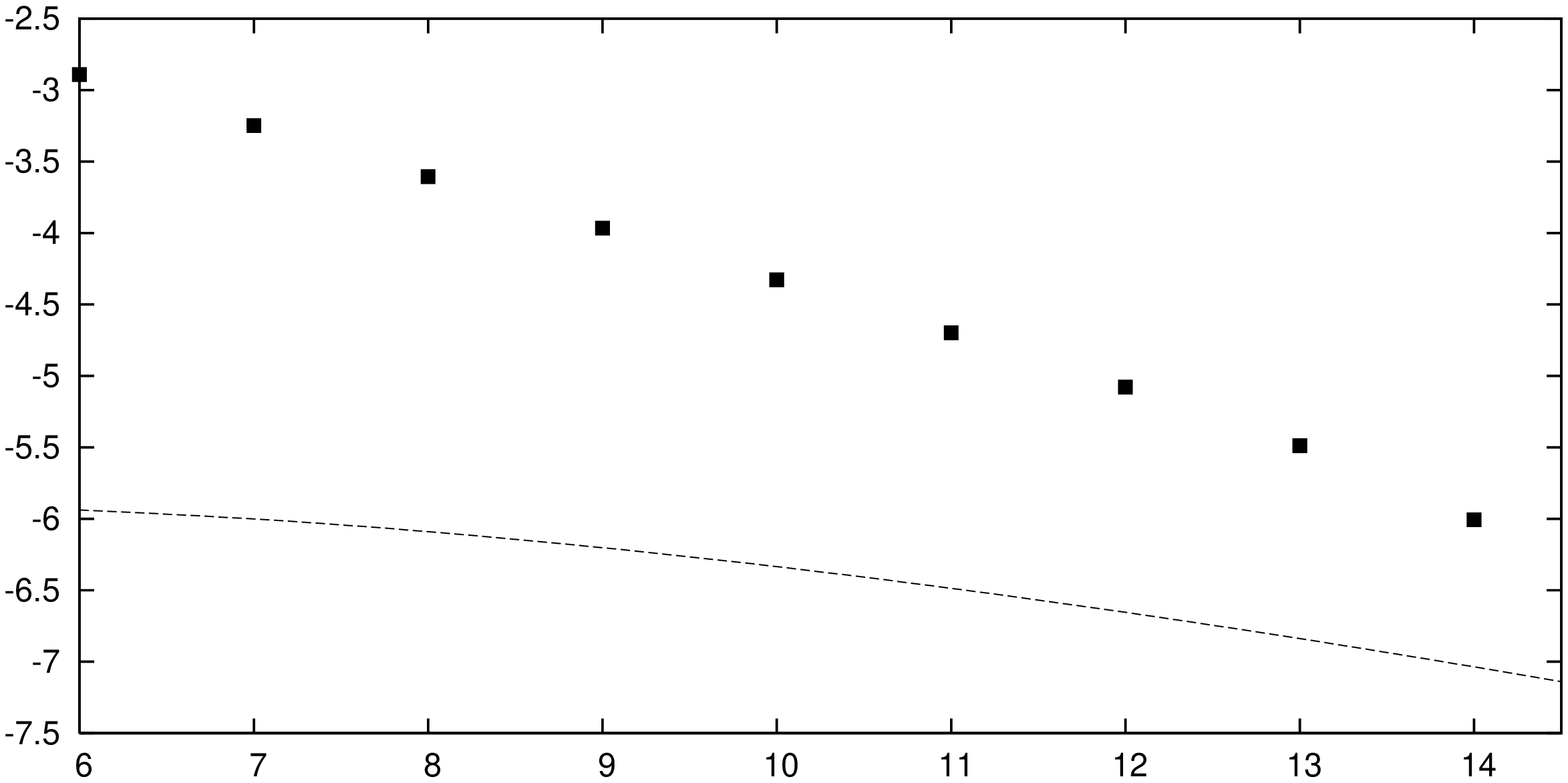}}
\caption{
Finite size corrections to $A_1$ one-particle levels in sectors
$I=0\dots 3$, $log_{10}\Delta e$ is plotted as a function of the
volume. Dots represent TCSA data, while the lines show the
$\mu$-term corresponding to the $A_1A_1\to A_1$ fusion.
\label{fig:A1_energia}}
\end{figure}

\begin{figure}
  \centering
\subfigure[$I=0$]{  \includegraphics[scale=0.3,angle=-90]{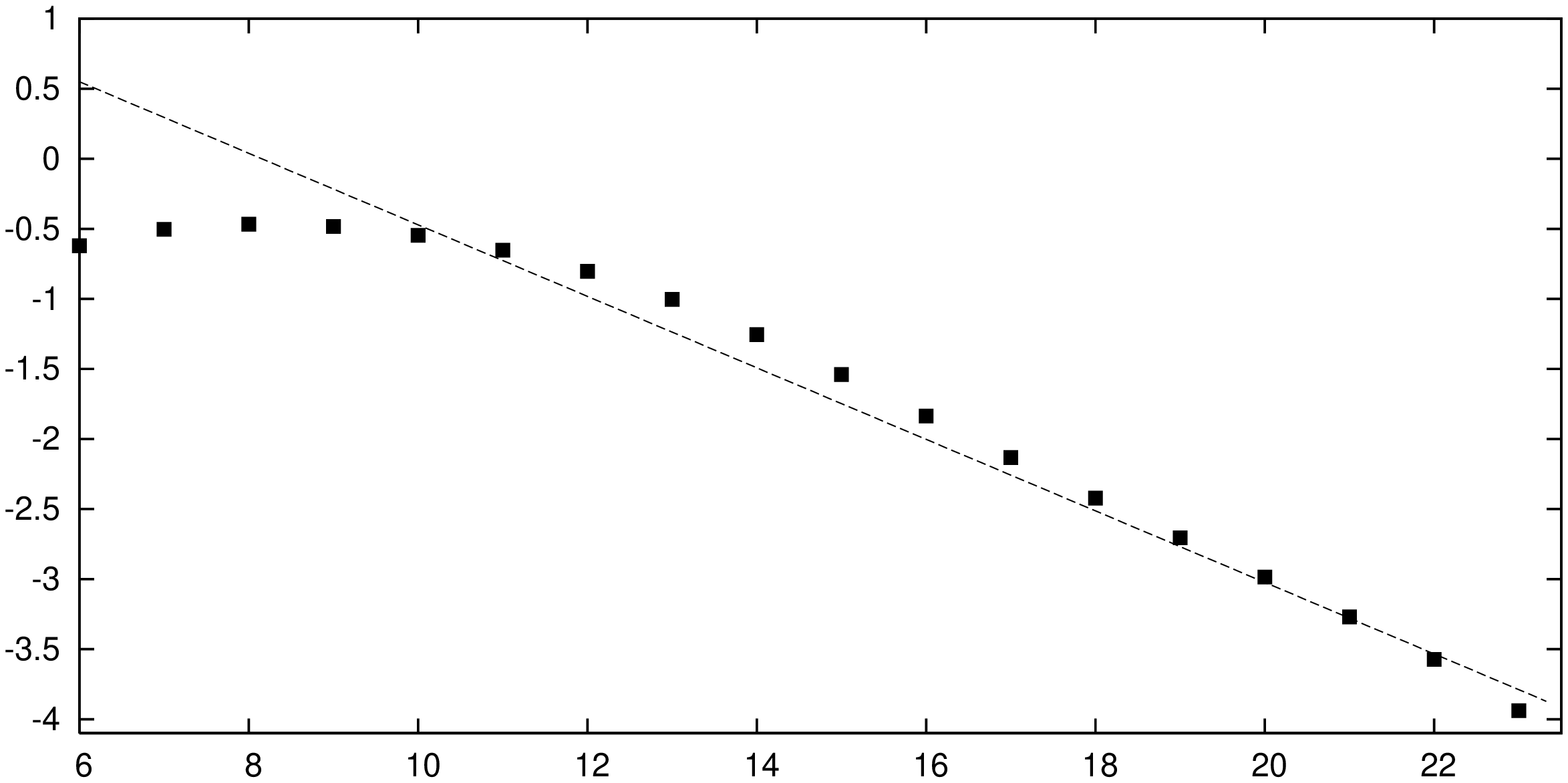}}
\subfigure[$I=1$]{ \includegraphics[scale=0.3,angle=-90]{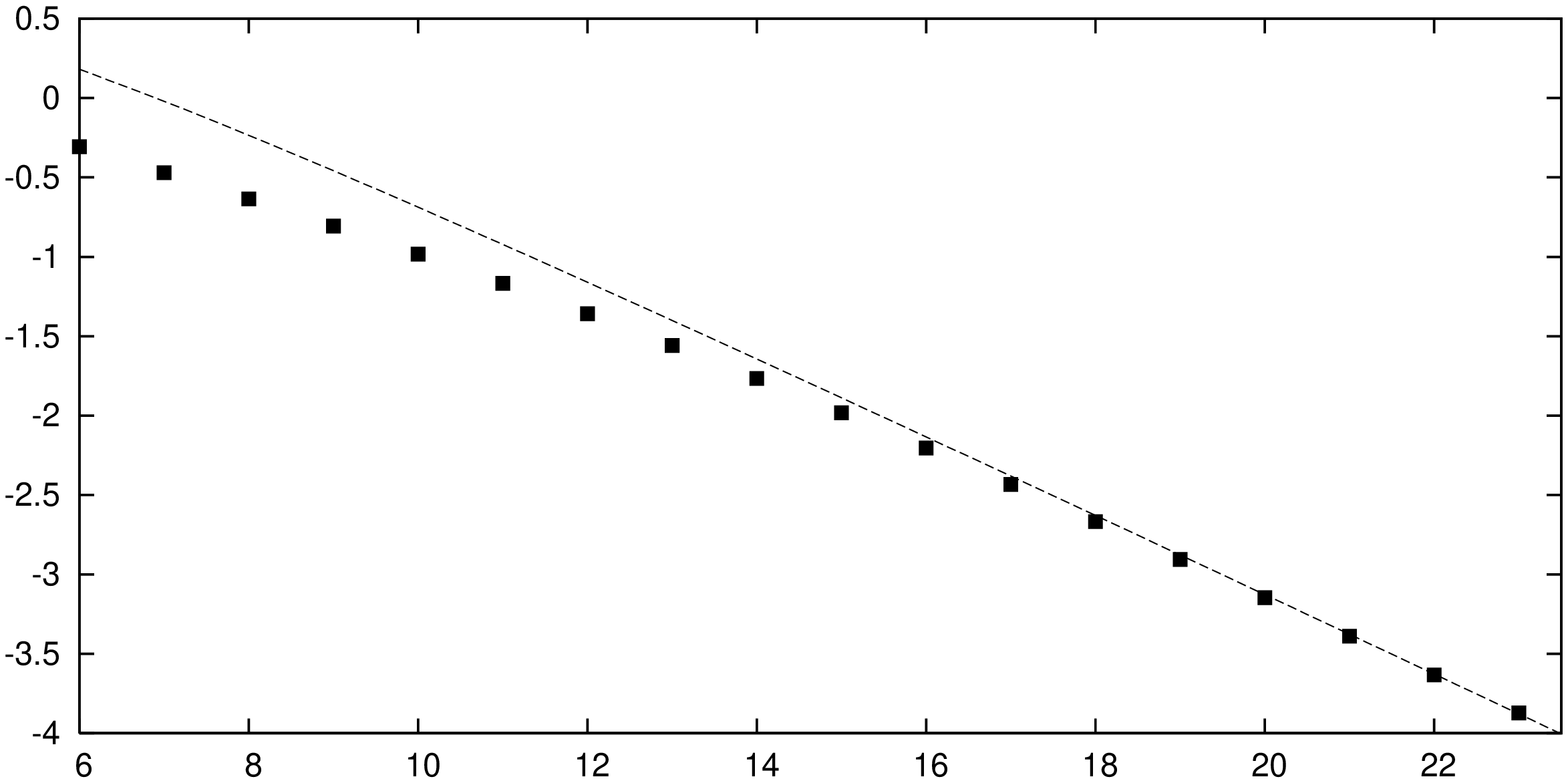}}

\subfigure[$I=2$]{ \includegraphics[scale=0.3,angle=-90]{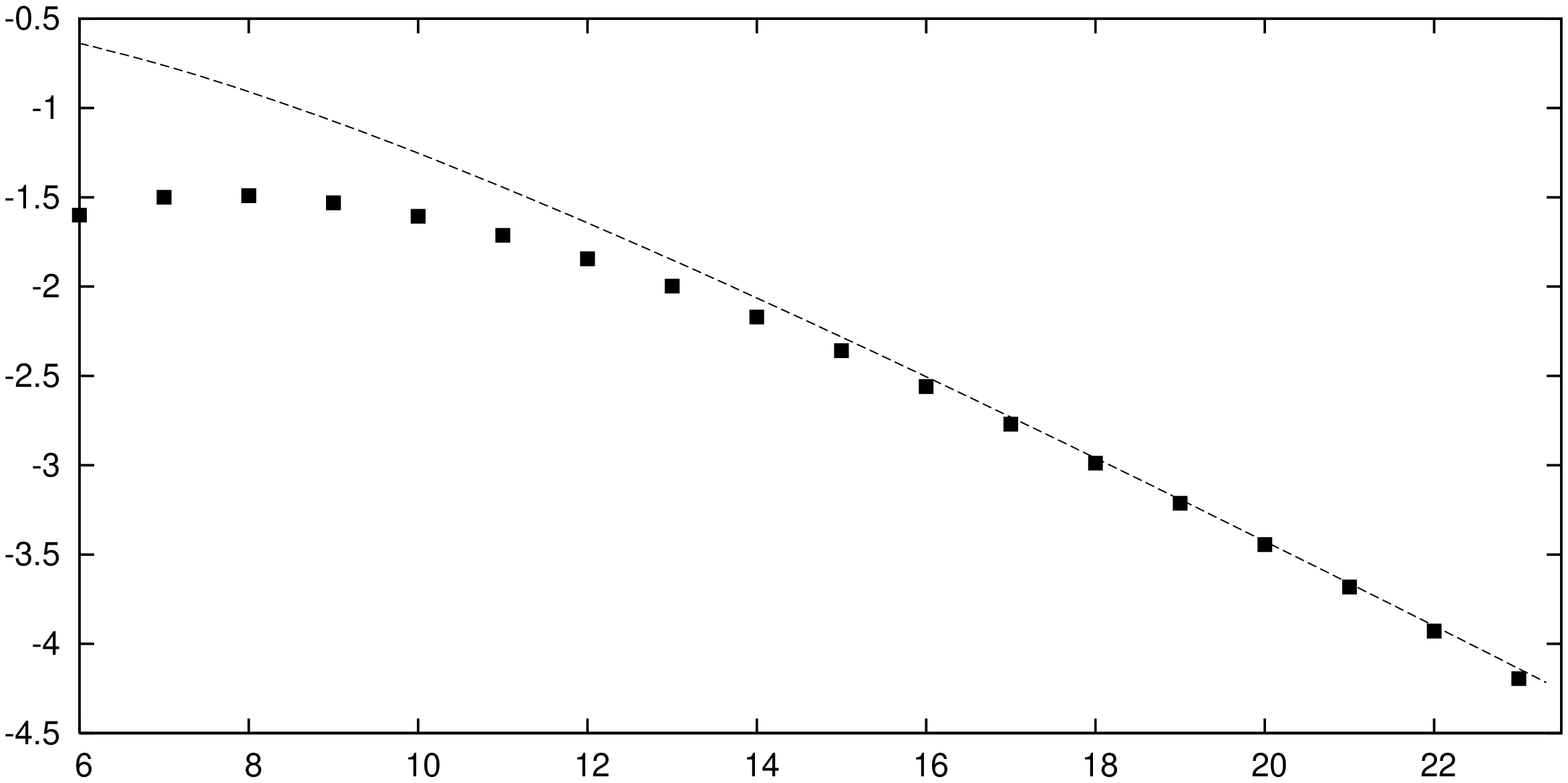}}
\subfigure[$I=3$]{\includegraphics[scale=0.3,angle=-90]{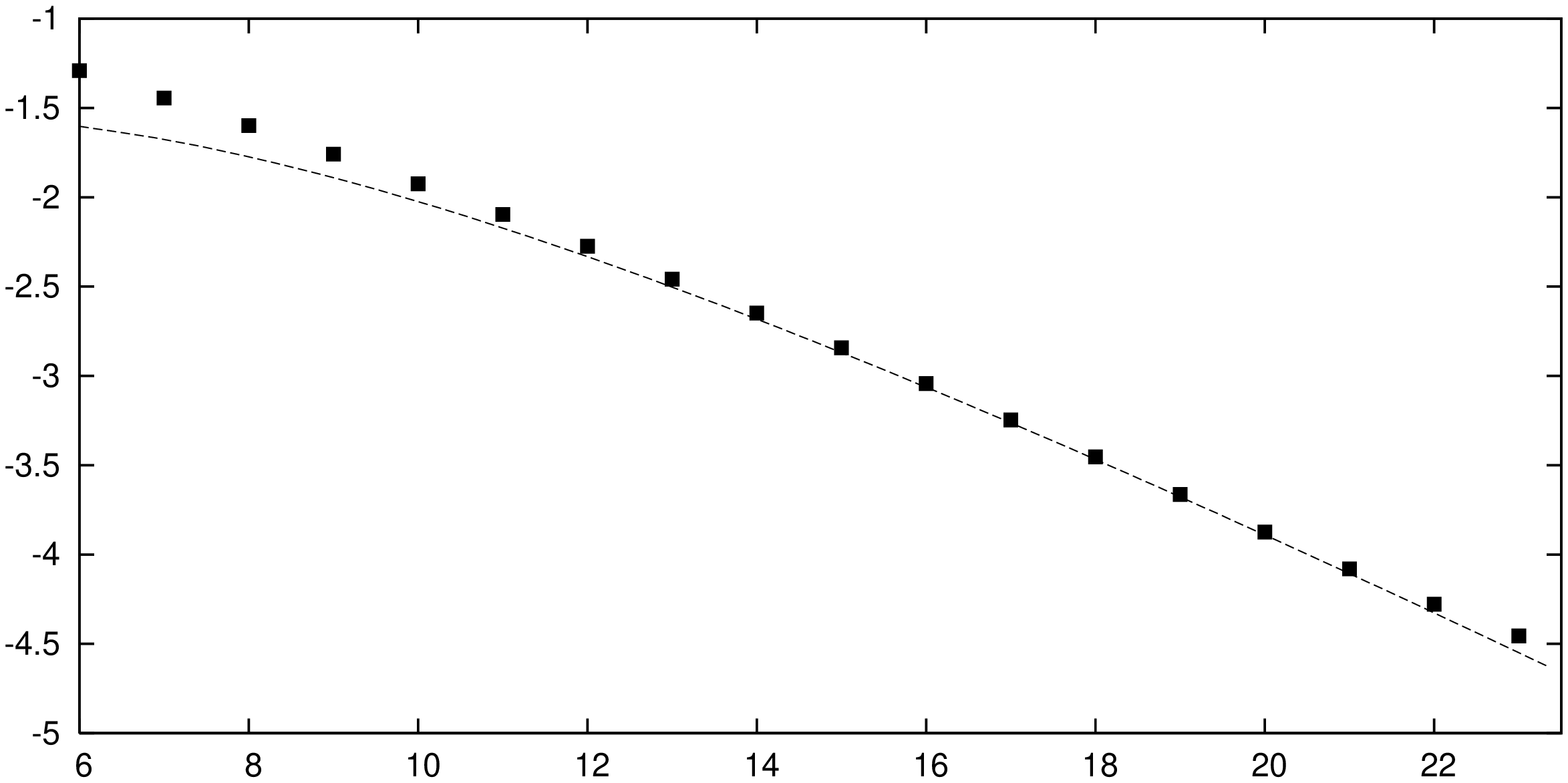}}
\caption{
Finite size corrections to $A_2$ one-particle levels in sectors
$I=0\dots 3$,  $log_{10}\Delta e$ is plotted as a function of the
volume. Dots represent TCSA data, while the lines show the
$\mu$-term corresponding to the $A_1A_1\to A_2$ fusion.
\label{fig:A2_energia}}
\end{figure}

 In figures
\ref{fig:A1_energia} and \ref{fig:A2_energia} 
$\textrm{log}(|\Delta e|)$ is plotted against the prediction of
\eqref{elso_korrekcio} for $l=6..16$ and $l=6..22$. (the sign of
$\Delta e$ was found to be in accordance with 
\eqref{elso_korrekcio} for both $A_1$ and $A_2$)

In the case of $A_1$ perfect agreement is observed for $l=10..18$
in the sectors $I=0$ and $I=1$. For $I=2$ and $I=3$ the energy corrections
become too small and therefore inaccessible to TCSA (note that the
prediction for $I=3$ is of order $10^{-6}$).

In the case of $A_2$ precise agreement is found for $l=14..22$ in all
four sectors.

\subsubsection{$A_5$}

Here we present an interesting calculation that determines the leading mass
corrections of $A_5$. The standard formulas are inapplicable in
this case, because $m_5$ lies above the two-particle
threshold. However, it is instructive to consider the composition of
$A_5$ under the bootstrap principle and to evaluate the $\mu$-term prediction.

There are two relevant fusions
\begin{eqnarray*}
  A_1A_3\to A_5 \quad \textrm{ with }\quad \mu_{13}^5=0.2079 \\
  A_2A_2\to A_5 \quad \textrm{ with }\quad \mu_{22}^5=0.6581 
\end{eqnarray*}

\begin{figure}
  \centering
\psfrag{igazi}{$m_5$}
\psfrag{l}{$l$}
\psfrag{e}{$e$}

\includegraphics[scale=1]{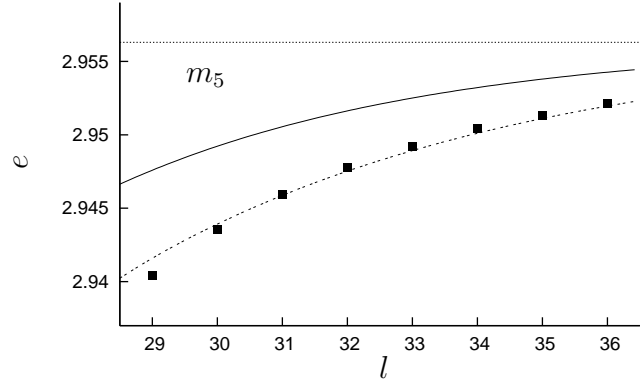}

\caption{
Finite size mass of $A_5$ as a function of the volume. The
squares represent the TCSA data which is compared to the leading
$\mu$-term (solid curve) and the solution of the quantization condition for the
``triple bound state`` $A_1A_1A_1$ (dotted curve). The straight line
shows the infinite volume mass.
\label{fig:A5}}
\end{figure}

Numerical evaluation of L\"uscher's formula for the $\mu$-term shows  that the
contribution of the second fusion is negligible for 
$l>30$. The first fusion on the other hand yields 
a significant discrepancy when compared to TCSA data. 
 This failure 
is connected to the two-particle threshold and it can be explained in
terms of the bound state quantization.
Experience with $A_3$ suggests that one should
first take into account the energy corrections of $A_3$ and consider the
$A_1A_3\to A_5$ fusion afterwards. $A_3$ can be split into $A_1A_1$
leading  to the ''triple bound state'' $A_1A_1A_1\to A_5$. In 
infinite volume one has (see also fig. \ref{fusion} c.)
\begin{equation*}
  \ket{\theta}_5 \sim \ket{\theta-2i\bar{u}_{11}^3,\theta,\theta+2i\bar{u}_{11}^3}_{111}
\end{equation*}
The finite volume realization of this identification is most easily
carried out in the $I=0$ sector with
\begin{equation*}
  \ket{\{0\}}_{5,L} \sim \ket{\{0,0,0\}}_{111,L}
\end{equation*}
Setting up the three-particle Bethe-Yang
equations with rapidities $(iu,0,-iu)$:
\begin{eqnarray*}
  e^{-m_1\sin(u)L} S_{11}(iu)S_{11}(2iu)&=&1\\
   S_{11}(iu)S_{11}(-iu)&=&1\\
  e^{m_1\sin(u)L} S_{11}(-iu)S_{11}(-2iu)&=&1
\end{eqnarray*}
The second equation is automatically satisfied due to unitarity and
real analyticity, whereas
the first and the third are equivalent and they serve as a
quantization condition for $u$. 
The finite volume mass of $A_5$ is given in terms of the solution by
\begin{equation}
\label{A5_pred}
  m_5(l)=2\cos(u)+1
\end{equation}
In the large $L$ limit the infinite volume mass is reproduced by $u\to
2\bar{u}_{11}^3$. Figure  \ref{fig:A5} demonstrates the agreement
between TCSA and the prediction of \eqref{A5_pred}.

The possibility of solving the quantization of the triple bound state
in a moving frame looks very appealing. In  the general case the
rapidities are expected to take the form $(\theta_1+iu,\theta_2,\theta_1-iu)$ where
$\theta_1$ and $\theta_2$ do not necessarily coincide. However, the numerical precision of
our TCSA data was not sufficient to check our predictions.

\subsection{Comparison with the TBA for excited states}

\label{excitedTBA_mu}

As a conclusion of this section \eqref{uminusubar} is compared to the lowest order results of the
TBA approach. The general discussion of excited states TBA equations
in diagonal scattering theories is not available. For simplicity we
restrict ourselves to the Lee-Yang model which was considered in the
original papers \cite{excited_TBA,Bazhanov:1996aq}. 

The exact TBA equations for moving one-particle states read
\begin{eqnarray}
\label{TBA_E}
  E&=&-im(\sinh\theta_0-\sinh\bar{\theta}_0)-
\int_{-\infty}^\infty\frac{d\theta}{2\pi} m\cosh(\theta) L(\theta)\\
\eps(\theta)&=&mR \cosh\theta
+\log\frac{S(\theta-\theta_0)}{S(\theta-\bar{\theta}_0)} 
-(\varphi\star L) (\theta)
\end{eqnarray}
where
\begin{equation*}
 S(\theta)=\frac{\sinh(\theta)+i\sin(\pi/3)}{\sinh(\theta)-i\sin(\pi/3)} \quad\quad
\varphi(\theta)=-i \frac{\partial S(\theta)}{\partial \theta}
\end{equation*}
and
\begin{equation*}
  L(\theta)=\log(1+e^{-\eps(\theta)})\quad\textrm{and}\quad 
(f\star g)(\theta)= \int_{-\infty}^\infty\frac{d\theta'}{2\pi}
f(\theta-\theta') g(\theta')
\end{equation*}
Here the volume is denoted by $R$ to avoid confusion with $L(\theta)$.
The complex rapidity $\theta_0$ satisfies the consistency equation 
\begin{equation}
\label{TBA_consist}
  \eps(\theta_0)=mR \cosh\theta_0+i\pi
-\log(S(2i\textrm{Im}\theta_0))-(\varphi\star L) (\theta)=i(2n+1)\pi
\end{equation}
The convolution term in \eqref{TBA_consist} can be neglected and one
obtains $\textrm{Im}\theta_0=\pi/6+\delta$ where 
$\delta$ is exponentially small. To zeroth order one also has
\begin{equation*}
mR\cosh(\textrm{Re}  \theta_0)=(4n+(1-\textrm{sign}\delta))\pi
\end{equation*}
which is the ordinary one-particle quantization condition with
$I=2n+\frac{1}{2}(1-\textrm{sign}\delta)$. 
Neglecting the contribution
of the integral in \eqref{TBA_E} and substituting $\theta_0=\theta+iu$
one has $E=2m\sin(u)\cosh(\theta)$. This is exactly the energy of an $AA$
bound state with the imaginary rapidities $\theta\pm iu$. Separating
the real and imaginary parts of \eqref{TBA_consist} and still neglecting
the convolution term (which is responsible for the F-term) one obtains equations \eqref{u_egyenlete} and
\eqref{kvantalasi_1}, thus proving the consistency of the two approaches.

\section{Multi-particle states}

\label{exponential:multi}

\subsection{Bethe-Yang quantization in the bound state picture}


Let us consider a scattering state
$\ket{\{I,I_1,\dots,I_n\}}_{cb_1\dots b_n,L}$ composed of $n+1$
particles, the first one being  $A_c$. The energy of this state is calculated as
\begin{equation}
\label{multi_elsotipp}
  E=\sum_{j=1}^n m_{i_j}
  \cosh(\bar{\theta}_j)+\dots
\end{equation}
where $(\bar{\theta},\bar{\theta}_1,\dots,\bar{\theta}_n)$ is
the solution of the Bethe-Yang equations
\begin{equation}
  \label{Bethe}
Q_j(\theta_1,\dots,\theta_n)=m_{i_j}\sinh(\theta_j)L+\sum_{k\ne
  j}\delta _{i_ji_k}(\theta_j-\theta_k) =2\pi I_j\quad \quad j=1\dots n
\end{equation}
 We determine the leading part of
the $\mu$-term associated to \eqref{multi_elsotipp} by considering $A_c$ as an $A_aA_a$ bound state
\textit{inside the multi-particle state}. Therefore we write
\begin{equation}
\label{multi_azonositas}
   \ket{\{I,I_1,\dots,I_n\}}_{cb_1\dots b_n,L}
\sim
 \ket{\{I/2,I/2,I_1,\dots,I_n\}}_{aab_1\dots b_n,L}
\end{equation}
The energy is then determined by analytic continuation
of the $n+2$ particle Bethe-Yang equations. They read
\begin{eqnarray}
\label{multi_Bethe_1}
  e^{-m_a\sin(u)\cosh(\theta)L} e^{im_a \cos(u)\sinh(\theta)L}
  S_{aa}(2iu) \prod_{j=1}^n S_{ab_j}(\theta+iu-\theta_j) &=&1\\
\label{multi_Bethe_2}
  e^{m_a\sin(u)\cosh(\theta)L} e^{im_a \cos(u)\sinh(\theta)L}
  S_{aa}(-2iu) \prod_{j=1}^n S_{ab_j}(\theta-iu-\theta_j) &=&1\\
\label{multi_Bethe_3}
e^{im_{b_j}\sinh(\theta_j)L} S_{ab_j}(\theta_j-\theta-iu)
S_{ab_j}(\theta_j-\theta+iu) \mathop{\prod_{k=1}^n}_{k\ne j}
S_{b_jb_k}(\theta_j-\theta_k) &=& 1 
\end{eqnarray}
The ordinary $n+1$ particle Bethe-equations are reproduced in
the $L\to\infty$ limit by multiplying \eqref{multi_Bethe_1} and
\eqref{multi_Bethe_2} and making use of the bootstrap equation
\begin{equation*}
  S_{cb_j}(\theta)=S_{ab_j}(\theta+i\bar{u}_{ac}^a)S_{ab_j}(\theta-i\bar{u}_{ac}^a)
\end{equation*}

We now proceed similar to the previous section and
derive a formula for the leading correction. 
The shift in the imaginary part of the rapidity can be calculated by
making use of \eqref{multi_Bethe_1} and \eqref{S-matrix_polus} as
\begin{equation}
\label{multi-particle_u}
\Delta u=  u-\bar{u}_{ac}^a=\frac{\left(\Gamma_{aa}^c\right)^2}{2}
  e^{-\mu\cosh(\bar{\theta})L} e^{im_c\sinh(\bar{\theta})L/2}
  \prod_{j=1}^n S_{ab_j}(\bar{\theta}+i\bar{u}_{ac}^a-\bar{\theta}_j)
\end{equation}
Multiplying \eqref{multi_Bethe_1} and \eqref{multi_Bethe_2}
\begin{eqnarray}
\label{mB1}
 e^{i2m_a \cos(u)\sinh(\theta)L}
 \prod_{j=1}^n
 S_{ab_j}(\theta-iu-\theta_j)S_{ab_j}(\theta+iu-\theta_j) &=&1\\
\label{mB2}
e^{im_{b_j}\sinh(\theta_j)L} S_{ab_j}(\theta_j-\theta-iu)
S_{ab_j}(\theta_j-\theta+iu) \mathop{\prod_{k=1}^n}_{k\ne j}
S_{b_jb_k}(\theta_j-\theta_k) &=& 1 
\end{eqnarray}
Let us define
\begin{equation*}
  S_{ab_j}(\theta-iu-\theta_j)S_{ab_j}(\theta+iu-\theta_j)\approx 
S_{cb_j}(\theta-\theta_j)e^{i \Delta u \bar{\varphi}_{cb_j}(\theta-\theta_j)}
\end{equation*}
where
\begin{equation*}
  \bar{\varphi}_{cb_j}(\theta)=
i\varphi_{cb_j}(\theta+i\bar{u}_{ac}^a)-i\varphi_{cb_j}(\theta-i\bar{u}_{ac}^a)
\quad\quad\textrm{with}\quad\quad \varphi_{ab}(\theta)=\delta_{ab}'(\theta)
\end{equation*}
Using $2m_a \cos(u)\approx m_c-2\mu_{aa}^c \Delta u$ the logarithm
of \eqref{mB1} and \eqref{mB2} can be written as
\begin{eqnarray*}
Q_0(\theta,\theta_1,\dots,\theta_n)  &=&\left(2\mu_{aa}^c
   \sinh(\theta)L- \sum_{j=1}^n\bar{\varphi}_{cb_j}(\theta-\theta_j)\right) \Delta u \\
Q_j(\theta,\theta_1,\dots,\theta_n)  &=& 
\bar{\varphi}_{cb_j}(\theta-\theta_j)  \Delta u 
\end{eqnarray*}
The lhs. can be expanded around the $n+1$ particle solution
$(\bar{\theta},\bar{\theta}_n,\dots,\bar{\theta}_n)$ 
to arrive at
\begin{equation}
\label{rapieltolodas}
  \begin{pmatrix}
    \theta- \bar{\theta} \\ \theta_1- \bar{\theta}_1 \\ 
\vdots \\ \theta_n- \bar{\theta}_n
  \end{pmatrix}
=
\left(\mathcal{J}^{(n+1)}\right)^{-1} 
\begin{pmatrix}
  2\mu_{aa}^c
   \sinh(\bar{\theta})L-\sum_{j=1}^n
   \bar{\varphi}_{cb_j}(\bar{\theta}-\bar{\theta}_j)\\
\bar{\varphi}_{cb_1}(\bar{\theta}-\bar{\theta}_1) \\
\vdots \\
\bar{\varphi}_{cb_n}(\bar{\theta}-\bar{\theta}_n) 
\end{pmatrix}\Delta u 
\end{equation}
where
\begin{equation*}
  \mathcal{J}^{(n+1)}_{kl}=\frac{\partial Q_k}{\partial \theta_l}
\end{equation*}

The final result for the energy correction reads 
\begin{equation}
\label{multi_mu}
  \Delta E=-2\mu_{aa}^c  \cosh(\bar{\theta}) \Delta u +
  \begin{pmatrix}
    m_c\sinh(\bar{\theta}) \\ m_{b_1}\sinh(\bar{\theta}_1) \\
    \vdots \\ m_{b_n}\sinh(\bar{\theta}_n)
  \end{pmatrix}
\left(\mathcal{J}^{n+1}\right)^{-1} 
\begin{pmatrix}
  2\mu_{aa}^c
   \sinh(\bar{\theta})L-\sum_{j=1}^n
   \bar{\varphi}_{cb_j}(\bar{\theta}-\bar{\theta}_j)\\
\bar{\varphi}_{cb_1}(\bar{\theta}-\bar{\theta}_1) \\
\vdots \\
\bar{\varphi}_{cb_n}(\bar{\theta}-\bar{\theta}_n) 
\end{pmatrix}\Delta u 
\end{equation}
with $\Delta u$ given by \eqref{multi-particle_u}.

Based on the previous section it is expected that there is a similar
contribution for every fusion leading to each one of the constituents of
the multi-particle state. 

We wish to remark that in the case of the Lee-Yang model  it is possible to derive \eqref{multi_mu} from
the excited state TBA, along the lines of subsection
\ref{excitedTBA_mu}. This carries over to other models with
diagonal scattering, provided that the excited state TBA equations are
available. The point of our calculation is, that the $\mu$-term can
be obtained without any reference to the TBA equations. Moreover, our
 arguments
can be applied directly to form factors as well (see section \ref{exponential:fv_ff}). 

\subsection{Multi-particle states -- Numerical analysis}

\begin{figure}
  \centering
\subfigure[$\ket{\{1,0\}}_{13}$] { \includegraphics[scale=0.3,angle=-90]{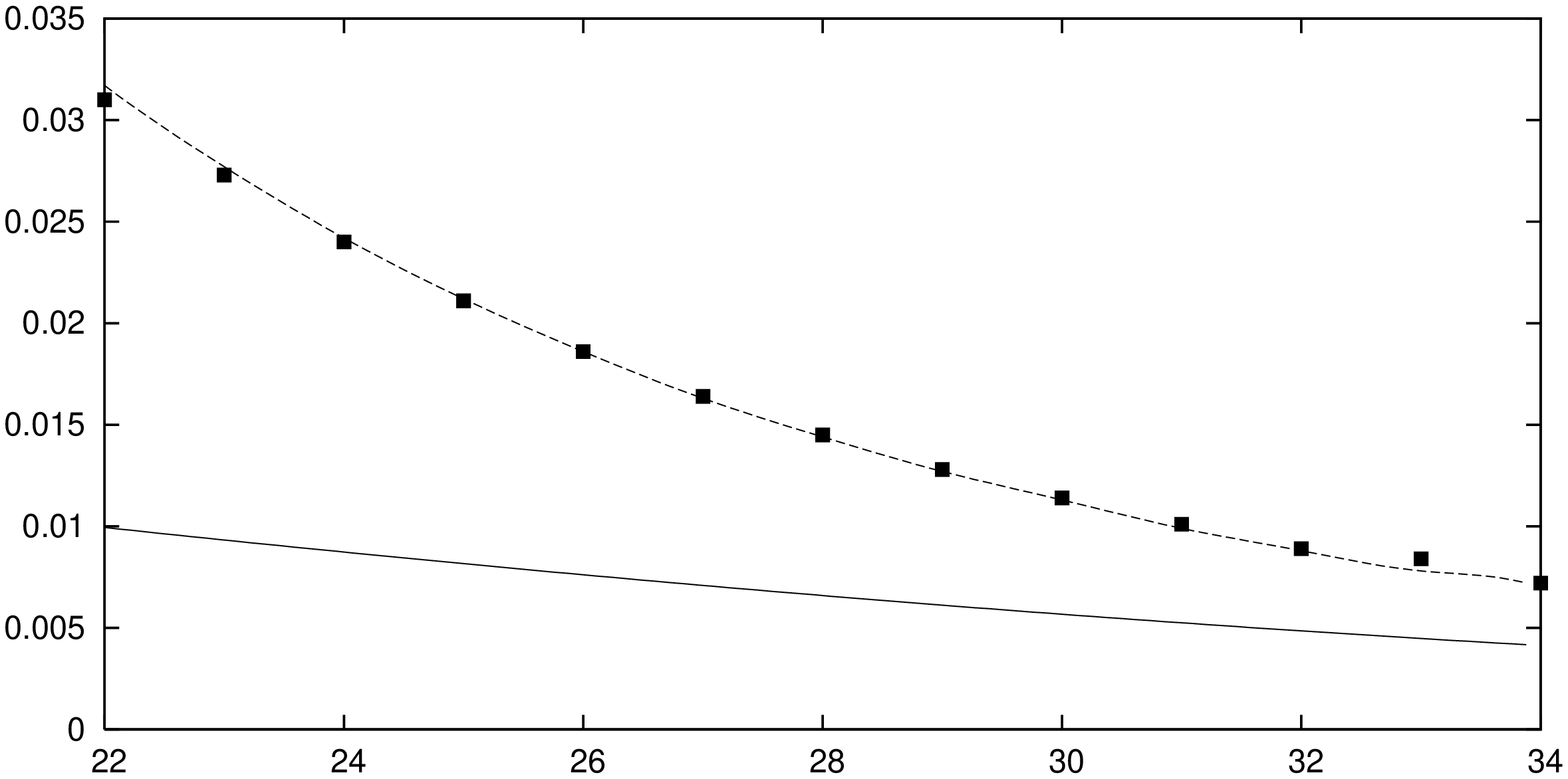}}
\subfigure[$\ket{\{0,2\}}_{13}$]{ \includegraphics[scale=0.3,angle=-90]{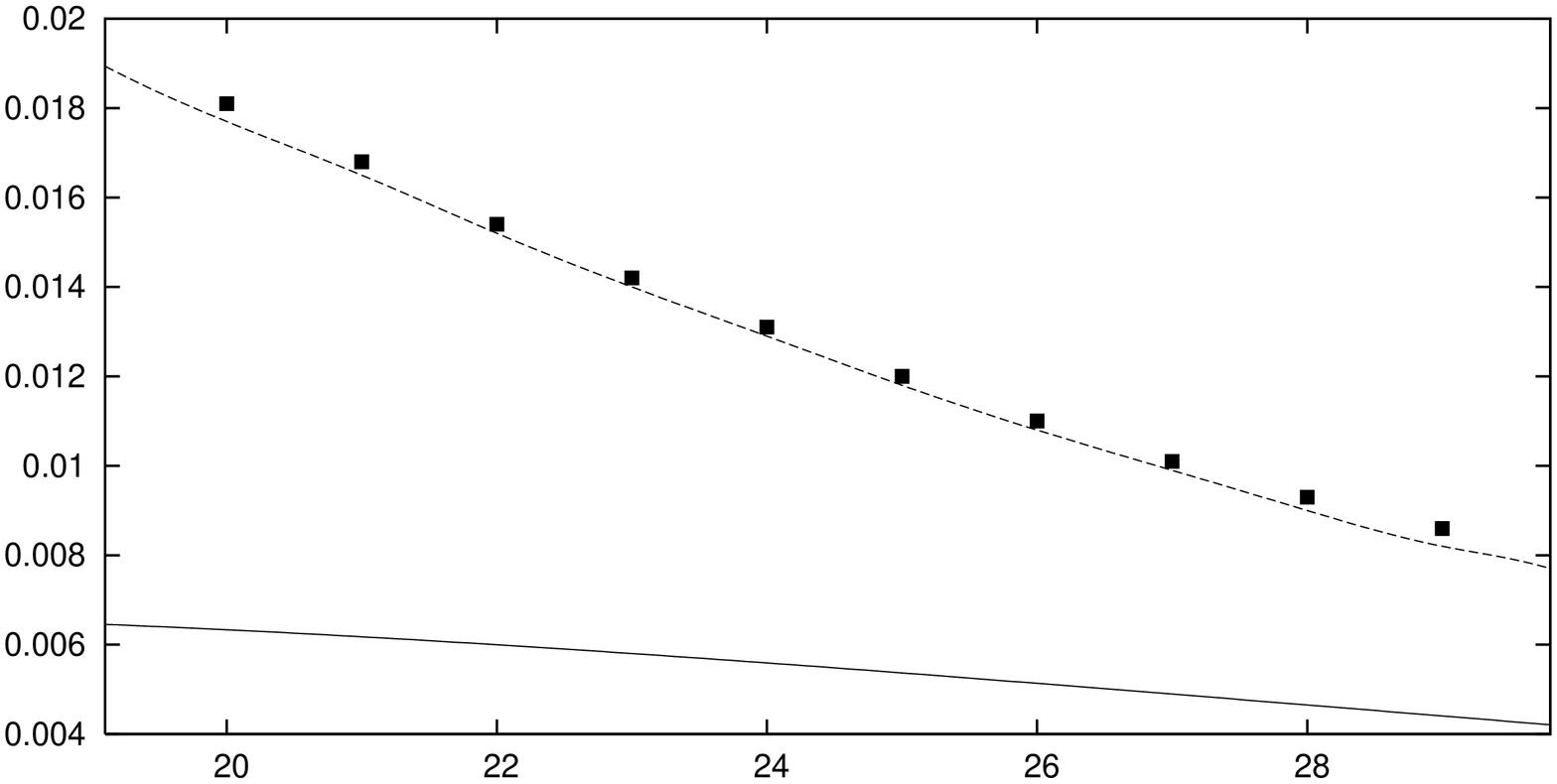}}

\subfigure[$\ket{\{0,1\}}_{13}$]{ \includegraphics[scale=0.3,angle=-90]{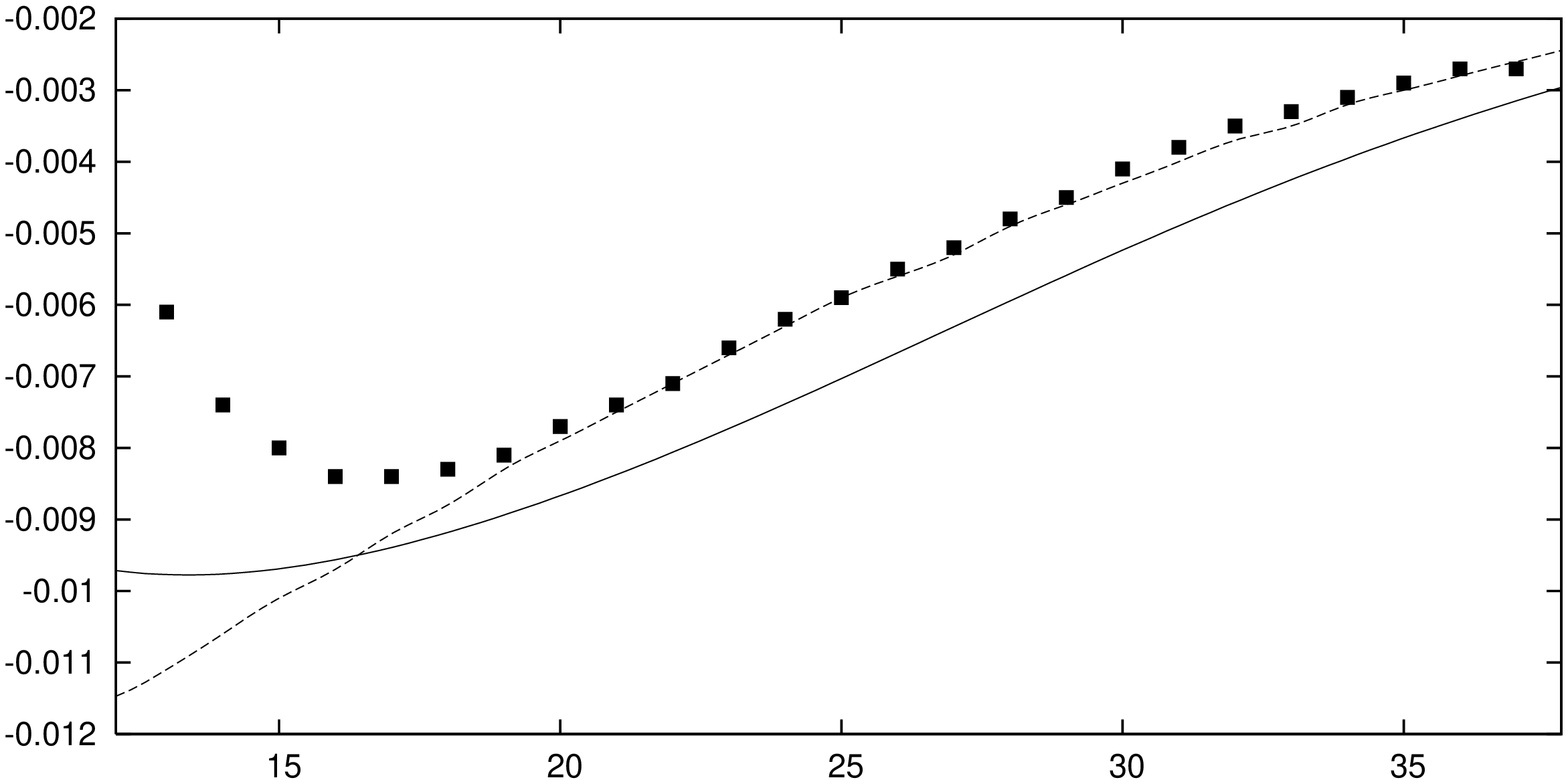}}
\subfigure[$\ket{\{1,1\}}_{13}$]{\includegraphics[scale=0.3,angle=-90]{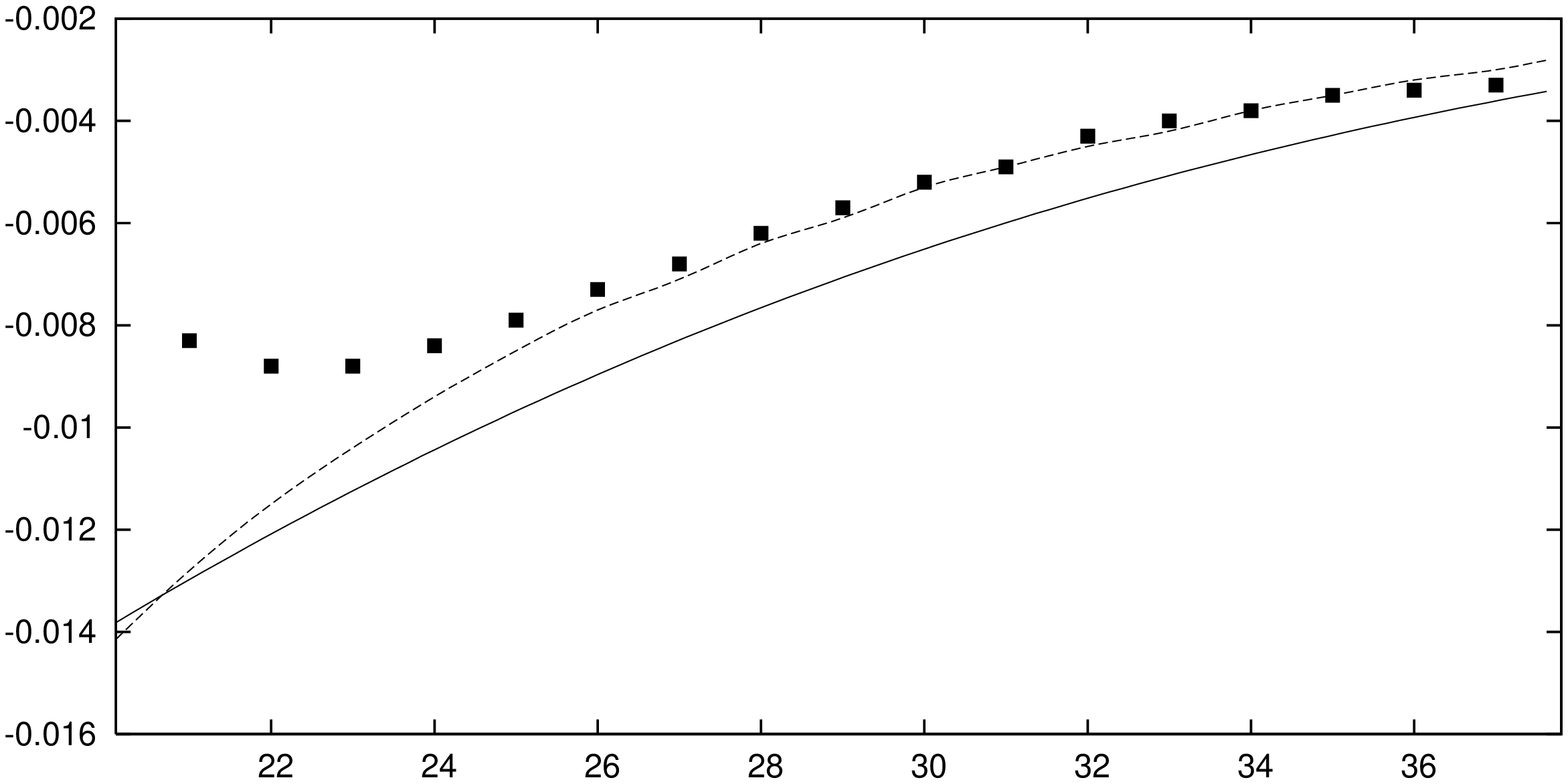}}
\caption{
Finite size corrections to $A_1A_3$ scattering states as a function of
the volume. Dots represent
TCSA data, the solid line shows the $\mu$-term corresponding to the
$A_1A_1\to A_3$ fusion. The dotted lines are obtained by the exact solution
of the quantization condition for the $A_1A_1A_1$ three-particle system.
 \label{fig:A1A3_enkorr}}
\end{figure}

We first consider finite size corrections to $A_1A_3$ states. They are
not the lowest lying two-particle states in the spectrum, but they possess
the largest $\mu$-term which is connected to the $A_1A_1\to A_3$
fusion. Given a particular state $\ket{\{I_1,I_3\}}_{13,L}$  the following procedure
is performed at each value of the volume:
\begin{itemize}
\item The two-particle Bethe-Yang equation for
  $\ket{\{I_1,I_3\}}_{13,L}$ is solved and the
  $\mu$-term is calculated according to \eqref{multi_mu}.
\item The exact three-particle Bethe-Yang equation is solved for $\ket{\{I_1,I_3/2,I_3/2\}}_{111,L}$
\end{itemize}
The results for different $A_1A_3$ levels are shown in figure
\ref{fig:A1A3_enkorr}. The situation is similar to 
 the case of the $A_3$ one-particle levels: 
the  bound state quantization yields a remarkably accurate
prediction, whereas
the single $\mu$-term
prediction only becomes correct in the $L \to\infty$ limit.

In table \ref{tab:en_A1A3_1} we present a numerical example for the
dissociation of the bound state inside the two-particle
state. In this case an $A_1A_3$ state turns into a conventional $A_1A_1A_1$
three-particle state at $l_c\approx 30$. 

\begin{figure}
  \centering
\subfigure[$\ket{\{-0.5,1.5\}}_{11}$]{  \includegraphics[scale=0.3,angle=-90]{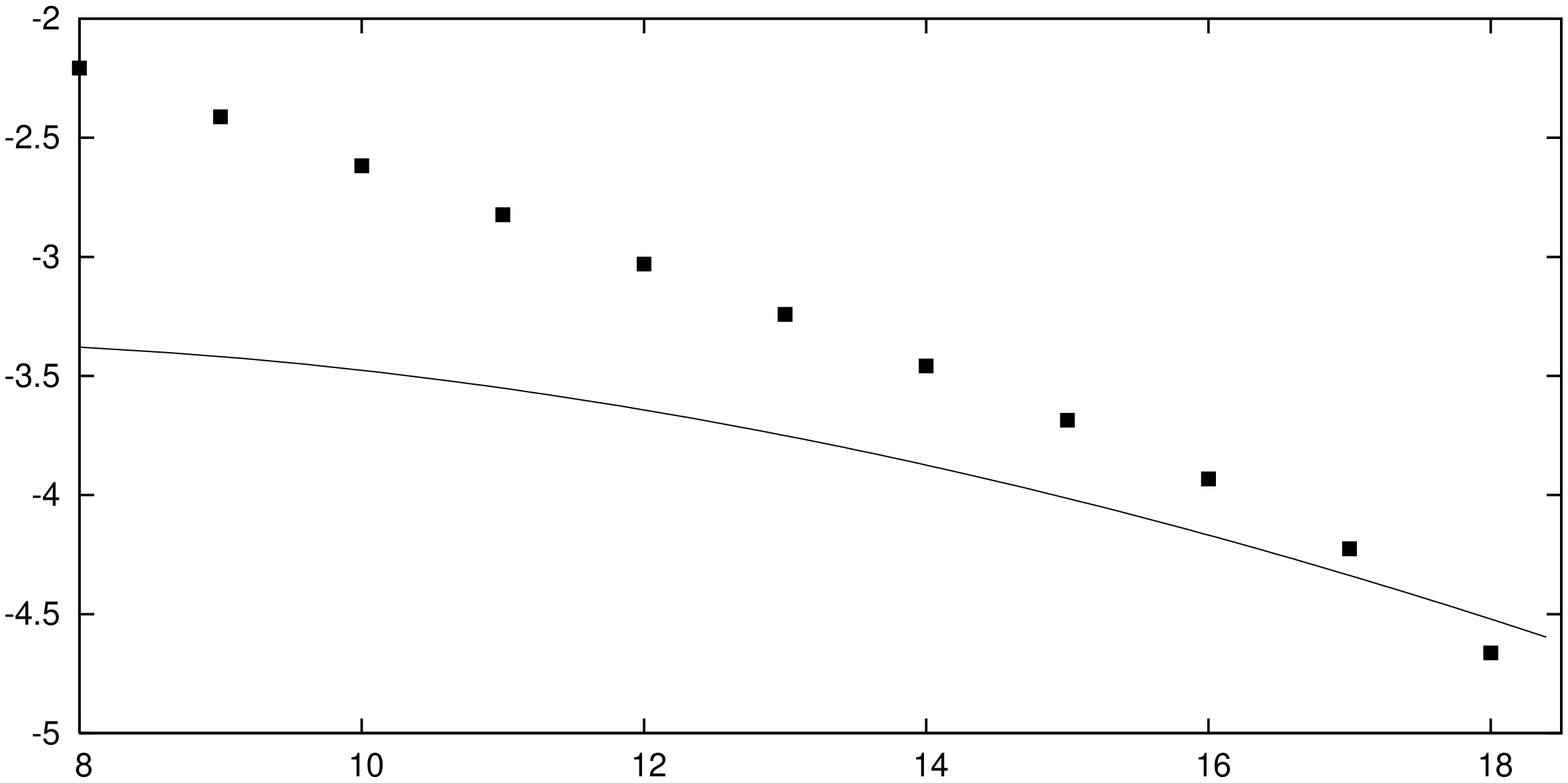}}
\subfigure[$\ket{\{-0.5,1.5\}}_{11}$]{ \includegraphics[scale=0.3,angle=-90]{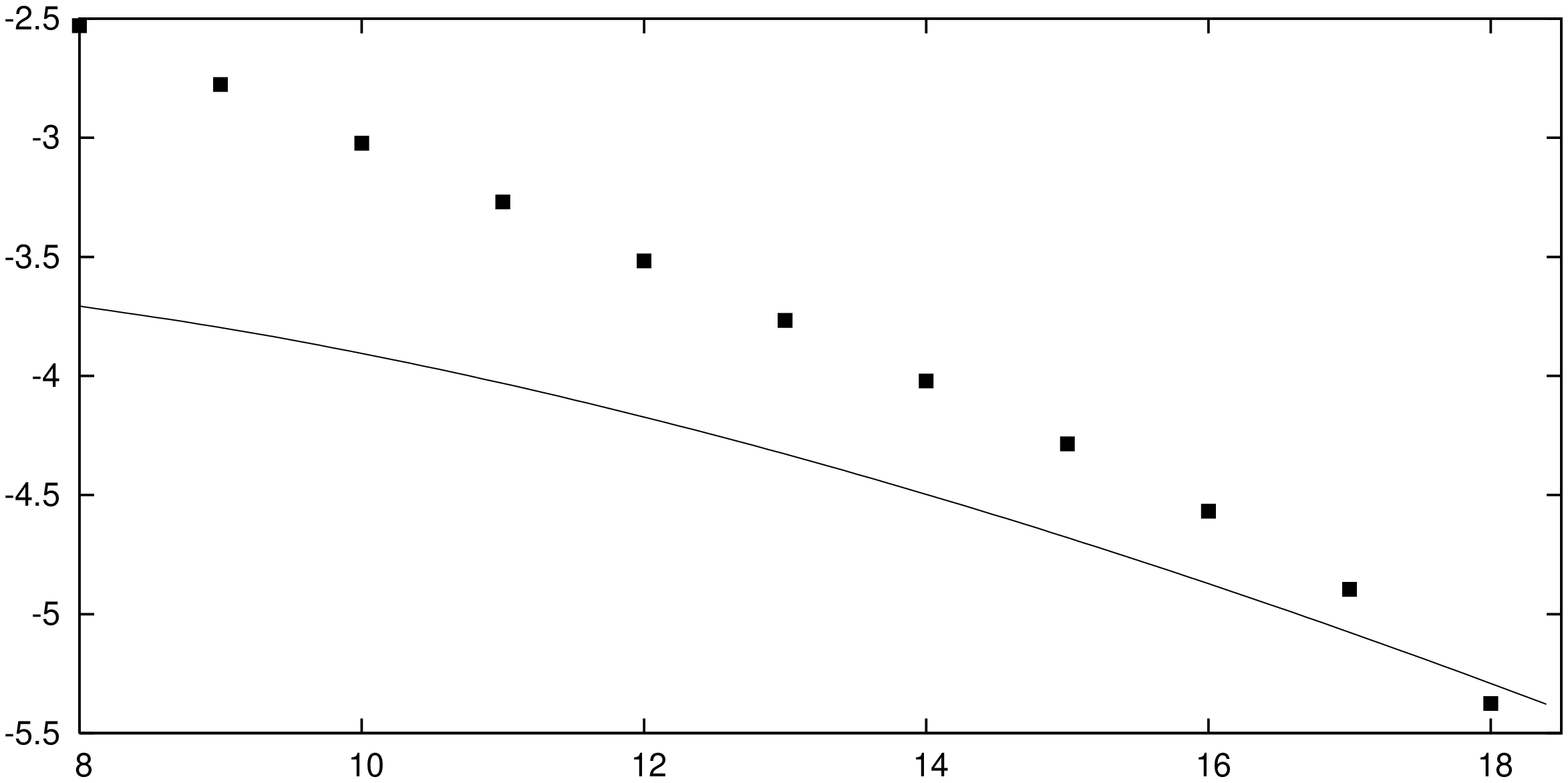}}

\subfigure[$\ket{\{0.5,1.5\}}_{11}$]{ \includegraphics[scale=0.3,angle=-90]{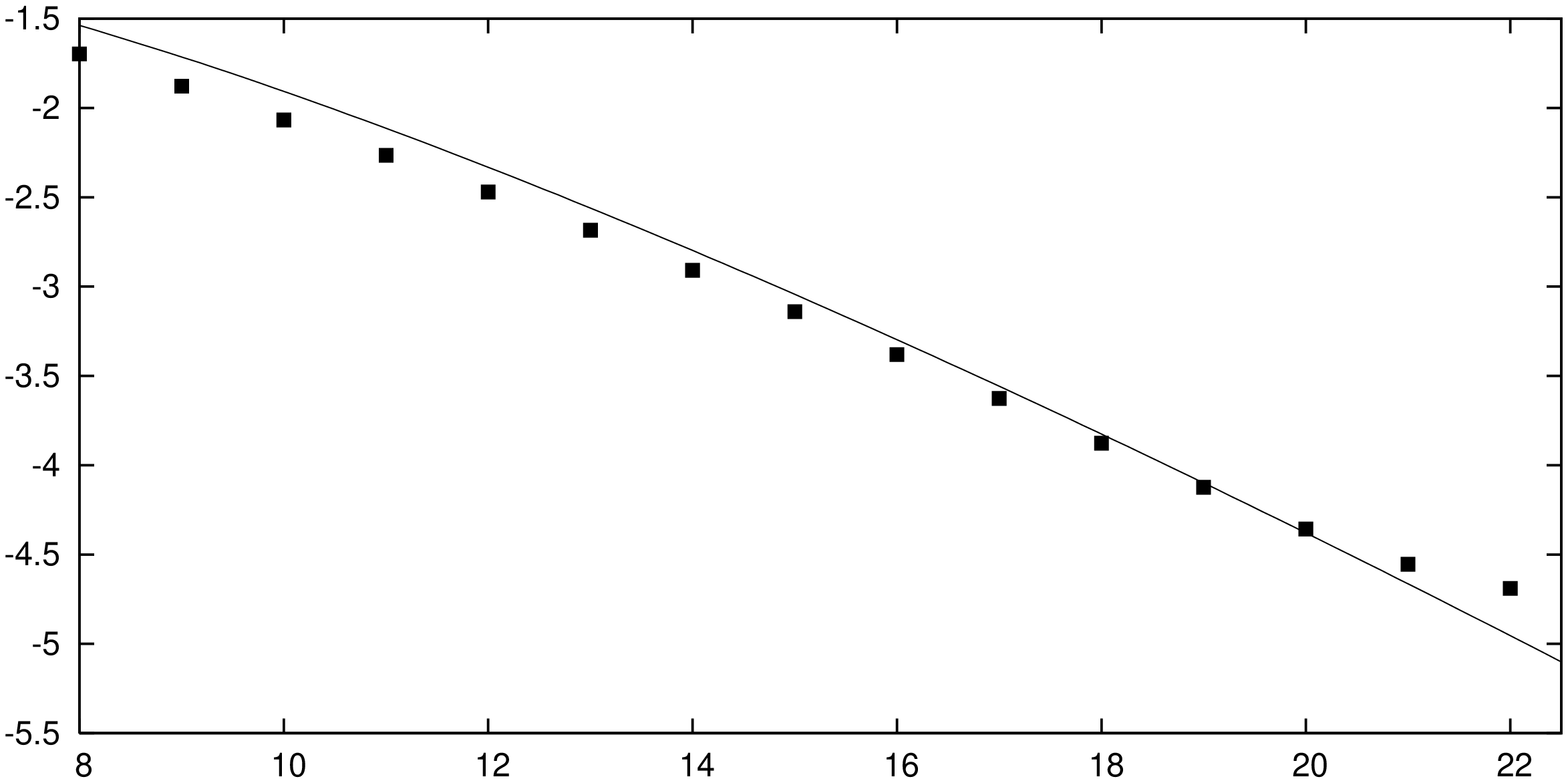}}
\subfigure[$\ket{\{0.5,2.5\}}_{11}$]{\includegraphics[scale=0.3,angle=-90]{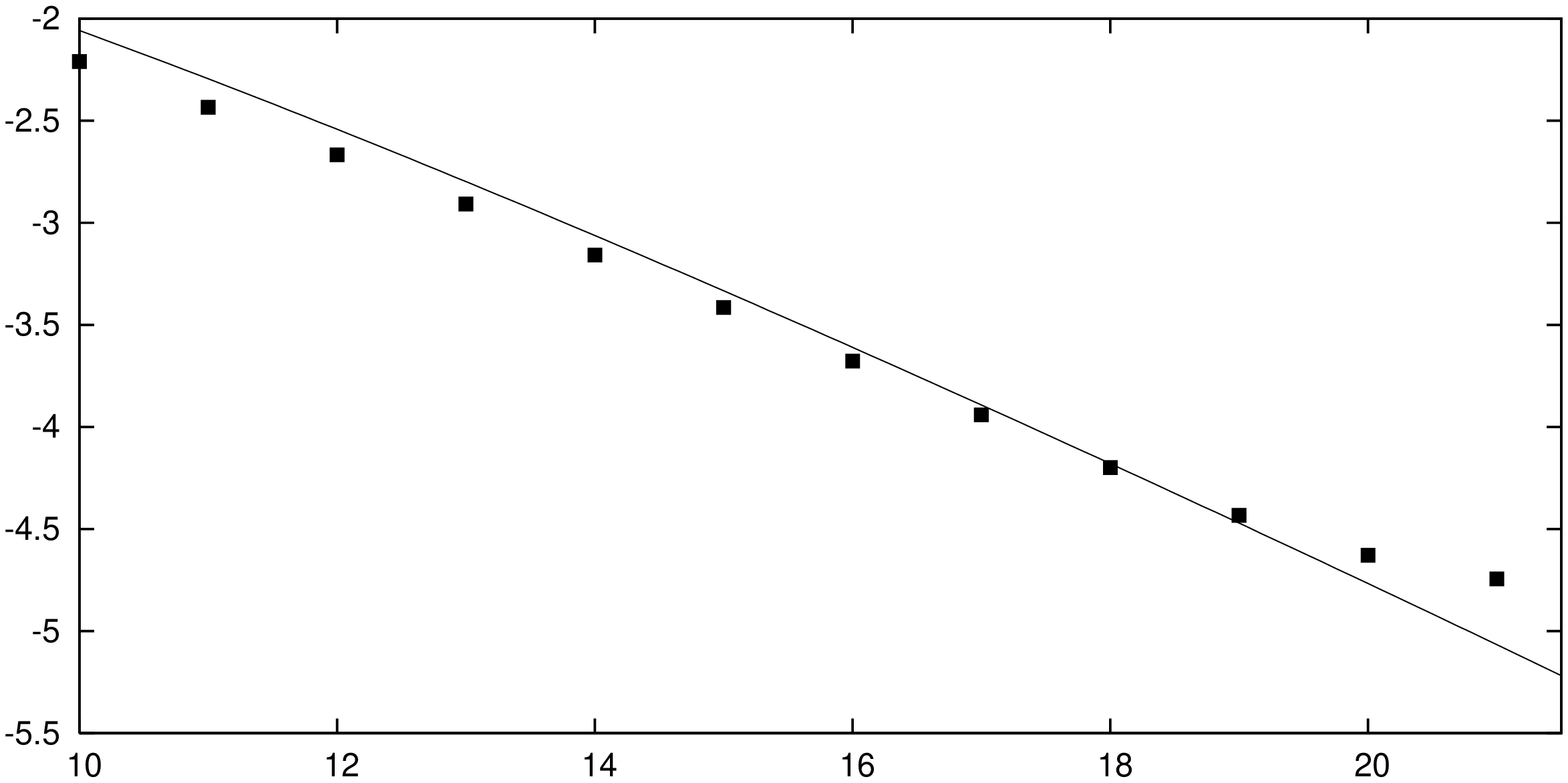}}
\caption{
Finite size corrections to $A_1A_1$ scattering states,  $log_{10}\Delta e$ is plotted as a function of the
volume. Dots represent
TCSA data, while the solid line show the sum of the two $\mu$-terms corresponding to the
$A_1A_1\to~A_1$ fusions. 
\label{fig:A1A1_enkorr}}
\end{figure}

\begin{figure}
  \centering
\subfigure[$\ket{\{0,1\}}_{12}$]{  \includegraphics[scale=0.3,angle=-90]{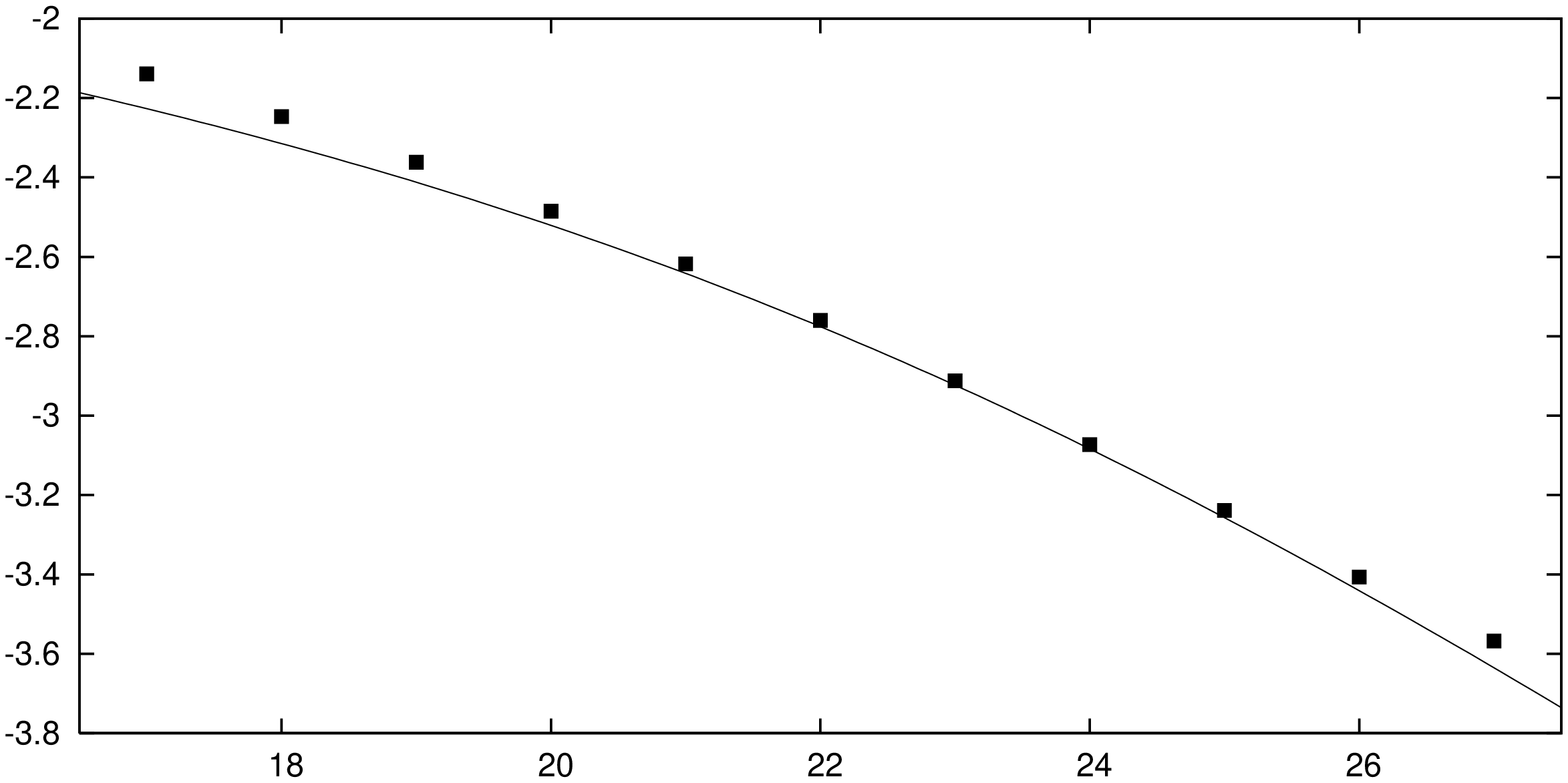}}
\subfigure[$\ket{\{1,1\}}_{12}$]{ \includegraphics[scale=0.3,angle=-90]{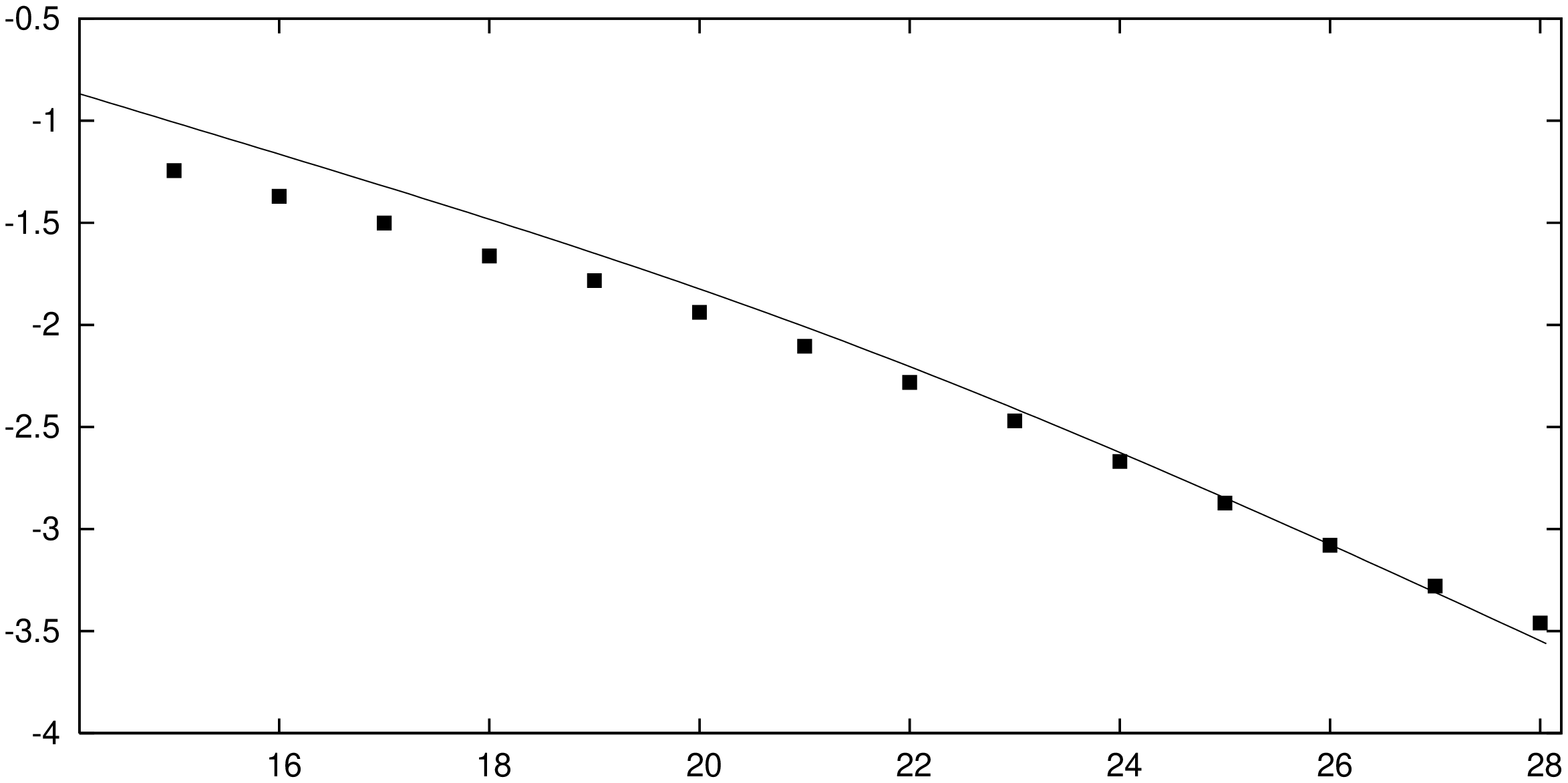}}
\caption{
Finite size corrections to $A_1A_2$ scattering states,  $log_{10}\Delta e$ is plotted as a function of the
volume. Dots represent
TCSA data, while the solid line show the $\mu$-term corresponding to the $A_1A_1\to A_2$ fusion. 
\label{fig:A1A2_enkorr}}
\end{figure}

Finite size corrections to $A_1A_1$ and $A_1A_2$
states are also investigated, the leading $\mu$-term given by the fusions $A_1A_1\to A_1$ and
$A_1A_1\to A_2$, respectively. In the former case we 
calculate separately the contribution associated to both $A_1$ particles and
 add them to get the total correction.
Results are exhibited in figures
\ref{fig:A1A1_enkorr} and \ref{fig:A1A2_enkorr} and formula
\eqref{multi_mu} is verified in both cases. 

\section{Finite volume form factors}

\label{exponential:fv_ff}

The connection between finite volume and infinite
volume form factors was derived in section \ref{fv_ff} as
\begin{eqnarray}
\nonumber
_{j_1\dots j_m,L}  \bra{\{I'_1,\dots,I'_m\}}
\mathcal{O}(0,0)\ket{\{I_1,\dots,I_n\}}_{i_1\dots i_n,L}=\\
\label{fvff}
\frac{F^{\mathcal{O}}(\bar{\theta}'_m+i\pi,\dots,\bar{\theta}'_1+i\pi,\bar{\theta}_1,\dots,\bar{\theta}_n)_{j_m\dots j_1i_1\dots i_n}}
{\sqrt{
\rho_{i_1\dots i_n}(\bar{\theta}_1,\dots,\bar{\theta}_n)
\rho_{j_1\dots j_m}(\bar{\theta}'_1,\dots,\bar{\theta}'_m)}}
+ O(e^{-\mu' L})
\end{eqnarray}
where the rapidities $\bar{\theta}$ are solutions of the corresponding Bethe-Yang
equations. Here we assume for simplicity, that there are no disconnected
terms present, ie. it is supposed that $\bar{\theta}_j\ne \bar{\theta}'_k$ whenever
$i_j=i_k$. 

Based on general arguments
it was shown in subsection \ref{estimate_of_residual} that $\mu'\ge \mu$ where $\mu$ is
determined by the pole of the S-matrix closest to the physical
line. A systematic finite volume perturbation theory (L\"uscher's method
applied to form factors)
is not available. However, it is expected that
the actual value of $\mu'$ depends on what diagrams contribute
to the form factor in question. Apart from the insertion of the local
operator they coincide with the diagrams determining the finite size
corrections of the multi-particle state. Therefore $\mu'$ is
associated to the bound state structure of the constituents 
of the multi-particle state. 

In this section we show that the leading correction term can be
obtained by the bound state quantization.

\subsection{Elementary one-particle form factors}

\eqref{fvff} yields a simple prediction for the elementary one-particle form factor:
\begin{equation}
\label{egyreszecskes_elsotipp}
  F^{\mathcal{O}}_c(I,L)\equiv\bra{0}\mathcal{O}(0,0)\ket{\{I\}}_{c,L}=\frac{F_c^\mathcal{O}}{\sqrt{EL}}+
O(e^{-\mu' L})
\end{equation}
where $E$
is the one-particle energy, and
$F_c^\mathcal{O}=F_c^\mathcal{O}(\theta)$ is the infinite volume
one-particle form factor, which is constant by 
Lorentz symmetry. 

The $\mu$-term associated to \eqref{egyreszecskes_elsotipp} is derived
by employing the bound state quantization. We gain some intuition from
the section \ref{exp:1particle} where it was
found that the bound state $A_aA_a$ may dissociate at a critical volume $L_c$.
For $L<L_c$ there is no one-particle level of type $A_c$ in the
given sector of the spectrum, however an $A_aA_a$ scattering state
appears instead. Finite volume form factors of this state are calculated using \eqref{fvff} as
\begin{equation}
\label{ketreszecskes_fazisnelkul}
    F^{\mathcal{O}}_c(I,L)=
\frac{F^{\mathcal{O}}({\theta}_1,{\theta}_2)_{aa}}
{\sqrt{\rho_{aa}({\theta}_1,{\theta}_2)}}
\quad \textrm{for } L<L_c
\end{equation}

The generalization to $L>L_c$ seems to be straightforward: one has to
 continue analytically \eqref{ketreszecskes_fazisnelkul} to the
solutions of the Bethe-Yang equation with imaginary rapidities
$\theta_{1,2}=\theta\pm iu$. However, note that equations
\eqref{egyreszecskes_elsotipp} and \eqref{ketreszecskes_fazisnelkul}
are valid up to a phase factor. 
In order to
continue analytically to imaginary rapidities we also need to fix
this phase\footnote{As we noted in earlier the phase of a (non-diagonal) infinite volume form factor is
  unphysical in the sense that it may be redefined by a complex
  rotation of the state vectors and 
physical quantities, e.g. correlation
functions, do not depend on such redefinitions.
 However, the bootstrap program uniquely assigns a phase
  to each form factor. }.

The two-particle form factor satisfies
\begin{equation*}
  F^{\mathcal{O}}(\theta_1,\theta_2)_{aa}=S_{aa}(\theta_1-\theta_2)F^{\mathcal{O}}(\theta_2,\theta_1)_{aa}
\end{equation*}
The simplest choice for the phase is therefore
\begin{equation}
\label{fazisvalasztas}
  F^{\mathcal{O}}(\theta_1,\theta_2)_{aa}=\sqrt{S_{aa}(\theta_1-\theta_2)}
\left|F^{\mathcal{O}}(\theta_1,\theta_2)_{aa}\right|
\end{equation}
This choice is dictated by CPT symmetry
\cite{maiani_testa_final_state_interaction,lellouch_luscher,k_to_pipi},
and it is respected by all known solutions of the 
form factor bootstrap axioms. 
There is a sign ambiguity caused by the square root, but it can
be fixed by demanding $(S_{aa}(0))^{1/2}=i$ and continuity.
Using \eqref{fazisvalasztas} 
\begin{equation*}
    F^{\mathcal{O}}_c(I,L)=
\frac{\sqrt{S_{aa}(\theta_2-\theta_1)} 
F^{\mathcal{O}}(\theta_1,\theta_2)_{aa}}
{\sqrt{\rho_{aa}(\theta_1,\theta_2)}}
\end{equation*}
and upon analytic continuation 
\begin{equation}
  \label{analitikus_elfolytatas}
    F^{\mathcal{O}}_c(I,L)=
\frac{\sqrt{S_{aa}(-2iu)} 
F^{\mathcal{O}}(\theta+iu,\theta-iu)_{aa}}
{\sqrt{\rho_{aa}(\theta+iu,\theta-iu)}}
\quad \textrm{for } L>L_c
\end{equation}

It is easy to see that the result
\eqref{egyreszecskes_elsotipp} is reproduced in the $L\to\infty$ limit.
First observe that 
\begin{equation*}
  F^{\mathcal{O}}(\theta+iu,\theta-iu)_{aa}\sim 
\frac{\Gamma_{aa}^c}{2(u-\bar{u}^c_{aa})}F_c^{\mathcal{O}}(\theta)
\end{equation*}
The residue of $\rho_{aa}$ is determined by $\varphi_{aa}(2iu)$
and it reads
\begin{equation*}
   \rho(\theta+iu,\theta-iu)_{aa}\sim
2m_aL \cos(u)\cosh(\theta) (-i) \frac{S_{aa}'(2iu)}{S_{aa}(2iu)}=
m_cL \cosh(\theta) \frac{1}{2(u-\bar{u}_{ac}^a)}
\end{equation*}
The singularities in the numerator and denominator of
\eqref{analitikus_elfolytatas} cancel  and  indeed
\begin{equation*}
  F^{\mathcal{O}}_c(I,L)\sim
\frac{F_c^{\mathcal{O}}}{ \sqrt{m_cL\cosh(\theta)}}
\end{equation*}
We emphasize that it is
crucial to include the extra normalization factor
$\sqrt{S_{aa}(-2iu)}$ to obtain a meaningful result.

Expression \eqref{analitikus_elfolytatas} can be developed into a
Taylor-series in $u-\bar{u}_{ac}^a$. 
First we use the exchange axiom to arrive at 
\begin{equation}
 \label{analitikus_elfolytatas_2}
    F^{\mathcal{O}}_c(I,L)=
\frac{
F^{\mathcal{O}}(\theta-iu,\theta+iu)_{aa}}
{\sqrt{S_{aa}(-2iu)\rho_{aa}(\theta+iu,\theta-iu)}}
\end{equation}
The form factor axioms imply that
\begin{equation*}
  \lim_{u\to \bar{u}_{ac}^a}F^{\mathcal{O}}(\theta-iu,\theta+iu)_{aa}  =
\frac{1}{\Gamma_{aa}^c}F_c^{\mathcal{O}}
\end{equation*}
The simple pole of $\varphi_{aa}(2iu)$ in $\rho_{aa}$ is canceled by
$S_{aa}(-2iu)$, therefore both the numerator and the denominator of
\eqref{analitikus_elfolytatas_2} have continuous limits as $u\to
\bar{u}_{ac}^a$. 

The form factor $F^{\mathcal{O}}(\theta-iu,\theta+iu)_{aa}$ only
depends on $u$ by Lorentz-symmetry. Therefore
\begin{equation*}
  F^{\mathcal{O}}(\theta-iu,\theta+iu)_{aa}=
\frac{1}{\Gamma_{aa}^c}
F_c^{\mathcal{O}}-2i\left(F^{\mathcal{O}}_{aa}\right)'(u-\bar{u}_{ac}^a)+
\dots
\end{equation*}
where
\begin{equation*}
  \left(F^{\mathcal{O}}_{aa}\right)'=
\frac{d}{d\theta}   F^{\mathcal{O}}(\theta,\theta')_{aa}
\Big|_{\theta-\theta'=-2i\bar{u}_{ac}^a}
\end{equation*}
Expanding the S-matrix element into a Laurent-series in the vicinity
of the pole 
\begin{eqnarray*}
S_{aa}(2iu)&=&\frac{\left(\Gamma_{aa}^c \right)^2}{2(u-\bar{u}_{ac}^a)}+
S_{aa}^{c,0}+\dots\\
S_{aa}(-2iu)&=&\frac{2(u-\bar{u}_{ac}^a)}{\left(\Gamma_{aa}^c
  \right)^2}-
\left(\frac{2(u-\bar{u}_{ac}^a)}{\left(\Gamma_{aa}^c \right)^2}\right)^2
S_{aa}^{c,0}+\dots
\end{eqnarray*}
Expanding the denominator:
\begin{eqnarray*}
&&  S(-2iu)\rho_{aa}(\theta+iu,\theta-iu)=\\
&&S(-2iu)E_1E_2L^2-i(E_1+E_2)L S'(2iu)\Big(S(-2iu)\Big)^2=\\
&&\frac{E_c L}{\left(\Gamma_{aa}^c \right)^2} +
\left(\frac{2E_1E_2L^2}{\left(\Gamma_{aa}^c \right)^2}
-\frac{4E_cL}{\left(\Gamma_{aa}^c \right)^4}S_{aa}^{c,0}\right) 
(u-\bar{u}_{ac}^a)+\dots
\end{eqnarray*}
where 
\begin{eqnarray*}
E_c=E_1+E_2=2m_a\cos(u)\cosh(\theta)
\end{eqnarray*}
Putting all this together
\begin{eqnarray}
 \nonumber
 && F^{\mathcal{O}}_c(I,L)=
\frac{F_c^{\mathcal{O}}}{ \sqrt{E_cL}}+\\
\nonumber
 &&+\left[\frac{-2i\Gamma_{aa}^c \left(F^{\mathcal{O}}_{aa}\right)'}{ \sqrt{E_cL}}
+
\frac{F_c^{\mathcal{O}}}{ \sqrt{E_cL}^3}\left(-E_1E_2L^2+
\frac{2E_cL}{\left(\Gamma_{aa}^c \right)^2}S_{aa}^{c,0} 
\right) \right] (u-\bar{u}_{ac}^a)+\dots
\end{eqnarray}
Note that in the preceding formulas $E_c$ does include the leading order correction to
the usual one-particle energy $E_c^0=\sqrt{m_c^2+(2\pi I)^2/L^2}$. Using 
\begin{equation*}
  E_c=E_c^0-2\frac{m_c\mu}{E_c^0}  (u-\bar{u}_{ac}^a)+O(e^{-2\mu L}) 
\quad\quad \textrm{and} \quad\quad E_1E_2=\frac{m_a^2}{m_c^2}E_c^2-\mu^2
\end{equation*}
the final result is given by
\begin{eqnarray}
 \nonumber
 && F^{\mathcal{O}}_c(I,L)=
\frac{F_c^{\mathcal{O}}}{\sqrt{E_c^0L}}
-\frac{2i\Gamma_{aa}^c \left(F^{\mathcal{O}}_{aa}\right)'}{ \sqrt{E_c^0L}} (u-\bar{u}_{ac}^a)+
\\
&&
+\frac{F_c^{\mathcal{O}}}{ \sqrt{E_c^0L}}
\left[
\frac{2S_{aa}^{c,0} }{\left(\Gamma_{aa}^c \right)^2}
+\frac{m_c\mu}{(E_c^0)^2}-
\left(\frac{m_a^2}{m_c^2}E_c^0-\frac{\mu^2}{E_c^0}\right)L
 \right](u-\bar{u}_{ac}^a) +O(e^{-2\mu L})
\label{ff_mu_tag_1}
\end{eqnarray}
with
\begin{equation*}
  u-\bar{u}_{ac}^a=\pm \frac{1}{2}\left(\Gamma_{aa}^c\right)^2
  e^{-\mu_{aa}^cL \sqrt{1+\left(\frac{\pi I}{m_aL\cos(\bar{u}_{ac}^a)}\right)^2}}  
\end{equation*}


\subsection{One-particle form factors -- Numerical analysis}

Let us introduce the dimensionless form factors as
\begin{equation*}
  f_i(I,l)=\frac{\bra{0}\eps(0,0)\ket{\{I\}}_{i,L}}{m_1}
\end{equation*}
We use the methods described in section \ref{numerics_howto}  to determine
$f_i(I,l)$ for $i=1,2,3$ and $I=0,1,2,3$. The numerical results are
compared to the exact infinite volume form factors.

\begin{figure}
  \centering
\small
\psfrag{Tspin0}{$\bra{0}\eps\ket{\{0\}}_3$}
\psfrag{Tspin1}{$\bra{0}\eps\ket{\{1\}}_3$}
\psfrag{Tspin2}{$\bra{0}\eps\ket{\{2\}}_3$}
\psfrag{Tspin3}{$\bra{0}\eps\ket{\{3\}}_3$}
\psfrag{l}{$l$}
\psfrag{exact}{$F_3$}
\psfrag{f3}{$|\bar{f}_3|$}
  \includegraphics[scale=0.5,angle=-90]{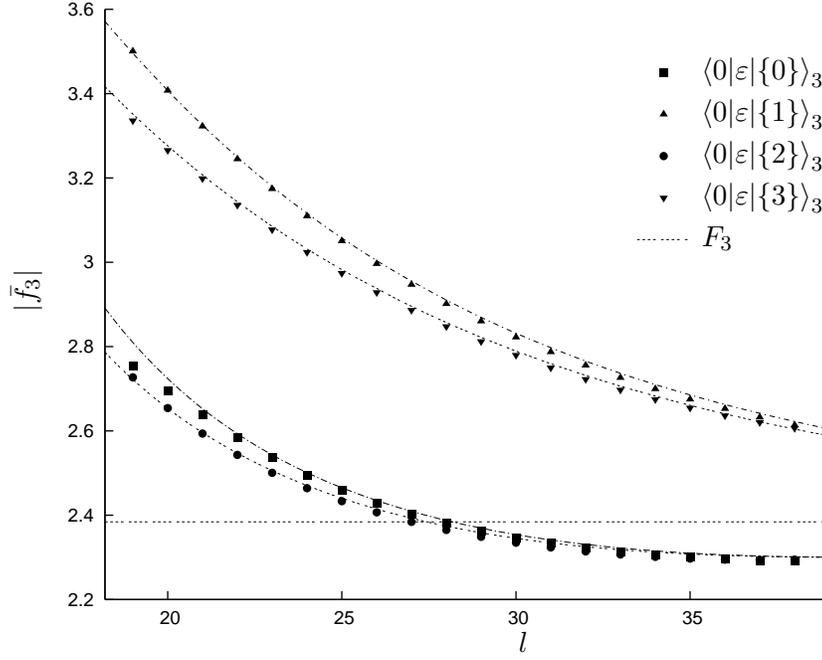}
\caption{
Elementary finite volume form factors of $A_3$ one-particle
levels. 
Here the normalization  \eqref{F3normalas} is applied to
obtain a finite $l\to\infty$ limit, which is given by the infinite
volume form factor $F_3=\bra{0}\eps\ket{A_3(\theta)}$. The TCSA 
data are plotted against the bound state prediction. The ordinary
evaluation of $\bar{f}_3$ is simply the constant $F_3$.
\label{fig:F3}}
\end{figure}

We start our investigation with  $f_3(I,l)$, for which
relatively large exponential corrections were found in subsection \ref{numerics_1p_ising}. 
It is convenient to consider
\begin{equation}
\label{F3normalas}
  \bar{f}_3(I,l)=\left(e_0l\right)^{1/2} f_3(I,L)\quad\textrm{with}
\quad \lim_{l\to\infty}\bar{f}_3(I,l)=F_3
\end{equation}
The numerical results are demonstrated in
fig. \ref{fig:F3}. 
 Note, that this is the same figure as
fig. \ref{fig:onepffising}, but here the interpretation  
of the huge deviations from $F_3$ is also provided. 

\begin{figure}
  \centering
\psfrag{Tspin0}{$I=0$}
\psfrag{Tspin1}{$I=1$}
\psfrag{Tspin2}{$I=2$}
\psfrag{Tspin3}{$I=3$}
\psfrag{l}{$l$}
\psfrag{f2}{}
  \includegraphics[scale=0.4,angle=-90]{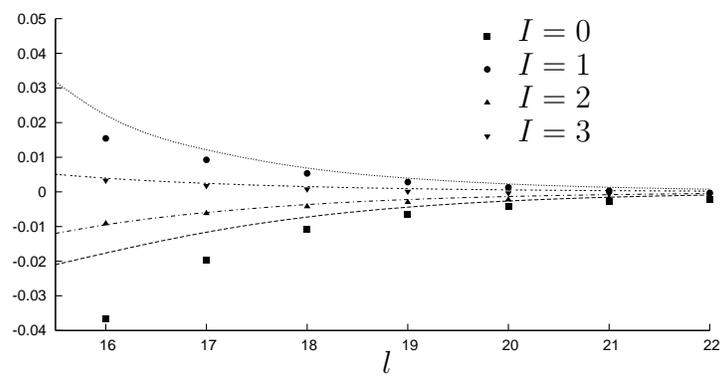}
\caption{
Finite size corrections to the elementary form factors of $A_2$. Dots
represent TCSA data, while the lines show the $\mu$-term
prediction corresponding to the $A_1A_1\to A_2$ fusion.
\label{fig:F2}}
\end{figure}

We also tried to verify the predictions  for $f_1$
and $f_2$. In the latter case reasonably good agreement was found
with TCSA, the results are demonstrated in fig. \ref{fig:F2}. In the case
of $f_1$ we encountered the unpleasant situation that the F-term
decays slower than the TCSA errors grow, thus making the observation of the
$\mu$-term impossible.

\begin{figure}
  \centering
\tiny
\psfrag{l}{$l$}
\psfrag{f13}{$f_{13}$}
\psfrag{f23}{$f_{23}$}
\psfrag{f33}{$f_{33}$}
\subfigure[$_{1}\bra{\{0\}}\eps\ket{\{3\}}_{3,L}$]{\includegraphics[scale=0.4,angle=-90]{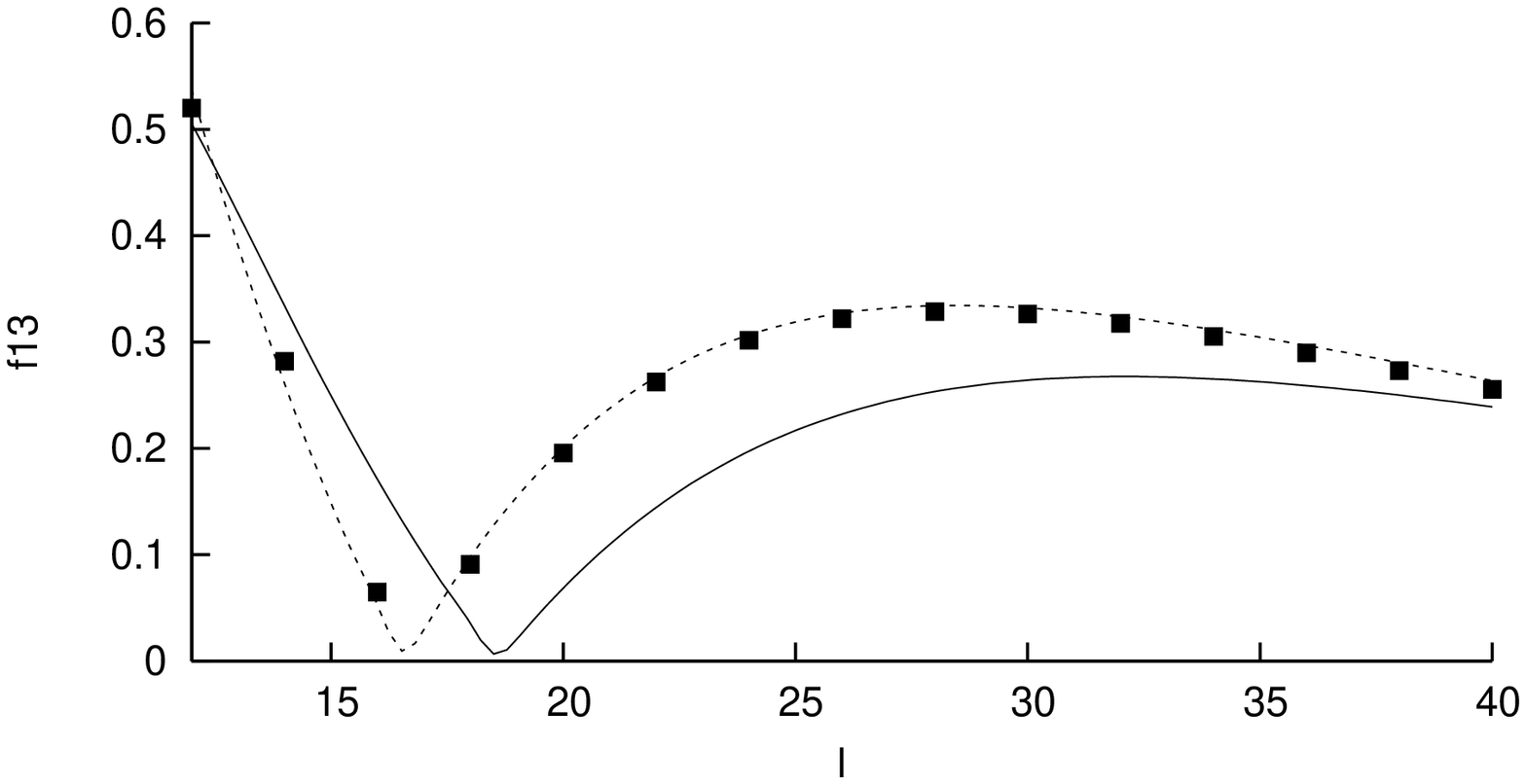}}
\subfigure[$_{1}\bra{\{0\}}\eps\ket{\{1\}}_{3,L}$]{ \includegraphics[scale=0.4,angle=-90]{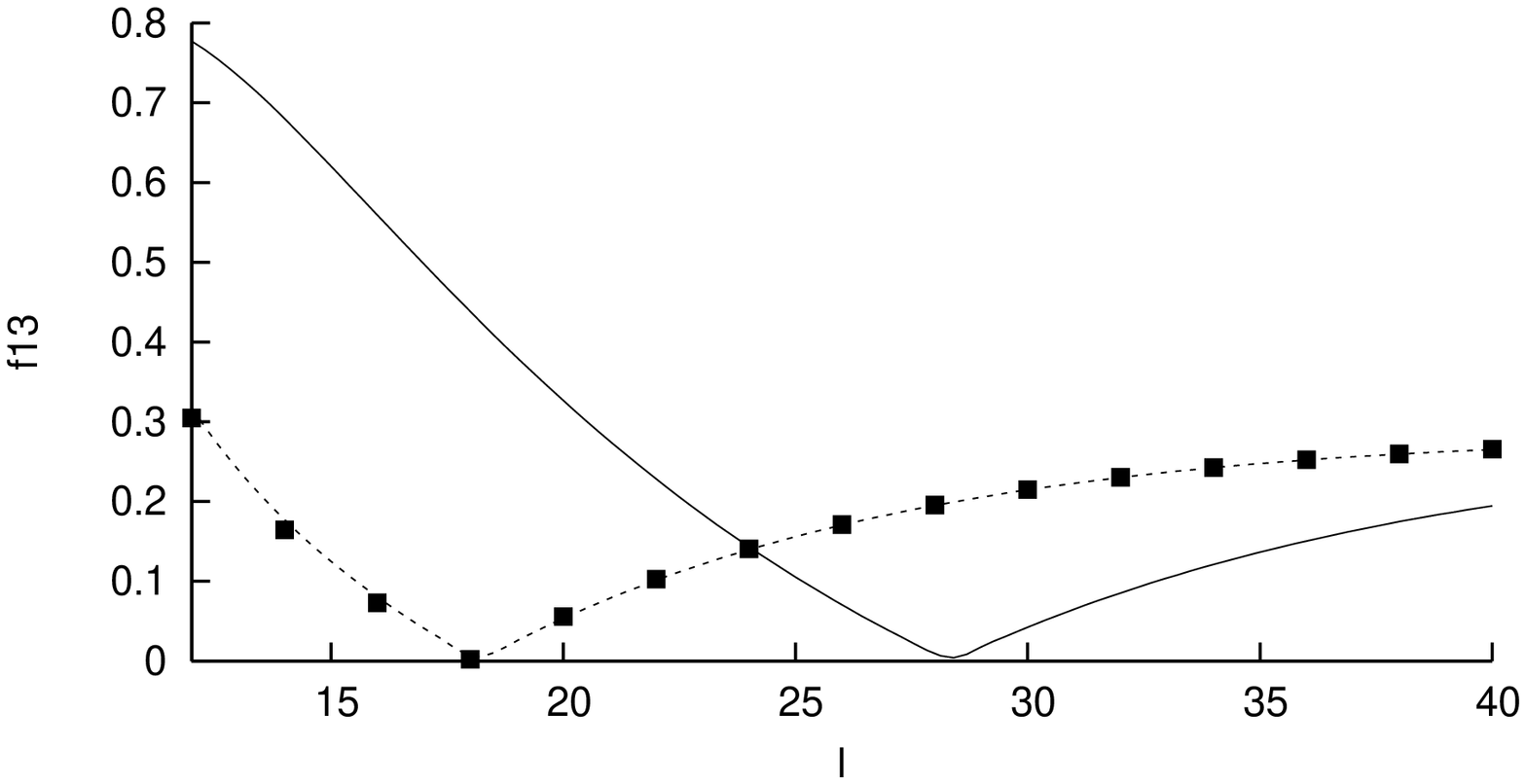}}

\subfigure[$_{2}\bra{\{-2\}}\eps\ket{\{3\}}_{3,L}$]{ \includegraphics[scale=0.4,angle=-90]{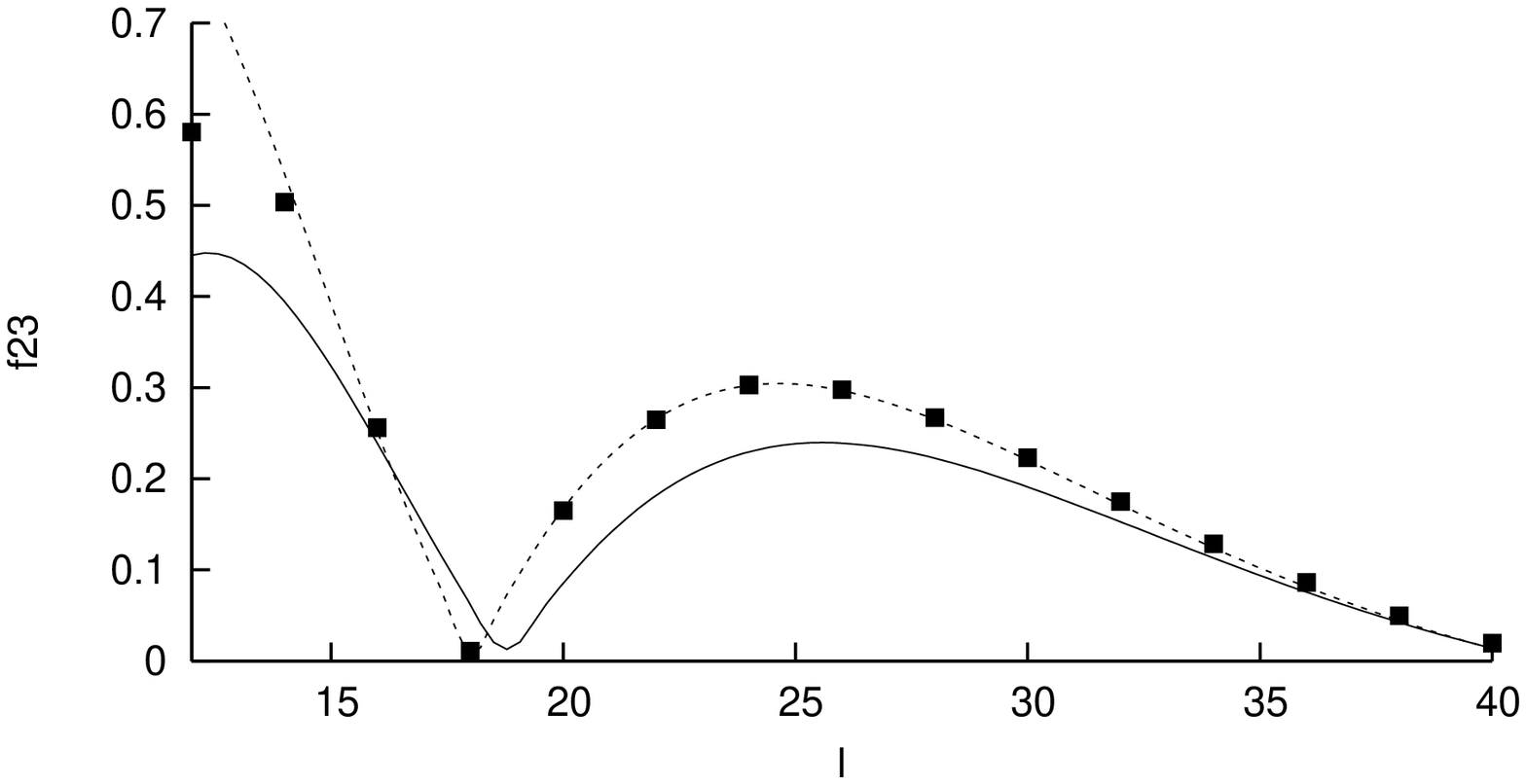}}
\subfigure[$_{3}\bra{\{-3\}}\eps\ket{\{3\}}_{3,L}$]{\includegraphics[scale=0.4,angle=-90]{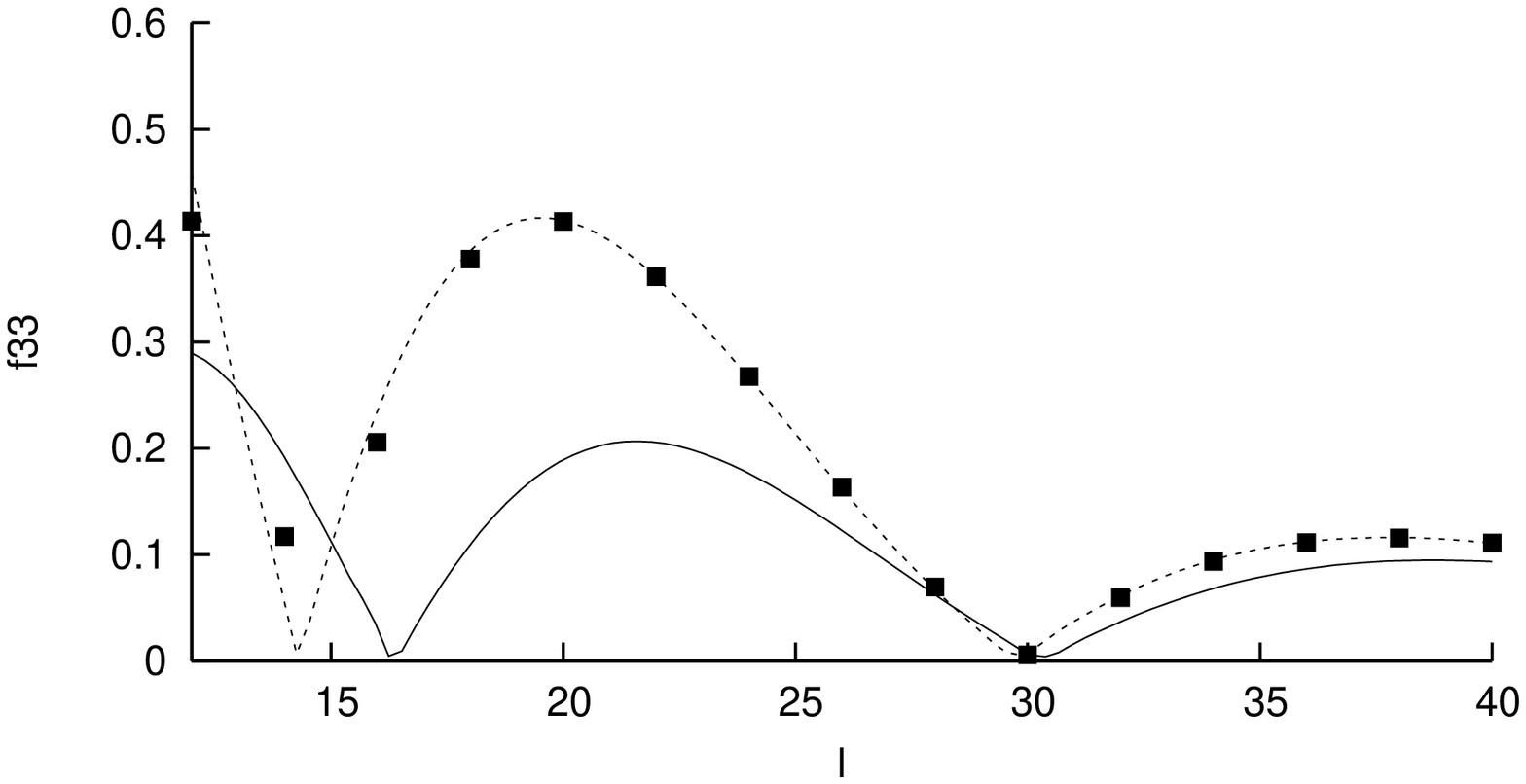}}
\caption{One-particle--one-particle form factors, dots correspond to
  TCSA data. The solid lines represent the ordinary evaluation of the finite volume form
  factors, while the dotted lines show the bound state
  prediction.  
\label{fig:Fx3}}
\end{figure}

It is straightforward to generalize \eqref{analitikus_elfolytatas} to
matrix elements between two different one-particle
 states. For $b\ne c$ one has for example
\begin{equation*}
_b\bra{\{I_b\}}\eps\ket{\{I_c\}}_{c,L}=\frac{F^{\eps}(\theta_b+i\pi,\theta+iu,\theta-iu)_{baa}}
{\sqrt{\rho_b(\theta_b)\rho_{aa}(\theta+iu,\theta-iu)}}
\end{equation*}
Numerical examples are presented in figures \ref{fig:Fx3} (a)-(c) for $c=3$ and
$b=1,2$. 

The most interesting case is the one shown in
fig. \ref{fig:Fx3} (d) where the matrix element between
two different $A_3$ one-particle states are investigated. This can be done by considering
both $A_3$ particles as the appropriate $A_1A_1$ bound states and then
calculating the finite volume form factor
$_3\bra{\{I\}}\eps\ket{\{I'\}}_{3,L}$ as
\begin{eqnarray*}
 _{11}\bra{\{I/2,I/2\}}\eps\ket{\{I'/2,I'/2\}}_{11,L}=
\frac{F^\eps(\theta+iu+i\pi,\theta-iu+i\pi,\theta'+iu',\theta'-iu')_{1111}}
{\sqrt{\rho_{11}(\theta'+iu',\theta'-iu')\rho_{11}(\theta+iu,\theta-iu)}}
\end{eqnarray*}
Once again we find complete agreement with the TCSA data.

\subsection{Elementary multi-particle form factors}

The generalization of \eqref{analitikus_elfolytatas} to multi-particle
states is straightforward, the only task is to find the 
appropriate phase factor.
Similar to the one-particle case one has
\begin{equation*}
  F^{\mathcal{O}}(\theta_1,\dots,\theta_m)_{b_1\dots b_m}=
\sqrt{\prod_{i<j} S_{b_ib_j}(\theta_i-\theta_j)}
\left|F^{\mathcal{O}}(\theta_1,\dots,\theta_m)_{b_1\dots b_m}\right|
\end{equation*}
A general $n$ particle finite volume form factor with real
rapidities can thus be
written as
\begin{eqnarray*}
  \sqrt{\frac{\prod_{i<j} S_{b_ib_j}(\theta_j-\theta_i)}{\rho^{n}
({\theta}_1,\dots,{\theta}_{n})_{b_1\dots
  b_{n}}}}
 F^{\mathcal{O}}(\theta_1,\dots,\theta_{n})_{b_1\dots b_{n}}
\end{eqnarray*}
Substituting the solution of the Bethe-equation for the state
$\ket{\{I/2,I/2,I_1,\dots,I_n\}}_{aab_1\dots b_n,L}$ and  
making use of the real analyticity condition
\begin{equation*}
 |S_{ab_j}(\theta_j-\theta-iu)S_{ab_j}(\theta_j-\theta+iu)|=1 
\end{equation*}
one gets
\begin{equation}
\label{multi_particle_form_factor} 
\bra{0}\mathcal{O}\ket{\{I/2,I/2,I_1,\dots,I_n\}}_{aab_1\dots b_n,L}=
\frac{\sqrt{S_{aa}(-2iu)}
\left| F^{\mathcal{O}}(\theta+iu,\theta-iu,\theta_1,\dots,\theta_{n})_{aab_1\dots b_{n}}
\right|
}{\sqrt{\rho^{(n+2)}
(\theta+iu,\theta-iu,\theta_1,\dots,\theta_{n})_{aab_1\dots
  b_{n}}}}
\end{equation}
up to a physically irrelevant phase.

It is easy to show once again that the ''naive'' result is reproduced in the
$L\to\infty$ limit. To do so, we first quote the dynamical pole equation of the infinite volume form factor:
\begin{equation*}
  F^{\mathcal{O}}({\theta}+iu,{\theta}-iu,{\theta}_1,\dots,{\theta}_{n})
_{aab_1\dots b_{n}}=\frac{\Gamma_{aa}^c}{2(u-\bar{u}_{ac}^a)}
F^{\mathcal{O}}({\theta},{\theta}_1,\dots,{\theta}_{n})
_{cb_1\dots b_{n}}+O(1)
\end{equation*}
The singularity of $\rho^{(n+2)}$ is given by
\begin{equation*}
\textrm{Res}_{u\to \bar{u}_{ac}^a} \rho^{(n+2)}
(\theta+iu,\theta-iu,\theta_1,\dots,\theta_n)_{aab_1\dots b_n}=\frac{1}{2}\rho^{(n+1)}
(\theta,\theta_1,\dots,\theta_n)_{cb_1\dots b_n}  
\end{equation*}
The ''naive'' formula is now recovered by inserting the last two
equations into \eqref{multi_particle_form_factor}. 

The leading exponential corrections can be obtained by plugging
\eqref{multi-particle_u} and \eqref{rapieltolodas} into
\eqref{multi_particle_form_factor} and expanding to first order in  $u-\bar{u}_{ac}^a$.
This procedure is straightforward
but quite lengthy, therefore we refrain from giving the details of the calculations.

\begin{figure}
  \centering
\psfrag{l}{$l$}
\psfrag{f111}{$f_{13}$}
\subfigure[$\bra{0}\eps\ket{\{1,1\}}_{13,L}$]{ \includegraphics[scale=0.5,angle=-90]{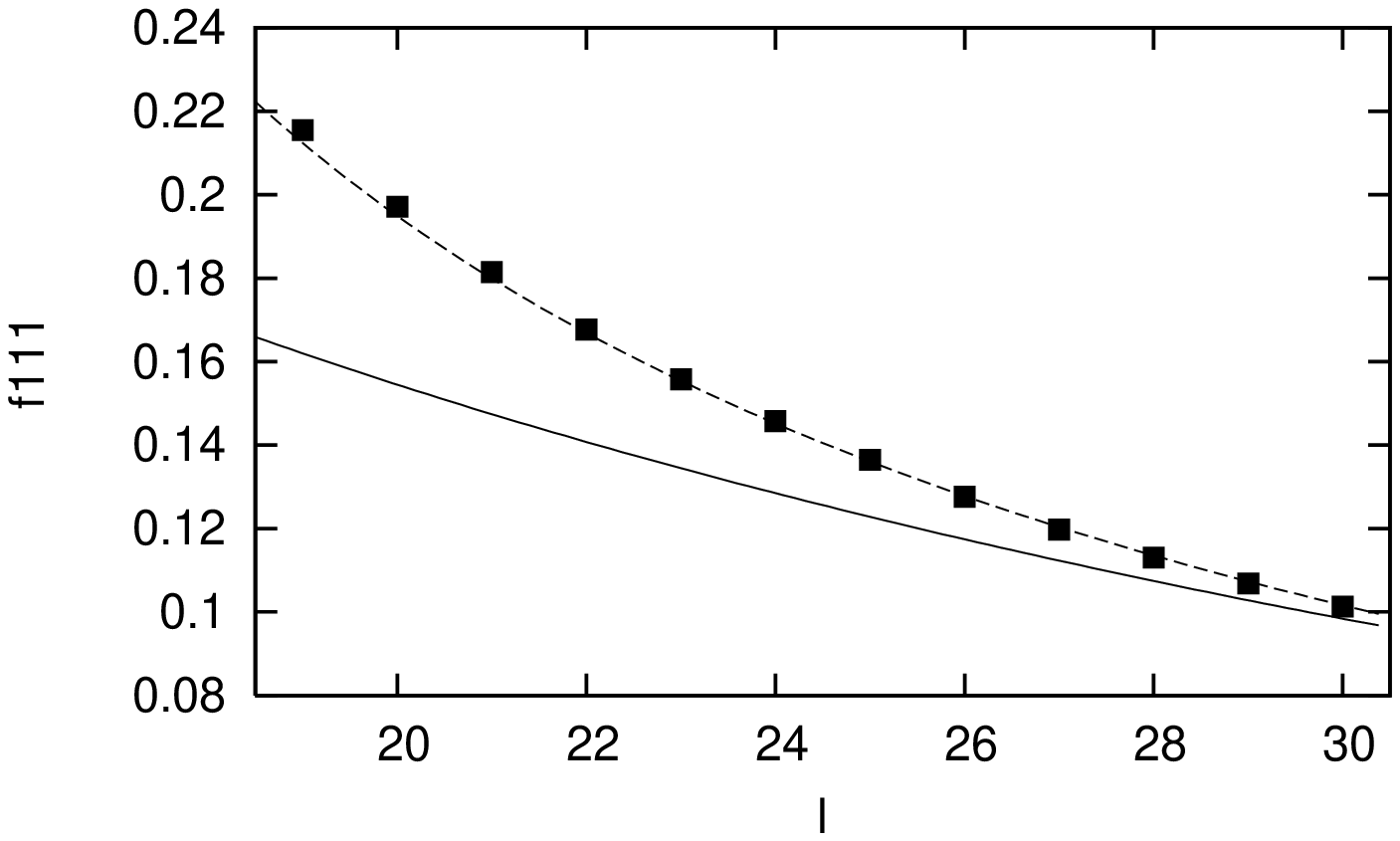}}
\subfigure[$\bra{0}\eps\ket{\{2,0\}}_{13,L}$]{ \includegraphics[scale=0.5,angle=-90]{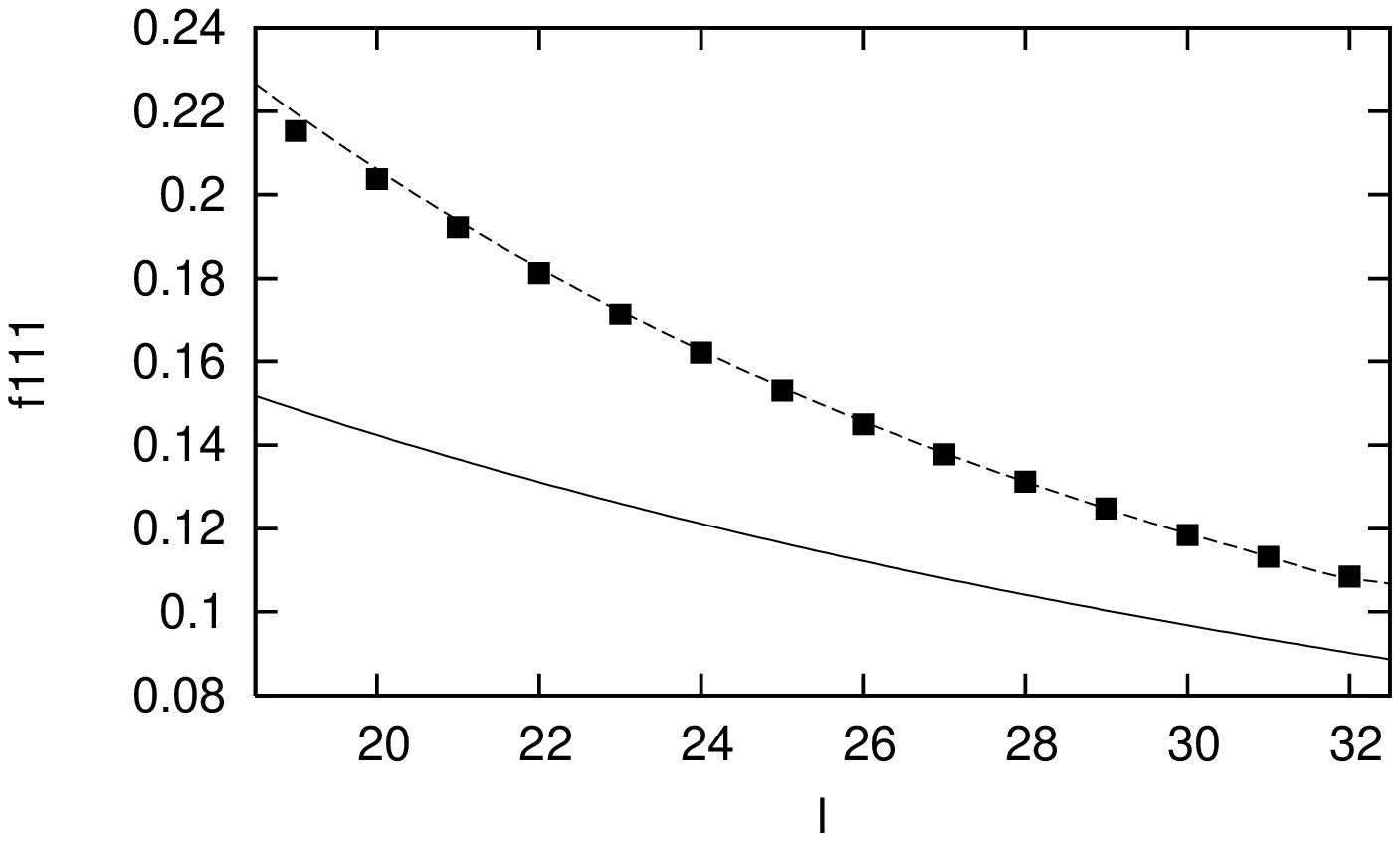}}
\caption{
Elementary form factors of $A_1A_3$ scattering states, dots correspond
to TCSA data. The solid lines are obtained by a ``naive``
evaluation of the finite volume form factors, while the dotted line
represents the bound state prediction. (in this case $A_1A_1A_1$ form
factors at the appropriate rapidities)
\label{fig:F13_durch_F111}}
\end{figure}

In fig. \ref{fig:F13_durch_F111} two examples are presented for the
evaluation of \eqref{multi_particle_form_factor} applied to $A_1A_3$
two-particle states. 

\begin{table}
  \centering
\small
  \begin{tabular}{|c||c|c|c||c|c|}
\hline
$l$ & $\theta_1$ & $\theta_2$  & $\theta_3$ & $e(l)$ (predicted) & $e(l)$ (TCSA) \\
\hline
\hline
22 & 0.44191 & 0.68235 & -0.58742 &  3.51877 & 3.51900 \\ 
\hline 
23 & 0.42574 & 0.64280 & -0.55190 &  3.46201 & 3.46219 \\ 
\hline 
24 & 0.60523 & 0.41155 & -0.51892 &  3.41239 & 3.41255 \\ 
\hline 
25 & 0.56934 & 0.39931 & -0.48831 &  3.36890 & 3.36907 \\ 
\hline 
26 & 0.38910 & 0.53475 & -0.45989 &  3.33071 & 3.33089 \\ 
\hline 
27 & 0.50092 & 0.38119 & -0.43350 &  3.29709 & 3.29729 \\ 
\hline 
28 & 0.46686 & 0.37634 & -0.40899 &  3.26743 & 3.26767 \\ 
\hline 
29 & 0.42947 & 0.37739 & -0.38622 &  3.24122 & 3.24164 \\ 
\hline 
30 & 0.38646 $+$ 0.02332 i & 0.38646 $-$ 0.02332 i & -0.36505 &  3.21801 & 3.21832 \\ 
\hline  
31 & 0.37059 $+$ 0.04042 i & 0.37059 $-$ 0.04042 i & -0.34537 &  3.19741 & 3.19775 \\ 
\hline 
32 & 0.35572 $+$ 0.05104 i & 0.35575 $-$ 0.05104 i & -0.32706 &  3.17908 & 3.17945 \\ 
\hline 
33 & 0.34182 $+$ 0.05891 i & 0.34182 $-$ 0.05891 i & -0.31002 &  3.16275 & 3.16313 \\ 
\hline 
34 & 0.32876 $+$ 0.06513 i & 0.32876 $-$ 0.06513 i & -0.29415 &  3.14816 & 3.14856 \\ 
\hline 
35 & 0.31650 $+$ 0.07022 i & 0.31650 $-$ 0.07022 i & -0.27936 &  3.13511 & 3.13547 \\ 
\hline 
36 & 0.30498 $+$ 0.07446 i & 0.30498 $-$ 0.07446 i & -0.26556 &  3.12341 & 3.12235 \\ 
\hline 
 \hline
 \end{tabular}
  \caption{An example for the dissociation of the $A_1A_1$ bound state
  inside a scattering state. $\ket{\{2,0\}}_{31,L}$ is identified with $\ket{\{1,1,0\}}_{111,L}$
  and the corresponding Bethe-Yang equations is solved.
For $l<30$ there is a real $A_1A_1A_1$ three-particle state in the spectrum,
  whereas at $l\approx 30$ two of the rapidities become complex and
  the two-particle state $A_1A_3$ emerges.}
  \label{tab:en_A1A3_1}
\end{table}

\chapter{Correlation functions at finite temperature}

\label{correlations}

\section{The form factor expansion for correlation functions}

The finite temperature correlation function of local operators
$\mathcal{O}_1$ and $\mathcal{O}_2$
is defined by
\begin{equation}
  \label{2point_funct}
\langle\mathcal{O}_1(x_1,t_1)\mathcal{O}_2(x_2,t_2)
\rangle^{R}=\frac{\text{Tr}\left(\mathrm{e}^{-RH}
\mathcal{O}_1(x_1,t_1)\mathcal{O}_2(x_2,t_2)
\right)}{\text{Tr}\left(\mathrm{e}^{-RH}\right)}  
\end{equation}
where $R=1/T$ is the temperature dependent extension of the Euclidean
time direction used in thermal quantum field theory and $H$ is the
Hamiltonian. 

In the following we briefly discuss the evaluation of the
correlation function at zero temperature using the form factor
expansion and show that there is no straightforward generalization of this
method to the finite temperature case.  To keep the exposition simple we assume that the spectrum
contains a single massive particle of mass $m$.



At zero temperature the two-point function of operators
$\mathcal{O}_1$ and $\mathcal{O}_2$ is given by the vacuum expectation
value
\begin{equation*}
\bra{0}\mathcal{O}_1(x_1,\tau_1)\mathcal{O}_2(x_2,\tau_2)\ket{0}
\end{equation*}
Inserting a complete set of asymptotic states and using Euclidean
invariance one obtains
\begin{eqnarray}
\nonumber
  \bra{0}\mathcal{O}_1(x_1,t_1)\mathcal{O}_2(x_2,t_2)\ket{0}=&\\
\sum_{n=0}^\infty \frac{1}{n!} \int d\theta_1\dots d\theta_n&
F^{\mathcal{O}_1}_n(\theta_1\dots\theta_n)
F^{\mathcal{O}_2}_n(\theta_1\dots\theta_n)^*\
 e^{-mr \sum_{i=1}^n \cosh(\theta_i)}
\label{ff_expansion}
\end{eqnarray}
where $r=\sqrt{(x_1-x_2)^2+(\tau_1-\tau_2)^2}$. The essence of the form factor approach 
\cite{karowski_weisz_elso_S,smirnov_ff,yurov_zam_Ising,cardy_mussardo_ff_corr_funct,zam_Lee_Yang}
is to determine the form factors appearing in  \eqref{ff_expansion}
as solutions of the form factor bootstrap program, and then numerically
integrate the resulting expressions to obtain the contributions to the
correlator. Due to the exponential damping factor and the vanishing
property of the form factor functions at $\theta_i=0$
the series \eqref{ff_expansion} is rapidly converging even for
relatively small values of $mr$, therefore in most cases it is
sufficient to consider only the first few terms in \eqref{ff_expansion}.

Switching on a non-zero temperature one faces severe problems. A
straightforward application of the form factor expansion results in
\begin{equation}
  \label{formal_2point}
  \langle\mathcal{O}_1(0,0)\mathcal{O}_2(0,\tau)\rangle^{R}=
\frac{1}{Z} \sum_{m,n}  \bra{m}\mathcal{O}_1\ket{n}
\bra{n}\mathcal{O}_2\ket{m} 
e^{-(R-\tau) E_\Psi} e^{-\tau E_{\Psi'}}
\end{equation}
where for simplicity we have chosen the displacement between the two
operators to lie in the imaginary time direction and the summation
over $m$ and $n$ runs over the complete set of asymptotic states of the theory.
Expression \eqref{formal_2point} is ill-defined in two ways:
\begin{itemize}
\item The partition function $Z$ is infinite if one considers the
  infinite volume theory.
\item One has to integrate over the poles of the form
  factors associated to the disconnected contributions, whenever the
  rapidities of two particles entering $m$ and $n$ approach
  each other.
\end{itemize}
One encounters similar problems already by the evaluation of the
thermal expectation value of local operators. The formal form factor
expansion  for the one-point function reads
\begin{equation}
  \label{formal_1point}
  \langle\mathcal{O}\rangle^{R}=\frac{\text{Tr}\left(\mathrm{e}^{-RH}\mathcal{O}\right)}{\text{Tr}\left(\mathrm{e}^{-RH}\right)}  = 
\frac{1}{Z} \sum_m \bra{m}\mathcal{O}\ket{m} e^{-RE_\Psi}
\end{equation}
The diagonal form factors entering the expression above are
ill-defined in the infinite volume theory due to disconnected
terms. 

In order to obtain well-defined expressions for the correlation
functions it is necessary to introduce a regularization scheme. 

A. LeClair and G. Mussardo
proposed an evaluation scheme \cite{leclair_mussardo} based on the
Thermodynamic Bethe Ansatz. It was proven by Saleur
\cite{saleurfiniteT} that it yields the correct results for
expectation values of local conserved charges. However it was also argued in
\cite{saleurfiniteT} that the LeClair-Mussardo formalism fails in the
case of the two-point function. The same conclusion was reached in
\cite{CastroAlvaredo:2002ud} studying massless limits of correlation
functions.
G. Delfino proposed a second, independent regularization scheme
\cite{delfinofiniteT} for the one-point function; these two approaches are presented and compared
in subsection \ref{ss:leclair_mussardo}. Mussardo showed using a toy model,
that the two methods yield different results 
\cite{mussardodifference}; the same conclusion is reached by comparing
the systematic low-temperature expansion of the two proposals (see \ref{ss:leclair_mussardo}). 

A third approach is to consider a theory in finite volume $L$, to evaluate
the form factor expansion using the well-defined finite volume form
factors, and to take the limit $L\to\infty$. This method is built on
first principles and it does not rely on any additional
assumptions; moreover it is expected to work for any n-point function.

In this work we develop a systematic low-temperature
expansion for the one-point function with a well-defined $L\to\infty$
limit; this calculation is presented in section
\ref{low_T_expansion}. We find complete agreement with the
LeClair-Mussardo proposal up to the third non-trivial order.
In section \ref{lowT_2p} we apply our method to the two-point
function. However, we only calculate the simplest terms; the systematic evaluation of the 
form factor expansion is out of the scope of the present work.

Finally we note, that it is also possible to numerically
evaluate the form-factor expansion \cite{Essler:2007jp,konik_heisenberg_spin_chains,james-2009} or
to develop a term by term regularization scheme based on the
infinite volume connected form factors \cite{konik-2001}. Eventually,
after the completion of this thesis there appeared an independent work by
Konik and Essler \cite{essler-2009} which uses (in addition to a novel
infinite volume regularization procedure) finite volume techniques
similar to the ones presented in this work.

\subsection{Comparison of the LeClair-Mussardo and the Delfino approaches}

\label{ss:leclair_mussardo}



 LeClair and Mussardo
proposed the following expression for the low temperature ($T\ll m$,
or equivalently $mR\gg1$) expansion for the one-point function \eqref{formal_1point}:\begin{equation}
\langle\mathcal{O}\rangle^{R}=\sum_{n=0}^{\infty}\frac{1}{n!}\frac{1}{(2\pi)^{n}}\int\left[\prod_{i=1}^{n}d\theta_{i}\frac{\mathrm{e}^{-\epsilon(\theta_{i})}}{1+\mathrm{e}^{-\epsilon(\theta_{i})}}\right]F_{2n}^{c}(\theta_{1},...,\theta_{n})\label{eq:leclmuss1pt}\end{equation}
where $F_{2n}^{c}$ is the connected diagonal form factor defined
in eqn. (\ref{eq:connected}) and $\epsilon(\theta)$ is the pseudo-energy
function, which is the solution of the thermodynamic Bethe Ansatz
equation
\begin{equation}
\epsilon(\theta)=mR\cosh(\theta)-\int\frac{d\theta'}{2\pi}\varphi(\theta-\theta')\log(1+\mathrm{e}^{-\epsilon(\theta')})\label{eq:TBA}\end{equation}
The solution of this equation can be found by successive iteration,
which results in \begin{eqnarray}
\epsilon(\theta) & = & mR\cosh(\theta)-\int\frac{d\theta'}{2\pi}\varphi(\theta-\theta')\mathrm{e}^{-mR\cosh\theta'}+\frac{1}{2}\int\frac{d\theta'}{2\pi}\varphi(\theta-\theta')\mathrm{e}^{-2mR\cosh\theta'}+\nonumber \\
 & + & \int\frac{d\theta'}{2\pi}\frac{d\theta''}{2\pi}\varphi(\theta-\theta')\varphi(\theta'-\theta'')\mathrm{e}^{-mR\cosh\theta'}\mathrm{e}^{-mR\cosh\theta''}+O\left(\mathrm{e}^{-3mR}\right)\label{eq:pseudoenergy2}\end{eqnarray}
Using this expression, it is easy to derive the following expansion
from (\ref{eq:leclmuss1pt})\begin{eqnarray}
\langle\mathcal{O}\rangle^{R} & = & \langle\mathcal{O}\rangle+\int\frac{d\theta}{2\pi}F_{2}^{c}\left(\mathrm{e}^{-mR\cosh{\theta}}-\mathrm{e}^{-2mR\cosh{\theta}}\right)\nonumber \\
 &  & +\frac{1}{2}\int\frac{d\theta_{1}}{2\pi}\frac{d\theta_{2}}{2\pi}\left(F_{4}^{c}(\theta_{1},\theta_{2})+2\Phi(\theta_{1}-\theta_{2})F_{2}^{c}\right)\mathrm{e}^{-mR\cosh{\theta_{1}}}\mathrm{e}^{-mR\cosh{\theta_{2}}}\nonumber \\
 &  & +O\left(\mathrm{e}^{-3mR}\right)\label{eq:TBA2c}\end{eqnarray}
where $\langle\mathcal{O}\rangle$ denotes the zero-temperature vacuum
expectation value. The above result can also be written in terms of
the symmetric evaluation (\ref{eq:Fs_definition}) as

\begin{eqnarray}
\langle\mathcal{O}\rangle^{R} & = & \langle\mathcal{O}\rangle+\int\frac{d\theta}{2\pi}F_{2}^{s}\left(\mathrm{e}^{-mR\cosh{\theta}}-\mathrm{e}^{-2mR\cosh{\theta}}\right)+\nonumber \\
 &  & \frac{1}{2}\int\frac{d\theta_{1}}{2\pi}\frac{d\theta_{2}}{2\pi}F_{4}^{s}(\theta_{1},\theta_{2})\mathrm{e}^{-mR(\cosh{\theta_{1}}+\cosh{\theta_{2}})}+O\left(\mathrm{e}^{-3mR}\right)\label{eq:TBA2s}\end{eqnarray}
where we used relations (\ref{eq:fs2fc2}) and (\ref{cs_four}). Obtaining the third order correction
from the LeClair-Mussardo expansion is a somewhat lengthy, but elementary
computation, which results in\begin{eqnarray}
 &  & \frac{1}{6}\int\frac{d\theta_{1}}{2\pi}\frac{d\theta_{2}}{2\pi}\frac{d\theta_{3}}{2\pi}F_{6}^{s}(\theta_{1},\theta_{2},\theta_{3})\mathrm{e}^{-mR(\cosh\theta_{1}+\cosh\theta_{2}+\cosh\theta_{3})}\nonumber \\
 &  & -\int\frac{d\theta_{1}}{2\pi}\frac{d\theta_{2}}{2\pi}F_{4}^{s}(\theta_{1},\theta_{2})\mathrm{e}^{-mR(\cosh\theta_{1}+2\cosh\theta_{2})}+\int\frac{d\theta_{1}}{2\pi}F_{2}^{s}\mathrm{e}^{-3mR\cosh\theta_{1}}\nonumber \\
 &  & -\frac{1}{2}\int\frac{d\theta_{1}}{2\pi}\frac{d\theta_{2}}{2\pi}F_{2}^{s}\varphi(\theta_{1}-\theta_{2})\mathrm{e}^{-mR(\cosh\theta_{1}+2\cosh\theta_{2})}\label{eq:lm3order}\end{eqnarray}
where we used eqns. (\ref{eq:fs2fc2}, \ref{cs_four}, \ref{eq:f6sa})
to express the result in terms of the symmetric evaluation.

Delfino's proposal for the one-point function reads:\begin{equation}
\langle\mathcal{O}\rangle_{D}^{R}=\sum_{n=0}^{\infty}\frac{1}{n!}\frac{1}{(2\pi)^{n}}\int\left[\prod_{i=1}^{n}d\theta_{i}\frac{\mathrm{e}^{-mR\cosh\theta_{i}}}{1+\mathrm{e}^{-mR\cosh\theta_{i}}}\right]F_{2n}^{s}(\theta_{1},...,\theta_{n})\label{eq:delfino1pt}\end{equation}
which gives the following result when expanded to second order: \begin{eqnarray}
\langle\mathcal{O}\rangle_{D}^{R} & = & \langle\mathcal{O}\rangle+\int\frac{d\theta}{2\pi}F_{2}^{s}\left(\mathrm{e}^{-mR\cosh{\theta}}-\mathrm{e}^{-2mR\cosh{\theta}}\right)+\nonumber \\
 &  & \frac{1}{2}\int\frac{d\theta_{1}}{2\pi}\frac{d\theta_{2}}{2\pi}F_{4}^{s}(\theta_{1},\theta_{2})\mathrm{e}^{-mR(\cosh{\theta_{1}}+\cosh{\theta_{2}})}+O\left(\mathrm{e}^{-3mR}\right)\label{eq:delfino2s}\end{eqnarray}
Note that the two proposals coincide with each other to this order,
which was already noted in \cite{delfinofiniteT}. However, this is
not the case in the next order.  Expanding (\ref{eq:delfino1pt}) results in \begin{eqnarray}
 &  & \frac{1}{6}\int\frac{d\theta_{1}}{2\pi}\frac{d\theta_{2}}{2\pi}\frac{d\theta_{3}}{2\pi}F_{6}^{s}(\theta_{1},\theta_{2},\theta_{3})\mathrm{e}^{-mR(\cosh\theta_{1}+\cosh\theta_{2}+\cosh\theta_{3})}\nonumber \\
 &  & -\int\frac{d\theta_{1}}{2\pi}\frac{d\theta_{2}}{2\pi}F_{4}^{s}(\theta_{1},\theta_{2})\mathrm{e}^{-mR(\cosh\theta_{1}+2\cosh\theta_{2})}+\int\frac{d\theta_{1}}{2\pi}F_{2}^{s}\mathrm{e}^{-3mR\cosh\theta_{1}}\label{eq:delfino3order}\end{eqnarray}
It can be seen that the two proposals differ at this order (the last
term of (\ref{eq:lm3order}) is missing from (\ref{eq:delfino3order})),
which was already noted by Mussardo using a toy model in \cite{mussardodifference},
but the computation presented above is model independent and shows the general
form of the discrepancy. 

\vspace{5cm}

\section{Low-temperature expansion for the one-point function}

\label{low_T_expansion}

We now evaluate the finite temperature expectations value in a finite,
but large volume $L$:

\begin{equation}
\langle\mathcal{O}\rangle_{L}^{R}=\frac{\text{Tr}_{L}\left(\mathrm{e}^{-RH_{L}}\mathcal{O}\right)}{\text{Tr}_{L}\left(\mathrm{e}^{-RH_{L}}\right)}\label{onepointRL}\end{equation}
where $H_{L}$ is the finite volume Hamiltonian, and $\mathrm{Tr}_{L}$
means that the trace is now taken over the finite volume Hilbert space.
For later convenience we introduce a new notation:\[
|\theta_{1},\dots,\theta_{n}\rangle_{L}=|\{ I_{1},\dots,I_{n}\}\rangle_{L}\]
where $\theta_{1},\dots,\theta_{n}$ solve the Bethe-Yang equations
for $n$ particles with quantum numbers $I_{1},\dots,I_{n}$ at the
given volume $L$. We can develop the low temperature expansion of
(\ref{onepointRL}) in powers of $\mathrm{e}^{-mR}$ using

\begin{eqnarray}
\text{Tr}_{L}\left(\mathrm{e}^{-RH_{L}}\mathcal{O}\right) & = & \langle\mathcal{O}\rangle_{L}+\sum_{\theta^{(1)}}\mathrm{e}^{-mR\cosh\theta^{(1)}}\langle\theta^{(1)}|\mathcal{O}|\theta^{(1)}\rangle_{L}\nonumber \\
 &  & +\frac{1}{2}\sum_{\theta_{1}^{(2)},\theta_{2}^{(2)}}{}^{'}\mathrm{e}^{-mR(\cosh\theta_{1}^{(2)}+\cosh\theta_{2}^{(2)})}\langle\theta_{1}^{(2)},\theta_{2}^{(2)}|\mathcal{O}|\theta_{1}^{(2)},\theta_{2}^{(2)}\rangle_{L}+\nonumber \\
 &  & +\frac{1}{6}\sum_{\theta_{1}^{(3)},\theta_{2}^{(3)},\theta_{3}^{(3)}}{}^{'}\mathrm{e}^{-mR(\cosh\theta_{1}^{(3)}+\cosh\theta_{2}^{(3)}+\cosh\theta_{3}^{(3)})}\langle\theta_{1}^{(3)},\theta_{2}^{(3)},\theta_{3}^{(3)}|\mathcal{O}|\theta_{1}^{(3)},\theta_{2}^{(3)},\theta_{3}^{(3)}\rangle_{L}\nonumber \\
 &  & +O(\mathrm{e}^{-4mR})\label{eq:nomexp}\end{eqnarray}
and\begin{eqnarray}
\text{Tr}_{L}\left(\mathrm{e}^{-RH_{L}}\right) & = & 1+\sum_{\theta^{(1)}}\mathrm{e}^{-mR\cosh(\theta^{(1)})}+\frac{1}{2}\sum_{\theta_{1}^{(2)},\theta_{2}^{(2)}}{}^{'}\mathrm{e}^{-mR(\cosh(\theta_{1}^{(2)})+\cosh(\theta_{2}^{(2)}))}\nonumber \\
 &  & +\frac{1}{6}\sum_{\theta_{1}^{(3)},\theta_{2}^{(3)},\theta_{3}^{(3)}}{}^{'}\mathrm{e}^{-mR(\cosh\theta_{1}^{(3)}+\cosh\theta_{2}^{(3)}+\cosh\theta_{3}^{(3)})}+O(\mathrm{e}^{-4mR})\label{eq:Zexp}\end{eqnarray}
The denominator of (\ref{onepointRL}) can then be easily expanded:\begin{eqnarray}
\frac{1}{\text{Tr}_{L}\left(\mathrm{e}^{-RH_{L}}\right)} & = & 1-\sum_{\theta^{(1)}}\mathrm{e}^{-mR\cosh\theta^{(1)}}+\left(\sum_{\theta^{(1)}}\mathrm{e}^{-mR\cosh\theta^{(1)}}\right)^{2}-\frac{1}{2}\sum_{\theta_{1}^{(2)},\theta_{2}^{(2)}}{}^{'}\mathrm{e}^{-mR(\cosh\theta_{1}^{(2)}+\cosh\theta_{2}^{(2)})}\nonumber \\
 &  & -\left(\sum_{\theta^{(1)}}\mathrm{e}^{-mR\cosh\theta^{(1)}}\right)^{3}+\left(\sum_{\theta^{(1)}}\mathrm{e}^{-mR\cosh\theta^{(1)}}\right)\sum_{\theta_{1}^{(2)},\theta_{2}^{(2)}}{}^{'}\mathrm{e}^{-mR(\cosh\theta_{1}^{(2)}+\cosh\theta_{2}^{(2)})}\nonumber \\
 &  & -\frac{1}{6}\sum_{\theta_{1}^{(3)},\theta_{2}^{(3)},\theta_{3}^{(3)}}{}^{'}\mathrm{e}^{-mR(\cosh\theta_{1}^{(3)}+\cosh\theta_{2}^{(3)}+\cosh\theta_{3}^{(3)})}+O(\mathrm{e}^{-4mR})\label{eq:Zinvexp}\end{eqnarray}
The primes in the multi-particle sums serve as a reminder that there
exist only states for which all quantum numbers are distinct. Since
we assumed that there is a single particle species, this means that
terms in which any two of the rapidities coincide are excluded. All
$n$-particle terms in (\ref{eq:nomexp}) and (\ref{eq:Zexp}) have
a $1/n!$ pre-factor which takes into account that different ordering
of the same rapidities give the same state; as the expansion contains
only diagonal matrix elements, phases resulting from reordering the
particles cancel. The upper indices of the rapidity variables indicate
the number of particles in the original finite volume states; this
is going to be handy when replacing the discrete sums with integrals
since it keeps track of which multi-particle state density is relevant. 

We also need an extension of the finite volume matrix elements to
rapidities that are not necessarily solutions of the appropriate Bethe-Yang
equations. The required analytic continuation is simply given by eqn.
(\ref{eq:diaggenrule})

\begin{equation}
\langle\theta_{1},\dots,\theta_{n}|\mathcal{O}|\theta_{1},\dots,\theta_{n}\rangle_{L}=\frac{1}{\rho_{n}(\theta_{1},\dots,\theta_{n})_{L}}\,\sum_{A\subset\{1,2,\dots n\}}F_{2|A|}^{s}(\{\theta_{i}\}_{i\in A})\rho_{n-|A|}(\{\theta_{i}\}_{i\notin A})_{L}+O(\mathrm{e}^{-\mu L})\label{eq:Fsextended}\end{equation}
where we made explicit the volume dependence of the $n$-particle
density factors. The last term serves as a reminder that this prescription
only defines the form factor to all orders in $1/L$ (i.e. up to residual
finite size corrections), but this is sufficient to perform the computations
in the sequel.

Using the leading behavior of the $n$-particle state density, contributions
from the $n$-particle sector scale as $L^{n}$, and for the series
expansions (\ref{eq:nomexp}), (\ref{eq:Zexp}) and (\ref{eq:Zinvexp})
it is necessary that $mL\ll\mathrm{e}^{mR}$. However if $mR$ is
big enough there remains a large interval \[
1\ll mL\ll\mathrm{e}^{mR}\]
where the expansions are expected to be valid. After substituting
these expansions into (\ref{onepointRL}) we will find order by order
that the leading term of the net result is $O(L^{0})$, and the corrections
scale as negative powers of $L$. Therefore in (\ref{onepointRL})
we can continue analytically to large $L$ and take the $L\rightarrow\infty$
limit. 

It is an interesting question why this limit exists, or to be more
precise, why do form factors with low number of particles
determine the leading terms of the low-temperature expansion. As a
matter of fact, the particle density $N/L$ has a non-vanishing $L\to\infty$ limit,
the thermal average is therefore dominated by thermodynamic
configurations where the particle number grows proportional with the
volume. However, the leading contributions 
come from disconnected terms and they
still involve form factors  with a low number of
particles, as it was pointed out in \cite{konik-2001}.

\subsection{Corrections of order $\mathrm{e}^{-mR}$}

Substituting the appropriate terms from (\ref{eq:Zinvexp}) and (\ref{eq:nomexp})
into (\ref{onepointRL}) gives the result\[
\langle\mathcal{O}\rangle_{L}^{R}=\langle\mathcal{O}\rangle_{L}+\sum_{\theta^{(1)}}\mathrm{e}^{-mR\cosh\theta^{(1)}}\left(\langle\theta^{(1)}|\mathcal{O}|\theta^{(1)}\rangle_{L}-\langle\mathcal{O}\rangle_{L}\right)+O(\mathrm{e}^{-2mR})\]
Taking the $L\to\infty$ limit one can replace the summation with
an integral over the states in the rapidity space: \[
\sum_{i}\to\int\frac{d\theta}{2\pi}\rho_{1}(\theta)\]
and using (\ref{eq:d1formula}) we can write

\begin{equation}
\rho_{1}(\theta)\left(\langle\theta|\mathcal{O}|\theta\rangle_{L}-\langle\mathcal{O}\rangle_{L}\right)=F_{2}^{s}+O(\mathrm{e}^{-\mu L})\label{1pbol1}\end{equation}
so we obtain\[
\langle\mathcal{O}\rangle^{R}=\langle\mathcal{O}\rangle+\int\frac{d\theta}{2\pi}F_{2}^{s}\mathrm{e}^{-mR\cosh{\theta}}+O(\mathrm{e}^{-2mR})\]
which coincides with eqn. (\ref{eq:TBA2s}) to this order.

\subsection{Corrections of order $e^{-2mR}$}

\label{e-2mr}

Substituting again the appropriate terms from (\ref{eq:Zinvexp})
and (\ref{eq:nomexp}) into (\ref{onepointRL}) gives the result\begin{eqnarray*}
\langle\mathcal{O}\rangle_{L}^{R} & = & \langle\mathcal{O}\rangle_{L}+\sum_{\theta^{(1)}}\mathrm{e}^{-mR\cosh\theta^{(1)}}\left(\langle\theta^{(1)}|\mathcal{O}|\theta^{(1)}\rangle_{L}-\langle\mathcal{O}\rangle_{L}\right)\\
 &  & -\left(\sum_{\theta_{1}^{(1)}}\mathrm{e}^{-mR\cosh\theta_{1}^{(1)}}\right)\left(\sum_{\theta_{2}^{(1)}}\mathrm{e}^{-mR\cosh\theta_{2}^{(1)}}\left(\langle\theta_{2}^{(1)}|\mathcal{O}|\theta_{2}^{(1)}\rangle_{L}-\langle\mathcal{O}\rangle_{L}\right)\right)\\
 &  & +\frac{1}{2}\sum_{\theta_{1}^{(2)},\theta_{2}^{(2)}}{}^{'}\mathrm{e}^{-mR(\cosh\theta_{1}^{(2)}+\cosh\theta_{2}^{(2)})}\left(\langle\theta_{1}^{(2)},\theta_{2}^{(2)}|\mathcal{O}|\theta_{1}^{(2)},\theta_{2}^{(2)}\rangle_{L}-\langle\mathcal{O}\rangle_{L}\right)+O(\mathrm{e}^{-3mR})\end{eqnarray*}
The $O(\mathrm{e}^{-2mR})$ terms can be rearranged as follows. We
add and subtract a term to remove the constraint from the two-particle
sum:\begin{eqnarray*}
 &  & +\frac{1}{2}\sum_{\theta_{1}^{(2)},\theta_{2}^{(2)}}\mathrm{e}^{-mR(\cosh\theta_{1}^{(2)}+\cosh\theta_{2}^{(2)})}\left(\langle\theta_{1}^{(2)},\theta_{2}^{(2)}|\mathcal{O}|\theta_{1}^{(2)},\theta_{2}^{(2)}\rangle_{L}-\langle\mathcal{O}\rangle_{L}\right)\\
 &  & -\frac{1}{2}\sum_{\theta_{1}^{(2)}=\theta_{2}^{(2)}}\mathrm{e}^{-2mR\cosh\theta_{1}^{(2)}}\left(\langle\theta_{1}^{(2)},\theta_{1}^{(2)}|\mathcal{O}|\theta_{1}^{(2)},\theta_{1}^{(2)}\rangle_{L}-\langle\mathcal{O}\rangle_{L}\right)\\
 &  & -\frac{1}{2}\sum_{\theta_{1}^{(1)}}\sum_{\theta_{2}^{(1)}}\mathrm{e}^{-mR(\cosh\theta_{1}^{(1)}+\cosh\theta_{2}^{(1)})}\left(\langle\theta_{1}^{(1)}|\mathcal{O}|\theta_{1}^{(1)}\rangle_{L}+\langle\theta_{2}^{(1)}|\mathcal{O}|\theta_{2}^{(1)}\rangle_{L}-2\langle\mathcal{O}\rangle_{L}\right)\end{eqnarray*}
The $\theta_{1}^{(2)}=\theta_{2}^{(2)}$ terms correspond to insertion
of some spurious two-particle states with equal Bethe quantum numbers
for the two particles ($I_{1}=I_{2}$). The two-particle Bethe-Yang
equations in this case degenerates to the one-particle case (as discussed
before, the matrix elements can be defined for these {}``states''
without any problems since we have the analytic formula (\ref{eq:Fsextended})
valid to any order in $1/L$). This also means that the density relevant
to the diagonal two-particle sum is $\rho_{1}$ and so for large $L$
we can substitute the sums with the following integrals \[
\sum_{\theta_{1,2}^{(1)}}\rightarrow\int\frac{d\theta_{1,2}}{2\pi}\rho_{1}(\theta_{1,2})\quad,\quad\sum_{\theta_{1}^{(2)}=\theta_{2}^{(2)}}\rightarrow\int\frac{d\theta}{2\pi}\rho_{1}(\theta)\quad,\quad\sum_{\theta_{1}^{(2)},\theta_{2}^{(2)}}\rightarrow\int\frac{d\theta_{1}}{2\pi}\frac{d\theta_{2}}{2\pi}\rho_{2}(\theta_{1.},\theta_{2})\]
Let us express the finite volume matrix elements in terms of form
factors using (\ref{eq:d1formula}) and (\ref{eq:d2formula}):\begin{eqnarray*}
 &  & \rho_{2}(\theta_{1},\theta_{2})\left(\langle\theta_{1}^{(2)},\theta_{2}^{(2)}|\mathcal{O}|\theta_{1}^{(2)},\theta_{2}^{(2)}\rangle_{L}-\langle\mathcal{O}\rangle_{L}\right)\\
 &  & -\rho_{1}\left(\theta_{1}\right)\rho_{1}\left(\theta_{2}\right)\left(\langle\theta_{1}|\mathcal{O}|\theta_{1}\rangle_{L}+\langle\theta_{2}|\mathcal{O}|\theta_{2}\rangle_{L}-2\langle\mathcal{O}\rangle_{L}\right)=F_{4}^{s}(\theta_{1},\theta_{2})+O(\mathrm{e}^{-\mu L})\end{eqnarray*}
Combining the above relation with (\ref{1pbol1}), we also have\[
\langle\theta,\theta|\mathcal{O}|\theta,\theta\rangle_{L}-\langle\mathcal{O}\rangle_{L}=\frac{2\rho_{1}\left(\theta\right)}{\rho_{2}(\theta,\theta)}F_{2}^{s}+O(\mathrm{e}^{-\mu L})\]
where we used that $F_{4}^{s}(\theta,\theta)=0$, which is just the
exclusion property mention after eqn. (\ref{eq:Fs_definition}). Note
that\[
\frac{\rho_{1}(\theta)^{2}}{\rho_{2}(\theta,\theta)}=1+O(L^{-1})\]
and therefore in the limit $L\rightarrow\infty$ we obtain\[
-\int\frac{d\theta}{2\pi}\mathrm{e}^{-2mR\cosh\theta}F_{2}^{s}+\frac{1}{2}\int\frac{d\theta_{1}}{2\pi}\frac{d\theta_{2}}{2\pi}F_{4}^{s}(\theta_{1},\theta_{2})\mathrm{e}^{-mR(\cosh\theta_{1}+\cosh\theta_{2})}\]
which is equal to the relevant contributions in the LeClair-Mussardo
expansion (\ref{eq:TBA2s}).

\subsection{Corrections of order $e^{-3mR}$}

In order to shorten the presentation, we introduce some further convenient
notations:\begin{eqnarray*}
 &  & E_{i}=m\cosh\theta_{i}\\
 &  & \langle\theta_{1},\dots,\theta_{n}|\mathcal{O}|\theta_{1},\dots,\theta_{n}\rangle_{L}=\langle1\dots n|\mathcal{O}|1\dots n\rangle_{L}\\
 &  & \rho_{n}(\theta_{1},\dots,\theta_{n})=\rho(1\dots n)\end{eqnarray*}
Summations will be shortened to\begin{eqnarray*}
\sum_{\theta_{1}\dots\theta_{n}} & \rightarrow & \sum_{1\dots n}\\
\sum_{\theta_{1}\dots\theta_{n}}{}^{'} & \rightarrow & \sum_{1\dots n}{}^{'}\end{eqnarray*}
Given these notations, we now multiply (\ref{eq:nomexp}) with (\ref{eq:Zinvexp})
and collect the third order correction terms:\begin{eqnarray*}
 &  & \frac{1}{6}\sum_{123}{}^{'}\mathrm{e}^{-R(E_{1}+E_{2}+E_{3})}\left(\langle123|\mathcal{O}|123\rangle_{L}-\langle\mathcal{O}\rangle_{L}\right)\\
 & - & \left(\sum_{1}\mathrm{e}^{-RE_{1}}\right)\frac{1}{2}\sum_{23}{}^{'}\mathrm{e}^{-R(E_{2}+E_{3})}\left(\langle23|\mathcal{O}|23\rangle_{L}-\langle\mathcal{O}\rangle_{L}\right)\\
 & + & \left\{ \left(\sum_{1}\mathrm{e}^{-RE_{1}}\right)\left(\sum_{2}\mathrm{e}^{-RE_{2}}\right)-\frac{1}{2}\sum_{12}{}^{'}\mathrm{e}^{-R(E_{1}+E_{2})}\right\} \left(\sum_{3}\mathrm{e}^{-RE_{3}}\right)\left(\langle3|\mathcal{O}|3\rangle_{L}-\langle\mathcal{O}\rangle_{L}\right)\end{eqnarray*}
To keep trace of the state densities, we avoid combining rapidity
sums. Now we replace the constrained summations by free sums with
the diagonal contributions subtracted:\begin{eqnarray*}
\sum_{12}{}^{'} & = & \sum_{12}-\sum_{1=2}\\
\sum_{123}{}^{'} & = & \sum_{123}-\left(\sum_{1=2,3}+\sum_{2=3,1}+\sum_{1=3,2}\right)+2\sum_{1=2=3}\end{eqnarray*}
where the diagonal contributions are labeled to show which diagonal
it sums over, but otherwise the given sum is free, e.g.\[
\sum_{1=2,3}\]
shows a summation over all triplets $\theta_{1}^{(3)},\theta_{2}^{(3)},\theta_{3}^{(3)}$
where $\theta_{1}^{(3)}=\theta_{2}^{(3)}$ and $\theta_{3}^{(3)}$
runs free (it can also be equal with the other two). We also make
use of the notation\[
F(12\dots n)=F_{2n}^{s}(\theta_{1},\dots,\theta_{n})\]
so the necessary matrix elements can be written in the form\begin{eqnarray}
\rho(123)\left(\langle123|\mathcal{O}|123\rangle_{L}-\langle\mathcal{O}\rangle_{L}\right) & = & F(123)+\rho(1)F(23)+\dots+\rho(12)F(3)+\dots\nonumber \\
\rho(122)\left(\langle122|\mathcal{O}|122\rangle_{L}-\langle\mathcal{O}\rangle_{L}\right) & = & 2\rho(2)F(12)+2\rho(12)F(3)+\rho(22)F(1)\nonumber \\
\rho(111)\left(\langle111|\mathcal{O}|111\rangle_{L}-\langle\mathcal{O}\rangle_{L}\right) & = & 3\rho(111)F(1)\nonumber \\
\rho(12)\left(\langle12|\mathcal{O}|12\rangle_{L}-\langle\mathcal{O}\rangle_{L}\right) & = & F(12)+\rho(1)F(2)+\rho(2)F(1)\nonumber \\
\rho(11)\left(\langle11|\mathcal{O}|11\rangle_{L}-\langle\mathcal{O}\rangle_{L}\right) & = & 2\rho(1)F(1)\nonumber \\
\rho(1)\left(\langle1|\mathcal{O}|1\rangle_{L}-\langle\mathcal{O}\rangle_{L}\right) & = & F(1)\label{eq:matelms}\end{eqnarray}
where we used that $F$ and $\rho$ are entirely symmetric in all
their arguments, and the ellipsis in the the first line denote two
plus two terms of the same form, but with different partitioning of
the rapidities, which can be obtained by cyclic permutation from those
displayed. We also used the exclusion property mentioned after eqn.
(\ref{eq:Fs_definition}).

We can now proceed by collecting terms according to the number of
free rapidity variables. The terms containing threefold summation
are\begin{eqnarray*}
 &  & \frac{1}{6}\sum_{123}\mathrm{e}^{-R(E_{1}+E_{2}+E_{3})}\left(\langle123|\mathcal{O}|123\rangle_{L}-\langle\mathcal{O}\rangle_{L}\right)-\frac{1}{2}\sum_{1}\sum_{2,3}\left(\langle23|\mathcal{O}|23\rangle_{L}-\langle\mathcal{O}\rangle_{L}\right)\\
 & + & \left(\sum_{1}\sum_{2}\sum_{3}-\frac{1}{2}\sum_{1,2}\sum_{3}\right)\left(\langle3|\mathcal{O}|3\rangle_{L}-\langle\mathcal{O}\rangle_{L}\right)\end{eqnarray*}
Replacing the sums with integrals\begin{eqnarray*}
\sum_{1} & \rightarrow & \int\frac{d\theta_{1}}{2\pi}\rho(1)\\
\sum_{1,2} & \rightarrow & \int\frac{d\theta_{1}}{2\pi}\frac{d\theta_{2}}{2\pi}\rho(12)\\
\sum_{1,2,3} & \rightarrow & \int\frac{d\theta_{1}}{2\pi}\frac{d\theta_{2}}{2\pi}\frac{d\theta_{3}}{2\pi}\rho(123)\end{eqnarray*}
and using (\ref{eq:matelms}) we get\begin{eqnarray*}
 &  & \frac{1}{6}\int\frac{d\theta_{1}}{2\pi}\frac{d\theta_{2}}{2\pi}\frac{d\theta_{3}}{2\pi}\mathrm{e}^{-R(E_{1}+E_{2}+E_{3})}\left(F(123)+3\rho(1)F(23)+3\rho(12)F(3)\right)\\
 & - & \frac{1}{2}\int\frac{d\theta_{1}}{2\pi}\frac{d\theta_{2}}{2\pi}\frac{d\theta_{3}}{2\pi}\mathrm{e}^{-R(E_{1}+E_{2}+E_{3})}\left(\rho(1)F(23)+2\rho(1)\rho(2)F(3)\right)\\
 & + & \int\frac{d\theta_{1}}{2\pi}\frac{d\theta_{2}}{2\pi}\frac{d\theta_{3}}{2\pi}\mathrm{e}^{-R(E_{1}+E_{2}+E_{3})}\left(\rho(1)\rho(2)F(3)-\frac{1}{2}\rho(12)F(3)\right)\end{eqnarray*}
where we reshuffled some of the integration variables. Note that all
terms cancel except the one containing $F(123)$ and writing it back
to its usual form we obtain\begin{equation}
\frac{1}{6}\int\frac{d\theta_{1}}{2\pi}\frac{d\theta_{2}}{2\pi}\frac{d\theta_{3}}{2\pi}F_{6}^{s}(\theta_{1},\theta_{2},\theta_{3})\mathrm{e}^{-mR(\cosh\theta_{1}+\cosh\theta_{2}+\cosh\theta_{3})}\label{eq:res3int}\end{equation}
It is also easy to deal with terms containing a single integral. The
only term of this form is\[
\frac{1}{3}\sum_{1=2=3}\mathrm{e}^{-R(E_{1}+E_{2}+E_{3})}\left(\langle123|\mathcal{O}|123\rangle_{L}-\langle\mathcal{O}\rangle_{L}\right)\]
When all rapidities $\theta_{1}^{(3)},\theta_{2}^{(3)},\theta_{3}^{(3)}$
are equal, the three-particle Bethe-Yang equations reduce to the one-particle
case\[
mL\sinh\theta_{1}^{(3)}=2\pi I_{1}\]
Therefore the relevant state density is that of the one-particle state:\begin{eqnarray}
\frac{1}{3}\int\frac{d\theta_{1}}{2\pi}\mathrm{e}^{-3RE_{1}}\rho(1)\left(\langle111|\mathcal{O}|111\rangle_{L}-\langle\mathcal{O}\rangle_{L}\right) & = & \int\frac{d\theta_{1}}{2\pi}\mathrm{e}^{-3RE_{1}}\rho(1)\frac{\rho(11)}{\rho(111)}F(1)\nonumber \\
 & \rightarrow & \int\frac{d\theta_{1}}{2\pi}\mathrm{e}^{-3mR\cosh\theta{}_{1}}F_{2}^{s}\label{eq:res1int}\end{eqnarray}
where we used that\[
\rho(1)\frac{\rho(11)}{\rho(111)}\rightarrow1\]
when $L\rightarrow\infty$.

The calculation of double integral terms is much more involved. We
need to consider\begin{eqnarray}
 &  & -\frac{1}{6}\left(\sum_{1=2,3}+\sum_{1=3,2}+\sum_{2=3,1}\right)\mathrm{e}^{-R(E_{1}+E_{2}+E_{3})}\left(\langle123|\mathcal{O}|123\rangle_{L}-\langle\mathcal{O}\rangle_{L}\right)\nonumber \\
 &  & +\frac{1}{2}\sum_{1}\sum_{2=3}\mathrm{e}^{-R(E_{1}+E_{2}+E_{3})}\left(\langle23|\mathcal{O}|23\rangle_{L}-\langle\mathcal{O}\rangle_{L}\right)\nonumber \\
 &  & +\frac{1}{2}\sum_{1=2}\sum_{3}\mathrm{e}^{-R(E_{1}+E_{2}+E_{3})}\left(\langle3|\mathcal{O}|3\rangle_{L}-\langle\mathcal{O}\rangle_{L}\right)\label{eq:dintstart}\end{eqnarray}
We need the density of partially degenerate three-particle states.
The relevant Bethe-Yang equations are\begin{eqnarray*}
mL\sinh\theta_{1}+\delta(\theta_{1}-\theta_{2}) & = & 2\pi I_{1}\\
mL\sinh\theta_{2}+2\delta(\theta_{2}-\theta_{1}) & = & 2\pi I_{2}\end{eqnarray*}
where we supposed that the first and the third particles are degenerate
(i.e. $I_{3}=I_{1}$), and used a convention for the phase-shift and
the quantum numbers where $\delta(0)=0$. The density of these degenerate
states is then given by\[
\bar{\rho}(13,2)=\det\left(\begin{array}{ll}
LE_{1}+\varphi(\theta_{1}-\theta_{2}) & -\varphi(\theta_{1}-\theta_{2})\\
-2\varphi(\theta_{1}-\theta_{2}) & LE_{2}+2\varphi(\theta_{1}-\theta_{2})\end{array}\right)\]
where we used that $\varphi(\theta)=\varphi(-\theta)$. Using the
above result and substituting integrals for the sums, we can rewrite
eqn. (\ref{eq:dintstart}) in the form\begin{eqnarray*}
 &  & -\frac{1}{6}\int\frac{d\theta_{1}}{2\pi}\frac{d\theta_{2}}{2\pi}\mathrm{e}^{-R(2E_{1}+E_{2})}\frac{\bar{\rho}(13,2)}{\rho(112)}\left(2\rho(1)F(12)+2\rho(12)F(1)+\rho(11)F(2)\right)+\dots\\
 &  & +\frac{1}{2}\int\frac{d\theta_{1}}{2\pi}\frac{d\theta_{2}}{2\pi}\mathrm{e}^{-R(E_{1}+2E_{2})}\rho(1)\rho(2)\frac{2\rho(2)}{\rho(22)}F(2)\\
 &  & +\frac{1}{2}\int\frac{d\theta_{1}}{2\pi}\frac{d\theta_{3}}{2\pi}\mathrm{e}^{-R(2E_{1}+E_{3})}\rho(1)\rho(3)\frac{1}{\rho(3)}F(3)\end{eqnarray*}
where the ellipsis denote two terms that can be obtained by cyclical
permutation of the indices $1,2,3$ from the one that is explicitly
displayed, and these three contributions can be shown to be equal
to each other by relabeling the integration variables: \begin{eqnarray}
 &  & -\frac{1}{2}\int\frac{d\theta_{1}}{2\pi}\frac{d\theta_{2}}{2\pi}\mathrm{e}^{-R(2E_{1}+E_{2})}\frac{\bar{\rho}(13,2)}{\rho(112)}\left(2\rho(1)F(12)+2\rho(12)F(1)+\rho(11)F(2)\right)\nonumber \\
 &  & +\frac{1}{2}\int\frac{d\theta_{1}}{2\pi}\frac{d\theta_{2}}{2\pi}\mathrm{e}^{-R(E_{1}+2E_{2})}\rho(1)\rho(2)\frac{2\rho(2)}{\rho(22)}F(2)\nonumber \\
 &  & +\frac{1}{2}\int\frac{d\theta_{1}}{2\pi}\frac{d\theta_{3}}{2\pi}\mathrm{e}^{-R(2E_{1}+E_{3})}\rho(1)\rho(3)\frac{1}{\rho(3)}F(3)\label{eq:dintreshuff}\end{eqnarray}
We first evaluate the terms containing $F(23)$ which results in\begin{equation}
-\int\frac{d\theta_{1}}{2\pi}\frac{d\theta_{2}}{2\pi}F_{4}^{s}(\theta_{1},\theta_{2})\mathrm{e}^{-mR(\cosh\theta_{1}+2\cosh\theta_{2})}\label{eq:dintres1}\end{equation}
using that \[
\frac{\bar{\rho}(13,2)}{\rho(112)}\rho(1)=1+O(L^{-1})\]
We can now treat the terms containing the amplitude $F(1)=F(2)=F(3)=F_{2}^{s}$.
Exchanging the variables $\theta_{1}\leftrightarrow\theta_{2}$ in
the second line and redefining $\theta_{3}\rightarrow\theta_{2}$
in the third line of eqn. (\ref{eq:dintreshuff}) results in\[
\frac{F_{2}^{s}}{2}\int\frac{d\theta_{1}}{2\pi}\frac{d\theta_{2}}{2\pi}\mathrm{e}^{-R(2E_{1}+E_{2})}\left\{ -\frac{\bar{\rho}(13,2)}{\rho(112)}\left(2\rho(12)+\rho(11)\right)+\frac{2\rho(1)^{2}\rho(2)}{\rho(11)}+\rho(1)\right\} \]
The combination of the various densities in this expression requires
special care. From the large $L$ asymptotics\[
\rho(i)\sim E_{i}L\quad,\quad\rho(ij)\sim E_{i}E_{j}L^{2}\quad,\quad\rho(ijk)\sim E_{i}E_{j}E_{k}L^{3}\quad,\quad\bar{\rho}(13,2)\sim E_{1}E_{2}L^{2}\]
it naively scales with $L$. However, it can be easily verified that
the coefficient of the leading term, which is linear in $L$, is exactly
zero. Without this, the large $L$ limit would not make sense, so
this is rather reassuring. We can then calculate the sub-leading term,
which requires tedious but elementary manipulations. The end result
turns out to be extremely simple\begin{equation}
-\frac{\bar{\rho}(13,2)}{\rho(112)}\left(2\rho(12)+\rho(11)\right)+\frac{2\rho(1)^{2}\rho(2)}{\rho(11)}+\rho(1)=-\varphi(\theta_{1}-\theta_{2})+O(L^{-1})\label{eq:anomdens}\end{equation}
so the contribution in the $L\rightarrow\infty$ limit turns out to
be just\begin{equation}
-\frac{1}{2}\int\frac{d\theta_{1}}{2\pi}\frac{d\theta_{2}}{2\pi}F_{2}^{s}\varphi(\theta_{1}-\theta_{2})\mathrm{e}^{-mR(2\cosh\theta_{1}+\cosh\theta_{2})}\label{eq:dintres2}\end{equation}
Summing up the contributions (\ref{eq:res3int}), (\ref{eq:res1int}),
(\ref{eq:dintres1}) and (\ref{eq:dintres2}) we obtain
\begin{eqnarray}
 &  & \frac{1}{6}\int\frac{d\theta_{1}}{2\pi}\frac{d\theta_{2}}{2\pi}\frac{d\theta_{3}}{2\pi}F_{6}^{s}(\theta_{1},\theta_{2},\theta_{3})\mathrm{e}^{-mR(\cosh\theta_{1}+\cosh\theta_{2}+\cosh\theta_{3})}\nonumber \\
 &  & -\int\frac{d\theta_{1}}{2\pi}\frac{d\theta_{2}}{2\pi}F_{4}^{s}(\theta_{1},\theta_{2})\mathrm{e}^{-mR(\cosh\theta_{1}+2\cosh\theta_{2})}+\int\frac{d\theta_{1}}{2\pi}F_{2}^{s}\mathrm{e}^{-3mR\cosh\theta_{1}}\nonumber \\
 &  & -\frac{1}{2}\int\frac{d\theta_{1}}{2\pi}\frac{d\theta_{2}}{2\pi}F_{2}^{s}\varphi(\theta_{1}-\theta_{2})\mathrm{e}^{-mR(\cosh\theta_{1}+2\cosh\theta_{2})}\label{eq:our3order}\end{eqnarray}
which agrees exactly with eqn. (\ref{eq:lm3order}).

This result gives an independent support
for the LeClair-Mussardo expansion. 
The calculation is model independent, and although we
only worked it out to order $\mathrm{e}^{-3mR}$, it is expected that it
coincides with the LeClair-Mussardo expansion to all orders. For a
complete proof we need a better understanding of its structure, which
is out of the scope of the present work.



\section{Low-temperature expansion for the two-point function}

\label{lowT_2p}

The method presented in the previous section has a straightforward extension to higher
point correlation functions. For example, a two-point correlation
function \[
\langle\mathcal{O}_{1}(x,t)\mathcal{O}_{2}(0)\rangle_{L}^{R}=\frac{\text{Tr}_{L}\left(\mathrm{e}^{-RH_{L}}\mathcal{O}_{1}(x,t)\mathcal{O}_{2}(0)\right)}{\text{Tr}_{L}\left(\mathrm{e}^{-RH_{L}}\right)}\]
can be expanded inserting two complete sets of states\begin{equation}
\text{Tr}_{L}\left(\mathrm{e}^{-RH_{L}}\mathcal{O}_{1}(x,t)\mathcal{O}_{2}(0)\right)=\sum_{m,n}\mathrm{e}^{-RE_{n}(L)}\langle n|\mathcal{O}(x,t)_1|m\rangle_{L}\langle m|\mathcal{O}_2(0)|n\rangle_{L}\label{eq:2ptexp}\end{equation}
The above expression can be evaluated along
the lines presented in the previous section, provided that the intermediate
state sums are properly truncated. The explicit evaluation
of expansion (\ref{eq:2ptexp}) has not yet been carried out.
As a first step, here we investigate the simplest terms and calculate
the first nontrivial contribution; the systematic evaluation of 
higher order terms is left
for further work. We wish to remark, that the material presented in
this section is unpublished.  

We define 
\begin{equation}
    \langle\mathcal{O}_{1}(x,t)\mathcal{O}_{2}(0)\rangle_{L}^{R}=
\frac{1}{Z} \sum_{N,M} I_{NM}
\label{Imn_def}
\end{equation}
where
\begin{eqnarray}
\nonumber
I_{NM}=\sum_{I_1\dots I_N}\sum_{J_1\dots J_M}
\bra{\{I_1\dots I_N\}}\mathcal{O}_1(0)\ket{\{J_1\dots J_M\}}_L \times\\
\bra{\{J_1\dots J_M\}}\mathcal{O}_2(0)\ket{\{I_1\dots I_N\}}_L
e^{i(P_1-P_2)x}e^{-E_1(R-t)}e^{-E_2t}
\end{eqnarray}
and $E_{1,2}$ and $P_{1,2}$ are the total energies and momenta of
the multi-particle states.

It is easy to see, that the terms
$N=0,M=0\dots\infty$ and $N=0\dots\infty,M=0$ add up to
zero-temperature correlation functions
\begin{equation*}
  \sum_{M=0}^\infty I_{0M}=
  \langle\mathcal{O}_{1}(x,t)\mathcal{O}_{2}(0)\rangle_{L}
\quad\quad
  \sum_{N=0}^\infty I_{N0}=
  \langle\mathcal{O}_{1}(x,R-t)\mathcal{O}_{2}(0)\rangle_{L}
\end{equation*}
The first nontrivial term is therefore $I_{11}$, which is given by
\begin{equation}
\label{I11}
  I_{11}=\sum_{I,J} 
\bra{\{I\}}  \mathcal{O}_1(0)\ket{\{J\}}_L\
\bra{\{J\}}\mathcal{O}_2(0)\ket{\{I\}}_L \
e^{i(p_1-p_2)x}e^{-E_1(R-t)}e^{-E_2t}
\end{equation}
where now $E_{1,2}=m\cosh(\theta_{1,2})$ and $p_{1,2}=m\sinh(\theta_{1,2})$ are finite size one-particle
energies and momenta. Upon \eqref{eq:genffrelation} and \eqref{eq:d1formula} the two-particle matrix elements are given by
\begin{eqnarray*}
  \bra{\{I\}}  \mathcal{O}_1(0)\ket{\{J\}}_L&=&
\frac{F_2^{\mathcal{O}_1}(\theta_1+i\pi,\theta_2)}
{\sqrt{\rho_1(\theta_1)\rho_1(\theta_2)}}+\delta_{IJ}
\vev{\mathcal{O}_1}\\
  \bra{\{J\}}  \mathcal{O}_2(0)\ket{\{I\}}_L&=&
\frac{F_2^{\mathcal{O}_2}(\theta_2+i\pi,\theta_1)}
{\sqrt{\rho_1(\theta_1)\rho_1(\theta_2)}}+\delta_{IJ}
\vev{\mathcal{O}_2}
\end{eqnarray*}
Substituting the above formulas into \eqref{I11} one obtains
\begin{eqnarray}
\label{I11b}
\nonumber
  I_{11}=\sum_{I,J} 
\frac{F_2^{\mathcal{O}_1}(\theta_1+i\pi,\theta_2)F_2^{\mathcal{O}_2}(\theta_2+i\pi,\theta_1)}
{\rho_1(\theta_1)\rho_1(\theta_2)}e^{i(p_1-p_2)x}e^{-E_1(R-t)}e^{-E_2t}\\
\nonumber
+\vev{\mathcal{O}_1} \sum_J 
\frac{F_2^{\mathcal{O}_2}(\theta_2+i\pi,\theta_2)}{\rho_1(\theta_2)}
e^{-E_2R}+
\vev{\mathcal{O}_2} \sum_J 
\frac{F_2^{\mathcal{O}_1}(\theta_1+i\pi,\theta_1)}{\rho_1(\theta_1)}
e^{-E_1R}\\
+\sum_{I}\vev{\mathcal{O}_1}\vev{\mathcal{O}_2} e^{-E_1R}
\end{eqnarray}
The first term in \eqref{I11b} can be transformed in the $L\to\infty$
limit into the well-defined
(singularity free) 
double integral
\begin{equation}
\label{double_integral}
  \int \frac{d\theta_1}{2\pi}\frac{d\theta_2}{2\pi}
F_2^{\mathcal{O}_1}(\theta_1+i\pi,\theta_2)F_2^{\mathcal{O}_2}(\theta_2+i\pi,\theta_1)
e^{i(\sinh(\theta_1)-\sinh(\theta_2))mx}
e^{-m(R-t)\cosh(\theta_1)-mt\cosh(\theta_2)}
\end{equation}
In the second and third terms of \eqref{I11b} one can  recognize the
first thermal corrections to the expectation values of $\mathcal{O}_2$
and $\mathcal{O}_1$, as given by \eqref{1pbol1}. The last term is of
$\mathcal{O}(L)$ and thus divergent; however, it gets canceled
by the $\mathcal{O}(L)$ term coming from the expansion
\begin{equation*}
  1/Z=1-\sum_{I} e^{-E_1R} +\mathcal{O}(e^{-2mR})
\end{equation*}

Putting everything together
\begin{eqnarray}
\nonumber
&&    \langle\mathcal{O}_{1}(x,t)\mathcal{O}_{2}(0)\rangle^{R}=
\vev{\mathcal{O}_{1}}^R\vev{\mathcal{O}_{2}}^R+\\
\label{eq:2p_funct_elso_hiphiphurra}
&&+\big(\langle\mathcal{O}_{1}(x,t)\mathcal{O}_{2}(0)\rangle-\vev{\mathcal{O}_{1}}\vev{\mathcal{O}_{2}}\big)+
\big(\langle\mathcal{O}_{1}(x,R-t)\mathcal{O}_{2}(0)\rangle-\vev{\mathcal{O}_{1}}\vev{\mathcal{O}_{2}}\big)\\
\nonumber
&&+ \int \frac{d\theta_1}{2\pi}\frac{d\theta_2}{2\pi}
F_2^{\mathcal{O}_1}(\theta_1+i\pi,\theta_2)F_2^{\mathcal{O}_2}(\theta_2+i\pi,\theta_1)
e^{i(\sinh(\theta_1)-\sinh(\theta_2))mx}
e^{-m(R-t)\cosh(\theta_1)-mt\cosh(\theta_2)}
\\
&&+\mathcal{O}(e^{-2mR})
\nonumber
\end{eqnarray}

Equation \eqref{eq:2p_funct_elso_hiphiphurra} is a new result of this
work and it can serve as a starting point to calculate higher
order terms. \footnote{We wish to note, that similar techniques are used in a recent work
\cite{essler-2009} which appeared after the completion of this
thesis. In addition to the results above,  the regularized finite contributions of
$I_{12}$ and $I_{21}$ are also determined in \cite{essler-2009}.
 }

\subsection{Comparison with the LeClair-Mussardo proposal}

Based on the TBA approach LeClair and Mussardo proposed the following
formula for the thermal two-point function (equation
(3.3) in \cite{leclair_mussardo})
\begin{eqnarray}
\vev{\mathcal{O}(x,t)\mathcal{O}(0,0)}^R=&&
\big(\vev{\mathcal{O}}^R\big)^2+
\sum_{N=1}^\infty \frac{1}{N!} \sum_{\sigma_i=\pm 1}
\int \frac{d\theta_1}{2\pi}\dots \frac{d\theta_N}{2\pi}
\left[\prod_{j=1}^N f_{\sigma_j}(\theta_j) 
\exp\Big(-\sigma_j(t\eps_j+ixk_j)\Big)\right]\nonumber\\
&&\times \big|\bra{0}\mathcal{O}\ket{\theta_1\dots\theta_N}_{\sigma_1\dots\sigma_N}\big|^2
  \label{eq:2point_leclair_mussardo}
\end{eqnarray}
where $f_{\sigma_j}(\theta_j)=1/(1+e^{-\sigma_j \eps(\theta_j)})$,  $\eps_j=\eps(\theta_j)/R$ and $k_j=k(\theta_j)$ with
$\eps(\theta)$ being the solution of the TBA equations \eqref{eq:TBA}
and $k(\theta)$ given by 
\begin{eqnarray*}
  k(\theta)&=&m\sinh(\theta)+\int d\theta' \delta(\theta-\theta')
  \rho_1(\theta')\\
2\pi \rho_1(\theta)(1+e^{\eps(\theta)})
&=&m\cosh(\theta)+\int d\theta' \varphi(\theta-\theta') \rho_1(\theta')
\end{eqnarray*}
The form factors appearing in \eqref{eq:2point_leclair_mussardo} are
defined by
\begin{equation*}
  \bra{0}\mathcal{O}\ket{\theta_1\dots\theta_N}_{\sigma_1\dots\sigma_N}=
F^{\mathcal{O}}_N(\theta_1-i\pi\tilde{\sigma}_1,\dots,\theta_N-i\pi\tilde{\sigma}_N)\quad\quad
\tilde{\sigma}_j=(1-\sigma_j)/2 \in \{0,1\}
\end{equation*}

We do not attempt here a thorough
analysis of \eqref{eq:2point_leclair_mussardo}; nevertheless we wish
to point out,
that to order $\mathcal{O}(e^{-mR})$ it coincides with
our result \eqref{eq:2p_funct_elso_hiphiphurra}. The double integral
\eqref{double_integral} is given for example by the $N=2$,\
$\sigma_{1,2}=(+1,-1)$ term in \eqref{eq:2point_leclair_mussardo} after
one substitutes the 0{\small th}-order approximations
\begin{equation*}
  \eps(\theta)=mR\cosh(\theta)+\mathcal{O}(e^{-mR})\quad\quad
 k(\theta)=m\sinh(\theta)+\mathcal{O}(e^{-mR})
\end{equation*}

The validity of the LeClair-Mussardo proposal was doubted by two
different authors \cite{saleurfiniteT,CastroAlvaredo:2002ud}.
We did not find a discrepancy at order $\mathcal{O}(e^{-mR})$;
it is a subject of future research to decide whether all terms in our
low-temperature expansion are properly reproduced by
\eqref{eq:2point_leclair_mussardo}.

\chapter{Conclusions}

In this thesis we investigated finite size effects in 1+1 dimensional
integrable quantum field theories;  we considered massive
integrable models with diagonal scattering.  There were two main subjects of
interest: finite volume form factors of local operators and finite
size effects to correlation functions.

\bigskip

In the first main part of the work we gave a description of finite volume form
factors in terms of the infinite volume form factors (solutions of the
form factor bootstrap) and the S-matrix of the theory. 

First we compared
the spectral representation of correlation functions in an infinite
system and in a finite volume $L$. We showed that the matrix elements in finite volume
essentially coincide with the infinite volume form factors up to a
nontrivial normalization factor, which is related to the particle
densities of the finite volume spectrum. We showed, that
our results capture all finite size corrections which scale as powers
of $1/L$. We also showed that the residual finite size effects are of
order $e^{-\mu L}$ with $\mu$ a universal characteristic mass scale of the
theory, which does not depend on the particular form factor in question.

As a second step, we conjectured a formula for generic matrix elements
without disconnected pieces. Together with the elementary case this is
a new result of the author,
although it 
can be considered as a generalization of partial results which
appeared in \cite{cikk_resonances} and independently in
\cite{lellouch_luscher,k_to_pipi}. 
Disconnected terms occur if there is a particle with a given rapidity,
which present in both the ''bra'' and the ''ket'' vector. We showed
that in finite volume this only happens in the case of diagonal form
factors and matrix elements of ''parity-symmetric'' states including
zero-momentum particles. 

The diagonal matrix elements require special care because of the presence
of various disconnected terms and the ambiguity of the infinite
volume diagonal form factors. We used form factor perturbation theory
to derive the first two cases (diagonal matrix elements of
one-particle and two-particle states) in terms of the symmetric evaluation of the infinite
volume diagonal form factors. Based on this result, we conjectured a
general formula, which we expressed with both the symmetric and
connected evaluations. The latter expression was used to prove that
our result coincides with a conjecture made independently by Saleur in
\cite{saleurfiniteT}. The rigorous proofs concerning the two simplest
cases (eqs. \eqref{eq:d1formula} and \eqref{eq:d2formula}) are new
results of the author, together with the determination of the general
relation 
between the symmetric and the connected form factors (Theorem 1).

To complete the description of finite volume matrix elements we also
conjectured a formula for the case of states with zero-momentum particles. 
The most general result concerns  situations with possibly more than one zero-momentum
particles present in both states; this only occurs in theories with
multiple particle species.  In \cite{fftcsa2}
we published the result for theories with only one particle type; the
generalization \eqref{eq:zero_mom_most_general} is an unpublished
result of this work.

\bigskip

Following the analytic study, in section \ref{s:numerics} we turned to numerical methods to
confirm our results and conjectures. We investigated the massive Lee-Yang model and the critical Ising
model in a magnetic field and considered form factors of the
perturbing field, and the energy operator, respectively. We used TCSA
as a numerical tool to obtain finite volume spectra and form factors.
First we identified finite volume states at different values of the volume
$L$: we
matched the numerical data with predictions of the Bethe-Yang
equations. We then calculated the finite volume form factors and
compared them to our analytic formulas (the infinite volume form
factors were already available in both models). In all cases
(including form factors with disconnected pieces) we
observed a satisfactory agreement in the scaling region, where both
the residual finite size effects and the truncation errors are
negligible. 

We stress that our numerical investigation not only confirms our
results about matrix elements in finite volume, but it is also the first direct test
of the infinite volume form factors. These functions are obtained  by
solving the form factor axioms of the bootstrap program and then selecting
the solutions with the desired symmetry and scaling properties. Although
the identification of scattering theories as 
perturbed conformal field theories is well-established, 
there were only a few direct tests of the individual form factor functions 
prior to our work; the usual tests in the literature proceed through
evaluation of correlation functions, sum-rules, etc.
In this work, on the other hand, we
directly compared the form factor functions at different values of the
rapidity parameters to TCSA data, thus providing evidence for the
applicability of the bootstrap approach to form factors.

\bigskip 

In the case of the Ising model we observed
that in some situations there is no scaling region, ie. the
residual finite size corrections remain relevant even in large volumes, which are out  of
the reach of our TCSA routines. The large exponential corrections  were explained by the presence
of a $\mu$-term with a small exponent. These findings served as
a motivation for section \ref{exponential}, where we investigated the
$\mu$-terms associated to finite size energies and finite volume form
factors. 

It is well-known that the $\mu$-term is always connected with the inner
structure of the particles, ie. their composition under the bootstrap procedure.
We applied a simple quantum mechanical picture of bound
state quantization in finite volume: the momenta of the constituents take
complex values but they are still determined by the analytical continuation of the
Bethe-Yang equations. We showed that if this principle is implemented
properly, the $\mu$-term can always be obtained by 
analytically
continuing the usual formulas to the complex rapidities of the
constituents. Moreover it was shown, that in some circumstances it is
possible for the constituents of a bound state to unbind. This phenomenon was
demonstrated numerically in case of the Ising model, where moving $A_3$
states with odd momentum quantum numbers dissociate in small volume into conventional
$A_1A_1$ scattering states.

\bigskip

In the second part of the work (section \ref{correlations}) we
investigated finite temperature corrections to vacuum expectation
values and correlation functions. We introduced finite volume as a
regulator of the otherwise ill-defined Boltzmann sum and 
established the spectral representation of $n$-point functions in terms of the finite volume
form factors. We showed that it is possible to derive a systematic
low temperature expansion with a well-defined $L\to\infty$ limit.  

In the case of the one-point function we calculated the first three
nontrivial terms and found complete agreement with the
LeClair-Mussardo approach. It was pointed out, that the Delfino proposal
differs from the LeClair-Mussardo formula exactly at this order; the
discrepancy is easily explained in light of our results. 
Any evaluation scheme of thermal averages based on a spectral
representation has to deal with infinities, which
are introduced by disconnected terms of the form factors and by
different contributions to the partition function itself. These
infinities cancel as expected, however a well-defined regularization
procedure is needed to obtain the left-over finite pieces. 
This point was missed in the proposals prior to our work. The
LeClair-Mussardo formula is believed to be correct to all orders, 
however a general proof (possibly based on a re-summation scheme of our
low-temperature expansion) 
is not available.

In our calculations the volume $L$  served as a cutoff to regulate the
divergent disconnected contributions. 
Those terms in the
low-temperature expansion which scale with positive powers of $L$ add
up to zero, as it is necessary to obtain a meaningful result. We stress that in order to obtain the correct
$\mathcal{O}(1)$ terms it was crucial to use the interacting densities
$\rho^{(n)}$, which differ from the free-theory densities by
sub-leading terms.

In this work we also considered the two-point function. We
only calculated the first non-trivial term, which is a new
(unpublished) result of
this work.  Our
leading-order calculation showed agreement with the LeClair-Mussardo
approach. 
A study of higher order terms  is left for future work, which will
hopefully shed more light on the  validity of  the LeClair-Mussardo
proposal.

\bigskip

There are several directions to generalize our results on finite
volume form factors. 

The extension to boundary field theories has
already been established \cite{Kormos:2007qx} together with the
evaluation of finite temperature expectation values of boundary
operators \cite{Takacs:2008ec}. 

The generalization to non-diagonal
scattering theories would be desirable, since these theories (for
example the $O(3)$ $\sigma$-model) serve as effective field theories
describing long-range interactions is real-word condensed matter
systems (Heisenberg spin chains, etc.). We expect that the principles
laid out in this work (finite volume matrix elements and infinite
volume form factors connected by a density factor) carry over to the
non-diagonal case.
However, non-diagonal scattering poses technical difficulties. One one hand,
the quantization of scattering states in
finite volume is more complicated, since it involves a diagonalization
of the transfer matrix.
On the other hand, infinite volume form factors (decomposed into the different
channels of the scattering) are also more difficult
to obtain. 

It is an interesting question whether some of our results can be
extended to massless scattering theories. These models can be used to
study renormalization group flows with a nontrivial conformal
IR fixed point \cite{zam_massless,saleur_massless2,Delfino:1994ea}.
Although the form factor
bootstrap for massless theories has already been established
\cite{Delfino:1994ea,Mejean:1996xg}, it is not clear 
what happens in a finite volume. 
In the absence of a mass-gap one does not have control over residual
finite size effects, which decay exponentially in a massive
theory. The situation is even worse in the case of thermal expectation
values and correlation functions: there is no suitable
variable (such as $e^{-mR}$ in a massive theory) to perform a low-temperature expansion.

Finally we would like to remark that an exact description of finite
volume form factors would be desirable, since they could be used to calculate correlations at
finite temperature in a different way, than the one presented
in this work. Instead of dealing with a theory defined in infinite
volume and finite temporal extent $R=1/T$ one can perform an Euclidean
rotation $(\tau,x)\to (-x,\tau)$ to deal with a zero-temperature
system defined in a finite volume $R$. Thermal corrections then become
finite size corrections which can be readily calculated by a conventional
zero-temperature form factor expansion. Obviously one needs the exact
finite volume form factors in this approach: the summation over form
factors not including the exponential corrections only reproduces the
infinite volume correlation functions.

As far as we know, the only exact results
available concern the form factors of the order and disorder operators
in Ising model with zero magnetic field \cite{Bugrij:2001nf,bugrij-2003-319,Fonseca:2001dc}. 
However, these calculations are rather model-specific and it is not clear at all how to approach
more complicated (interacting) models. A TBA-like integral equation
similar to the one describing excited state energies
\cite{excited_TBA}  would be desirable; however, it is not clear how
to solve this problem.

\cleardoublepage
\addcontentsline{toc}{chapter}{Bibliography}
\bibliography{doktoribib}

\begin{thebibliography}{100}

\bibitem{Fisher:1972zza}
M.~E. Fisher and M.~N. Barber, \emph{{Scaling Theory for Finite-Size Effects in
  the Critical Region}}, Phys. Rev. Lett. \textbf{28} (1972) 1516-1519.

\bibitem{Cardy:1986ie}
J.~L. Cardy, \emph{{Operator Content of Two-Dimensional Conformally Invariant
  Theories}}, Nucl. Phys. \textbf{B270} (1986) 186-204.

\bibitem{casimir}
H.~B.~G. Casimir, \emph{{On the Attraction Between Two Perfectly Conducting
  Plates}}, Indag. Math. \textbf{10} (1948) 261-263.

\bibitem{luscher_1particle}
M.~L{\"u}scher, \emph{{Volume Dependence of the Energy Spectrum in Massive
  Quantum Field Theories. 1. Stable Particle States}}, Commun. Math. Phys.
  \textbf{104} (1986) 177.

\bibitem{luscher_2particle}
M.~L{\"u}scher, \emph{{Volume Dependence of the Energy Spectrum in Massive
  Quantum Field Theories. 2. Scattering States}}, Commun. Math. Phys.
  \textbf{105} (1986) 153-188.

\bibitem{luscher_szorodas_1+1}
M.~L{\"u}scher and U.~Wolff, \emph{{How to calculate the elastic scattering
  matrix in two dimensional quantum field theories by numerical simulation}},
  Nucl. Phys. \textbf{B339} (1990) 222-252.

\bibitem{luscher_unstable}
M.~L{\"u}scher, \emph{{Signatures of unstable particles in finite volume}},
  Nucl. Phys. \textbf{B364} (1991) 237-254.

\bibitem{cikk_resonances}
B.~Pozsgay and G.~Takacs, \emph{{Characterization of resonances using finite
  size effects}}, Nucl. Phys. \textbf{B748} (2006) 485-523,
  \texttt{hep-th/0604022}.

\bibitem{Affleck_Kondo}
I.~Affleck, \emph{{Conformal Field Theory Approach to the Kondo Effect}}, Acta
  Phys. Polon. \textbf{B26} (1995) 1869-1932, \texttt{cond-mat/9512099}.

\bibitem{FLS}
P.~Fendley, F.~Lesage and H.~Saleur, \emph{{A unified framework for the Kondo
  problem and for an impurity in a Luttinger liquid}}, J. Statist. Phys.
  \textbf{85} (1996) 211, \texttt{cond-mat/9510055}.

\bibitem{CS1}
J.~S. Caux, H.~Saleur and F.~Siano, \emph{{The two-boundary sine-Gordon
  model}}, Nucl. Phys. \textbf{B672} (2003) 411-461, \texttt{cond-mat/0306328}.

\bibitem{CS2}
J.~S. Caux, H.~Saleur and F.~Siano, \emph{{The Josephson current in Luttinger
  liquid-superconductor junctions}}, Phys. Rev. Lett. \textbf{88} (2002)
  106402, \texttt{cond-mat/0109103}.

\bibitem{saleur00}
H.~Saleur, \emph{Lectures on Non Perturbative Field Theory and Quantum Impurity
  Problems: Part II}, ~~~, \texttt{cond-mat/0007309}.

\bibitem{saleur98}
H.~Saleur, \emph{Lectures on Non Perturbative Field Theory and Quantum Impurity
  Problems: Part I}, ~~~, \texttt{cond-mat/9812110}.

\bibitem{zam_integrable}
A.~B. Zamolodchikov, \emph{{Integrable field theory from conformal field
  theory}}, Adv. Stud. Pure Math. \textbf{19} (1989) 641-674.

\bibitem{zam_int1}
A.~B. Zamolodchikov, \emph{{Higher Order Integrals of Motion in Two-Dimensional
  Models of the Field Theory with a Broken Conformal Symmetry}}, JETP Lett.
  \textbf{46} (1987) 160-164.

\bibitem{zam_int2}
A.~B. Zamolodchikov, \emph{{Renormalization Group and Perturbation Theory Near
  Fixed Points in Two-Dimensional Field Theory}}, Sov. J. Nucl. Phys.
  \textbf{46} (1987) 1090.

\bibitem{smirnov_ff}
F.~A. Smirnov, \emph{{Form-factors in completely integrable models of quantum
  field theory}}, Adv. Ser. Math. Phys. \textbf{14} (1992) 1-208.

\bibitem{zam_Lee_Yang}
A.~B. Zamolodchikov, \emph{{Two point correlation function in scaling Lee-Yang
  model}}, Nucl. Phys. \textbf{B348} (1991) 619-641.

\bibitem{Korepin:1998vp}
V.~E. Korepin and N.~A. Slavnov, \emph{{The determinant representation for
  quantum correlation functions of the sinh-Gordon model}}, J. Phys.
  \textbf{A31} (1998) 9283-9295, \texttt{hep-th/9801046}.

\bibitem{Korepin:1998rj}
V.~E. Korepin and T.~Oota, \emph{{The determinant representation for a
  correlation function in scaling Lee-Yang model}}, J. Phys. \textbf{A31}
  (1998) L371-L380, \texttt{hep-th/9802003}.

\bibitem{Oota:1998tr}
T.~Oota, \emph{{Two-point correlation functions in perturbed minimal models}},
  J. Phys. \textbf{A31} (1998) 7611-7626, \texttt{hep-th/9804050}.

\bibitem{zam_tba}
A.~B. Zamolodchikov, \emph{{Thermodynamic Bethe Ansatz in relativistic models.
  Scaling three state Potts and Lee-Yang models}}, Nucl. Phys. \textbf{B342}
  (1990) 695-720.

\bibitem{LeClair:1995uf}
A.~LeClair, G.~Mussardo, H.~Saleur and S.~Skorik, \emph{{Boundary energy and
  boundary states in integrable quantum field theories}}, Nucl. Phys.
  \textbf{B453} (1995) 581-618, \texttt{hep-th/9503227}.

\bibitem{excited_TBA}
P.~Dorey and R.~Tateo, \emph{{Excited states by analytic continuation of TBA
  equations}}, Nucl. Phys. \textbf{B482} (1996) 639-659,
  \texttt{hep-th/9607167}.

\bibitem{Bazhanov:1996aq}
V.~V. Bazhanov, S.~L. Lukyanov and A.~B. Zamolodchikov, \emph{{Quantum field
  theories in finite volume: Excited state energies}}, Nucl. Phys.
  \textbf{B489} (1997) 487-531, \texttt{hep-th/9607099}.

\bibitem{Destri:1992qk}
C.~Destri and H.~J. d~Vega, \emph{{New thermodynamic Bethe ansatz equations
  without strings}}, Phys. Rev. Lett. \textbf{69} (1992) 2313-2317.

\bibitem{Destri:1994bv}
C.~Destri and H.~J. De~Vega, \emph{{Unified approach to thermodynamic Bethe
  Ansatz and finite size corrections for lattice models and field theories}},
  Nucl. Phys. \textbf{B438} (1995) 413-454, \texttt{hep-th/9407117}.

\bibitem{sachdev}
A.~LeClair, F.~Lesage, S.~Sachdev and H.~Saleur, \emph{Finite temperature
  correlations in the one-dimensional quantum Ising model}, Nuclear Physics B
  \textbf{482} (1996) 579.

\bibitem{leclair_mussardo}
A.~Leclair and G.~Mussardo, \emph{{Finite temperature correlation functions in
  integrable QFT}}, Nucl. Phys. \textbf{B552} (1999) 624-642,
  \texttt{hep-th/9902075}.

\bibitem{saleurfiniteT}
H.~Saleur, \emph{{A comment on finite temperature correlations in integrable
  QFT}}, Nucl. Phys. \textbf{B567} (2000) 602-610, \texttt{hep-th/9909019}.

\bibitem{lukyanovfiniteT}
S.~Lukyanov, \emph{Finite temperature expectation values of local fields in the
  sinh-Gordon model}, Nuclear Physics B \textbf{612} (2001) 391.

\bibitem{delfinofiniteT}
G.~Delfino, \emph{{One-point functions in integrable quantum field theory at
  finite temperature}}, J. Phys. \textbf{A34} (2001) L161-L168,
  \texttt{hep-th/0101180}.

\bibitem{mussardodifference}
G.~Mussardo, \emph{{On the finite temperature formalism in integrable quantum
  field theories}}, J. Phys. \textbf{A34} (2001) 7399-7410,
  \texttt{hep-th/0103214}.

\bibitem{CastroAlvaredo:2002ud}
O.~A. Castro-Alvaredo and A.~Fring, \emph{{Finite temperature correlation
  functions from form factors}}, Nucl. Phys. \textbf{B636} (2002) 611-631,
  \texttt{hep-th/0203130}.

\bibitem{esslerfiniteT}
F.~H.~L. Essler and R.~M. Konik.
\newblock \emph{Applications of Massive Integrable Quantum Field Theories to
  Problems in Condensed Matter Physics}, (2004).

\bibitem{tsvelikfiniteTcorr}
B.~L. Altshuler, R.~M. Konik and A.~M. Tsvelik, \emph{Low temperature
  correlation functions in integrable models: Derivation of the large distance
  and time asymptotics from the form factor expansion}, Nuclear Physics B
  \textbf{739} (2006) 311.

\bibitem{Essler:2007jp}
F.~H.~L. Essler and R.~M. Konik, \emph{{Finite-temperature lineshapes in gapped
  quantum spin chains}}, Phys. Rev. \textbf{B78} (2008) 100403,
  \texttt{arxiv:}\texttt{0711.2524}\texttt{[cond-mat.str-el]}.

\bibitem{konik_heisenberg_spin_chains}
R.~M.~K. A.~J. A.~James, F. H. L.~E, \emph{Finite Temperature Dynamical
  Structure Factor of Alternating Heisenberg Chains}, Phys. Rev. \textbf{B78}
  (2008) 094411.

\bibitem{james-2009}
A.~J.~A. James, W.~D. Goetze and F.~H.~L. Essler.
\newblock \emph{Finite Temperature Dynamical Structure Factor of the
  Heisenberg-Ising Chain}, (2009).

\bibitem{Smirnov:1998kv}
F.~A. Smirnov, \emph{{Quasi-classical study of form factors in finite volume}},
  ~ (1998)~, \texttt{hep-th/9802132}.

\bibitem{Mussardo:2003ji}
G.~Mussardo, V.~Riva and G.~Sotkov, \emph{{Finite-volume form factors in
  semiclassical approximation}}, Nucl. Phys. \textbf{B670} (2003) 464-478,
  \texttt{hep-th/0307125}.

\bibitem{Bugrij:2001nf}
A.~I. Bugrij, \emph{{Form factor representation of the correlation function of
  the two dimensional Ising model on a cylinder}}, ~ (2001)~,
  \texttt{hep-th/0107117}.

\bibitem{bugrij-2003-319}
A.~I. Bugrij and O.~Lisovyy, \emph{Spin matrix elements in 2D Ising model on
  the finite lattice}, Physics Letters A \textbf{319} (2003) 390.

\bibitem{Fonseca:2001dc}
P.~Fonseca and A.~Zamolodchikov, \emph{{Ising field theory in a magnetic field:
  Analytic properties of the free energy}}, ~ (2001)~, \texttt{hep-th/0112167}.

\bibitem{Koubek:1993ke}
A.~Koubek and G.~Mussardo, \emph{{On the operator content of the sinh-Gordon
  model}}, Phys. Lett. \textbf{B311} (1993) 193-201, \texttt{hep-th/9306044}.

\bibitem{Koubek:1994gk}
A.~Koubek, \emph{{Form-factor bootstrap and the operator content of perturbed
  minimal models}}, Nucl. Phys. \textbf{B428} (1994) 655-680,
  \texttt{hep-th/9405014}.

\bibitem{Delfino:2008ia}
G.~Delfino, \emph{{On the space of quantum fields in massive two-dimensional
  theories}}, Nucl. Phys. \textbf{B807} (2009) 455-470,
  \texttt{arxiv:}\texttt{0806.1883}\texttt{[hep-th]}.

\bibitem{Belavin:2003pu}
A.~A. Belavin, V.~A. Belavin, A.~V. Litvinov, Y.~P. Pugai and A.~B.
  Zamolodchikov, \emph{{On correlation functions in the perturbed minimal
  models M(2,2n+1)}}, Nucl. Phys. \textbf{B676} (2004) 587-614,
  \texttt{hep-th/0309137}.

\bibitem{Zamolodchikov:1986gt}
A.~B. Zamolodchikov, \emph{{Irreversibility of the Flux of the Renormalization
  Group in a 2D Field Theory}}, JETP Lett. \textbf{43} (1986) 730-732.

\bibitem{Delfino:1996nf}
G.~Delfino, P.~Simonetti and J.~L. Cardy, \emph{{Asymptotic factorisation of
  form factors in two- dimensional quantum field theory}}, Phys. Lett.
  \textbf{B387} (1996) 327-333, \texttt{hep-th/9607046}.

\bibitem{lellouch_luscher}
L.~Lellouch and M.~L{\"u}scher, \emph{{Weak transition matrix elements from
  finite-volume correlation functions}}, Commun. Math. Phys. \textbf{219}
  (2001) 31-44, \texttt{hep-lat/0003023}.

\bibitem{sachrajda}
C.~J.~D. Lin, G.~Martinelli, C.~T. Sachrajda and M.~Testa, \emph{{K --> pi pi
  decays in a finite volume}}, Nucl. Phys. \textbf{B619} (2001) 467-498,
  \texttt{hep-lat/0104006}.

\bibitem{Mussardo:1992uc}
G.~Mussardo, \emph{{Off critical statistical models: Factorized scattering
  theories and bootstrap program}}, Phys. Rept. \textbf{218} (1992) 215-379.

\bibitem{Yang:1967bm}
C.-N. Yang, \emph{{Some exact results for the many body problems in one
  dimension with repulsive delta function interaction}}, Phys. Rev. Lett.
  \textbf{19} (1967) 1312-1314.

\bibitem{Baxter:1972hz}
R.~J. Baxter, \emph{{Partition function of the eight-vertex lattice model}},
  Annals Phys. \textbf{70} (1972) 193-228.

\bibitem{zam_zam}
A.~B. Zamolodchikov and A.~B. Zamolodchikov, \emph{{Factorized S-matrices in
  two dimensions as the exact solutions of certain relativistic quantum field
  models}}, Annals Phys. \textbf{120} (1979) 253-291.

\bibitem{Coleman:1978kk}
S.~R. Coleman and H.~J. Thun, \emph{{On the prosaic origin of the double poles
  in the Sine-Gordon S matrix}}, Commun. Math. Phys. \textbf{61} (1978)~31.

\bibitem{Goebel:1986na}
C.~J. Goebel, \emph{{On the sine-Gordon S matrix}}, Prog. Theor. Phys. Suppl.
  \textbf{86} (1986) 261-273.

\bibitem{Braden:1990wx}
H.~W. Braden, E.~Corrigan, P.~E. Dorey and R.~Sasaki, \emph{{Multiple poles and
  other features of affine Toda field theory}}, Nucl. Phys. \textbf{B356}
  (1991) 469-498.

\bibitem{Beane:2007qr}
S.~R. Beane, W.~Detmold and M.~J. Savage, \emph{{n-Boson Energies at Finite
  Volume and Three-Boson Interactions}}, Phys. Rev. \textbf{D76} (2007) 074507,
  \texttt{arxiv:}\texttt{0707.1670}\texttt{[hep-lat]}.

\bibitem{Detmold:2008gh}
W.~Detmold and M.~J. Savage, \emph{{The Energy of n Identical Bosons in a
  Finite Volume at $O(L^{-7})$}}, Phys. Rev. \textbf{D77} (2008) 057502,
  \texttt{arxiv:}\texttt{0801.0763}\texttt{[hep-lat]}.

\bibitem{Luu:2008fg}
T.~Luu, \emph{{Three fermions in a box}}, ~ (2008)~,
  \texttt{arxiv:}\texttt{0810.2331}\texttt{[hep-lat]}.

\bibitem{klassen_melzer}
T.~R. Klassen and E.~Melzer, \emph{{On the relation between scattering
  amplitudes and finite size mass corrections in QFT}}, Nucl. Phys.
  \textbf{B362} (1991) 329-388.

\bibitem{bethe_original}
H.~Bethe, \emph{Zur Theorie der Metalle. I. Eigenwerte und Eigenfunktionen der
  linearen Atomkette.}, Zeitschrift f{\"u}r Physik \textbf{A71} (1931) 205-226.

\bibitem{Polyakov:1970xd}
A.~M. Polyakov, \emph{{Conformal symmetry of critical fluctuations}}, JETP
  Lett. \textbf{12} (1970) 381-383.

\bibitem{CFT_alap}
A.~A. Belavin, A.~M. Polyakov and A.~B. Zamolodchikov, \emph{{Infinite
  conformal symmetry in two-dimensional quantum field theory}}, Nucl. Phys.
  \textbf{B241} (1984) 333-380.

\bibitem{Ginsparg:1988ui}
P.~H. Ginsparg, \emph{{Applied conformal field theory}}, ~ (1988)~,
  \texttt{hep-th/9108028}.

\bibitem{Cardy:2008jc}
J.~Cardy, \emph{{Conformal Field Theory and Statistical Mechanics}}, ~ (2008)~,
  \texttt{arxiv:}\texttt{0807.3472}\texttt{[cond-mat.stat-mech]}.

\bibitem{CFT}
A.~M. Polyakov, A.~A. Belavin and A.~B. Zamolodchikov, \emph{{Infinite
  Conformal Symmetry of Critical Fluctuations in Two-Dimensions}}, J. Statist.
  Phys. \textbf{34} (1984) 763.

\bibitem{cardy_perturbing_cfts}
A.~W.~W. Ludwig and J.~L. Cardy, \emph{{Perturbative Evaluation of the
  Conformal Anomaly at New Critical Points with Applications to Random
  Systems}}, Nucl. Phys. \textbf{B285} (1987) 687-718.

\bibitem{mussardo_bosonic_S}
G.~Mussardo and P.~Simon, \emph{{Bosonic-type S-matrix, vacuum instability and
  CDD ambiguities}}, Nucl. Phys. \textbf{B578} (2000) 527-551,
  \texttt{hep-th/9903072}.

\bibitem{zam_potts}
A.~B. Zamolodchikov, \emph{{Integrals of Motion in Scaling Three State Potts
  Model Field Theory}}, Int. J. Mod. Phys. \textbf{A3} (1988) 743-750.

\bibitem{zam_rsos}
A.~B. Zamolodchikov, \emph{{Thermodynamic Bethe ansatz for RSOS scattering
  theories}}, Nucl. Phys. \textbf{B358} (1991) 497-523.

\bibitem{zam_massless}
A.~B. Zamolodchikov, \emph{{From tricritical Ising to critical Ising by
  thermodynamic Bethe ansatz}}, Nucl. Phys. \textbf{B358} (1991) 524-546.

\bibitem{saleur_massless2}
P.~Fendley, H.~Saleur and A.~B. Zamolodchikov, \emph{{Massless flows, 2. The
  Exact S matrix approach}}, Int. J. Mod. Phys. \textbf{A8} (1993) 5751-5778,
  \texttt{hep-th/9304051}.

\bibitem{klassen_melzer_tba1}
T.~R. Klassen and E.~Melzer, \emph{{Purely Elastic Scattering Theories and
  their Ultraviolet Limits}}, Nucl. Phys. \textbf{B338} (1990) 485-528.

\bibitem{klassen_melzer_tba2}
T.~R. Klassen and E.~Melzer, \emph{{The Thermodynamics of purely elastic
  scattering theories and conformal perturbation theory}}, Nucl. Phys.
  \textbf{B350} (1991) 635-689.

\bibitem{Yang:1952be}
C.-N. Yang and T.~D. Lee, \emph{{Statistical theory of equations of state and
  phase transitions. I: Theory of condensation}}, Phys. Rev. \textbf{87} (1952)
  404-409.

\bibitem{Lee:1952ig}
T.~D. Lee and C.-N. Yang, \emph{{Statistical theory of equations of state and
  phase transitions. II: Lattice gas and Ising model}}, Phys. Rev. \textbf{87}
  (1952) 410-419.

\bibitem{Cardy:1985yy}
J.~L. Cardy, \emph{{Conformal invariance and the Yang-Lee edge singularity in
  two dimensions}}, Phys. Rev. Lett. \textbf{54} (1985) 1354-1356.

\bibitem{Parisi:1980ia}
G.~Parisi and N.~Sourlas, \emph{{Critical behavior of branched polymers and the
  Lee-Yang edge singularity}}, Phys. Rev. Lett. \textbf{46} (1981) 871.

\bibitem{Cardy_Mussardo__Lee_Yang}
J.~L. Cardy and G.~Mussardo, \emph{{S Matrix of the Yang-Lee Edge Singularity
  in Two- Dimensions}}, Phys. Lett. \textbf{B225} (1989) 275.

\bibitem{zamE8}
A.~B. Zamolodchikov, \emph{{Integrals of Motion and S Matrix of the (Scaled)
  T=T(c) Ising Model with Magnetic Field}}, Int. J. Mod. Phys. \textbf{A4}
  (1989) 4235.

\bibitem{phonebook}
V.~A. Fateev, \emph{{The Exact relations between the coupling constants and the
  masses of particles for the integrable perturbed conformal field theories}},
  Phys. Lett. \textbf{B324} (1994) 45-51.

\bibitem{yurovzam}
V.~P. Yurov and A.~B. Zamolodchikov, \emph{{Truncated Conformal Space Approach
  to Scaling Lee-Yang model}}, Int. J. Mod. Phys. \textbf{A5} (1990) 3221-3246.

\bibitem{Karowski:1978vz}
M.~Karowski and P.~Weisz, \emph{{Exact Form-Factors in (1+1)-Dimensional Field
  Theoretic Models with Soliton Behavior}}, Nucl. Phys. \textbf{B139} (1978)
  455.

\bibitem{karowski_weisz_elso_S}
B.~Berg, M.~Karowski and P.~Weisz, \emph{{Construction of Green Functions from
  an Exact S Matrix}}, Phys. Rev. \textbf{D19} (1979) 2477.

\bibitem{delfino_mussardo}
G.~Delfino and G.~Mussardo, \emph{{The Spin spin correlation function in the
  two-dimensional Ising model in a magnetic field at T = T(c)}}, Nucl. Phys.
  \textbf{B455} (1995) 724-758, \texttt{hep-th/9507010}.

\bibitem{ising_ff2}
G.~Delfino, P.~Grinza and G.~Mussardo, \emph{{Decay of particles above
  threshold in the Ising field theory with magnetic field}}, Nucl. Phys.
  \textbf{B737} (2006) 291-303, \texttt{hep-th/0507133}.

\bibitem{nonintegrable}
G.~Delfino, G.~Mussardo and P.~Simonetti, \emph{{Non-integrable Quantum Field
  Theories as Perturbations of Certain Integrable Models}}, Nucl. Phys.
  \textbf{B473} (1996) 469-508, \texttt{hep-th/9603011}.

\bibitem{resonances}
G.~Delfino, P.~Grinza and G.~Mussardo, \emph{{Decay of particles above
  threshold in the Ising field theory with magnetic field}}, Nucl. Phys.
  \textbf{B737} (2006) 291-303, \texttt{hep-th/0507133}.

\bibitem{balogtba}
J.~Balog, \emph{{Field theoretical derivation of the TBA integral equation}},
  Nucl. Phys. \textbf{B419} (1994) 480-506.

\bibitem{qism}
N.~B. V.E.~Korepin and A.~Izergin.
\newblock {\em Quantum inverse scattering method and correlation functions}.
\newblock Cambridge University Press, (1993).

\bibitem{takacs_watts}
H.~Kausch, G.~Takacs and G.~Watts, \emph{{On the relation between Phi(1,2) and
  Phi(1,5) perturbed minimal models and unitarity}}, Nucl. Phys. \textbf{B489}
  (1997) 557-579, \texttt{hep-th/9605104}.

\bibitem{yurovzam_fermionic_TFCSA}
V.~P. Yurov and A.~B. Zamolodchikov, \emph{{Truncated fermionic space approach
  to the critical 2-D Ising model with magnetic field}}, Int. J. Mod. Phys.
  \textbf{A6} (1991) 4557-4578.

\bibitem{ising_ff1}
G.~Delfino and P.~Simonetti, \emph{{Correlation Functions in the
  Two-dimensional Ising Model in a Magnetic Field at $T=T_c$}}, Phys. Lett.
  \textbf{B383} (1996) 450-456, \texttt{hep-th/9605065}.

\bibitem{vevs}
V.~Fateev, S.~L. Lukyanov, A.~B. Zamolodchikov and A.~B. Zamolodchikov,
  \emph{{Expectation values of local fields in Bullough-Dodd model and
  integrable perturbed conformal field theories}}, Nucl. Phys. \textbf{B516}
  (1998) 652-674, \texttt{hep-th/9709034}.

\bibitem{isingff}
G.~Delfino, \texttt{http://www.sissa.it/\textasciitilde{}delfino/isingff.html}.

\bibitem{guidamagnoli}
R.~Guida and N.~Magnoli, \emph{{Vacuum expectation values from a variational
  approach}}, Phys. Lett. \textbf{B411} (1997) 127-133,
  \texttt{hep-th/9706017}.

\bibitem{onepff}
G.~M. D.~Fioravanti and P.~Simon, \emph{Universal Amplitude Ratios of The
  Renormalization Group: Two-Dimensional Tricritical Ising Model}, Phys.Rev.
  \textbf{E63} (2001) 016103, \texttt{cond-mat/0008216}.

\bibitem{takacs_extrapolation}
Z.~Bajnok, L.~Palla, G.~Takacs and F.~Wagner, \emph{{The k-folded sine-Gordon
  model in finite volume}}, Nucl. Phys. \textbf{B587} (2000) 585-618,
  \texttt{hep-th/0004181}.

\bibitem{Rummukainen:1995vs}
K.~Rummukainen and S.~A. Gottlieb, \emph{{Resonance scattering phase shifts on
  a nonrest frame lattice}}, Nucl. Phys. \textbf{B450} (1995) 397-436,
  \texttt{hep-lat/9503028}.

\bibitem{fftcsa1}
B.~Pozsgay and G.~Takacs, \emph{{Form factors in finite volume I: form factor
  bootstrap and truncated conformal space}}, Nucl. Phys. \textbf{B788} (2008)
  167-208, \texttt{arxiv:}\texttt{0706.1445}\texttt{[hep-th]}.

\bibitem{Janik:2007wt}
R.~A. Janik and T.~Lukowski, \emph{{Wrapping interactions at strong coupling --
  the giant magnon}}, Phys. Rev. \textbf{D76} (2007) 126008,
  \texttt{arxiv:}\texttt{0708.2208}\texttt{[hep-th]}.

\bibitem{Heller:2008at}
M.~P. Heller, R.~A. Janik and T.~Lukowski, \emph{{A new derivation of L\"uscher
  F-term and fluctuations around the giant magnon}}, JHEP \textbf{06} (2008)
  036, \texttt{arxiv:}\texttt{0801.4463}\texttt{[hep-th]}.

\bibitem{Bajnok:2008bm}
Z.~Bajnok and R.~A. Janik, \emph{{Four-loop perturbative Konishi from strings
  and finite size effects for multiparticle states}}, Nucl. Phys. \textbf{B807}
  (2009) 625-650, \texttt{arxiv:}\texttt{0807.0399}\texttt{[hep-th]}.

\bibitem{Hatsuda:2008na}
Y.~Hatsuda and R.~Suzuki, \emph{{Finite-Size Effects for Multi-Magnon States}},
  JHEP \textbf{09} (2008) 025,
  \texttt{arxiv:}\texttt{0807.0643}\texttt{[hep-th]}.

\bibitem{bpt1}
Z.~Bajnok, L.~Palla and G.~Takacs, \emph{{Boundary states and finite size
  effects in sine-Gordon model with Neumann boundary condition}}, Nucl. Phys.
  \textbf{B614} (2001) 405-448, \texttt{hep-th/0106069}.

\bibitem{bpt_semiclass}
Z.~Bajnok, L.~Palla and G.~Takacs, \emph{{(Semi)classical analysis of
  sine-Gordon theory on a strip}}, Nucl. Phys. \textbf{B702} (2004) 448-480,
  \texttt{hep-th/0406149}.

\bibitem{maiani_testa_final_state_interaction}
L.~Maiani and M.~Testa, \emph{{Final state interactions from Euclidean
  correlation functions}}, Phys. Lett. \textbf{B245} (1990) 585-590.

\bibitem{k_to_pipi}
C.~J.~D. Lin, G.~Martinelli, C.~T. Sachrajda and M.~Testa, \emph{{K --> pi pi
  decays in a finite volume}}, Nucl. Phys. \textbf{B619} (2001) 467-498,
  \texttt{hep-lat/0104006}.

\bibitem{yurov_zam_Ising}
V.~P. Yurov and A.~B. Zamolodchikov, \emph{{Correlation functions of integrable
  2-D models of relativistic field theory. Ising model}}, Int. J. Mod. Phys.
  \textbf{A6} (1991) 3419-3440.

\bibitem{cardy_mussardo_ff_corr_funct}
J.~L. Cardy and G.~Mussardo, \emph{Form-factors of descendant operators in
  perturbed conformal field theories}, Nucl. Phys. \textbf{B340} (1990)
  387-402.

\bibitem{konik-2001}
R.~M. Konik.
\newblock \emph{Haldane Gapped Spin Chains: Exact Low Temperature Expansions of
  Correlation Functions}, (2001).

\bibitem{essler-2009}
F.~H.~L. Essler and R.~M. Konik, \emph{Finite Temperature Dynamical
  Correlations in Massive Integrable Quantum Field Theories}, ~ (2009)~,
  \texttt{arxiv:}\texttt{0907.0779}\texttt{[cond-mat]}.

\bibitem{fftcsa2}
B.~Pozsgay and G.~Takacs, \emph{{Form factors in finite volume II:disconnected
  terms and finite temperature correlators}}, Nucl. Phys. \textbf{B788} (2008)
  209-251, \texttt{arxiv:}\texttt{0706.3605}\texttt{[hep-th]}.

\bibitem{Kormos:2007qx}
M.~Kormos and G.~Takacs, \emph{{Boundary form factors in finite volume}}, Nucl.
  Phys. \textbf{B803} (2008) 277-298,
  \texttt{arxiv:}\texttt{0712.1886}\texttt{[hep-th]}.

\bibitem{Takacs:2008ec}
G.~Takacs, \emph{{Finite temperature expectation values of boundary
  operators}}, Nucl. Phys. \textbf{B805} (2008) 391-417,
  \texttt{arxiv:}\texttt{0804.4096}\texttt{[hep-th]}.

\bibitem{Delfino:1994ea}
G.~Delfino, G.~Mussardo and P.~Simonetti, \emph{{Correlation functions along a
  massless flow}}, Phys. Rev. \textbf{D51} (1995) 6620-6624,
  \texttt{hep-th/9410117}.

\bibitem{Mejean:1996xg}
P.~Mejean and F.~A. Smirnov, \emph{{Form factors for principal chiral field
  model with Wess- Zumino-Novikov-Witten term}}, Int. J. Mod. Phys.
  \textbf{A12} (1997) 3383-3396, \texttt{hep-th/9609068}.

\end{thebibliography}
\bibliographystyle{pozsi}

\end{document}